%% file: DoubleDoubleMain.tex
\documentclass[11pt] {article} 
\usepackage[margin=1in]{geometry}
\usepackage[english]{babel}  
\usepackage[T1]{fontenc}
\usepackage{amsmath,amssymb,graphicx, latexsym, theorem, enumerate,bbm,bm}
{\theorembodyfont{\rmfamily}\newtheorem{lemma}{Lemma}[section]}
\newenvironment{proof}{\noindent\textit{Proof.}\enspace}{\hfill$\square$\medskip}
\usepackage{booktabs, multirow, nicefrac, float, xspace, comment}
\usepackage[section]{placeins}
\usepackage{xcolor}
\usepackage{slashed}
\usepackage{cite}
\usepackage[normalem]{ulem}

\usepackage{hyperref}
 
\usepackage{tikz}
\usetikzlibrary{positioning, backgrounds, scopes, 3d, calc, shapes.geometric, arrows.meta, decorations.markings, decorations.pathmorphing}

\usepackage[compat=1.1.0]{tikz-feynman}
\tikzfeynmanset{compat=1.1.0, warn luatex=false}

\usepackage{tikz-cd}
\usepackage{pgfplots}
\pgfplotsset{compat=1.17}

\usepackage{graphicx}
\usepackage{tikz,color}
\usepackage{feynmp-auto}
\usepackage{amsmath}
\usepackage{mathtools}
\usepackage{float} 
\numberwithin{equation}{section} 
\definecolor{darkred}{rgb}{0.5,0.0,0.0}
\definecolor{darkblue}{rgb}{0.0,0.0,0.9}
\definecolor{darkerblue}{rgb}{0.0,0.0,0.5}
\definecolor{darkgreen}{rgb}{0.0,0.5,0.0}
\definecolor{darkpurple}{rgb}{0.5, 0.2, 0.8}
\definecolor{branchone}{rgb}{0.368,0.507,0.710}   % Mathematica ColorData[97][1] blue
\definecolor{branchtwo}{rgb}{0.881,0.611,0.142}    % Mathematica ColorData[97][2] orange
\definecolor{branchthree}{rgb}{0.560,0.692,0.195}  % Mathematica ColorData[97][3] green
\definecolor{branchfour}{rgb}{0.58,0.40,0.74}   % medium purple

\DeclareMathOperator{\Ai}{Ai} 
\DeclareMathOperator{\Bi}{Bi}

\newcommand{\Ein}{\operatorname{Ein}}
\newcommand{\Seff}{ S_\text{eff}}

\def \CC {\mathbb{C}}
\def \R {\mathbb{R}} 
\def \RR {\mathbb{R}} 
\newcommand{\DV}{\Delta_V}          % vacuum-bubble energy shift
\newcommand{\DL}{\Delta_L}          % localized core correction
\newcommand{\opO}{{\mathcal O}}     % fluctuation operator
\newcommand{\prop}{\Pi}             % propagator (replaces \Delta for propagators)
\def \cC {\mathcal{C}}

\def \cJ {\mathcal{J}}
\def \cI {\mathcal{I}}

\def \cO {\mathcal{O}}
\def \cG {\mathcal{G}}
\def \cN {\mathcal{N}}
\def \cP {\mathcal{P}}
\def \cS {\mathcal{S}}

\newcommand{\fS}{\mathfrak{S}}
\newcommand{\fSbr}{\fS_{\text{br}}}
\newcommand{\fSC}{\fS_{C}}
\newcommand{\re}{\operatorname{Re}}
\newcommand{\im}{\operatorname{Im}}
\newcommand{\VN}{ {\mathcal V}_N }
\newcommand{\VP}{ {\mathcal V}_P }
\newcommand{\fv}{\Upsilon}

\newcommand{\Tr}{ {\operatorname{Tr}}}
\newcommand{\Shat} {\widehat{S}}
\newcommand{\GR}{\Gamma_{\mathbb{R}}}

\newcommand{\xii}{ x_2^{\text{s}} }
\newcommand{\sig}{\sigma}           % elliptic modulus (renamed from k to avoid clash with winding number k)

\input{tikzfigs}

\newcommand{\fSp}{{\color{stokesplus}\fS_{+}}}
\newcommand{\fSm}{{\color{stokesminus}\fS_{-}}}
\newcommand{\SA}{{\color{darkred}\bm{\mathcal{S}}}}

\newcommand{\LT}{T_3}
\newcommand{\LF}{T_4}
\newcommand{\Ln}{T_n}

\newcommand{\ntwoScale}{\kappa_2}
\newcommand{\nthreeScale}{\kappa_3}
\newcommand{\nfourScale}{\kappa_4}
\newcommand{\nnScale}{\kappa_n}
\newcommand{\amin}{\alpha_0}
\newcommand{\umin}{u_0}
\newcommand{\ximin}{\xi_0}

\newcommand{\sn}{\operatorname{sn}}

\begin{document}

\title{The Double Well Done Doubly-Well}

\author{Aur\'elien Dersy and Matthew D.\ Schwartz\\
\textit{Department of Physics, Harvard University, Cambridge, MA 02138, USA}}

\maketitle

\begin{abstract}
The symmetric double-well potential is one of the simplest quantum-mechanical systems in which perturbative and non-perturbative physics are deeply entangled. The expansion of any of its energy levels in inverse powers of the inter-well separation is non-analytic, with factorially growing coefficients, while the splitting between parity eigenstates is exponentially small and invisible to perturbation theory at any finite order. Resurgence ties the two features together, organizing the exact spectrum into a single tightly-constrained trans-series. This paper gives a self-contained, first-principles account of how this trans-series can be understood from two complementary approaches: exact WKB and the Euclidean path integral, both developed in a common notation with explicit calculations through the four-instanton level and three-loop order, whenever possible. In exact WKB, Stokes phenomena encoded in the Delabaere--Dillinger--Pham relations control the analytic continuation of the wavefunction past turning points. Then the quantization condition expressed in terms of Voros symbols determines the full trans-series. The DDP relations are local and do not require knowing the global topology of the energy surface, but that surface is an elliptic curve. In the path integral, elliptic curves enter differently: the classical saddle points are doubly-periodic elliptic functions of Euclidean time, and the Stokes phenomenon plays out within the finite-dimensional manifold of quasi-zero modes rather than through analytic continuation of the wavefunction. A Lefschetz thimble decomposition determines which saddles contribute, and the resulting partition function trans-series is much simpler than the energy trans-series: at each instanton order the $T$-dependence is a polynomial fixed by the quasi-zero-mode thimble integrals. Together, the two approaches deploy a shared mathematical infrastructure of elliptic curves and Stokes phenomena in complementary ways to achieve extraordinary computational depth, thereby showing that the double well is an ideal setting to explore and understand resurgence.

\end{abstract}  
\newpage
\setcounter{tocdepth}{2}
\tableofcontents 
\newpage

% ============================================================
% Main body
%  =============================================================
\input{sections/introduction}

\input{sections/history}
\input{sections/zero_dimensions}

\input{sections/exact_wkb}
\input{sections/path_integral}
\input{sections/conclusions}

% ============================================================
% Acknowledgments
% ============================================================
\section*{Acknowledgments}
We thank G\"ok\c{c}e Ba\c{s}ar, Arindam Bhattacharya, Jordan Cotler, Gerald Dunne, and Marco Serone for helpful discussions. We also thank Claude Opus~4.7 (Anthropic) and GPT Codex~5.5 (OpenAI) for assistance with computations, writing, and figure generation. AD and MDS take full responsibility for the correctness of the work. This work was supported by the U.S.\ Department of Energy, Office of Science, under grant DE-SC0013607.

% ============================================================
% Appendices
% ============================================================
\appendix
\input{sections/appendix_perturbative}
\input{sections/appendix_Pn_decomposition}
\input{sections/appendix_Weber_v2}
\input{sections/appendix_Lame_propagator}

\input{sections/appendix_loop_corrections}
\input{sections/appendix_transseries}

% ============================================================
% Bibliography
% ============================================================
\bibliographystyle{JHEP}
\bibliography{sections/biblio}

\end{document}

%% file: tikzfigs.tex
\usetikzlibrary{arrows}
\usetikzlibrary{decorations.pathmorphing}
\usetikzlibrary{calc,math,patterns}
\usetikzlibrary{decorations.markings,arrows.meta}
\usetikzlibrary{decorations.pathreplacing}

\tikzset{squiggly/.style={decorate, decoration={snake, amplitude=.4mm, segment length=2mm, post length=0mm}}}

\definecolor{stokesplus}{rgb}{0.560181, 0.691569, 0.194885}
\definecolor{stokesminus}{rgb}{0.922526, 0.385626, 0.209179}

\newcommand{\StokesGraphAiry}{%
\begin{tikzpicture}[scale=0.5]
    \def\R{2.2}

    \draw[stokesminus, thick] (0,0) -- (\R,0) node[pos=0.75, above] {$-$};
    \draw[stokesplus, thick] (0,0) -- (120:\R) node[pos=0.75, above left] {$+$};
    \draw[stokesplus, thick] (0,0) -- (-120:\R) node[pos=0.75, below left] {$+$};

    \draw[squiggly, thick] (0,0) -- (-\R,0);

    \fill (0,0) circle (2.5pt);
\end{tikzpicture}%
}

\newcommand{\SHOStokesWithPotential}{%
\begin{tikzpicture}
    \begin{scope}[shift={(-3,-1)}, scale=0.8]
        \draw[->] (-2.5,0) -- (2.5,0) node[right] {$x$};
        \draw[->] (0,-0.5) -- (0,2.5) ;

        \draw[blue, thick, domain=-2.4:2.4, samples=50] plot (\x, {0.5*\x*\x});
        
        \node[blue] at (2.5,2.5) {$V$};

        \draw[red, dashed, thick] (-2.3,1) -- (2.3,1) node[right] {$E$};

        \fill (-1.414,1) circle (2.5pt) node[below left] {$x_1$};
        \fill (1.414,1) circle (2.5pt) node[below right] {$x_2$};

        \draw[thick, dotted, purple, decoration={markings, mark=at position 0.25 with {\arrow[purple]{Triangle[length=2.5mm, width=1.5mm]}}, mark=at position 0.5 with {\arrow[purple]{Triangle[length=2.5mm, width=1.5mm]}}, mark=at position 0.75 with {\arrow[purple]{Triangle[length=2.5mm, width=1.5mm]}}}, postaction={decorate}] (3.0,0.3) arc[start angle=40, end angle=180-40, radius=4];
        \fill[purple] (3.0,0.3) circle (1.8pt);
        \fill[purple] (-3.0,0.3) circle (1.8pt);
    \end{scope}
\end{tikzpicture}%
}

\newcommand{\SHOMonodromyF}{%
\begin{tikzpicture}[scale=1.2]
    \draw[squiggly, thick] (-1,0) -- (1,0);

    \draw[stokesplus, thick] (-1,0) -- ++(-1,0) node[left] {$+$};
    \draw[stokesminus, thick] (-1,0) -- ++(60:1) node[above] {$-$};
    \draw[stokesminus, thick] (-1,0) -- ++(-60:1) node[below] {$-$};

    \draw[stokesminus, thick] (1,0) -- ++(1,0) node[right] {$-$};
    \draw[stokesplus, thick] (1,0) -- ++(120:1) node[above] {$+$};
    \draw[stokesplus, thick] (1,0) -- ++(-120:1) node[below] {$+$};

    \fill (-1,0) circle (1.5pt) node[below left, font=\scriptsize] {$x_1$};
    \fill (1,0) circle (1.5pt) node[below right, font=\scriptsize] {$x_2$};

    \draw[thick, dotted, purple, decoration={markings, mark=at position 0.12 with {\arrow[purple]{Triangle[length=2.5mm, width=1.5mm]}}, mark=at position 0.37 with {\arrow[purple]{Triangle[length=2.5mm, width=1.5mm]}}, mark=at position 0.62 with {\arrow[purple]{Triangle[length=2.5mm, width=1.5mm]}}, mark=at position 0.87 with {\arrow[purple]{Triangle[length=2.5mm, width=1.5mm]}}}, postaction={decorate}]
        ({1.5*cos(45)},{0.55*sin(45)}) arc[x radius=1.5, y radius=0.55, start angle=45, end angle=375];
    \fill[purple] ({1.5*cos(45)},{0.55*sin(45)}) circle (1.8pt);
    \fill[purple] ({1.5*cos(375)},{0.55*sin(375)}) circle (1.8pt);
\end{tikzpicture}%
}

\newcommand{\SHOPathC}{%
\begin{tikzpicture}
    \def\xa{-3}
    \def\xb{-0.5}
    \def\xc{0.5}
    \def\xd{3}

    \draw[squiggly, thick] (\xa,0) .. controls ++(0.6,0) .. ++(1.3,0.9);
    \draw[squiggly, thick] (\xb,0) .. controls ++(-0.6,0) ..++(-1.3,-0.9);
    
    \draw[stokesplus, thick] (\xa,0) -- ++(-1,0) node[left] {$+$};
    \draw[stokesminus, thick] (\xa,0) -- ++(60:1) node[above] {$-$};
    \draw[stokesminus, thick] (\xa,0) -- ++(-60:1) node[below] {$-$};
    \draw[stokesplus, thick] (\xb,0) -- ++(120:1) node[above] {$+$};
    \draw[stokesplus, thick] (\xb,0) -- ++(-120:1) node[below] {$+$};
    \draw[stokesminus, thick] (\xb,0) -- ++(0:1) node[below] {$-$};

    \draw[thick, dotted, purple, decoration={markings, mark=at position 0.25 with {\arrow[purple]{Triangle[length=2.5mm, width=1.5mm]}}, mark=at position 0.5 with {\arrow[purple]{Triangle[length=2.5mm, width=1.5mm]}}, mark=at position 0.75 with {\arrow[purple]{Triangle[length=2.5mm, width=1.5mm]}}}, postaction={decorate}] (0.3,0.15) arc[start angle=70, end angle=180-70, radius=6];
    \fill[purple] (0.3,0.15) circle (1.8pt);
    \fill[purple] (-4.0,0.15) circle (1.8pt);

    \fill (\xa, 0) circle (1.5pt) node at ($(\xa,0)+(240:0.3)$) {\scalebox{0.8} {$x_1$}};
    \fill (\xb, 0) circle (1.5pt) node at ($(\xb,0)+(-60:0.3)$) {\scalebox{0.8} {$x_2$}};
\end{tikzpicture}%
}

\newcommand{\DWpotential}{%
\begin{tikzpicture}
        \draw[->] (-2.5,0) -- (2.5,0) node[right] {$x$};
        \draw[->] (0,-0.5) -- (0,2.5) ;

       \draw[blue, thick, domain=-1.6:1.6, samples=50] plot (\x, { 1.0*(\x*\x -1.0)*(\x*\x -1.0)});         
        \draw[red, dashed, thick] (-2.3,0.5) -- (2.3,0.5) node[right] {$E$};

        \fill (-1.30656,0.5) circle (2.5pt) node[below left,scale=0.8] {$x_1$};
        \fill (-0.541196, 0.5) circle (2.5pt) node[below right,scale=0.8] {$x_2$};
        \fill (0.541196, 0.5) circle (2.5pt) node[below left,scale=0.8] {$x_3$};
        \fill (1.30656,0.5) circle (2.5pt) node[below right,scale=0.8] {$x_4$};

\end{tikzpicture}%
}

\newcommand{\StokesGraphDoubleWellA}{%
\begin{tikzpicture}
    \def\xa{-2}
    \def\xb{-0.5}
    \def\xc{0.5}
    \def\xd{2}

    \draw[thick, brown] (\xb,0) -- (\xc,0);

    \draw[squiggly, thick] (\xa,0) -- (\xb,0);
    \draw[squiggly, thick] (\xc,0) -- (\xd,0);

    \draw[stokesplus, thick] (\xa,0) -- ++(-1,0) node[left] {$+$};
    \draw[stokesminus, thick] (\xa,0) -- ++(60:1) node[above] {$-$};
    \draw[stokesminus, thick] (\xa,0) -- ++(-60:1) node[below] {$-$};
    \draw[stokesplus, thick] (\xb,0) -- ++(120:1) node[above] {$+$};
    \draw[stokesplus, thick] (\xb,0) -- ++(-120:1) node[below] {$+$};

    \draw[stokesminus, thick] (\xc,0) -- ++(60:1) node[above] {$-$};
    \draw[stokesminus, thick] (\xc,0) -- ++(-60:1) node[below] {$-$};
    \draw[stokesminus, thick] (\xd,0) -- ++(1,0) node[right] {$-$};
    \draw[stokesplus, thick] (\xd,0) -- ++(120:1) node[above] {$+$};
    \draw[stokesplus, thick] (\xd,0) -- ++(-120:1) node[below] {$+$};

    \fill (\xa, 0) circle (1.5pt) node at ($(\xa,0)+(135:0.3)$) {\scalebox{0.8} {$x_1$}};
    \fill (\xb, 0) circle (1.5pt) node at ($(\xb,0)+(45:0.3)$) {\scalebox{0.8} {$x_2$}};
    \fill (\xc, 0) circle (1.5pt) node at ($(\xc,0)+(135:0.3)$) {\scalebox{0.8} {$x_3$}};
    \fill (\xd, 0) circle (1.5pt) node at ($(\xd,0)+(45:0.3)$) {\scalebox{0.8} {$x_4$}};
\end{tikzpicture}%
}

\newcommand{\StokesGraphDoubleWellMisumi}{%
\begin{tikzpicture}
    \def\xa{-3}
    \def\xb{-0.5}
    \def\xc{0.5}
    \def\xd{3}

    \draw[squiggly, thick] (\xa,0) .. controls ++(0.6,0) .. ++(1.3,0.9);
    \draw[squiggly, thick] (\xb,0) .. controls ++(-0.6,0) ..++(-1.3,-0.9);
    
    \draw[squiggly, thick] (\xc,0) .. controls ++(0.6,0) .. ++(1.3,0.9);
    \draw[squiggly, thick] (\xd,0) .. controls ++(-0.6,0) ..++(-1.3,-0.9);   
    
    \draw[stokesplus, thick] (\xa,0) -- ++(-1,0) node[left] {$+$};
    \draw[stokesminus, thick] (\xa,0) -- ++(60:1) node[above] {$-$};
    \draw[stokesminus, thick] (\xa,0) -- ++(-60:1) node[below] {$-$};
    \draw[stokesplus, thick] (\xb,0) -- ++(120:1) node[above] {$+$};
    \draw[stokesplus, thick] (\xb,0) -- ++(-120:1) node[below] {$+$};

    \draw[stokesminus, thick] (\xc,0) -- ++(60:1) node[above] {$-$};
    \draw[stokesminus, thick] (\xc,0) -- ++(-60:1) node[below] {$-$};
    \draw[stokesminus, thick] (\xd,0) -- ++(1,0) node[right] {$-$};
    \draw[stokesplus, thick] (\xd,0) -- ++(120:1) node[above] {$+$};
    \draw[stokesplus, thick] (\xd,0) -- ++(-120:1) node[below] {$+$};

    \draw[stokesminus, thick] (\xb,0) .. controls ++(0.6,0.0) .. ++(1.2,0.9) node[above] {$$};
    \draw[stokesplus, thick] (\xc,0) .. controls ++(-0.6,0.0) .. ++(-1.2,-0.9) node[above] {$$};

    \draw[thick, dotted, purple, decoration={markings, mark=at position 0.25 with {\arrow[purple]{Triangle[length=2.5mm, width=1.5mm]}}, mark=at position 0.5 with {\arrow[purple]{Triangle[length=2.5mm, width=1.5mm]}}, mark=at position 0.75 with {\arrow[purple]{Triangle[length=2.5mm, width=1.5mm]}}}, postaction={decorate}] (3.5,0.15) arc[start angle=70, end angle=180-70, radius=10];
    \fill[purple] (3.5,0.15) circle (1.8pt);
    \fill[purple] (-3.5,0.15) circle (1.8pt);

    \fill (\xa, 0) circle (1.5pt) node at ($(\xa,0)+(210:0.3)$) {\scalebox{0.8} {$x_1$}};
    \fill (\xb, 0) circle (1.5pt) node at ($(\xb,0)+(60:0.3)$) {\scalebox{0.8} {$x_2$}};
    \fill (\xc, 0) circle (1.5pt) node at ($(\xc,0)+(210:0.3)$) {\scalebox{0.8} {$x_3$}};
    \fill (\xd, 0) circle (1.5pt) node at ($(\xd,0)+(-30:0.3)$) {\scalebox{0.8} {$x_4$}};
\end{tikzpicture}%
}

\newcommand{\StokesGraphDoubleWellMisumiB}{%
\begin{tikzpicture}
    \def\xa{-3}
    \def\xb{-0.5}
    \def\xc{0.5}
    \def\xd{3}

    \draw[squiggly, thick] (\xa,0) .. controls ++(0.6,0) .. ++(1.3,0.9);
    \draw[squiggly, thick] (\xb,0) .. controls ++(-0.6,0) ..++(-1.3,-0.9);
    
    \draw[squiggly, thick] (\xc,0) .. controls ++(0.6,0) .. ++(1.3,0.9);
    \draw[squiggly, thick] (\xd,0) .. controls ++(-0.6,0) ..++(-1.3,-0.9);   
    
    \draw[stokesplus, thick] (\xa,0) -- ++(-1,0) node[left] {$+$};
    \draw[stokesminus, thick] (\xa,0) -- ++(60:1) node[above] {$-$};
    \draw[stokesminus, thick] (\xa,0) -- ++(-60:1) node[below] {$-$};
    \draw[stokesplus, thick] (\xb,0) -- ++(120:1) node[above] {$+$};
    \draw[stokesplus, thick] (\xb,0) -- ++(-120:1) node[below] {$+$};

    \draw[stokesminus, thick] (\xc,0) -- ++(60:1) node[above] {$-$};
    \draw[stokesminus, thick] (\xc,0) -- ++(-60:1) node[below] {$-$};
    \draw[stokesminus, thick] (\xd,0) -- ++(1,0) node[right] {$-$};
    \draw[stokesplus, thick] (\xd,0) -- ++(120:1) node[above] {$+$};
    \draw[stokesplus, thick] (\xd,0) -- ++(-120:1) node[below] {$+$};

    \draw[stokesminus, thick] (\xb,0) .. controls ++(0.6,0.0) .. ++(1.2,-0.9) node[above] {$$};
    \draw[stokesplus, thick] (\xc,0) .. controls ++(-0.6,0.0) .. ++(-1.2,0.9) node[above] {$$};

    \draw[thick, dotted, purple, decoration={markings, mark=at position 0.25 with {\arrow[purple]{Triangle[length=2.5mm, width=1.5mm]}}, mark=at position 0.5 with {\arrow[purple]{Triangle[length=2.5mm, width=1.5mm]}}, mark=at position 0.75 with {\arrow[purple]{Triangle[length=2.5mm, width=1.5mm]}}}, postaction={decorate}] (3.5,0.15) arc[start angle=70, end angle=180-70, radius=10];
    \fill[purple] (3.5,0.15) circle (1.8pt);
    \fill[purple] (-3.5,0.15) circle (1.8pt);

    \fill (\xa, 0) circle (1.5pt) node at ($(\xa,0)+(210:0.3)$) {\scalebox{0.8} {$x_1$}};
    \fill (\xb, 0) circle (1.5pt) node at ($(\xb,0)+(-60:0.3)$) {\scalebox{0.8} {$x_2$}};
    \fill (\xc, 0) circle (1.5pt) node at ($(\xc,0)+(-210:0.3)$) {\scalebox{0.8} {$x_3$}};
    \fill (\xd, 0) circle (1.5pt) node at ($(\xd,0)+(-30:0.3)$) {\scalebox{0.8} {$x_4$}};
\end{tikzpicture}%
}

\newcommand{\CubicPotentialPeriods}{%
\begin{tikzpicture}[xscale=1.0, yscale=1.0, baseline=(redline)]
  \coordinate (redline) at (0,0);
  \draw[domain=-2.1:2.1, smooth, variable=\x, blue, thick]
    plot (\x, {0.25*(\x*\x*\x - 3*\x)});

  \draw[red, dashed, thick] (-2.2, 0) -- (2.5, 0)
    node[below right] {$y=0$};

  \draw[<->, thick] (1,-0.6) ++(200:0.45 and 0.20)
    arc (200:340:0.45 and 0.30)
    node[midway, below] {$\omega$};

  \draw[<->, thick] (-1,0.6) ++(20:0.45 and 0.20)
    arc (20:160:0.45 and 0.30)
    node[midway, above] {$\omega'$};

  \fill (-1.732, 0) circle (2pt) node[above left] {$e_1$};
  \fill (0, 0) circle (2pt) node[above] {$e_2$};
  \fill (1.732, 0) circle (2pt) node[above left] {$e_3$};
\end{tikzpicture}%
}

\newcommand{\DoubleWellPeriods}{%
\begin{tikzpicture}[xscale=1.5, yscale=2, baseline=(redline)]
  \coordinate (redline) at (0,0.4);
  \draw[domain=-1.5:1.5, smooth, variable=\x, blue, thick]
    plot (\x, {0.5*(\x*\x-1)*(\x*\x-1)});
  \draw[red, dashed, thick] (-1.5, 0.4) -- (1.5, 0.4)
    node[below right] {$\varepsilon$};

  \draw[<->, thick] (-1,-0.03) ++(200:0.35 and 0.16)
    arc (200:340:0.35 and 0.25)
    node[midway, below] {$\omega_P$};

  \draw[<->, thick] (0,0.5) ++(20:0.38 and 0.20)
    arc (20:160:0.38 and 0.20)
    node[midway, above] {$\omega_N$};

  \fill (-1.376, 0.4) circle (1.5pt) node[above left] {$x_1$};
  \fill (-0.325, 0.4) circle (1.5pt) node[below] {$x_2$};
  \fill (0.325, 0.4) circle (1.5pt) node[below] {$x_3$};
  \fill (1.376, 0.4) circle (1.5pt) node[above right] {$x_4$};
\end{tikzpicture}%
}

\newcommand{\InvertedDoubleWellPeriods}{%
\begin{tikzpicture}[xscale=1.5, yscale=2, baseline=(redline)]
  \coordinate (redline) at (0,-0.4);
  \draw[domain=-1.5:1.5, smooth, variable=\x, blue, thick]
    plot (\x, {-0.5*(\x*\x-1)*(\x*\x-1)});
  \draw[red, dashed, thick] (-1.5, -0.4) -- (1.5, -0.4)
    node[above right] {$\varepsilon$};

  \draw[<->, thick] (-1,0.03) ++(160:0.35 and 0.16)
    arc (160:20:0.35 and 0.25)
    node[midway, above] {$\omega_P$};

  \draw[<->, thick] (0,-0.5) ++(340:0.38 and 0.20)
    arc (340:200:0.38 and 0.20)
    node[midway, below] {$\omega_N$};
\end{tikzpicture}%
}

\newcommand{\InvertedDoubleWellAxes}{%
\begin{tikzpicture}[xscale=1.5, yscale=2, baseline=(current bounding box.center)]
  \draw[domain=-1.5:1.5, smooth, variable=\x, blue, thick]
    plot (\x, {-0.5*(\x*\x-1)*(\x*\x-1)});
  \draw[red, dashed, thick] (-1.5, -0.05) -- (1.5, -0.05)
    node[above right] {$\varepsilon$};

  \draw[<->, thick, green!60!black] (1,0.03) ++(160:0.35 and 0.16)
    arc (160:20:0.35 and 0.25)
    node[midway, above] {\color{green!60!black}$\omega_P$};

  \draw[<->, thick, red!70!black] (0,-0.5) ++(340:0.38 and 0.20)
    arc (340:200:0.38 and 0.20)
    node[midway, below] {\color{red!70!black}$\omega_N$};

  \draw[->] (-1.9, -0.6) -- (-1.9, 0.15) node[above] {$U(x)$};
  \draw (-1.93, 0) -- (-1.87, 0) node[left, anchor=east, font=\scriptsize] {$0$};
  \draw (-1.93, -0.5) -- (-1.87, -0.5) node[left, anchor=east, font=\scriptsize] {$-\tfrac{1}{8}$};

  \draw[<->] (-1.5, -1.05) -- (1.5, -1.05) node[right] {$x$};
  \draw (-1, -1.02) -- (-1, -1.08) node[below, font=\scriptsize] {$-1$};
  \draw (0, -1.02) -- (0, -1.08) node[below, font=\scriptsize] {$0$};
  \draw (1, -1.02) -- (1, -1.08) node[below, font=\scriptsize] {$1$};
\end{tikzpicture}%
}

\newcommand{\PropagatorSHO}{%
\begin{tikzpicture}[baseline=(current bounding box.center)]
\begin{axis}[
  width=4.5cm,
  height=3.5cm,
  axis lines=none,
  xlabel={},
  ylabel={},
  xmin=-0.55, xmax=0.55,
  ymin=0.95, ymax=1.09,
  xtick=\empty,
  ytick=\empty,
]
\addplot[blue, thick, domain=-0.5:0.5, samples=100]
  {cosh(0.5 - abs(x)) / (2*sinh(0.5))};
\end{axis}
\end{tikzpicture}%
}

\newcommand{\FeynmanQuarticVertex}{%
\begin{tikzpicture}[baseline=-0.5ex, scale=1.0]
  \draw[line width=0.8pt] (-0.35,0) circle (0.35cm);
  \draw[line width=0.8pt] (0.35,0) circle (0.35cm);
  \fill (0,0) circle (1.8pt);
\end{tikzpicture}%
}

\newcommand{\FeynmanCubicBubble}{%
\begin{tikzpicture}[baseline=-0.5ex, scale=1.0]
  \draw[line width=0.8pt] (0,0) circle (0.5cm);
  \fill (0,0.5) circle (1.8pt);
  \fill (0,-0.5) circle (1.8pt);
  \draw[line width=0.8pt] (0,0.5) -- (0,-0.5);
\end{tikzpicture}%
}

\newcommand{\FeynmanCubicSunset}{%
\begin{tikzpicture}[baseline=-0.5ex, scale=1.0]
  \fill (-0.5,0) circle (1.8pt);
  \fill (0.5,0) circle (1.8pt);
  \draw[line width=0.8pt] (-0.5,0) -- (0.5,0);
  \draw[line width=0.8pt] (-0.5,0.35) circle (0.35cm);
  \draw[line width=0.8pt] (0.5,0.35) circle (0.35cm);
\end{tikzpicture}%
}

\newcommand{\FeynmanQuarticVertexInst}{%
\begin{tikzpicture}[baseline=-0.5ex, scale=1.0]
  \draw[line width=0.8pt, double, double distance=1.2pt] (-0.35,0) circle (0.35cm);
  \draw[line width=0.8pt, double, double distance=1.2pt] (0.35,0) circle (0.35cm);
  \fill (0,0) circle (1.8pt);
\end{tikzpicture}%
}

\newcommand{\FeynmanCubicBubbleInst}{%
\begin{tikzpicture}[baseline=-0.5ex, scale=1.0]
  \draw[line width=0.8pt, double, double distance=1.2pt] (0,0) circle (0.5cm);
  \fill (0,0.5) circle (1.8pt);
  \fill (0,-0.5) circle (1.8pt);
  \draw[line width=0.8pt, double, double distance=1.2pt] (0,0.5) -- (0,-0.5);
\end{tikzpicture}%
}

\newcommand{\FeynmanCubicSunsetInst}{%
\begin{tikzpicture}[baseline=-0.5ex, scale=1.0]
  \fill (-0.5,0) circle (1.8pt);
  \fill (0.5,0) circle (1.8pt);
  \draw[line width=0.8pt, double, double distance=1.2pt] (-0.5,0) -- (0.5,0);
  \draw[line width=0.8pt, double, double distance=1.2pt] (-0.5,0.35) circle (0.35cm);
  \draw[line width=0.8pt, double, double distance=1.2pt] (0.5,0.35) circle (0.35cm);
\end{tikzpicture}%
}

\newcommand{\FeynmanCubicJacobian}{%
\begin{tikzpicture}[baseline=-0.5ex, scale=1.0]
  \draw[line width=0.8pt, fill=white] (-0.5,0) circle (1.8pt);
  \draw[line width=0.8pt, double, double distance=1.2pt] (-0.5,0) -- (0.15,0);
  \fill (0.15,0) circle (1.8pt);
  \draw[line width=0.8pt, double, double distance=1.2pt] (0.5,0) circle (0.35cm);
\end{tikzpicture}%
}

\newcommand{\StokesGraphAiryWithArrow}[2]{%
\begin{tikzpicture}[scale=0.38, baseline=(current bounding box.center)]
    \def\R{2.2}

    \draw[stokesminus, line width=1.0pt] (0,0) -- (\R,0) node[pos=0.75, above] {\scriptsize $-$};
    \draw[stokesplus, line width=1.0pt] (0,0) -- (120:\R) node[pos=0.75, above left] {\scriptsize $+$};
    \draw[stokesplus, line width=1.0pt] (0,0) -- (-120:\R) node[pos=0.75, below left] {\scriptsize $+$};

    \draw[squiggly, line width=1.0pt] (0,0) -- (-\R,0);

    \fill (0,0) circle (3pt);

    \draw[blue, line width=1.0pt, ->, >=stealth] (0,0) -- (#1:\R);

    \node[blue, font=\footnotesize, anchor=south] at (0,2.8) {#2};

    \useasboundingbox (-3.2,-3.2) rectangle (3.2,3.2);
\end{tikzpicture}%
}

%% file: sections/introduction.tex
% \!TEX root = ../DoubleDoubleMain.tex
\section{Introduction}
The symmetric double well potential in quantum mechanics has provided deep insights into quantum theory for many decades. Its exploration by Sidney Coleman in his timeless Erice lectures~\cite{ColemanErice}, from which the title of this paper is taken, led generations of students to appreciate the importance of both instantons and levity in physics. On one hand the double well is phenomenologically relevant. For example, the simplest molecule $\mathrm{H}_2^+$ has an electron in a double well and more broadly, the local stability of matter relies on double-well physics. It also generalizes to quantum field theory: many features of quantum chromodynamics, such as the structure of the vacuum and the role of instantons can be fruitfully explored by analogy to the double well. On the other hand, it is mathematically rich. For example, it connects geometry (Picard--Lefschetz theory), algebra (elliptic curves), and analysis (Borel resummation). Indeed, the double well provides an almost ideal use case for some of these mathematical tools, demonstrating their utility in a concrete physical setting where the resurgent trans-series can be computed and explored systematically to high order. 
The goal of the current paper is to present a self-contained rigorous treatment of the double well from two complementary perspectives: Exact WKB and the Euclidean path integral. We review and build on the extensive literature on both methods, provide a pedagogical introduction, push the calculations to high order, and compile, relate and cross check results from the two approaches.

The focus of this paper is the quantum-mechanical energy spectrum of a particle subject to the one-dimensional symmetric double-well potential
\begin{equation}
V(z) = \frac{\omega^2}{8R^2}(z^2-R^2)^2\quad = \quad
\begin{tikzpicture}[baseline=1ex, scale=0.8]
\draw[->] (-1.8,0) -- (1.8,0) node[right] {\small $z$};
\draw[->] (0,-0.2) -- (0,1.4);
\draw[thick, blue, domain=-1.5:1.5, samples=80] plot (\x, {0.125*((\x)^2-1)^2*8});
\fill[red] (-1,0) circle (2pt);
\fill[red] (1,0) circle (2pt);
\fill[orange] (0,1) circle (2pt);
\node[above right, scale=0.7] at (0,1) {$\frac{1}{8}R^2\omega^2$};
\node[below, scale=0.7] at (-1,0) {$-R$};
\node[below, scale=0.7] at (1,0) {$R$};
\end{tikzpicture}
\label{eq:Vdefined}
\end{equation}
In the limit $R\to\infty$ at fixed $\omega$ the wells separate and the barrier becomes infinite. Then $V(\pm R+y)= \frac{1}{2}\omega^2y^2$ so that the potential separates into two infinitely-separated simple harmonic oscillators. In this limit the spectrum is doubly degenerate. As the wells are moved in from infinity the states mix and the degeneracy is lifted. The dimensionless scaling parameter around the $R=\infty$ limit is $\hbar/(m \omega R^2)$. For $H_2^+$ the dimensionless parameter is  $a_0/R = \hbar/(m_e \alpha c R)$ with $R$ the separation between the two nuclei and $a_0$ the Bohr radius. Setting $m_e=c=\alpha R=1$ lets us expand simply in $\hbar$. To explore this limit in the double well case and expand in $\hbar$ it is conventional to set $\omega=R=m=1$ so that $V(z)=\frac{1}{8}(z^2-1)^2$.

The energies can be computed using textbook Rayleigh-Schr\"odinger (time-independent) perturbation theory~\cite{Rayleigh1894,Schrodinger1926}: start from $\hbar=0$ where the spectrum and energies are known and expand around that limit (see Appendix~\ref{appendix:benderWu}). For example, the ground state energy of $\mathrm{H}_2^+$ at large internuclear separation $R$ (in atomic units) is given by the perturbative series~\cite{DamburgPRL1984,CizekDamburg1986}
\begin{equation}
  E_0(R) = -\frac{1}{2} - \frac{1}{R}- \frac{9}{4R^4} - \frac{15}{2R^6} - \frac{213}{4R^7} + \cdots
  \label{eq:H2pert}
\end{equation}
where $-1/2$ is the hydrogen atom ground state energy, $R^{-1}$ is the repulsion term, and the $R^{-4}$ term comes from the polarizability of the hydrogen atom. For the double well, the perturbative series for the $N$th energy level takes the form 
\begin{equation}
E_N(\hbar) = \sum_{k=1}^\infty e_k(N) \hbar^k
\end{equation}
with $e_1 = N+\tfrac{1}{2}$, $e_2 = -\tfrac{3}{4}(N+\tfrac{1}{2})^2-\tfrac{1}{16}$, and so on.

The perturbation series in $\hbar$ for both $H_2^+$ and the double well are straightforward to compute. Both series have two shortcomings. First the series are asymptotic, with $e_k \sim k!$ at large $k$ with zero radius of convergence. Second, the splittings between even and odd parity states are invisible to perturbation theory at any finite order. These issues have the same source: missing physics. The missing physics is associated with tunneling between the wells. Tunneling contributes non-perturbative effects proportional to $e^{-S_I/\hbar}$ where $S_I$ is the instanton action, given by the integral of the classical momentum through the classically forbidden region between the wells. Including tunneling effects, the result for $H_2^+$ can be written as ~\cite{DamburgPRL1984,CizekDamburg1986,Holstein1952,HolsteinHerring}
\begin{equation}
  E_0(R) = -\frac{1}{2}- \frac{1}{R} -  \frac{9}{4R^4} + \cdots - \frac{2}{e} R \, e^{-R} \Big(1 + \frac{1}{2R} + \cdots \Big) + \frac{4 R^2}{e^2} e^{-2R} \Big(R+ 2 \ln 2R + 2\gamma_E \pm i \pi+  \cdots \Big) + \cdots \,.
\end{equation}
This is an example of \emph{trans-series}. For the double-well we write
\begin{equation}
  E_N = \sum e_k \hbar^k + e^{-\frac{S_I}{\hbar}} \sum e_k^{(1)} \hbar^k + e^{-\frac{2S_I}{\hbar}}  \sum  (e_k^{(2)} + c_k^{(2)} \ln \hbar ) \hbar^k +\cdots
\end{equation}
which should be thought of as a formal collection of coefficients. We call the series at each order in $e^{-S_I/\hbar}$ the $n$-instanton sector. Each instanton sector by itself is asymptotic, so one cannot simply sum the terms.  In addition, the coefficients are generally complex, while the energies are real. Remarkably, though, the coefficients in the different instanton sectors are all related. For example, the imaginary parts of sector $n$ are completely determined by sectors $n-2$ and lower. The relationships between different sectors are called resurgence relations and are a consequence of the underlying analytic structure of the energy as a function of $\hbar$. These relations are efficiently encoded through the Alien Calculus of \'Ecalle~\cite{ecalle1981fonctions1,ecalle1981fonctions2,ecalle1985fonctions3}. 

A central tool in analyzing trans-series is the Borel transform. A Borel transform is a misleadingly simple operation: replace the asymptotic series $\sum e_k \hbar^k$ with a convergent one $B(t)= \sum(e_k/k!)t^k$. From $B(t)$ one then constructs a function of $\hbar$ by Borel resummation
\begin{equation}
  \mathcal{S}[B] = \frac{1}{\hbar}\int_0^\infty dt\, e^{-t/\hbar} B(t)  \,.
  \label{eq:Borel_resummation}
\end{equation}
Since $\cS[t^k]=k! \hbar^k$, if this integral exists, it (trivially) has the asymptotic series $\sum e_k \hbar^k$. If $B(t)$ has singularities on the integration contour, then it is said to be non-Borel resummable, and one must deform the contour to integrate. Doing so leads to ambiguities depending on how the deformation is done. These ambiguities are the key to relating the different instanton sectors in the trans-series and reconstructing the real $E_N$ from its expansion. 

Taking a step back, one can understand the structure of Borel transforms by noting that well-defined quantities, like $E_N$, have a valid non-perturbative description. When described directly from the path integral point of view, one needs to compute Laplace integrals such as
\begin{equation}
  Z(\hbar) = \int d^d z \, e^{-S(z)/\hbar} \,.
\end{equation}
Comparing to Eq.~\eqref{eq:Borel_resummation} one sees that $t$ is naturally identified with the Euclidean action. The Borel transform is given by the Jacobian $B(t) \sim |\partial S/\partial z|^{-1}$, so that singularities in $B(t)$ arise from critical points of the action. This connects obstructions to Borel resummation (singularities at real positive $t$) with semi-classical saddle points (instantons).

The central observation which demystifies the entire Borel resummation program is that each instanton sector of the trans-series comes from the saddle-point expansion around some critical point $z_\star$ of the action. Saddle point expansions are how series arise in quantum mechanics and quantum field theory. The prefactor $\exp(-n S_I/\hbar)$ is due to the action at the saddle $n S_I = S(z_\star)$. A saddle-point expansion must be done so that the Gaussian integrals are convergent, meaning $S''(z_\star)>0$ along every integration direction. If the saddle point is a local maximum in some real direction it is a minimum in the corresponding imaginary direction. Thus it is typically necessary to complexify the integration contour to remain along steepest descent contours, also called Lefschetz thimbles\footnote{Conventionally one discusses exponential integrals where the integrand is $\exp(\lambda g(z))$ and the Lefschetz thimbles are steepest descent paths. When $g(z)=-S(z)$ this corresponds to steepest \emph{ascent} paths of the action.}. 
Expanding the action around the saddle gives a Gaussian integral at quadratic order, convergent in affine coordinates along the steepest descent directions. Higher-order terms in the expansion produce additional integrals which converge against the same Gaussian kernel, giving generically an asymptotic perturbative series. Alternatively, we can keep the full action and integrate along the full thimble, which generally bends through complex space. This integral is also convergent, and, somewhat remarkably, gives the Borel resummation of the asymptotic series. The quantity we are ultimately interested in is the exact answer ($Z$ or the energies $E_N$), while the one we often have access to is the series. So Borel resummation allows us (in principle) to compute the exact answer from the series.

To actually complete the program of reconstructing the partition function or the energies from the trans-series, one needs the trans-series! For the double well, there are two powerful and complementary methods for computing it: Exact WKB and the Euclidean path integral. Why and how the trans-series arises is different in the two methods. The goal of this paper is to present the two methods together, emphasizing the commonalities and differences, and building an explicit bridge between them. We will see that the two methods are in fact closely related, with the same complex classical solutions playing a central role in both.

Exact WKB is a method to make the Wentzel, Kramers and Brillouin~\cite{Wentzel1926,Kramers1926,Brillouin1926} (WKB) approach systematic. Recall that WKB approximates a wavefunction with energy $E$ as 
%$\psi(x) =(P(x))^{-1/2}\exp{\tfrac{i}{\hbar}\int^x P(x')dx'}$ 
$\psi(x) =\big[1/\sqrt{P(x)} \,\big]\exp{\tfrac{i}{\hbar}\int^x P(x')dx'}$ 
where $P(x)=\sqrt{2E-2V(x)}$ is the classical momentum. This approximation runs into problems at the turning points where $V(x)=E$ since $P(x)=0$ there and $\psi(x)$ blows up. The Exact WKB approach gets around these singular points by literally getting around the singular points, via analytic continuation in the complex plane. The Exact WKB approach was developed by Voros~\cite{Voros1983} and further refined by Delabaere and Pham~\cite{DelabaerePham1999,DelabaereDillingerPham1993}. Its starting point is the observation that in the neighborhood of a turning point, after shifting and rescaling, the Schr\"odinger equation reduces to
\begin{equation}
  f'' (x) - x f (x) = 0 \,.
\end{equation}
This is Airy's equation with two independent solutions, $\Ai (x)$ and $\Bi (x)$.  The two solutions have different asymptotic behavior, like $e^{-x}$ or $e^{+x}$ but in three angular wedges rather than two. Then Exact WKB constructs a wavefunction which behaves like $e^{-x}$ at large $x$, analytically continues it through the 3-wedge structure around each turning point, and then matches it to the $e^{+x}$ as $x\to -\infty$ behavior required for normalizability. Doing this matching carefully leads to a quantization condition
\begin{equation} 
  1 + \VP \pm i \sqrt{\VN} = 0 \label{quant}
\end{equation}
where the $\mathcal{V}_j$ are called Voros symbols, given by
\begin{equation}
  \VP = \exp \left( \frac{1}{\hbar}\oint_{\gamma_P} d x \, P(E,x) \right),\quad
  \VN = \exp \left(\frac{1}{\hbar} \oint_{\gamma_N} d x \, P(E,x) \right)
\end{equation}
where $P(E,x) = \sum_k P_k(E,x)\hbar^k$ is the quantum momentum, computable perturbatively from $V(x)$ through the Riccati (WKB) expansion. The contours $\gamma_P$ and $\gamma_N$ associated with the symbols $\VP$ and $\VN$ are respectively the perturbative and non-perturbative cycles, surrounding adjacent pairs of turning points. 

Each integral in the expansion of Eq.~\eqref{quant} is a contour integral over an algebraic function such as $P_0(E,x) = \sqrt{2E-2V(x)}$. The integrals over $P_0$ give closed form elliptic functions, the classical perturbative and non-perturbative actions $S_P^0$ and $S_N^0$. Remarkably, all subsequent integrals can be derived from these algebraically. This is possible because the integrals are all associated with a genus 1 elliptic curve and satisfy a second order Picard-Fuchs equation. At each order in $\hbar$, the integrals $S^k = \oint P_k(E,x) dx$ are linear combinations of $S^0$ and $\partial_E S^0$ with coefficients that are rational functions of $E$. Thus $\VP$ and $\VN$ can be algorithmically computed to arbitrary order in $\hbar$. 

To go from the expansions of $\VN$ and $\VP$ to the energies, one must solve Eq.~\eqref{quant} around the leading solution $E_N = \hbar (N + \tfrac{1}{2})$. For the perturbative spectrum this is straightforward to carry out, reproducing the Rayleigh-Schr\"odinger perturbation theory. However, for the leading non-perturbative corrections a complication arises: the rational functions of $E$ in the period integrals have poles at $E=0$. Thus setting $E=\hbar\kappa$ mixes orders in perturbation theory. The result is that, at each order in $\hbar$, terms to all orders in the WKB expansion are relevant. With a careful calculation these terms can be resummed leading to 
\begin{equation}
E_{N} = \hbar\left(N+\frac{1}{2}\right) + e^{-S_I/\hbar}
\frac{\hbar}{\sqrt{2\pi}}\,
  \frac{1}{\Gamma\!\left(N+1\right)}
  \left(\frac{8}{\hbar}\right)^{N+1/2} + \cdots \,.
  \label{eq:EN_intro}
\end{equation}
Higher order corrections in $\hbar$ require subleading resummation of the pole towers using the Picard--Fuchs coefficients, leading to digamma and polygamma functions in the intermediate steps, though remarkable cancellations simplify the final expressions. In that way, Exact WKB can provide corrections in increasing powers of $e^{-n S_I/\hbar}$ ($n$-instanton sectors), with the leading-order $\hbar$ terms relatively easy to get and the higher-order corrections increasingly taxing.

The second approach to computing the spectrum of the double well uses the Euclidean path integral and the partition function
\begin{equation}
  Z (T) = \operatorname{Tr} (e^{- H T/\hbar}) = \int_{x (T) = x (0)} \mathcal{D} x \,  e^{- S(x)/\hbar} \,.
\end{equation}
The integral is over real paths $x (t)$ with periodic boundary conditions. The energies are buried within the partition function. They can be extracted through its Laplace transform called the resolvent
\begin{equation}
  G (E) = \operatorname{Tr} \frac{1}{H - E} = \frac{1}{\hbar}\int_0^{\infty} d T\, e^{E T/\hbar}\, Z (T) \,,
\end{equation}
as poles of $G(E)$ give the energies. If one computes $Z$ in the saddle point approximation around the perturbative saddles $x=\pm 1$, the leading result is $Z=1/\sinh(T/2)$ with resolvent $G(E) = -\psi(\frac{1}{2}-E/\hbar)$ where $\psi$ is the digamma function. The poles of $G$ are at $E_N = \hbar(N + \tfrac{1}{2})$, reproducing the leading order SHO spectrum. Including fluctuations around the perturbative saddle gives corrections to the energies, and the resulting energies are exactly those of Rayleigh-Schr\"odinger perturbation theory.

The key to getting the full trans-series with the path integral is observing that there are other relevant stationary points of the action, associated with instantons. The traditional approach to incorporating these into the partition function is through the dilute instanton gas (DIG) approximation. In this approximation one takes $T\to \infty$ and treats the sum over paths in $Z$ as splitting into approximate integrals over paths in each instanton sector. Unfortunately, this approach is not systematic and cannot produce the energies of the excited states. The first problem with the DIG is that the action decreases when instanton centers are displaced. This feature is often described as an attractive force between instantons. The negative eigenvalue associated with the curvature of the action in these quasi-zero mode directions makes the associated functional determinant imaginary. This implies there must be some implicit analytic continuation being used in the path integral which we have not yet identified. A second problem with the instanton gas picture is conceptual:  why are we summing over configurations with different numbers of instantons in the first place? The path integral, no matter where we expand around, sums over all real paths so any particular saddle point expansion would already include the sum over all modes. 

A rigorous way to see the emergence of an instanton sum is not with the DIG but with Picard-Lefschetz theory. In this approach, the path integral is decomposed into a sum over Lefschetz thimbles, each associated with a critical point of the action. To begin this program, we must first find the exact saddles at finite $T$. These saddles are known and can be enumerated and written in closed form because the symmetric well corresponds to a genus 1 elliptic curve for which there is a rich mathematical literature. The exact saddles are Weierstrass elliptic functions $x_{k,k'}$ labeled by integer pairs where $0\le k' < k$ are integers counting the number of times the solution winds around the two cycles of a torus. With the exact saddles, there is an exact decomposition of the partition function as
\begin{equation}
  Z = \sum_{k,k'} \eta_{k,k'} Z_{k,k'},\quad
  Z_{k,k'} = \int_{\mathcal{J}_{k,k'}} \mathcal{D} x \, e^{- S(x)/\hbar}\,,
\end{equation}
where $\mathcal{J}_{k,k'}$ is the complex Lefschetz thimble associated with the saddle $x_{k,k'}$ and $\eta_{k,k'}$ is the intersection number between the thimble and the original integration contour. This intersection number determines the weight with which each thimble contributes. A nontrivial result is that $\eta_{k,k'}=0$ for $k'\ne 0$, so that only the thimbles passing through real saddle points contribute.

Each $Z_{k,k'}$ can be computed in a systematic expansion around the real saddle $x_{k,k'}$. The leading behavior is $e^{-2kS_I/\hbar}$. Thus the $2k$-instanton sector of the trans-series for $Z$ is exactly given by the perturbative series for $Z_{k,0}$. The one-loop contribution to $Z_{k,k'}$ can also be computed in closed form. This is possible because the operator for fluctuations around the instanton $\cO_{k,k'} = -\partial_t^2 + V''(x_{k,k'}(t))$ is a Lam\'e operator, whose spectrum and determinant is known. The determinant itself can be expressed in terms of the classical period integrals $S_P^0$ and $S_N^0$ that appeared in Exact WKB and are known in closed form. The operator $\cO_{k,k'}$ has one zero mode which must be removed as a collective coordinate. In addition there are quasi-zero modes, associated with the relative motion of the instanton centers, with exponentially small negative eigenvalues $\lambda_j \sim -e^{-T}$ of $\cO_{k,k'}$. These can be removed as well and integrated over separately.  The result is that the path integral around the $(k,k')$ saddle can be written as
\begin{equation}
   Z_{k,k'}
  =   \sqrt{\frac{\|\dot x_{k,k'}\|^2}{2\pi\hbar}} \;T\;
  e^{-\frac{1}{\hbar} S_{k,k'}}\,
(\det\nolimits_\perp \opO_{k,k'})^{-1/2}\,
\, e^{T \DV + 2 k \DL}\, Y_{k,k'} \,
 \cI_{k,k'}
 \label{ZcZ}
\end{equation}
where the prefactor and $T$ are from the collective-coordinate zero mode, $S_{k,k'}$ is the action of the saddle, $(\det\nolimits_\perp \opO_{k,k'})^{-1/2}$ is the functional determinant over transverse modes,
$T \DV + 2 k \DL$ encode connected Feynman diagram loops, and $\cI_{k,k'}$ is the thimble integral along the $(2k-1)$-dimensional quasi-zero mode manifold with $Y_{k,k'}$ the appropriate Jacobian. 

Up to terms exponentially small at large $T$, both $\DV$ and $\DL$ are $T$-independent, $S_{k,k'}=2kS_I$, and $\cI_{k,k'}$ is a polynomial of degree $2k-1$ in $T$. 
The large $T$ behavior is enough to get the ground state energy, but to get the energies of the excited states, exponentially subleading corrections are needed. For example, the real one-instanton saddle has antisymmetric boundary conditions and contributes to the twisted partition function $\widetilde{Z} = \Tr[\cP \exp(-H T/\hbar)]$, with $\cP$ the parity operator, rather than to $Z$. Its action is $\widetilde S_1 = S_I - 8 e^{-T} +\cdots$, and so
\begin{equation}
  \widetilde Z\sim e^{-\widetilde S_1/\hbar} = e^{-S_I/\hbar}\sum_{N=0}^\infty \frac{1}{\Gamma(N+1)}\left(\frac{8}{\hbar}\right)^N e^{-N T}  \,.
\end{equation}
This structure is the same as in Eq.~\eqref{eq:EN_intro}. Matching to the spectral decomposition of $\widetilde{Z}$ gives the one-instanton contribution to the $N$th energy level as the coefficient of $e^{-N T}$ in this expansion in exact agreement with the Exact WKB result, but without the need for any resummation.

One of the most important payoffs of the path integral approach is a geometric picture of resurgence, encoded in the thimble integral $\cI_{k,k'}$ of Eq.~\eqref{ZcZ}. For the $n=2k$ real saddle, $\cI_{k,k'}$ is an integral over an $(n-1)$-dimensional manifold within $\CC^{n-1}$, resulting in a Meijer $G$-function (which reduces to a polynomial in $T$ at large $T$). Beyond producing $\cI_{k,k'}$, this manifold encodes the thimble geometry and Stokes structure of the other instanton sectors related to the $n=2k$ saddle by resurgence. The imaginary parts in the trans-series then have a geometric origin as concrete middle-dimensional contours. In contrast, in Exact WKB the resurgent structure is only accessible through the energies and the alien calculus. With the path integral, the resurgence structure of the energy trans-series is induced from that of the partition function, which exposes the underlying thimble geometry.

The paper is organized as follows. In Section~\ref{sec:history}, we review the historical development of resurgence in the context of anharmonic oscillators and double-well potentials. In Section~\ref{sec:zerodim}, we warm up with the zero-dimensional double-well integral, illustrating Borel summation, Stokes phenomena, and thimble decomposition in a setting where everything can be computed exactly.  In Section~\ref{sec:exactWKB}, we develop the Exact WKB method, starting with a review of Airy functions and Stokes phenomena, then deriving the quantization condition for the double well in terms of Voros symbols. We work out the Picard--Fuchs structure of the period integrals, the resummation of colliding-turning-point singularities into gamma and polygamma functions, and the full energy trans-series through four-instanton corrections, including alien calculus constraints and the resurgence structure. In Section~\ref{sec:pathintegral}, we turn to the path integral approach, beginning with the simple harmonic oscillator, then treating instantons and the elliptic curve structure of saddle points at finite $T$, and finally performing the Lefschetz thimble decomposition and quasi-zero-mode integrals, starting with the $n=2$, $n=3$ and $n=4$ sectors as examples, before moving on to discuss general $n$ results. We conclude in Section~\ref{sec:conclusions} with a discussion of results and outlook. Several appendices collect technical details: perturbative calculations and Bender--Wu recursion (Appendix~\ref{appendix:benderWu}), the $P_n$ decomposition and Picard--Fuchs algorithm (Appendix~\ref{appendix:PicardFuchs}), the Weber equation and nonperturbative resummation of the pole towers (Appendix~\ref{appendix:Weber}), the spectrum of the Lam\'e equation (Appendix~\ref{appendix:Lame}), higher-loop perturbative corrections on an instanton background (Appendix~\ref{appendix:loop_corrections}) and explicit trans-series coefficients (Appendix~\ref{appendix:terms}).

%% file: sections/history.tex
% \!TEX root = ../DoubleDoubleMain.tex
\section{Historical development of resurgence}
\label{sec:history}

The realization that perturbation theory in quantum mechanics generically produces divergent series has a long history, intertwined with the development of instanton methods, Borel summation, and what is now called resurgence. This section provides a brief overview of this development, focusing on the anharmonic oscillator and double-well potential as the central examples. See the reviews by Dorigoni~\cite{Dorigoni2014}, Mari\~no~\cite{MarinoLectures}, Aniceto, Ba\c{s}ar, and Schiappa~\cite{AnicetoSchiappa2018}, and Sauzin~\cite{Sauzin2007} for more comprehensive introductions to resurgence.

The mathematical foundations for understanding divergent series were laid in the 19th century, though not without resistance. In an 1826 letter to Holmboe, Abel declared that ``divergent series are the invention of the devil, and it is shameful to base on them any demonstration whatsoever''~\cite{Abel1828}. Yet divergent series refused to disappear from physics. In work read in 1857 and published in 1864~\cite{Stokes1857}, Stokes showed that the asymptotic expansion of the Airy function, introduced by Airy in 1838 in connection with the intensity of light near a caustic, changes discontinuously across certain lines in the complex plane. This ``Stokes phenomenon'' was an early indication that asymptotic series contain hidden information about exponentially small terms invisible to the power series expansion.

The crisis came to a head in celestial mechanics.  The central problem of 19th century astronomy was the stability of the solar system: could the perturbation series used to compute planetary orbits be trusted for all time? In 1885, King Oscar~II of Sweden and Norway offered a prize for a rigorous solution to the $N$-body problem in the form of a convergent series, to be awarded on his 60th birthday in January~1889~\cite{BarrowGreen1994}.  Poincar\'e won the prize with a memoir on the restricted three-body problem that appeared to establish stability by showing that certain invariant manifolds join smoothly.  But after the prize had been awarded, while the memoir was being prepared for publication in \textit{Acta Mathematica}, Poincar\'e discovered a fundamental error~\cite{BarrowGreen1994}: the stable and unstable manifolds of an unstable periodic orbit do \emph{not} join smoothly---they intersect transversely.  The corrected memoir~\cite{Poincare1890}, published in 1890 after the original printing was recalled, contained a discovery that would reshape dynamical systems: these transverse \emph{homoclinic intersections} force the manifolds to cross infinitely many times, weaving together in a pattern of such complexity that Poincar\'e, writing a decade later in his \textit{Les m\'ethodes nouvelles de la m\'ecanique c\'eleste}, declared he would ``not even attempt to draw it.''  This homoclinic tangle is the geometric origin of what we now call chaotic dynamics.

\begin{figure}[t]
\centering
\includegraphics[width=\textwidth]{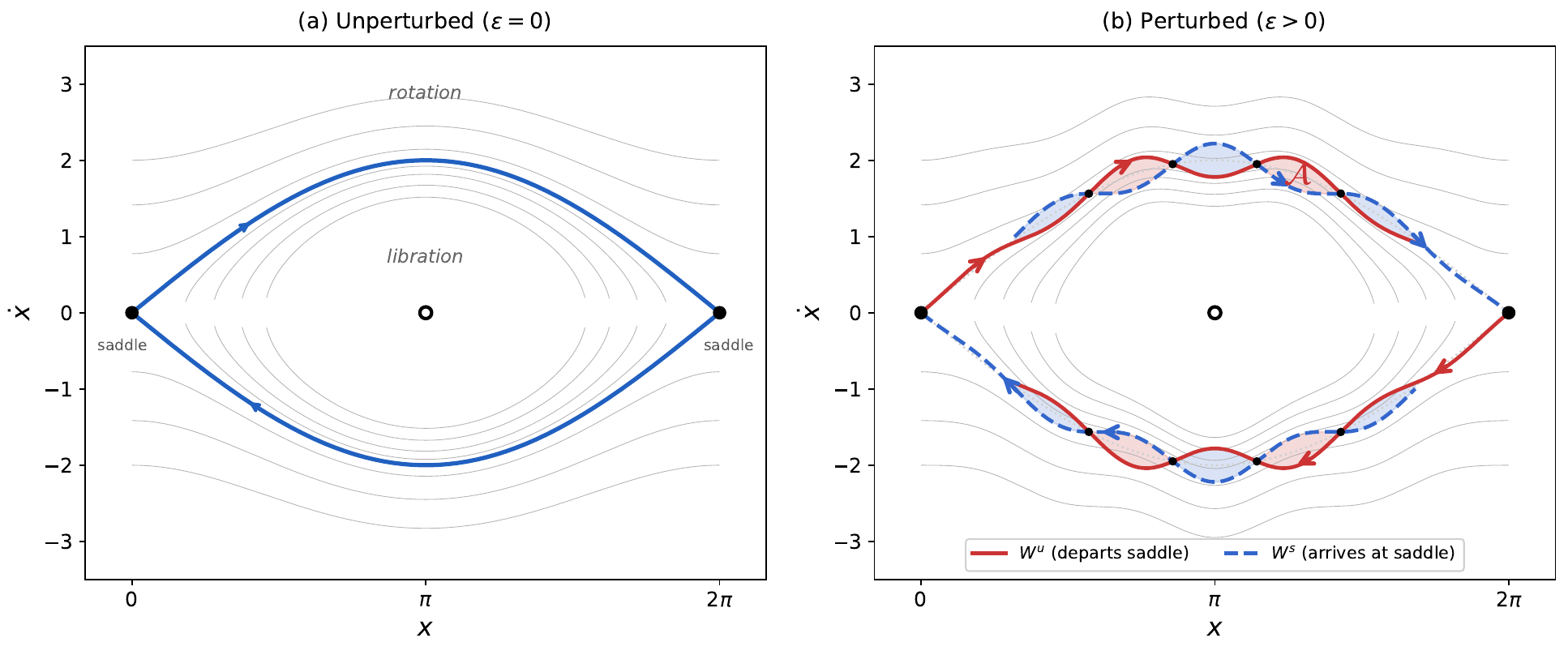}
\caption{Phase portrait of the pendulum $\ddot{x} = \sin x$. The pendulum down at rest is $x=\pi$ and the unstable equilibrium saddle is at $x=0$ or $x=2\pi$.
 (a)~Unperturbed: the separatrix (blue) divides libration from rotation.  The stable and unstable manifolds of each saddle point coincide exactly.  (b)~With rapid periodic forcing ($\varepsilon > 0$): the unstable manifold $W^u$ (solid red, departing each saddle) and stable manifold $W^s$ (dashed blue, arriving at each saddle) split apart and cross transversely at homoclinic points (black dots), creating lobes of area $\mathcal{A}$.  This area---the flux of initial conditions transported across the old separatrix per forcing cycle---is exponentially small: $\mathcal{A}(\varepsilon) \sim e^{-\pi/(2\varepsilon)}$.}
\label{fig:separatrix_lobes}
\end{figure}

Unlike Poincar\'e, we will attempt to draw it (see Fig.\ref{fig:separatrix_lobes}). For a concrete illustration consider the driven pendulum~\cite{HolmesMarsdenScheurle1988,DelshamsSeara1992}. This simple classical system already has a resurgent structure. Let $x$ be the angle measured from the top of the pendulum. The equations of motion are
\begin{equation}
\ddot{x} = \sin x + \frac{\mu}{\varepsilon^2}\,\sin\!\Big(\frac{t}{\varepsilon}\Big),
\label{eq:forced_pendulum}
\end{equation}
where $\varepsilon \ll 1$ controls the forcing frequency and $\mu$ its strength; the $1/\varepsilon^2$ scaling keeps the time-averaged effective potential fixed as $\varepsilon \to 0$.  For the unforced system ($\mu=0$), whether the pendulum oscillates back and forth (libration) or rotates over the top depends on the initial conditions. The two possibilities are clearly separated by the separatrix curve $\dot{x} = \pm 2\sin(x/2)$ in phase space. The separatrix passes through the saddle at $x=0$, where the pendulum is upside down: this is an unstable fixed point. The \emph{unstable manifold} $W^u$ is the set of trajectories that depart from this saddle as $t\to -\infty$ and the \emph{stable manifold} $W^s$ is the set of trajectories that arrive at the saddle as $t\to +\infty$: the initial conditions from which the pendulum asymptotically approaches the inverted position. Without forcing, $W^u$ and $W^s$ coincide: they are both the separatrix, because a pendulum falling from the top traces a path that, running backwards in time, returns to the top. With forcing turned on, $W^u$ and $W^s$ split apart and cross transversely, creating the homoclinic lobes shown in Fig.~\ref{fig:separatrix_lobes}.  The lobe area $\mathcal{A}$ has a trans-series expansion~\cite{Sauzin1995,Sauzin2007} as a function of $\varepsilon$:
\begin{equation}
\mathcal{A}(\varepsilon) \sim
e^{-\frac{\pi}{2 \varepsilon}}\left[ a_0 + a_1\varepsilon + a_2\varepsilon^2 + \cdots + \left(\frac{2 \varepsilon }{\pi}\right)^n n!
+ e^{-\frac{\pi}{2 \varepsilon}}\Big(b_0 + b_1\varepsilon + \cdots\Big)\right]
+ \cdots
\label{eq:lobe_transseries}
\end{equation}
The power series within each sector diverges factorially, with the large-order growth of $a_n\sim (2/\pi)^n \, n!$. The Borel transform of this series has its first singularity at $|t|=\pi/2$. This is the same as the distance to the nearest complex-time singularity of the unperturbed separatrix solution $x_0(t) = 4\arctan(e^t) $, which has logarithmic branch points at $t = \pm i\pi/2$, and sets the scale for the exponential suppression of the first subleading sector in the trans-series. This is a purely classical realization of the same resurgent structure that governs quantum tunneling, with $\varepsilon$ playing the role of $\hbar$ and the complex-time singularity distance $\pi/2$ playing the role of the instanton action $S_I$.

Poincar\'e also made a foundational contribution of a different kind~\cite{Poincare1886}: he formalized the definition of an asymptotic expansion, showing that a series could be everywhere divergent yet still approximate a function to any desired accuracy by truncating at an optimal order.  Around the same time, Stieltjes~\cite{Stieltjes1886} studied what he called ``semi-convergent'' series: series whose terms decrease for a while before eventually growing without bound. Borel~\cite{Borel1899} then introduced his summation method in 1899, providing a way to assign meaningful values to certain divergent series.  Together, these developments legitimized the long-used but previously suspect series of the astronomers and physicists as valid mathematical objects, and provided the framework on which all later work on divergent series, including resurgence, would be built.  Their relevance to quantum mechanics, however, was not immediately apparent.

When Schr\"odinger developed his perturbation theory in 1926~\cite{Schrodinger1926}, it was immediately applied to problems like the anharmonic oscillator, but the question of convergence was not seriously investigated. Physicists computed the first few terms and moved on. The first clear physical argument that such series must diverge came from Dyson's 1952 paper~\cite{Dyson1952} which argued that perturbation theory in quantum electrodynamics should be divergent. His reasoning was physical: if the perturbation series in the coupling $e^2$ converged, it would define an analytic function in a neighborhood of $e^2 = 0$. But for $e^2 < 0$, like charges attract, the vacuum becomes unstable to pair production, and the theory is pathological. This singularity at negative coupling implies the series cannot converge. Though not mathematically rigorous, Dyson's argument captured a profound insight: \emph{instabilities and the divergence of perturbation theory are intimately connected}. Similar arguments apply to the anharmonic oscillator $V(x) = \frac{1}{2}x^2 + gx^4$: for $g < 0$ the potential is unbounded below, so analyticity in $g$ at the origin is impossible.

The first quantitative understanding of large-order behavior came from Bender and Wu~\cite{BenderWu, BenderWu2}, who studied the ground state of the quartic anharmonic oscillator in a landmark series of papers beginning in 1969. They computed the perturbation coefficients $E_n$ in $E(g) = \sum_n E_n g^n$ to high order and discovered that they grow factorially:
\begin{equation}
E_n \sim (-1)^{n+1} \frac{\sqrt{6}}{\pi^{3/2}}\, 3^n\, \Gamma(n + 1/2) \left(1 + O(1/n)\right) \,.
\end{equation}
This factorial growth means the series has zero radius of convergence. However, because of the alternating sign the series is Borel summable.

By the late 1970s, several atomic and molecular problems were recognized to have divergent perturbation series tied to exponentially small non-perturbative effects.  The Stark effect in hydrogen~\cite{HerbstSimon1978,GraffiGrecchi1978} has a perturbation series in the electric field strength that diverges because any nonzero field makes the potential unbounded below, turning the bound state into a resonance; the Borel sum of the perturbative series has an imaginary ambiguity exactly canceled by the exponentially small ionization width.  The Mathieu equation~\cite{Harrell1978} describes band structure in a periodic cosine potential, with exponentially small band gaps of the same resurgent origin.  The hydrogen molecular ion $\mathrm{H}_2^+$~\cite{DamburgPRL1984,CizekDamburg1986} is the richest of these examples: in the dissociation limit it is a genuine two-center double-well problem, and its even/odd level splitting admits a full trans-series computed by Damburg, \v{C}\'{\i}\v{z}ek, Silverstone and collaborators using Borel summation, dispersion relations, and the Holstein--Herring method.

The quartic double well $V(x)=\frac{1}{8}(x^2-1)^2$ is a simpler 1D model sharing the same resurgent structure, and it has the virtue that a full multi-instanton trans-series can be derived from first principles without the technical overhead of Coulomb two-center coordinates.  Br\'ezin, Parisi, and Zinn-Justin~\cite{BrezinParisiZJ} extended the Bender--Wu analysis to it in 1977, showing that the perturbation series around either minimum is \emph{not} Borel summable (its Borel transform has singularities on the positive real axis) and connecting this non-summability to the real instantons mediating tunneling between the wells.  The instanton contribution to the energy splitting was understood through path integral methods building on Langer's theory of metastable decay~\cite{Langer1967} and the discovery of gauge theory instantons by Belavin, Polyakov, Schwarz, and Tyupkin~\cite{BPST}, whose one-instanton functional determinant (with its collective coordinates, fermion zero modes, and the axial anomaly) was computed by 't~Hooft~\cite{tHooft1976}.  Coleman's 1977 Erice lectures~\cite{ColemanErice} gave the influential pedagogical treatment of the dilute instanton gas framework for quantum mechanics.

Bogomolny~\cite{Bogomolny} and Zinn-Justin~\cite{ZinnJustin1981} then went further, computing the correlated instanton--anti-instanton $[I\bar I]$ contribution and discovering the mechanism now named after them: the imaginary ambiguity from Borel-resumming the perturbative series is exactly cancelled by an ambiguity in the $[I\bar I]$ sector, rendering the total energy real. This Bogomolny--Zinn-Justin (BZJ) cancellation is the hallmark of resurgence: the perturbative and non-perturbative sectors are not independent but are locked together by consistency. The path integral approach to these results relied on the \emph{dilute instanton gas} approximation, in which one sums over configurations of widely separated instantons and anti-instantons; multi-instanton contributions at the two-instanton level involve characteristic $\log g$ terms arising from the quasi-zero mode associated with the instanton--anti-instanton separation.

Zinn-Justin and Jentschura~\cite{ZinnJustinJentschura1, ZinnJustinJentschura2} organized these results into a generalized quantization condition that extends Bohr--Sommerfeld quantization~\cite{BohrSommerfeld1} to include instanton effects. Bohr--Sommerfeld quantization, from 1916, predates the Schr\"odinger equation by a decade; it determines bound-state energies by requiring the classical action around a periodic orbit to be quantized: $\oint p_E(x)\, dx = 2\pi\hbar(N + \frac{1}{2})$, where $p_E = \sqrt{2(E-V)}$ is the classical momentum. The modern understanding is that this is the leading WKB approximation, connecting the spectrum to \emph{period integrals} over classical trajectories. For the double well, the key insight is that there is a second period: the instanton action $S_I = \int_{-x_0}^{x_0} |p_E(x)|\, dx$, the integral of momentum through the classically forbidden region between the two minima. Zinn-Justin's generalized quantization condition incorporates both periods (Ref.~\cite{ZinnJustinJentschura} Eq.~(5)):
\begin{equation}
\frac{1}{\sqrt{2\pi}}\, \Gamma\!\Big(\tfrac{1}{2} - B(E,g)\Big)\, \Big(\!-\tfrac{2}{g}\Big)^{\!B(E,g)} e^{-A(E,g)/2} = \pm i
\label{eq:ZJ_quantization}
\end{equation}
where $g$ plays the role of $\hbar$, $B(E,g)$ is a perturbative function whose leading approximation $B = N + \frac{1}{2}$ reproduces the Bohr--Sommerfeld condition, and $A(E,g)$ has leading term $2S_I/g$. Both $B$ and $A$ are formal power series in $g$ with polynomial coefficients in $E$. The poles of $\Gamma(\frac{1}{2} - B)$ at $B = N + \frac{1}{2}$ generate the perturbative spectrum; expanding around these poles produces all multi-instanton sectors, including the $\log g$ terms characteristic of the $[I\bar I]$ contribution. This quantization condition was initially conjectured from path integral reasoning and extensive numerical checks; the systematic derivation from Exact WKB came later.

Trying to go beyond the dilute instanton-gas approximation has been challenging. For example, Bogomolny \cite{Bogomolny}, Zinn-Justin \cite{ZinnJustin1981}, Balitsky and Yung \cite{Balitsky:1986qn} and others used analytic continuation to justify getting a real integral out of the bion configuration. Another approach, by Richard and Rouet, enumerates all real and \textit{complex} saddles at finite $T$ and reproduces the dilute instanton gas result after summing over the contributions from all saddle points \cite{RichardRouet, Richard:1981gn}. Although Richard and Rouet's result reproduces the dilute instanton gas, they do not explain why one should sum over every possible saddle. 
Not long after, Carlitz and Nicole~\cite{Carlitz:1984ab} touched on this oversight with a saddle-selection rule based on deformation of the elapsed-time contour, but restricted to real classical energies.
Around the same time, complex-saddle methods were applied to related systems such as the octic double well and the sine-Gordon potential~\cite{Lapedes:1981tz, Mottola:1982mz, Millard:1984qt}.
Decades later, Nekrasov~\cite{Nekrasov2018} independently rediscovered the complex saddle-point classification of Richard and Rouet in the context of supersymmetric integrable systems. In Nekrasov's framework, the saddles are classified by pairs of winding numbers $(m,n)$ around the cycles of the spectral curve, and the Lefschetz thimbles are identified with critical points of an effective superpotential. 
% However, the explicit computation of fluctuation determinants, quasi-zero-mode integrals, and the resulting trans-series was left for future work.
Returning to the double well, Behtash and collaborators~\cite{Behtash:2018voa} introduced the notion of ``critical points at infinity'' (complex saddles whose location is pushed to infinity in the $T\to\infty$ limit while their action stays finite) and appreciated that such saddles are essential and that the integration contour must be deformed into the complex plane.  At finite $T$, however, these saddles in fact sit only partway out in the complex plane and additional saddles, lying further beyond them, must also be included in the sum.  Their thimble decomposition also required an \textit{ad hoc} factor of $\frac{1}{2}$ to achieve cancellation of imaginary ambiguities between the perturbative and instanton--anti-instanton sectors. We now know the answer to which saddles need to be summed over: all saddles with non-zero intersection number between their associated unstable Lefschetz thimble and the original (real) path integral contour (see Section~\ref{sec:pathintegral} below).

The mathematical framework unifying the BZJ mechanism was developed by \'Ecalle beginning in 1981~\cite{ecalle1981fonctions1, ecalle1981fonctions2, ecalle1985fonctions3}. His theory of \emph{resurgent functions} showed that the divergent series and the exponentially small corrections are not independent but are related by precise algebraic structures. The perturbative series ``knows about'' the non-perturbative physics: the singularities of its Borel transform encode the instanton contributions, and the discontinuities across these singularities are computed via \emph{alien calculus} (``calcul \'etranger''), which reconstructs the full trans-series. The alien derivatives are so named because they extract information from ``foreign'' singularities in the Borel plane invisible to the original asymptotic expansion. More strongly, alien calculus is constructive: an alien derivative at a Borel singularity relates the late-order behavior of one trans-series sector to the leading behavior of the next, so the perturbative series determines all the higher instanton coefficients. The BZJ cancellation follows automatically.

In parallel with the path integral developments, the WKB method was being made exact. Balian, Parisi, and Voros~\cite{BalianParisiVoros1978} showed in 1978 that discrepancies between the WKB series and exact eigenvalues could be explained by complex classical trajectories. Voros's 1983 paper~\cite{Voros1983} made WKB fully exact by working in the complex plane and using Borel resummation. Silverstone~\cite{Silverstone1985} revisited the WKB connection formula problem using Borel summation, showing that the Stokes phenomenon is encoded in the singularities of the Borel transform. The rigorous mathematical foundation was provided by Delabaere, Dillinger, and Pham~\cite{DelabaereDillingerPham1993,DelabaerePham1999}, who used \'Ecalle's resurgence theory to prove the Exact WKB results. Their work justified Zinn-Justin's quantization condition, conjectured from path integral reasoning, by showing that the period integrals (both the real period giving perturbation theory and the complex period giving instanton contributions) must be Borel resummed, and that the resulting exact quantization condition reproduces the full trans-series.

The 2000s saw Zinn-Justin and Jentschura~\cite{ZinnJustinJentschura,ZinnJustinJentschura1,ZinnJustinJentschura2} work out detailed multi-instanton expansions using their generalized Bohr--Sommerfeld conditions.  Independently, \'Alvarez~\cite{Alvarez2004} derived the same multi-instanton coefficients from a purely differential-equation approach (Langer--Cherry uniform asymptotics matching to parabolic cylinder functions), confirming the results without any path-integral input. On the Exact WKB side, Kawai and Takei~\cite{KawaiTakei2005} developed the theory systematically as a branch of algebraic analysis, building on the foundational work of Voros, Delabaere, and Pham, and providing a rigorous framework for the Borel resummation of WKB series and the associated Stokes geometry.

A major impetus for the modern synthesis came from Witten's application of Picard--Lefschetz theory to path integrals in 2010~\cite{Witten2010,Witten2010Morse}. The mathematical foundations (Lefschetz thimbles as steepest-descent contours attached to critical points of a holomorphic function) go back to Picard and Lefschetz \cite{Picard, Lefschetz} around the turn of the 20th century, and were developed for oscillatory integrals by Pham \cite{Pham1965, Pham1967} in the 1960s. But it was Witten's 2010 work on the analytic continuation of Chern--Simons gauge theory that brought these ideas into the mainstream of modern physics. The central insight is that a real oscillatory integral can be rigorously defined by deforming the integration contour into the complexified field space, decomposing it as a sum over Lefschetz thimbles, one for each relevant saddle point. The integer coefficients in this decomposition are topological intersection numbers that determine which saddles contribute. The thimble perspective led to practical advances: it was used to put quantum tunneling rates in quantum field theory on a rigorous footing directly in the Minkowski path integral~\cite{AFFS2016,AFFS2017}, bypassing the traditional but \textit{ad hoc} analytic continuation of the potential. Applied to the electroweak vacuum, this framework enabled the first complete calculation of the lifetime of the Standard Model~\cite{AFS2018,ChigusaMoroiShoji2017}. For the double well, the thimble approach provides a systematic answer to a question that the path integral had left open since Coleman's work: when there are multiple saddle points, which ones should be included in the semiclassical expansion, and with what weight?

The 2010s saw a ``resurgence renaissance'' in which these ideas were brought together and extended to quantum field theory and to a broad range of one-dimensional quantum-mechanical systems. Building on the work of Nekrasov and of Behtash and collaborators discussed above, Dunne and \"Unsal~\cite{DunneUnsal2014,DunneUnsal2016} connected resurgence and trans-series to compactified gauge theories, showing that correlated instanton--anti-instanton pairs (``bions'') play the same role in field theory as they do in quantum mechanics: their contribution cancels the Borel ambiguity of the perturbative expansion, precisely as the Bogomolny--Zinn-Justin mechanism requires. The same authors~\cite{DunneUnsal2014WKB} showed that uniform WKB reproduces the full multi-instanton trans-series and verified the resurgence relations directly in the cosine potential. Picard--Lefschetz theory was used to analytically continue path integrals~\cite{BasarDunneUnsal}, to address the sign problem of Minkowski path integrals~\cite{TanizakiKoike}, to push thimble calculations to all orders in integrable quantum-mechanical systems~\cite{Fujimori_2023}, and to connect the path integral approach to Exact WKB via complex bions~\cite{Sueishi2020}. Beyond the cosine and $\mathrm{H}_2^+$ examples already discussed, resurgent trans-series have been computed for $\mathcal{PT}$-symmetric anharmonic oscillators~\cite{Kamata2024PT}, the Hofstadter butterfly~\cite{GuXuHofstadter}, asymmetric and higher-degree anharmonic oscillators~\cite{Serone2017,Serone2021,BucciottiReisSerone}, quasi-exactly solvable potentials via complex bions~\cite{Kozcaz:2016wvy}, and in closed form for cubic, double-well, and cosine systems~\cite{vanSpaendonck:2023znn}. (The Razavy potential~\cite{Razavy1980,BaradaranPanahi2017} is a notable exception: quasi-exactly solvable, with a finite spectrum that yields no continuous trans-series.) The lesson across these examples is that the trans-series structure of the symmetric double well (perturbative plus one-instanton splitting plus two-instanton with $\ln\hbar$) is generic to any 1D system with a real instanton between degenerate minima, while the specific coefficients are model-dependent and must be computed separately for each potential.

%% file: sections/zero_dimensions.tex
\section{Warm-up: the double well in zero dimensions}
\label{sec:zerodim}

Before tackling the full quantum mechanical double well, it is instructive to study a zero-dimensional analog: a simple integral over the real line with a double-well ``action.'' This toy model exhibits all the essential features of resurgence---multiple saddle points, asymptotic series, Borel resummation, and the reconstruction of the exact answer from thimble decompositions---in a context where everything can be computed exactly. We begin with a brief review of the Borel transform and its importance for trans-series and then discuss the 0D double-well example.

\subsection{Borel transforms \label{sec:borel}}
Expansions in quantum mechanics or quantum field theory are generically asymptotic, with coefficients often growing as $n!$ at large $n$. To extract meaningful information from such series, we use the \emph{Borel transform}. Given an asymptotic series 
\begin{equation}
    f(g) = \sum_{n=0}^{\infty} c_n g^{n+a}
    \label{eq:fdef}
\end{equation}
its Borel transform is defined by
\begin{equation}
\mathcal{B}[f](t) = \sum_{n=0}^{\infty} \frac{c_n}{\Gamma(n+a+1)} t^{n+a}\,.
\label{eq:Boreldef}
\end{equation}
The division by $\Gamma(n+a+1) \sim n!$ tames the factorial growth, often producing a convergent series with finite radius of convergence. Upon summing this convergent series, one can analyze the singularity structure of the Borel transform.  If the singularity nearest to the origin of $\mathcal{B}[f](t)$ is at $t = A$, the radius of convergence of the Borel series is $|A|$, so by Cauchy--Hadamard $\limsup_{n\to\infty} |c_n/\Gamma(n{+}a{+}1)|^{1/n} = 1/|A|$. Equivalently, $c_n \sim n!/A^n$ at large $n$, up to a power-law prefactor determined by the nature of the singularity at $t=A$. Thus the location of the leading Borel singularity directly controls the dominant large-order behavior of $f(g)$. 

The original function can be recovered (at least formally) by the \emph{inverse Borel transform}:
\begin{equation}
\mathcal{S}[f](g) = \frac{1}{g}\int_0^{\infty} dt\, e^{-t/g} \mathcal{B}[f](t)\,.
\label{eq:Borelinverse}
\end{equation}
It is easy to check that plugging Eq.~\eqref{eq:Boreldef} into Eq.~\eqref{eq:Borelinverse} reproduces Eq.~\eqref{eq:fdef}. When $\mathcal{B}(t)$ has singularities on the positive real axis, the integral must be defined by deforming the contour above or below the singularity, yielding the \emph{lateral Borel resummations} $\mathcal{S}_{\pm}$. The difference between these is purely imaginary and exponentially small in $g$.

In physics, we are often interested in functions defined as Laplace integrals of the form
\begin{equation}
    f(g) = \int d z e^{-S(z)/g}
\end{equation}
or the multidimensional generalization. The asymptotic series for such functions can be computed in the saddle point approximation by expanding $z$ around any saddle point $z^\star$.  A fundamental result, which can be proven using steepest descent analysis (see, e.g., \cite{Dorigoni2014,AnicetoSchiappa2018}), is that the Borel resummation of the asymptotic series around a saddle point reproduces the integral along the \emph{Lefschetz thimble} (the steepest descent contour) passing through that saddle. Critically this thimble is often different from both the original integration contour, which is typically real, and the saddle-point contour, which is linear. The importance of this theorem and its subtleties can be appreciated after some examples.

\subsection{Double well in \texorpdfstring{$D=0$}{D=0}}

Consider the 0D Euclidean path integral
\begin{equation}
Z(\hbar) = \int_{-\infty}^{\infty} dz\, e^{-S(z)/\hbar} \,.
\label{eq:0dintegral}
\end{equation}
There is no kinetic term so the action is just the potential, which we take to be that of the double well:
\begin{equation}
S(z)  = V(z) = \frac{1}{8}(z^2-1)^2\quad = \quad
\begin{tikzpicture}[baseline=1ex, scale=0.8]
\draw[->] (-1.8,0) -- (1.8,0) node[right] {\small $z$};
\draw[->] (0,-0.2) -- (0,1.4);
\draw[thick, blue, domain=-1.5:1.5, samples=80] plot (\x, {0.125*((\x)^2-1)^2*8});
\fill[red] (-1,0) circle (2pt);
\fill[red] (1,0) circle (2pt);
\fill[orange] (0,1) circle (2pt);
\node[above right, scale=0.7] at (0,1) {$\frac{1}{8}$};
\node[below, scale=0.7] at (-1,0) {$-1$};
\node[below, scale=0.7] at (1,0) {$1$};
\end{tikzpicture}
\label{eq:0daction}
\end{equation}
The integral can be evaluated in terms of Bessel functions:
\begin{equation}
Z(\hbar) = \frac{\pi}{2}\, e^{-\frac{1}{16\hbar}} \left[ I_{-1/4}\left(\frac{1}{16\hbar}\right) + I_{1/4}\left(\frac{1}{16\hbar}\right) \right]\,.
\label{eq:0dexact}
\end{equation}
This is the exact answer. Can we recover it from the perturbative series?

\begin{figure}[t]
\centering
\begin{tikzpicture}[scale=1.2]
\draw[very thick, ->] (-2.5,0) -- (2.5,0);
\draw[very thick, ->] (0,-2) -- (0,2);
% Corner label for complex z plane (top right)
\node at (1.5,1.5) {$z$};
\draw (1.7,1.3) -- (1.3,1.3) -- (1.3,1.7);
% Original contour (real line) - dark green
\draw[very thick, green!50!black, ->] (-2.2,0) -- (0,0);
\draw[very thick, green!50!black] (0,0) -- (2.2,0);
% Thimble J_- from -infty (hugs real axis, curves near origin down to -i*infty) - blue
\draw[very thick, blue!70!black, dashed, ->] (-2.2,-0.08) -- (-0.5,-0.08);
\draw[very thick, blue!70!black, dashed] (-0.5,-0.08) .. controls (-0.18,-0.08) and (-0.15,-0.18) .. (-0.15,-0.5);
\draw[very thick, blue!70!black, dashed, ->] (-0.15,-0.5) -- (-0.15,-2.0);
\node[blue!70!black] at (-1.5,-0.32) {$\mathcal{J}_{-1}$};
% Thimble J_+ from +i*infty down to origin then curving right to +infty - red
\draw[very thick, red!70!black, dashed, ->] (0.2,2.0) -- (0.2,0.5);
\draw[very thick, red!70!black, dashed] (0.2,0.5) .. controls (0.2,0.18) and (0.18,0.08) .. (0.5,0.08);
\draw[very thick, red!70!black, dashed, ->] (0.5,0.08) -- (2.2,0.08);
\node[red!70!black] at (1.5,0.32) {$\mathcal{J}_{+1}$};
% Thimble J_0 along imaginary axis - green (one arrow in negative region, one in positive)
\draw[very thick, green!70!black, dashed,->] (0.1,-2.0) -- (0.1,0.0);
\draw[very thick, green!70!black, dashed,->] (0.1,0.0) -- (0.1,2.0);
\node[green!70!black] at (-0.35,1.6) {$\mathcal{J}_0$};
% Saddle points with distinct colors (shades of blue, red, green)
\fill[blue!70!black] (-1,0) circle (2.5pt);
\fill[red!70!black] (1,0) circle (2.5pt); 
\fill[green!70!black] (0,0) circle (2.5pt);
\end{tikzpicture}
\hfill
\includegraphics[width=0.48\textwidth]{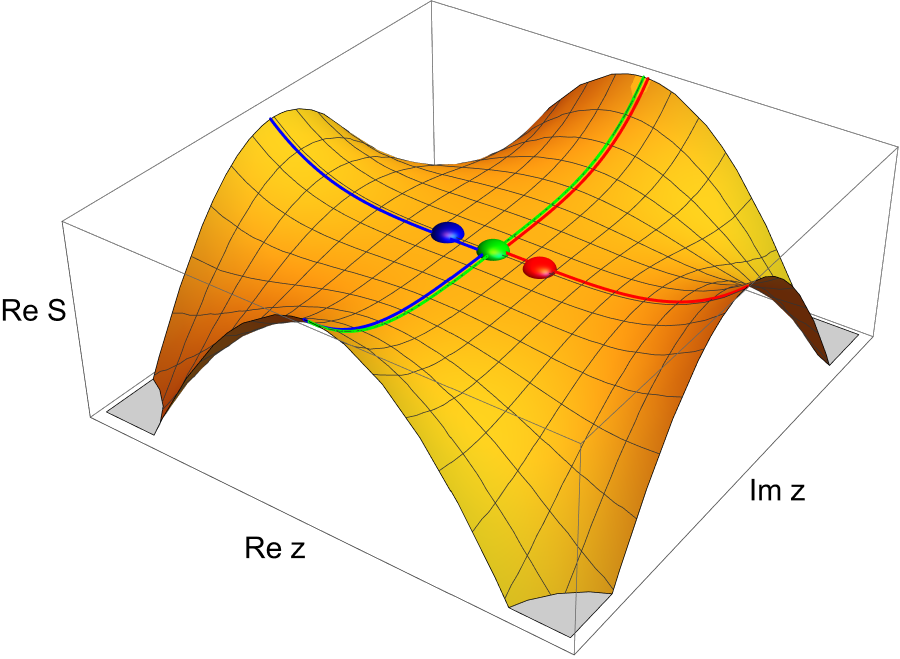}
\caption{Left: The complex $z$-plane with three saddle points: minima at $z = \pm 1$ and a maximum at $z = 0$. The original integration contour is the real line. Each thimble passes through its corresponding saddle point with matching color: $\mathcal{J}_{-1}$ (blue) from $-\infty$ curving down to $-i\infty$, $\mathcal{J}_{+1}$ (red) from $+i\infty$ curving right to $+\infty$, and $\mathcal{J}_0$ (green) along the imaginary axis. Right: The real part of the action $S(z) = \frac{1}{8}(z^2-1)^2$ as a function of complex $z$, with the three saddle points and their thimbles shown.}
\label{fig:0dpotential}
\end{figure}
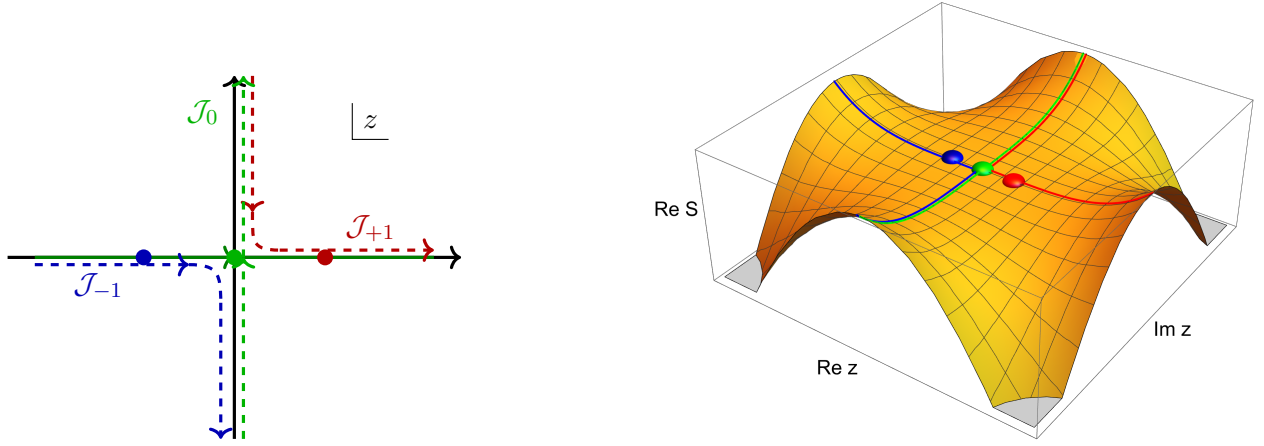

The analog of Rayleigh-Schr\"odinger perturbation theory in the 0D case is the perturbative expansion of $Z(\hbar)$. Because the action has 3 saddle
points, two minima at $z=\pm1$ and a local maximum at $z=0$, there are 3 different series expansions we can compute. 
For the saddles at $z = \pm 1$, setting $z = \pm 1 + x$ and expanding gives
\begin{equation}
Z^{(\pm)}(\hbar) = \int_{- \infty}^{\infty} d x\, e^{- \frac{x^2}{2\hbar}} \sum_n
  \frac{1}{n!} \left( - \frac{\frac{1}{2}x^3 + \frac{1}{8} x^4}{\hbar} \right)^n
  =\sum_{n=0}^{\infty} (2\hbar)^{n+1/2} \frac{\Gamma(\frac{1}{2}+2n)}{n!}\,.
\label{eq:0dseriespm}
\end{equation}
This is an ordinary series, up to the overall factor of $\hbar^{1/2}$, which is a normalization convention. It is asymptotic with coefficients that are factorially divergent as $c_n = 2^{n+1/2}\Gamma(\frac{1}{2}+2n)(n!)^{-1} \sim n!$. 

The saddle-point approximation around $z=0$ is a little different. First of all, at $z=0$ the action is nonzero. This leads to a non-perturbative factor $e^{-S(0)/\hbar} = e^{-1/(8\hbar)}$ multiplying each term in the saddle-point approximation. Since $z=0$ is a local maximum, $\text{Re}(S)$ increases in the imaginary direction, not in the real direction, so we must expand along $z=\pm i y$. The choice of sign is arbitrary. The result is then a trans-series: 
\begin{equation}
Z^{(0)}(\hbar) =
    \pm e^{- \frac{1}{8 \hbar}} \sum_{n = 0}^{\infty} \int_{- \infty}^{\infty} d (i y)\,
e^{- \frac{y^2}{4 \hbar}} \frac{1}{n!} \left( - \frac{y^4}{8 \hbar} \right)^n
= \pm i\, e^{-\frac{1}{8\hbar}} \sum_{n=0}^{\infty} \sqrt{2} (2\hbar)^{n+1/2}(-1)^n \frac{\Gamma(\frac{1}{2}+2n)}{n!}\,.
\label{eq:0dseries0}
\end{equation}
This series is also asymptotic, with the coefficients growing as $n!$ at large $n$, but now alternating in sign. This is a general phenomenon, as we will see, at the heart of resurgence: saddle points with higher action give alternating-sign series, and those with lower action give non-alternating series.

The key to combining these separate expansions into the full trans-series and then reconstructing the non-perturbative function is to use the Borel transform, as introduced in Section~\ref{sec:borel}. Since the original contour can be decomposed into a sum of thimbles, the exact answer is recovered by summing the Borel resummations of all contributing saddles. In our case, the integration contours used in the saddle-point expansions are straight lines -- along the real axis for $z=\pm1$ and the imaginary axis for $z=0$. However, the Lefschetz thimbles (steepest descent trajectories) labeled $\mathcal{J}_{\pm1,0}$ are not linear. These can be seen in Fig.~\ref{fig:0dpotential}.  By observation, the original real contour can be seen to be a concatenation of these Lefschetz thimble paths:
\begin{equation}
\mathbb{R} = \sum \eta_j \mathcal{J}_{j}= \mathcal{J}_{-1}  + \mathcal{J}_0+ \mathcal{J}_{+1}\,.
\end{equation}
The exact integral is then
\begin{equation}
Z(\hbar) =\int_{\mathbb{R}} dz\, e^{-S(z)/\hbar}
= \sum_{j=-1,0,1} \eta_j \int_{\mathcal{J}_j} dz\, e^{-S(z)/\hbar}
\end{equation}
with intersection numbers  $\eta_{-1}=\eta_0=\eta_1=1$. The exact integral can be recovered from the series expansions by
\begin{equation}
Z(\hbar) = \mathcal{S}[Z^{(+)}](\hbar) + \mathcal{S}[Z^{(-)}](\hbar) +  \mathcal{S}[Z^{(0)}](\hbar) = \mathcal{S}[Z^{\text{tot}}](\hbar) \,,
\end{equation}
where $\mathcal{S}$ denotes Borel resummation and
\begin{equation}
    Z^{\text{tot}}(\hbar) =2\sum_{n=0}^{\infty} (2\hbar)^{n+1/2} \frac{\Gamma(\frac{1}{2}+2n)}{n!}
    \pm i\, e^{-\frac{1}{8\hbar}} \sum_{n=0}^{\infty} \sqrt{2}(2\hbar)^{n+1/2}(-1)^n \frac{\Gamma(\frac{1}{2}+2n)}{n!}
\end{equation}
is the full trans-series. The $\pm i$ sign ambiguity arises from a choice of contour deformation to avoid the Stokes point at $z=0$ where the thimbles intersect. 

To verify that this procedure is correct, we next explicitly compute the Borel transforms and resummations for the separate saddle point expansions and verify that their sum reproduces the full non-perturbative function. The Borel transform of the series in Eq.~\eqref{eq:0dseriespm} is
\begin{equation}
\mathcal{B}[Z^{(-)}](t) =\sum_{n=0}^\infty \frac{2^{n+1/2}\Gamma(\frac{1}{2}+2n)}{n!\,\Gamma(n+\frac{3}{2})} t^{n+1/2} =  \sqrt{2 - 2\sqrt{1-8t}}\,.
\label{eq:0dBorelpm}
\end{equation}
This has a branch point at $t=1/8$, corresponding to the action $S(0)=1/8$ at the unstable saddle. The lateral Borel resummation is
\begin{equation}\label{eq:lateral_borel_saddle1}
    \mathcal{S}_\pm[Z^{(-)}](\hbar) = \frac{1}{\hbar}\int_0^{\infty_\pm} dt\, e^{-t/\hbar} \sqrt{2 - 2\sqrt{1-8t}}
\end{equation}
where $\infty_\pm$ indicates deforming the contour above (+) or below ($-$) the branch cut starting at $t=1/8$. This integral can be evaluated by changing variables to $u = \sqrt{1-8t}$, giving 
\begin{equation}
    \mathcal{S}_\pm[Z^{(-)}](\hbar) =\frac{1}{2}\, e^{-\frac{1}{16\hbar}} \!\left[e^{\mp i\pi/4}\, K_{1/4}\!\left(\frac{1}{16\hbar}\right) + \pi\, I_{1/4}\!\left(\frac{1}{16\hbar}\right)\right] \,.
\end{equation}
We can confirm this by direct integration along the Lefschetz thimble $\mathcal{J}_{-1}$. As shown in Fig.~\ref{fig:0dpotential}, the thimble through $z=-1$ runs from $-\infty$ along the real axis to the origin, then down to $-i\infty$ along the imaginary axis. Parametrizing the contour piecewise with $z=x$ for $x\in(-\infty,0]$ and $z=iy$ for $y\in[0,-\infty)$:
\begin{align}
    \int_{\mathcal{J}_{-1}} dz\, e^{-S(z)/\hbar} &= \int_{-\infty}^{0} dx\, e^{-\frac{(x^2-1)^2}{8\hbar}} + \int_{0}^{-i\infty} d(iy)\, e^{-\frac{((iy)^2-1)^2}{8\hbar}} \nonumber\\
    &= \int_{-\infty}^{0} dx\, e^{-\frac{(x^2-1)^2}{8\hbar}} + i\int_{0}^{-\infty} dy\, e^{-\frac{(y^2+1)^2}{8\hbar}}\,.
\end{align}
These integrals can be evaluated using the substitutions $u = x^2-1$ and $v = y^2+1$, yielding Bessel functions. The result is 
\begin{equation}
    \int_{\mathcal{J}_{-1}} dz\, e^{-S(z)/\hbar} = \frac{1}{4} \pi  e^{-\frac{1}{16 \hbar}} \left[I_{1/4}\left(\frac{1}{16 \hbar}\right)+I_{-1/4}\left(\frac{1}{16 \hbar}\right)\right] -\frac{i e^{-\frac{1}{16 \hbar}} K_{1/4}\left(\frac{1}{16 \hbar}\right)}{2 \sqrt{2}} \,,
\end{equation}
which, after using the identity $K_{1/4}(x) = \frac{\pi\sqrt{2}}{2}[I_{-1/4}(x) - I_{1/4}(x)]$, confirms the equivalence between Borel resummation and thimble integration. Note that the choice of $\mathcal{J}_{-1}$ is related to the choice of contour deformation in Eq.~\eqref{eq:lateral_borel_saddle1}. If we deform the contour below the branch cut, we would instead curve $\mathcal{J}_{-1}$ up to $+i \infty$ in the complex plane. The analysis of the saddle at $z=+1$ is similar, but the thimble $\mathcal{J}_{+1}$ approaches the origin from the opposite side. Explicitly,
\begin{equation}
    \int_{\mathcal{J}_{+1}} dz\, e^{-S(z)/\hbar} = \frac{1}{4} \pi  e^{-\frac{1}{16 \hbar}} \left[I_{1/4}\left(\frac{1}{16 \hbar}\right)+I_{-1/4}\left(\frac{1}{16 \hbar}\right)\right] -\frac{i e^{-\frac{1}{16 \hbar}} K_{1/4}\left(\frac{1}{16 \hbar}\right)}{2 \sqrt{2}} \,.
\end{equation}

For Eq.~\eqref{eq:0dseries0} the Borel transform is
\begin{equation}
\mathcal{B}[e^{\frac{1}{8\hbar}}Z^{(0)}](t) = \pm i \sum_{n=0}^\infty \frac{2^{n+1}(-1)^n\Gamma(\frac{1}{2}+2n)}{\Gamma(n+\frac{3}{2})\, n!} t^{n+1/2} = \pm 2i\sqrt{\sqrt{1+8t}-1}\,.
\label{eq:0dBorel0}
\end{equation}
This has no singularity on the positive real axis, so Borel resummation is unambiguous:
\begin{equation}
    \mathcal{S}[Z^{(0)}](\hbar) = \pm i\, e^{-\frac{1}{8\hbar}} \frac{1}{\hbar}\int_0^{\infty} dt\, e^{-t/\hbar} \cdot 2\sqrt{\sqrt{1+8t}-1} = \pm \frac{i}{\sqrt{2}}\, \, e^{-\frac{1}{16\hbar}} K_{1/4}\left(\frac{1}{16\hbar}\right)\,.
\end{equation}
This can also be verified by direct integration along the thimble $\mathcal{J}_0$, which runs along the imaginary axis from $-i\infty$ to $+i\infty$, where the orientation is again defined by the choice of sign in Eq.~\eqref{eq:0dBorel0}:
\begin{equation}
    \int_{\mathcal{J}_0} dz\, e^{-S(z)/\hbar} = i\int_{-\infty}^{\infty} dy\, e^{-\frac{(y^2+1)^2}{8\hbar}} = \frac{i}{\sqrt{2}}\,  e^{-\frac{1}{16\hbar}} K_{1/4}\left(\frac{1}{16\hbar}\right)\,.
\end{equation}

Finally, we confirm that the sum reproduces the exact function. Adding the contributions,
the two minima contribute the $I_{\pm 1/4}\left(\frac{1}{16\hbar}\right)$  functions along with an imaginary $K_{1/4}\left(\frac{1}{16\hbar}\right)$, while the unstable saddle contributes a compensating imaginary $K_{1/4}\left(\frac{1}{16\hbar}\right)$ such that the sum of all thimbles reproduces Eq.~\eqref{eq:0dexact}. The imaginary ambiguity from $Z^{(0)}$ reflects that different choices of thimble orientation give results differing by exponentially small imaginary terms that cancel in physical observables.

This zero-dimensional example illustrates the key features that will reappear in the quantum mechanical double well: (i) multiple saddle points with different actions, (ii) asymptotic series around each saddle whose Borel transforms have singularities at the actions of other saddles, (iii) reconstruction of the exact answer through thimble decomposition, and (iv) cancellation of imaginary ambiguities between different saddle contributions.

\subsection{Alien calculus and resurgence}\label{sec:alien_calc}
We saw in the $D=0$ double well that the full path integral can be reconstructed by combining the asymptotic series around each saddle separately. That the imaginary parts must cancel implies that these series are not completely independent. A central idea of resurgence is that you can actually reconstruct the full function from the expansion around a single saddle. The general framework for doing this is \emph{alien calculus}, developed by \'Ecalle~\cite{ecalle1981fonctions1, ecalle1981fonctions2, ecalle1985fonctions3}. For pedagogical introductions, see the reviews by Sauzin~\cite{Sauzin2007}, Dorigoni~\cite{Dorigoni2014}, and Aniceto, Ba\c{s}ar, and Schiappa~\cite{AnicetoSchiappa2018}.

The key object is the \emph{alien derivative} $\Delta_\omega$, which extracts information about singularities of the Borel transform at $t = \omega$. For a function $f(\hbar)$ with Borel transform $\mathcal{B}[f](t)$, the alien derivative at $\omega$ measures the discontinuity across the branch cut starting at $t = \omega$:
\begin{equation}
\Delta_\omega f \sim \text{Disc}_{t=\omega} \mathcal{B}[f](t)\,.
\end{equation}
More precisely, for a ``simple'' resurgent function the Borel transform has the local form
\begin{equation}
\mathcal{B}[f](\omega + s) = -\frac{1}{2\pi i}\,\varphi(s)\,\log s + \text{regular}(s),
\end{equation}
with $\varphi(s)$ analytic at $s=0$ (the \emph{minor}). The alien derivative is then defined by $\mathcal{B}[\Delta_\omega f](s) = \varphi(s)$: it strips off the logarithmic branch and returns the analytic coefficient. Since $\text{Disc}\,\log s = -2\pi i$, this is equivalent to the recentered discontinuity $\mathcal{B}[\Delta_\omega f](s) = \text{Disc}_{t=\omega}\,\mathcal{B}[f](\omega + s)$.

For the 0D double well, consider again the perturbative series $Z^{(+)}(\hbar)$ around $z = +1$ from Eq.~\eqref{eq:0dseriespm}. Its Borel transform $\mathcal{B}[Z^{(+)}](t) = \sqrt{2 - 2\sqrt{1-8t}}$ has a branch point at $t = 1/8$. Near this singularity:
\begin{equation}
\mathcal{B}[Z^{(+)}](t) \sim  \sqrt{2} - 2i \sqrt{t - \frac{1}{8}}  + \cdots \quad \text{as } t \to \frac{1}{8}\,.
\end{equation}
The discontinuity across the branch cut for $t > 1/8$ is
\begin{equation}
\text{Disc}_{t=\frac{1}{8}} \mathcal{B}[Z^{(+)}](t) = \mathcal{B}[Z^{(+)}](t + i\epsilon) - \mathcal{B}[Z^{(+)}](t - i\epsilon) = \pm 2i\sqrt{\sqrt{8t}-1}\,.
\end{equation}
Remarkably, this is precisely the expression in Eq.~\eqref{eq:0dBorel0}. That is,
\begin{equation}
\text{Disc}_{t=\frac{1}{8}} \mathcal{B}[Z^{(+)}](t) =  \mathcal{B}[Z^{(0)}](t - \frac{1}{8})\,.
\end{equation}
This is the resurgence relation: the discontinuity of the perturbative Borel transform at the instanton action encodes the instanton series itself. In the alien-calculus language above, the shift $t\to t-\omega$ is built into $\Delta_\omega$, so this reads:
\begin{equation}
\Delta_{\frac{1}{8}} Z^{(+)} =  Z^{(0)}\,.
\end{equation}

The alien derivative is directly connected to the Lefschetz thimble picture. When performing lateral Borel resummation $\mathcal{S}_+$ versus $\mathcal{S}_-$ (integrating above or below a singularity), the difference corresponds to picking up the thimble $\mathcal{J}_\omega$ associated with the saddle at action $\omega$:
\begin{equation}
\mathcal{S}_+ f - \mathcal{S}_- f = \eta_\omega \int_{\mathcal{J}_\omega} dz\, e^{-S(z)/\hbar}\,.
\end{equation}
Thus the alien derivative $\Delta_\omega f$ gives the Stokes constant---the coefficient with which the thimble $\mathcal{J}_\omega$ contributes when crossing a Stokes line. Alien calculus provides an algebraic way to compute the intersection numbers $\eta_j$ directly from the singularity structure of the Borel transform, without explicitly constructing the thimbles.

The full trans-series can now be reconstructed iteratively. Starting from $Z^{(+)}$ alone:
\begin{enumerate}
\item The Borel singularity at $t = 1/8$ reveals the instanton action $S(0) = 1/8$.
\item The alien derivative $\Delta_{1/8} Z^{(+)} = Z^{(0)}$ gives the instanton series.
\item Since $\mathcal{B}[Z^{(0)}](t)$ has no singularities on the positive real axis, there are no further alien derivatives to compute.
\end{enumerate}
The full answer is then
\begin{equation}
Z(\hbar) = \mathcal{S}_+[Z^{(+)}](\hbar) + \mathcal{S}_+[Z^{(-)}](\hbar) + \mathcal{S}[Z^{(0)}](\hbar)
\end{equation}
where $\mathcal{S}[Z^{(0)}]$ requires no lateral choice since $\mathcal{B}[Z^{(0)}]$ has no singularities on the positive real axis. For either choice of lateral resummation, $\mathcal{S}_\pm[Z^{(+)}]$ and $\mathcal{S}_\pm[Z^{(-)}]$ each acquire imaginary contributions from the Borel singularity at $t = 1/8$, which cancel exactly against $Z^{(0)}$, ensuring the total partition function is real.

This demonstrates the key insight of resurgence: perturbation theory is not merely incomplete, it contains hidden information about the full non-perturbative structure of the theory. The divergent tail of the perturbative series, which grows as $n!$, encodes the locations and contributions of non-perturbative saddles. From the Borel transform of a single asymptotic expansion, one can in principle reconstruct the entire trans-series and thereby obtain a non-perturbative definition of the theory.

\subsection{Action-Borel correspondence}

For the 0D double well, there is an alternative and more direct way to reconstruct the full function from a single saddle, using the Action-Borel correspondence introduced in~\cite{Bhattacharya:2024hhh}. The basic
idea is that if we have a function defined through an action
\begin{equation}
f_S(\hbar) = \int dz\, e^{-S(z)/\hbar}\,,
\end{equation}
with $S(z)\geq 0$, then we can compute the Borel transform through a change of variables. Writing
\begin{equation}
f_S(\hbar) = \frac{1}{\hbar}\int_0^\infty dt\, e^{-t/\hbar} \left[\hbar \int dz\, \delta(t - S(z))\right]\,,
\end{equation}
we can read off the Borel transform as
\begin{equation}
\mathcal{B}\left[\frac{1}{\hbar}f_S\right](t) = \int dz\, \delta(t - S(z)) = \frac{d}{dt}\int dz\, \Theta(t - S(z))\,.
\end{equation}
This manipulation identifies the Borel variable $t$ with the action $S$. Integrating and using the fact that $\mathcal{B}[\frac{1}{\hbar}f] = \frac{d}{dt}\mathcal{B}[f]$, we find that in one dimension
\begin{equation}
\frac{d}{dt}\mathcal{B}[f_S](t) = \sum_{z_i | S(z_i)=t} \left|\frac{1}{S'(z_i)} \right |\  \rightarrow  \ \mathcal{B}(t) = \sum_{z_i | S(z_i)=t} \pm z_i\,,
\label{eq:actionBorel}
\end{equation}
where the sum is over all roots of $S(z) = t$ and the signs are $\pm z_i = \text{sgn}(z_i'(t))z_i$. This correspondence is therefore a duality: not only is the Borel variable $t$ equal to the action, but the action variable $z$ equals the Borel transform. Some examples can be found in~\cite{Bhattacharya:2024hhh}. Singularities in $\mathcal{B}(t)$ arise precisely at the critical values of $S(z)$---the saddle point actions---providing a direct geometric interpretation of Borel singularities.

The power of this correspondence is that it works in reverse: given the Borel transform, we can reconstruct the action. To see how this works for the double well, suppose we only have the asymptotic series $Z^{(+)}$ (or $Z^{(-)}$)  around one of the minima, as in Eq.~\eqref{eq:0dseriespm}. Its Borel transform is in Eq.~\eqref{eq:0dBorelpm}, $\mathcal{B}[Z^{(+)}](t) = \sqrt{2-2\sqrt{1-8t}}$ and from its discontinuity we can reconstruct the Borel transform of $Z^{(0)}$ as $\mathcal{B}[Z^{(0)}](t) = \theta(t-1/8) 2 \sqrt{1-\sqrt{8t}} $. In the $t > \frac{1}{8}$ region we recover the full Borel transform by adding $\mathcal{B}[Z]=\mathcal{B}[Z^{(0)}]+\mathcal{B}[Z^{(+)}]+\mathcal{B}[Z^{(-)}]= 2\sqrt{1+\sqrt{8t}}$. Anticipating two domains for a stable action we can solve $\mathcal{B}[Z](t = S) = 2z$ and recover $S(z) = \frac{1}{8} - \frac{1}{4}z^2 + \frac{1}{8}z^4$, which is the correct action. In this way, we can reconstruct the full function from the perturbative expansion around a single saddle, providing a concrete realization of the resurgence program.

\subsection{Riemann surface picture}
\label{sec:0dRiemann}
\begin{figure}[t]
  \centering
  \begin{tikzpicture}
    \node[anchor=south west,inner sep=0] (img) at (1,0)
      {\includegraphics[width=0.7\textwidth,trim=40 60 40 40,clip]{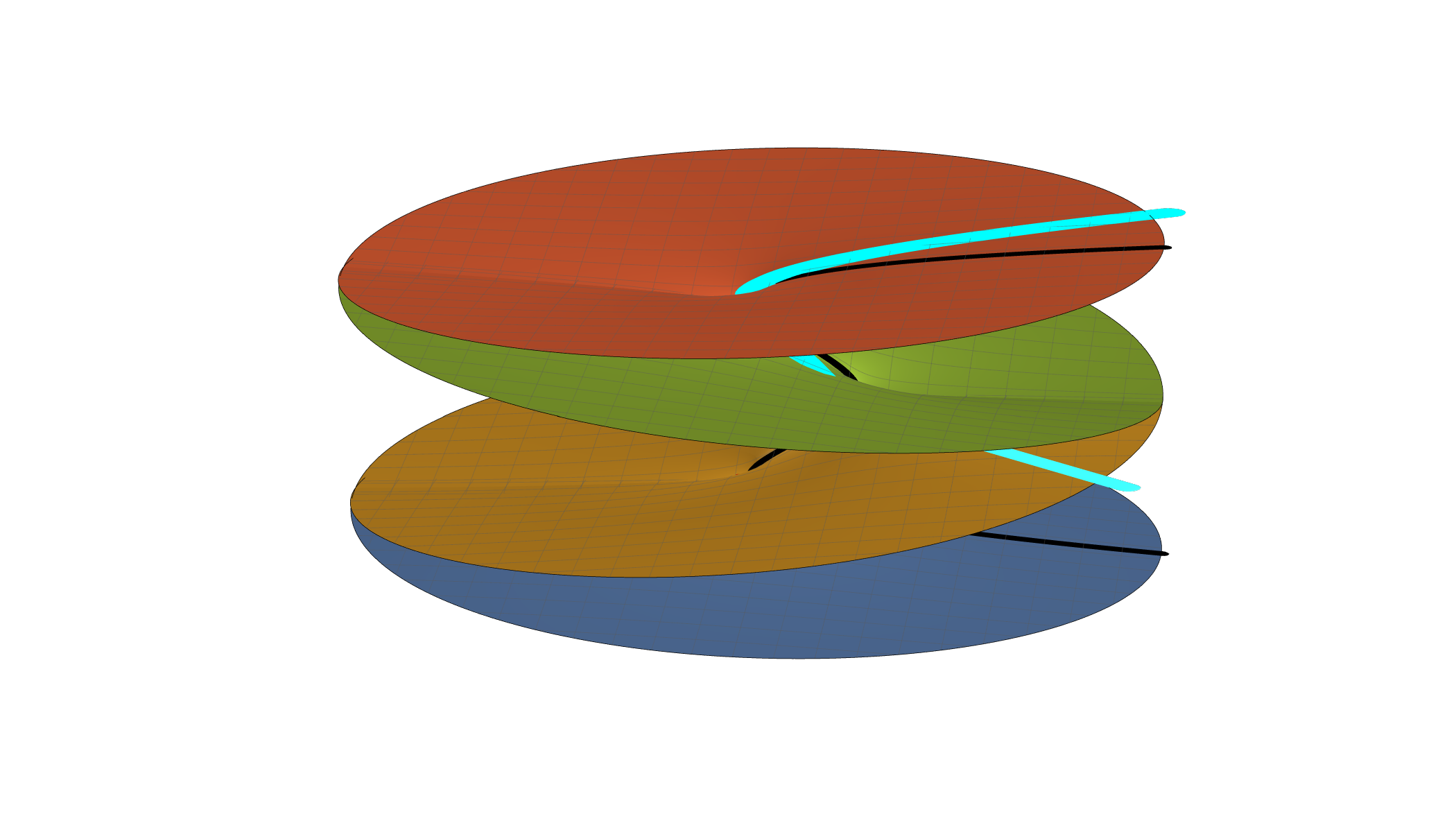}};
    % Axis legend shifted into empty space at lower-left, clear of the plot
    \coordinate (O) at (-0.6,1);
    \draw[->,thick] (O) -- ++(1.8,0) node[right] {$\re\, t$};
    \draw[->,thick] (O) -- ++(1.0,1.2) node[above right] {$\im\, t$};
    \draw[->,thick] (O) -- ++(0,1.9) node[above] {$\re\, B$};
  \end{tikzpicture}
  \caption{Four-sheeted Riemann surface for the 0D double well, defined by $t=\tfrac{1}{8}(B^2-1)^2$. The four sheets are shown in different colors. They meet pairwise at three branch points $(t,B)=(0,+1),(\tfrac{1}{8},0),(0,-1)$, which are the saddle points of the action $S(z)$. The black curves show the real branches $B_j(t)$ for real $t$. The cyan curve is the thimble $\mathcal{J}_{+1}$ through the $z=+1$ saddle at $(t,B)=(0,+1)$, where sheets 3 and 4 meet. The upper arm, on sheet 4, corresponds to $z\in[1,\infty)$: the action is real and grows, so the thimble stays on the real-$B$ surface out to $t=\infty$. The lower arm, on sheet 3, corresponds to $z\in[0,1]$ and reaches the local maximum $(t,B)=(\tfrac{1}{8},0)$, where it runs into the $z=0$ saddle. To continue with $\re S$ still increasing, the thimble turns 90\textdegree{} onto the imaginary-$z$ axis. Along that branch $S(i\lambda)=(1+\lambda^2)^2/8$ is still real, so $t$ stays real, but $B=i\lambda$ is pure imaginary: $\re B$ levels off at zero and the whole motion is hidden in $\im B$, which this plot does not resolve. The cyan curve is drawn tilting off into imaginary $t$ purely as an artistic cue that the thimble has left the real-$B$ surface.}
  \label{fig:0d_riemann_surface}
\end{figure}
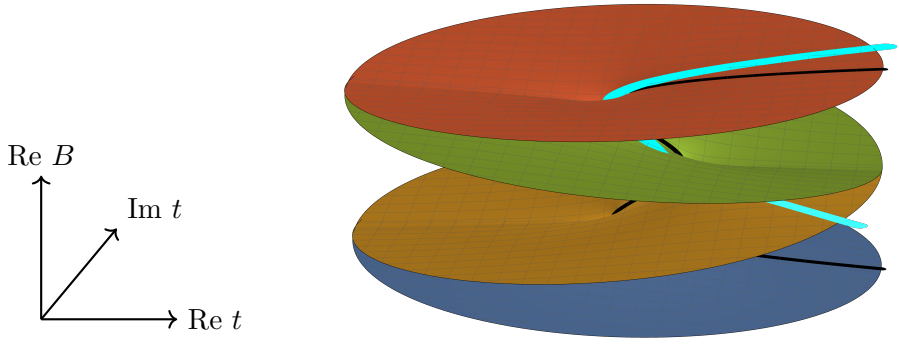

The different Borel transforms arising from the different asymptotic expansions around the different saddle points are not independent. They are different branches of a single Borel transform which naturally lives on a Riemann surface. 
For the double well, the action is
\begin{equation}
S(z) = \frac{1}{8}(z^2-1)^2
  \quad = \quad
\begin{tikzpicture}[baseline=1.5ex, scale=1.1]
  \clip (-2.4,-0.3) rectangle (2.4,1.5);
  % Colored regions
  \fill[branchone, opacity=0.12] (-2.4,-0.3) rectangle (-1,1.5);
  \fill[branchtwo, opacity=0.12] (-1,-0.3) rectangle (0,1.5);
  \fill[branchthree, opacity=0.12] (0,-0.3) rectangle (1,1.5);
  \fill[branchfour, opacity=0.12] (1,-0.3) rectangle (2.4,1.5);
  % Region labels
  \node[branchone!80!black, scale=0.7, font=\bfseries] at (-1.7,1.3) {$z_1$};
  \node[branchtwo!80!black, scale=0.7, font=\bfseries] at (-0.5,1.3) {$z_2$};
  \node[branchthree!80!black, scale=0.7, font=\bfseries] at (0.5,1.3) {$z_3$};
  \node[branchfour!80!black, scale=0.7, font=\bfseries] at (1.7,1.3) {$z_4$};
  % Axes
  \draw[->] (-2.3,0) -- (2.3,0) node[above left] {\scriptsize $z$};
  \draw[->] (0,-0.2) -- (0,1.4);
  % Curve: scale S(z) vertically by factor 8 so max at z=0 is 1.0
  \draw[thick, black!80, variable=\z, domain=-2.0:2.0, samples=120]
    plot ({\z}, {1.0*(\z*\z-1)*(\z*\z-1)});
  % Saddle points
  \fill[red!70!black] (-1,0) circle (1.8pt);
  \fill[red!70!black] (1,0) circle (1.8pt);
  \fill[orange!80!black] (0,1.0) circle (1.8pt);
  % Labels
  \node[scale=0.65] at (-1.08,-0.2) {$-1$};
  \node[scale=0.65] at (1.08,-0.2) {$+1$};
  \node[above right, scale=0.65] at (0.08,1.0) {$\tfrac{1}{8}$};
\end{tikzpicture}
  \label{eq:0d_Splot}
\end{equation}
Mapping $S\to t$ and $z\to B$ according to the action-Borel correspondence, we then see that the Borel transform $B(t)$ is a multi-valued function on a 4-sheeted Riemann surface defined by
\begin{equation}
  t \;=\; \frac{1}{8}(B^2-1)^2\,.
  \label{eq:0d_Bt}
\end{equation}
Solving $(z^2-1)^2 = 8t$ gives the four branches,
\begin{equation}
  B_4 = -B_1 = \sqrt{1{+}\sqrt{8t}} \,,\qquad
  B_3 = -B_2 = \sqrt{1{-}\sqrt{8t}}
  \;\;=\;\;
\begin{tikzpicture}[baseline=0ex, scale=0.85]
  \clip (-0.6,-1.5) rectangle (4.5,1.8);
  % axes
  \draw[->] (-0.2,0) -- (3.0,0) node[right, scale=0.8] {$t{=}S$};
  \draw[->] (0,-1.4) -- (0,1.4) node[above, scale=0.8] {$\re\, B$};
  % dashed line at t=1/8 (x=2)
  \draw[dashed, gray] (2,-1.5) -- (2,1.5);
  % branch point markers
  \fill[red!70!black] (0,0.8) circle (2.5pt);
  \fill[red!70!black] (0,-0.8) circle (2.5pt);
  \fill[red!70!black] (2,0) circle (2.5pt);
  \node[left, scale=0.6] at (-0.05,0.8) {$1$};
  \node[left, scale=0.6] at (-0.05,-0.8) {$-1$};
  \node[below right, scale=0.6] at (2,-0.05) {$\frac{1}{8}$};
  %% B_4 (purple): real for all t>0, domain t in [0, 0.27]
  \draw[thick, branchfour, variable=\t, domain=0.0:0.175, samples=100]
    plot ({16*\t}, {0.8*sqrt(1+sqrt(8*\t))});
  \node[branchfour, scale=0.6] at (2.6,1.0) {$B_4$};
  %% B_1 (steel blue): = -B_4
  \draw[thick, branchone, variable=\t, domain=0.0:0.175, samples=100]
    plot ({16*\t}, {-0.8*sqrt(1+sqrt(8*\t))});
  \node[branchone, scale=0.6] at (2.6,-1.0) {$B_1$};
  %% B_3 (forest green): real for 0 < t < 1/8 only
  \draw[thick, branchthree, variable=\t, domain=0.0:0.124, samples=100]
    plot ({16*\t}, {0.8*sqrt(1-sqrt(8*\t))});
  \node[branchthree, scale=0.6, above right] at (1.5,0.35) {$B_3$};
  %% B_2 (burnt orange): = -B_3
  \draw[thick, branchtwo, variable=\t, domain=0.0:0.124, samples=100]
    plot ({16*\t}, {-0.8*sqrt(1-sqrt(8*\t))});
  \node[branchtwo, scale=0.6, below right] at (1.5,-0.35) {$B_2$};
\end{tikzpicture}
  \label{eq:0d_Bj}
\end{equation}
numbered from bottom to top at $t = 0^+$.  The plot shows just the real parts of the four branches: the Borel transforms become imaginary at the branch points.  These four sheets describe the Riemann surface $\Sigma$ over the $t$-plane, with branch points at the critical values of $S$. At $t = 0$, sheets 1 and 2 and sheets 3 and 4 coalesce, while at $t = \frac{1}{8}$, sheets 2 and 3 coalesce. The four branches are related by analytic continuation around the branch points: $B_1 \leftrightarrow B_2$ and $B_3 \leftrightarrow B_4$ around $t = 0$, and $B_2 \leftrightarrow B_3$ around $t = 1/8$. Figure~\ref{fig:0d_riemann_surface} shows the full Riemann surface in three dimensions, with the two steepest-ascent arms of the $\mathcal{J}_{+1}$ thimble traced in cyan.

At a saddle $z_\star$  the steepest descent contour follows the two directions along which $\operatorname{Re} S$ increases away from $t_\star = S(z_\star)$. These two arms correspond to the two sheets $j,j'$ that coalesce at $t_\star = S(z_\star)$. Under the change of variables $t = S(z)$, both arms map to the same ray $t\in[t_\star,\infty)$, but with opposite orientations: the contour first traverses one arm from $t=\infty$ inward to $t_\star$, passes through the saddle, and then traverses the other arm from $t_\star$ outward to $t=\infty$. Re-parametrising both arms by increasing $t$, the inward-traversed arm acquires a sign reversal, giving
\begin{equation}
  \int_{\mathcal{J}} dz\, e^{-S(z)/\hbar} \;=\; \int_{t_\star}^{\infty}\!dt\, e^{-t/\hbar}\,\bigl[B'_{j'}(t) - B'_j(t)\bigr]\,.
\end{equation}
Integrating by parts, the boundary terms vanish (at $t = t_\star$  $B_{j'} = B_j$, and at $t = \infty$ the exponential goes to zero).  This gives
\begin{align}
  \int_{\mathcal{J}} dz\, e^{-S(z)/\hbar} &= \frac{1}{\hbar}\int_{t_\star}^{\infty}\!dt\, e^{-t/\hbar}\,\bigl[B_{j'}(t) - B_j(t)\bigr]\,\\
  &=  \frac{e^{-t_\star/\hbar}}{\hbar}\int_0^\infty\! d t\, e^{-t/\hbar}\, \bigl[B_{j'}(t + t_\star) - B_j(t + t_\star)\bigr]\,,
  \label{eq:0d_thimble_Borel_t}
\end{align}
 where we shifted $t\to t+t_\star$ in the second step. From here, we can read off that the Borel transform is $\theta(t-t^\star)\left[B_{j'}(t) - B_j(t)\right]$. This is the same Borel transform computed from the asymptotic series around the $z_\star$ saddle.

To check, consider the minima at $z = 1$  where  $t_\star = 0$. At this point sheets $3$ and $4$ coalesce and so 
\begin{equation}
  Z^{(+)}(\hbar)=
   \int_{\mathcal{J}_{+1}} dz\, e^{-S(z)/\hbar}
   = \frac{1}{\hbar}\int_0^{\infty}\!dt\, e^{-t/\hbar}\,\bigl[B_4(t) - B_3(t)\bigr]\,.
  \label{eq:0d_Zplus_Bj}
\end{equation}
From the explicit formulas in Eq.~\eqref{eq:0d_Bj}
\begin{equation}
  B_4(t) - B_3(t) \;=\; \sqrt{1{+}\sqrt{8t}} - \sqrt{1{-}\sqrt{8t}} \;=\; \sqrt{2-2\sqrt{1-8t}}\,,
\end{equation}
which is precisely $\mathcal{B}[Z^{(+)}](t)$ from Eq.~\eqref{eq:0dBorelpm}: the Borel transform of the series is the difference of the two coalescing branches.

For the unstable saddle at $z = 0$ we have $t_\star = 1/8$  and sheets 2 and  3 coalesce there. In this case,
\begin{equation}
   \int_{\mathcal{J}_0} dz\, e^{-S(z)/\hbar} 
   =\frac{e^{-\frac{1}{8\hbar}}}{\hbar}\int_0^{\infty}\!d\tau\, e^{-\tau/\hbar}\,\bigl[B_3(\tau{+}\frac{1}{8}) - B_2(\tau{+}\frac{1}{8})\bigr]
  \label{eq:0d_Z0_Bj}
\end{equation}
  and
  \begin{equation}
    B_3(\tau{+}\frac{1}{8}) - B_2(\tau{+}\frac{1}{8})=  2i\sqrt{\sqrt{1+8\tau}-1} \,,
  \end{equation}
 which matches Eq.~\eqref{eq:0dBorel0}.

%% file: sections/exact_wkb.tex
% \!TEX root = ../DoubleDoubleMain.tex
\section{Exact WKB}
\label{sec:exactWKB}
Although the simple double well $0D$ integral in Section~\ref{sec:zerodim} demonstrates the basic principles of resurgence, to compute the trans-series in quantum mechanics requires significantly more sophisticated methodology. In this section, the Exact WKB method is discussed. Section~\ref{sec:pathintegral} discusses the complementary path integral approach.

In the WKB method, the energy levels of a particle in a double-well potential are quantized by demanding the wavefunction be normalizable, dying off exponentially both as $x \rightarrow \infty$ and $x\to -\infty$.  If we want to compute an approximate solution, we could start from large $x$ where the solution to the Schr{\"o}dinger equation should be of the form $\psi(x)\sim e^{-x}$ which is exponentially small, with zero admixture of the complementary $e^{x}$ solution. If we then analytically continue this form to large negative $x$ we still have $e^{-x}$ which is now exponentially growing as $x\to -\infty$. On the other hand,  energy eigenstates for a polynomial potential must be smooth (in fact $C^{\infty}$) functions. The way to reconcile these two behaviors involves understanding asymptotic expansions: it is possible for a function to be analytic around $z = 0$ with an infinite radius of convergence, but to have asymptotic expansions that change abruptly at $| z | = \infty$. A simple example is $2\cosh (z) = e^{- z} + e^z$. Its asymptotic behavior for $\operatorname{Re} (z) \rightarrow \infty$ is $e^z$ while its behavior as $\text{Re} (z) \rightarrow - \infty$ is $e^{- z}$. The transition is along a line in the complex $z$ plane where $\text{Re}(z) = 0$. More generally the region of the complex plane where the asymptotic expansion abruptly changes is called a {\bf{Stokes line}}.

In quantum mechanics, we can take any potential $V (z)$ and expand locally around a turning point $z_i$ where $V (z_i) = E$. Then $V (z_i + z) - E = V' (z_i) z$ is linear. Thus, near a turning point the Schr{\"o}dinger equation reduces to Airy's equation
\begin{equation}
  f'' (z) - z f (z) = 0 \,.
  \label{eq:airy}
\end{equation}
The two independent solutions to this equation are $\Ai (z)$ and $\Bi(z)$ which are analytic functions, $\Ai (z)$ being the solution which decays exponentially at large $z$, while $\Bi (z)$ grows exponentially. These Airy functions, like $\cosh z$, are analytic around $z = 0$ with an infinite radius of convergence, but have asymptotic expansions at $|z| = \infty$ whose form changes abruptly for certain values of $\arg z$. So to overcome the challenge with WKB, matching across turning points analytically while imposing asymptotic boundary conditions, we begin with a thorough exploration of Airy functions and the associated identification of the relevant Stokes lines. 

\subsection{Airy functions}
\label{sec:airy_functions}
The most familiar place where Airy functions appear in nature is the rainbow. When sunlight scatters off water droplets in the atmosphere, most of the light is scattered in random directions. However, at certain special angles, many rays pile up, creating a bright caustic---the rainbow. For a single water droplet, classical ray optics predicts that light entering at impact parameter $b$ scatters at deflection angle $\Theta(b)$. The deflection function has an extremum at the rainbow angle $\Theta_R$ where $\frac{d\Theta}{db} = 0$. At this angle, an infinite number of nearby rays pile up, leading to a divergence in the classical cross section.

\begin{figure}[t]
\centering
\includegraphics[width=0.95\textwidth]{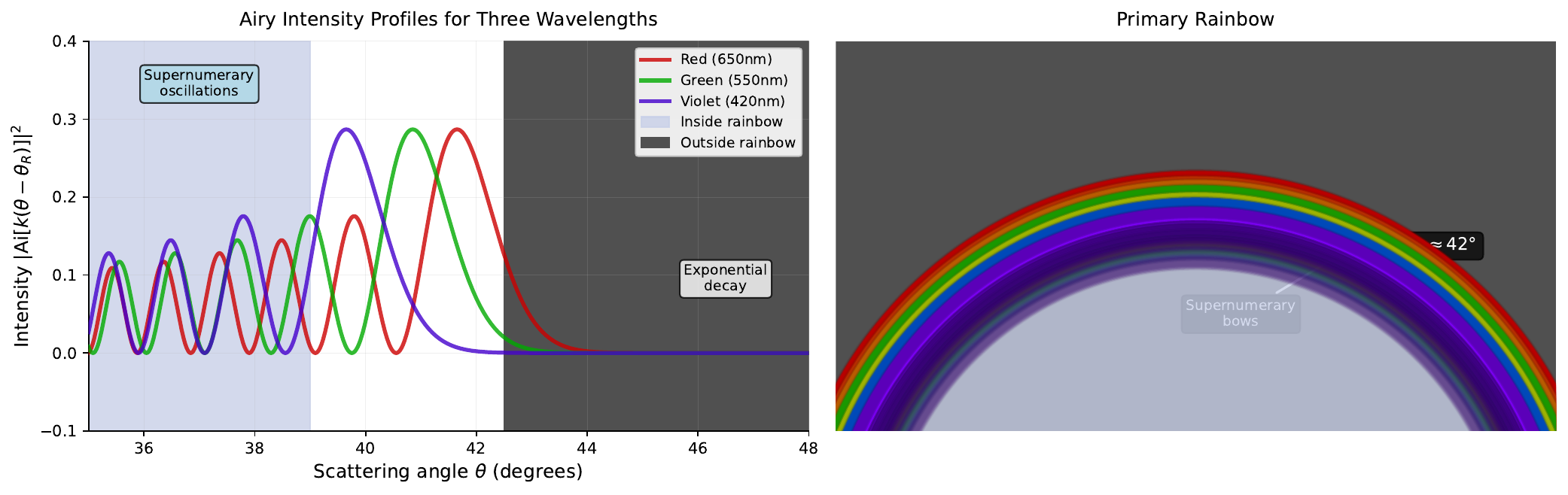}
\caption{Left: The Airy function $\Ai(z)$ showing the transition from oscillatory behavior for $z < 0$ to exponential decay for $z > 0$. Right: Double rainbow with intensity computed from $I(\theta) \propto |\Ai[k^{2/3}(\theta - \theta_R)]|^2$ for each wavelength. The primary rainbow appears at $\sim 42^\circ$ and secondary at $\sim 51^\circ$, with supernumerary bows visible as brightness variations inside each bow.}
\label{fig:rainbow}
\end{figure}

This problem was first investigated systematically by George Biddell Airy in 1838 \cite{Airy} (see also \cite{Nussenzveig} for a modern account), who sought to explain the supernumerary bows---faint colored fringes appearing just inside the primary rainbow. Airy recognized that the classical geometric optics prediction breaks down near the caustic and that one must account for the wave nature of light. Near the rainbow angle, the wavefront emerging from the droplet acquires a cubic shape: in coordinates aligned with the caustic ray, the wavefront is approximately $y \sim c\, x^3$ for some constant $c$. Applying Huygens' principle, the amplitude at observation angle $\theta$ relative to the caustic ray is obtained by integrating the secondary-wave contributions along the wavefront. The path difference to the observer from a point at position $x$ on the wavefront is $\delta(x) \approx x \sin\theta + c\, x^3 \cos\theta \approx x\, \theta + c\, x^3$ for small angles, so the amplitude is
\begin{equation}
  A(\theta) \propto \int_{-\infty}^\infty dx\, e^{ik(\theta\, x + c\, x^3)} \,.
\end{equation}
Rescaling $u = (3kc)^{1/3} x$ and defining $z = k^{2/3}\theta/(3c)^{1/3}$, this becomes
\begin{equation}
  A(\theta) \propto \frac{1}{2\pi}\int_{-\infty}^\infty du\, e^{i(u^3/3 + z u)} = \Ai(z) \,.
  \label{airyintegral}
\end{equation}
The parameter $z$ is proportional to the deviation from the rainbow angle, so the intensity is
\begin{equation}
  I(\theta) \propto |\Ai(z)|^2 \propto \left|\Ai\left[k^{2/3}(\theta - \Theta_R)\right]\right|^2 \,.
  \label{eq:rainbowintensity}
\end{equation}
This formula predicts not only the bright primary rainbow but also the supernumerary bows: the oscillations of $\Ai(z)$ for $z < 0$ correspond to alternating bright and dark fringes just inside the main bow, as shown in Fig.~\ref{fig:rainbow}. These oscillations are precisely analogous to the quantum mechanical oscillations that occur in the classically allowed region of a potential well. The characteristic behavior of the Airy function -- oscillations for $z<0$ and exponential decay for $z > 0$ -- directly maps to the bright rainbow and the dark region beyond it.

The series expansion of the Airy function is
\begin{equation}
  \Ai (z) = \sum_{n = 0}^{\infty} \left[ 3^{- 2 / 3} \frac{z^{3 n}}{9^n
  n! \Gamma \left( n + \frac{2}{3} \right)} - 3^{- 4 / 3} \frac{z^{3 n +
  1}}{9^n n! \Gamma \left( n + \frac{4}{3} \right)} \right]
\end{equation}
which is convergent for all $z$. We can also write
\begin{equation}
  \Ai (z) = \frac{1}{\pi} \sqrt{\frac{z}{3}} K_{1 / 3} \left(
  \frac{2}{3} z^{3 / 2} \right)
\end{equation}
where $K$ is a modified Bessel function of the second kind. The asymptotic expansions
of $\Ai (z)$ for large positive and negative $z$ are
\begin{equation}
  \Ai (z) \xrightarrow[z \rightarrow \infty]{} \frac{1}{2\sqrt{\pi}}
  z^{- \frac{1}{4}} e^{- \frac{2}{3} z^{3 / 2}}, \quad \Ai (z)
  \xrightarrow[z \rightarrow - \infty]{} \frac{1}{\sqrt{\pi}} |z|^{- \frac{1}{4}} \sin \left( \frac{2}{3} |z|^{3 / 2}
  + \frac{\pi}{4} \right)  \,. \label{eq:Aiasymp}
\end{equation}
As expected from our quantum mechanics intuition, the Airy function transitions from exponential behavior in the forbidden region of the linear potential barrier, where $V (z) > E$ on the right side of the barrier, to oscillatory behavior, when $E > V (z)$ for $z < 0$.

The other independent solution to Airy's equation, Eq.~\eqref{eq:airy}, is $\Bi(z)$, which can be written as
\begin{equation}
  \Bi(z) = \sqrt{\frac{z}{3}} \left[ I_{1/3}\left(\frac{2}{3}z^{3/2}\right) + I_{-1/3}\left(\frac{2}{3}z^{3/2}\right) \right]
\end{equation}
where $I$ is a modified Bessel function of the first kind. Like $\Ai(z)$, the function $\Bi(z)$ also transitions between exponential and oscillatory behavior, but with opposite character:
\begin{equation}
  \Bi(z) \xrightarrow[z \rightarrow \infty]{} \frac{1}{\sqrt{\pi}} z^{-\frac{1}{4}} e^{\frac{2}{3}z^{3/2}}, \quad
  \Bi(z) \xrightarrow[z \rightarrow -\infty]{} \frac{1}{\sqrt{\pi}} |z|^{-\frac{1}{4}} \cos\left(\frac{2}{3}|z|^{3/2} + \frac{\pi}{4}\right)  \,.
\end{equation}
For $z > 0$, $\Bi(z)$ grows exponentially while $\Ai(z)$ decays; for $z < 0$, both oscillate but with a $\pi/2$ phase difference.

To understand the change in the form of the asymptotic expansion of $\Ai (z)$ it is helpful to explore its integral representation, cf.\ Eq.~\eqref{airyintegral}, extended to complex $z$ values as
\begin{equation}
  \Ai (z) = \frac{1}{2 \pi} \int_{\gamma_A} d u\, e^{- S (u)}
\end{equation}
where
\begin{equation}
  S (u) = - i \left( \frac{1}{3} u^3 + z u \right) \,
\end{equation}
and $\gamma_A$ is the Airy integration contour\footnote{We are free to deform this contour as we desire, provided that the end points remain in the same convergent regions. In particular, for small $\arg z$ values we only need to slightly curve the real line.} that goes between the asymptotic regions $u \,  e^{\mp i \pi/3}$ as $u\rightarrow \infty$. This way of writing $\Ai (z)$ is very powerful, since the analytic continuation is simple: just do the integral with different values of $z$. Note that because of the $u^3$ behavior of the action at large $u$ the integral is convergent within 3 wedges $A, B$ and $C$ of angular size $\frac{\pi}{3}$ in the complex plane centered on $\theta_A = \frac{5 \pi}{6}, \theta_B = \frac{\pi}{6}$ and $\theta_C = - \frac{\pi}{2}$:
\begin{equation}
\begin{array}{l}
\begin{tikzpicture}[scale=0.5]
    \def\R{3.8}
    \draw[->] (-3.5,0) -- (4.2,0) node[right] {$\text{Re}(u)$};
    \draw[->] (0,-3.5) -- (0,3.5) node[above] {$\text{Im}(u)$};
    \foreach \angle in {30, 150, 270} {
        \draw[thick, red, -, domain=\angle-30:\angle+30]
            plot ({\R*cos(\x)}, {\R*sin(\x)});
    }
    \foreach \angle in {30, 150, 270} {
        \draw[thick, blue, domain=0:2.5, variable=\r, samples=50]
            plot ({\r*cos(\angle)}, {\r*sin(\angle)});
    }
    % Theta labels on blue lines
    \node at (2.2,1.6) {$\theta_B$};
    \node at (-2.2,1.6) {$\theta_A$};
    \node at (0.6,-2.8) {$\theta_C$};
    % A, B, C labels on red arcs (outside)
    \node[red] at (4.5,1.5) {$B$};
    \node[red] at (-4.5,1.5) {$A$};
    \node[red] at (-3,-3.0) {$C$};
    \fill (0,0) circle (2pt);
    % \node at (0.3,0.) {$0$};
\end{tikzpicture}
  \end{array}
\end{equation}

If we consider more generally integrals defined by a contour $\gamma$
\begin{equation}
  I_{\gamma} (z) = \frac{1}{2 \pi} \int_{\gamma} d u\, e^{- S (u)}
\end{equation}
then the contours $\gamma$ fall into one of 6 relative homology classes, determined only by which two distinct regions $A, B$ or $C$ the integral starts and ends in. Any contour going between the same boundary regions gives the same result. At real $z$ the Airy function $\Ai(z)$ has an integration contour along the real line which is just barely integrable, as its endpoints are on the boundary of the allowed asymptotic region. For complex $z$ values we need to slightly curve the real line so that it starts and ends in the convergent regions $A$ and $B$ respectively (see Fig.~\ref{fig:contours}). So in general we have
\begin{equation}
  \Ai(z) = I_{A \to B} \,.
\end{equation}
The other two independent contour integrals give linear combinations of $\Ai$ and $\Bi$:
\begin{equation}
  I_{A \to C} = \frac{1}{2}\Ai(z) - \frac{i}{2}\Bi(z)\,, \qquad
  I_{B \to C} = -\frac{1}{2}\Ai(z) - \frac{i}{2}\Bi(z) \,.
\end{equation}
Thus,
\begin{equation}
  \Bi(z) = i\left(I_{A \to C} + I_{B \to C}\right) \,.
\end{equation}
Note that $\Bi$ is a sum of two contours that both end at $C$, so they cannot be joined into a single contour.

\subsection{Stokes lines}
\label{sec:stokes_auto}

The integral representation allows us to determine the asymptotic behavior. The Airy integrand is highly oscillatory except near saddle points, so its dominant behavior will be determined by the points of stationary phase. For the Airy action, these are at $u = \pm i \sqrt{z}$ with action $S_{\pm} = \pm \frac{2}{3} z^{3 / 2}$ and behavior $e^{\mp \frac{2}{3} z^{3 / 2}}$. As $z$ is moved around in $\mathbb{C}$, these saddle points move around as well, switching places when the branch cut, conventionally placed on the negative real line, is crossed. If we avoid the branch cut with $- \pi < \arg z < \pi$ there are then two well-defined integrals, defined not through asymptotically convergent regions $A, B, C$ but rather as thimbles, steepest-descent trajectories passing through the saddles. For $z > 0$ these integrals give in the saddle point approximation
\begin{align}
  I_+ (z) &= \frac{1}{2\pi}\int_{- \infty}^{\infty} d u\, e^{- \frac{2}{3} z^{3 / 2} - u^2 \sqrt{z}+\frac{i}{3} u^3} = \frac{1}{2\sqrt{\pi}} z^{- 1 / 4} e^{-
  \frac{2}{3} z^{3 / 2}}
  \sum_{n=0}^{\infty} \left(-\frac{1}{z^{3/2}}\right)^n c_n
  \label{Iplusdef}
  \\
  I_- (z) &= \frac{1}{2\pi}\int_{- i \infty}^{i \infty} d u\,
  e^{\frac{2}{3} z^{3 / 2} + u^2 \sqrt{z}+\frac{i}{3} u^3} = \frac{i}{2\sqrt{\pi}} z^{- 1 / 4}
  e^{\frac{2}{3} z^{3 / 2}}
  \sum_{n=0}^{\infty} \left(\frac{1}{z^{3/2}}\right)^n c_n
  \label{Iminusdef}
\end{align}
where
\begin{equation}
  c_n \;=\; \frac{(6n)!}{576^n\,(3n)!\,(2n)!} \;=\; 1,\; \frac{5}{48},\; \frac{385}{4608},\; \ldots
  \;\;\underset{n\to\infty}{\sim}\;\;
  \left(\frac{3}{4}\right)^n n!\,.
\end{equation}
Note that $I_+$ is real for real $z$, while $I_-$ is purely imaginary for real $z > 0$. The subscript $\pm$ in $I_\pm$ refers to the saddle point location $u = \pm i\sqrt{z}$, which is opposite to the sign of the exponent $e^{\mp \frac{2}{3}z^{3/2}}$. 

\begin{figure}[t]
  \centering
  \noindent\begin{tabular}{@{}*{6}{p{0.1583\textwidth}@{}}}
    \centering\StokesGraphAiryWithArrow{-140}{$\theta = -\frac{2\pi}{3} - \varepsilon$} &
    \centering\StokesGraphAiryWithArrow{-100}{$\theta = -\frac{2\pi}{3} + \varepsilon$} &
    \centering\StokesGraphAiryWithArrow{-20}{$\theta = -\varepsilon$} &
    \centering\StokesGraphAiryWithArrow{20}{$\theta = \varepsilon$} &
    \centering\StokesGraphAiryWithArrow{100}{$\theta = \frac{2\pi}{3} - \varepsilon$} &
    \centering\StokesGraphAiryWithArrow{140}{$\theta = \frac{2\pi}{3} + \varepsilon$}
  \end{tabular}

  \vspace{0.2cm}

  \includegraphics[width=\textwidth]{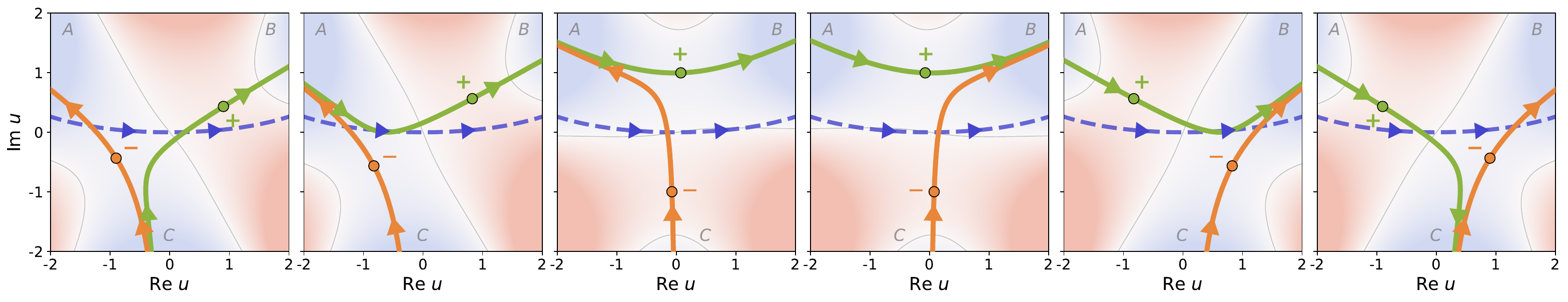}

  \caption{Top: Stokes diagrams showing the phase $\theta = \arg z$ in each region. Bottom: The thimble structure computed numerically by steepest descent flow, with the Airy contour (dashed blue), $\mathcal{J}_+$ (green), and $\mathcal{J}_-$ (orange). The asymptotically convergent regions are shaded in blue and denoted as $A,B,C$.}
  \label{fig:contours}
\end{figure}

To see which asymptotic behavior applies for which integral $I_{\gamma}$ we need to decompose $\gamma$ into a concatenation of thimbles. For any fixed $z$, each thimble belongs to one of the 6 relative homology classes. However, unlike the Airy contour which is always defined as $A \rightarrow B$, which class the thimble is in depends on the phase of $z$. Thimbles for different phases are shown in the bottom panels of Fig. \ref{fig:contours}. We can then see, for example, that for real $z > 0$, the Airy contour $A \rightarrow B$ is the same as the thimble passing through $+ i \sqrt{z}$, so its asymptotic expansion is $e^{- \frac{2}{3} z^{3 / 2}}$. However, as the phase of $z$ passes $\frac{2 \pi}{3}$ we find that neither thimble goes between $A \rightarrow B$ anymore. Instead, the Airy contour follows an $A \rightarrow C$ thimble and then $C \rightarrow B$. Then the asymptotic behavior of Ai in this region will be given by the sum $I_+ + I_- \sim  e^{- \frac{2}{3} z^{3/2}} + i e^{\frac{2}{3} z^{3 / 2}}$ which is oscillatory.

In an asymptotic expansion, only the leading asymptotic behavior is kept. Which behavior dominates is determined by the sign of $\operatorname{Re} (z^{3 / 2})$. On a \textbf{Stokes line}, the real part of the two asymptotic forms is maximally different. The domains between Stokes lines are called \textbf{Stokes regions}. For the Airy integral, we sketch the Stokes lines as
\begin{equation}
\StokesGraphAiry
\label{eq:stokes-lines}
\end{equation}
The solid lines are the Stokes lines at $\arg z = 0, \pm \frac{2\pi}{3}$, and the wavy line is the branch cut at $\arg z = \pi$. These Stokes lines are in the complex $z$ plane and the $+$ or $-$ indicate which asymptotic expansion applies exclusively when $|z| \to \infty$ along that line. At $\arg z = 0$ (positive real $z$), $z^{3/2} > 0$ so $e^{+\frac{2}{3}z^{3/2}}$ grows, meaning $I_-$ is dominant. At $\arg z = \pm \frac{2\pi}{3}$, $z^{3/2}$ is real and negative, so $e^{-\frac{2}{3}z^{3/2}}$ grows, meaning $I_+$ is dominant. Note that $\pm$ here are a convention with no physical meaning. 

Now let us return to the quantum mechanics problem where we have a potential barrier at $z=0$, with $z>0$ the forbidden region and $z<0$ the allowed region. We want to match from a function which is exponentially decaying for $z > 0$ across the barrier. So in the $z > 0$ region the function should be $\psi \propto I_+ \sim e^{- \frac{2}{3}z^{3 / 2}}$, which is the green contour in either of the  middle two panels of Fig.~\ref{fig:contours}. There cannot be any admixture of $I_-$ for it to be normalizable. As we analytically continue past $\arg z = \frac{2 \pi}{3}$ the $+$ thimble flips asymptotic regions. This means that a well-defined integral such as $\Ai (z) = I_{A\to B}$ that had a component of $I_+$ will now have the sum of $I_+$ and $I_-$, which, as we observed, becomes oscillatory. This is the expected behavior in the classically allowed region. If we had started with $I_{B\to C}$ instead, which matches onto the $-$ thimble in the $0<\theta<\frac{2\pi}{3}$ region, then since the $-$ thimble does not flip when the $\theta=\frac{2\pi}{3}$ Stokes line is crossed, its asymptotic expansion also does not change.

To determine how the asymptotic behavior  changes across Stokes lines more systematically, suppose we have some physical\footnote{Here physical means defined by an integral like a path integral with a fixed contour or a differential equation with boundary conditions, in contrast to being defined through a perturbative expansion, which would be associated with a thimble and a saddle point.} function $\psi (z)$ like $\Ai (z)$ or a wavefunction, or a path integral. In any asymptotic region, we can write $\psi (z) = c_+ \psi_+ (z) + c_- \psi_- (z)$ where $\psi_{\pm} (z)$ are integrals along thimbles. This decomposition is local, applying near any turning point (the linear potential giving rise to the Airy integral only has one turning point; the multiple turning point case will be considered shortly). Then as $z$ is analytically continued across a Stokes line, the thimbles may get associated with different asymptotically convergent regions. Then the coefficients $c_{\pm}$ will change. For example, with the Airy function we observed that if $\vec{c} = (1, 0)$ to begin with, on the other side of the $\frac{2 \pi}{3}$ Stokes line it is $\vec{c} = (1, 1)$. With $I_{B\to C}$, which corresponds to $\vec{c} = (0,-1)$, when the same Stokes line is crossed, it doesn't change. Combining these observations, we conclude that when crossing the $+$ Stokes line counterclockwise the asymptotic behavior changes according to $\vec{c} \to \fSp\cdot \vec{c}$ where
\begin{equation}
  \fSp = \left(\begin{array}{cc}
    1 & 0\\
    1 & 1
  \end{array}\right) \,.
\end{equation}
One can check that the same behavior occurs when crossing the other $+$ Stokes line at $\theta = -\frac{2\pi}{3}$. When traversing a $-$ Stokes line, the corresponding matrix is\footnote{Some references, e.g.~\cite{Sueishi2020, BucciottiReisSerone}, exhibit connection matrices with an extra factor of $i$ in the off-diagonal terms. This stems from a convention that the asymptotic expansions $I_\pm$ are real. Our convention instead treats the two saddle-point contributions on the same footing --- each $I_\pm$ is the integral over its thimble with no relative phase inserted by hand.} 
\begin{equation}
  \fSm = \left(\begin{array}{cc}
    1 & -1\\
    0 & 1
  \end{array}\right) \,.
\end{equation}
These matrices are known as \textbf{connection matrices} (or \textbf{Stokes matrices}). When crossing a Stokes line clockwise around the turning point, the matrices $\fS_{\pm}^{- 1}$ are invoked.

For the Airy function, with our convention, the branch cut is at  $\theta = \pi$ on the principal sheet. Traversing it counterclockwise corresponds to moving from the 6th panel in Fig.~\ref{fig:contours} to the first, whereby the $-$ thimble becomes the $+$ thimble and the $+$ thimble becomes the $-$ thimble with opposite orientation. We can represent this transition with the connection matrix
\begin{equation}
  \fSbr = \left(\begin{array}{cc}
    0 & 1 \\
    -1 & 0
  \end{array}\right) \,. \label{Sbr}
\end{equation}
This is a $90^\circ$ rotation matrix. As a check, referring back to the diagram in Eq.~\eqref{eq:stokes-lines} and the Fig.~\ref{fig:contours}, if we go in a full circle, we pick up a trivial monodromy $\fSm \fSp \fSbr\fSp = 1$, confirming that analytic continuation is self-consistent (or providing an alternative way to compute $\fSbr$).

For an example use of these automorphisms, let's use them to continue the Airy function from positive real $z$ to negative real $z$. For positive $z$, we want exponential damping only so that the asymptotic expansion of $\Ai$ must be $I_+$, so $\vec{c} = (1, 0)$ as before. Note that crossing the $\theta = 0$ Stokes line does nothing at $\fSm \cdot (1, 0) = (1, 0)$, so it doesn't matter if we had started above or below the real $z$ axis. We then analytically continue counterclockwise. When we cross the Stokes line at $\theta = \frac{2\pi}{3}$ we get $\fSp \cdot \vec{c} = (1, 1)$. This form holds until the negative real $z$ axis. Thus, when we get to $z < 0$, we have $I_+ + I_-$. Combining the two exponentials $e^{\pm \frac{2}{3}i|z|^{3/2}}$ with the phase shift gives the oscillatory asymptotic expansion of the Airy function:
\begin{equation}
  I_+ + I_- \underset{z \to -\infty}{\sim} \frac{e^{-i\pi/4}}{2\sqrt{\pi}} |z|^{-1/4} \left( e^{\frac{2}{3}i|z|^{3/2}} + ie^{-\frac{2}{3}i|z|^{3/2}} \right) = \frac{1}{\sqrt{\pi}} |z|^{-1/4} \cos\left(\frac{2}{3}|z|^{3/2} - \frac{\pi}{4}\right) \,.
\end{equation}
In this way we have analytically continued a function from the forbidden region to the allowed region avoiding the turning point at $z = 0$. Note that although we have placed the branch cut at $z < 0$, we can either push it out of the way when we get there, or cross it. Crossing it gives $\fSbr\cdot (1, 1) = (1, -1)$. Although Ai is an analytic function, this sign flip correctly accounts for its asymptotic expansion $\Ai(z) \sim I_+(z) - I_-(z)$ for $z$ below the negative real axis.

In summary, the Stokes automorphisms allow us to track the abrupt changes in the asymptotic expansion of a function as a Stokes line is crossed. As we noted, on a Stokes line one asymptotic expansion is maximally dominant over the other. As the Stokes line is crossed, the subdominant component picks up an admixture of the dominant component, but the dominant component does not change. For instance crossing the $\theta=\frac{2\pi}{3}$ Stokes line where $\psi_+$ dominates takes us from $\psi_+(x) \rightarrow \psi_+(x) + \psi_-(x)$. This arrangement is in a sense minimally disruptive: the mixing occurs where the smallest possible contribution is added to the maximum possible contribution. We have presented the discussion of Stokes automorphisms without mentioning Borel resummation. However, as we discuss next, this is the same phenomenon seen in lateral Borel resummation: as the integration contour is continued across a branch cut in the Borel plane, the resummed expression picks up an exponentially subdominant contribution.

\subsection{Stokes automorphisms in the Borel plane} \label{sec:BorelThimbles}
Recall from Section~\ref{sec:0dRiemann} that there is a natural way to interpret saddle points of the action as branch points of a Riemann surface.  In this picture, thimbles move along the Riemann surface, passing from one sheet to another at the branch point. The Riemann surface picture connects the Borel transform and the action. In the Airy case, we first rescale the coordinates so the Borel transform is easier to see, then we use the geometric Riemann-surface picture to connect the Stokes matrices to lateral Borel resummation.

As discussed above, both the Airy function $\Ai(z)$ and the thimble integrals $I_\pm(z)$ are defined by the same integrand $e^{-S(u)}$ with $S(u) = -i(u^3/3 + zu)$; they differ only in the choice of contour. Changing variables to  $u = i\sqrt{z}\,w$ and defining $\zeta = \frac{2}{3}z^{3/2}$, these integrals then have the form
\begin{equation}
  I_\gamma(\zeta) \;=\; K\!\int_{\gamma} dw\;e^{-\zeta\, S_w(w)}\,,\qquad
  K \;\equiv\; \frac{i}{2\pi}\Bigl(\frac{3\zeta}{2}\Bigr)^{\!1/3}
  \label{eq:rescaled_integral}
\end{equation}
where the rescaled action is
\begin{equation}
  S_w(w) \;=\;  \tfrac{3}{2} w -\tfrac{1}{2} w^3
  \quad = \quad
\begin{tikzpicture}[baseline=-0.5ex, scale=0.8]
  \clip (-2.4,-1.2) rectangle (2.4,1.2);
  % Translucent shaded bands for the three root regions (1,2,3 from left to right)
  \fill[branchone, opacity=0.12] (-2.4,-1.2) rectangle (-1,1.2);
  \fill[branchtwo, opacity=0.12] (-1,-1.2) rectangle (1,1.2);
  \fill[branchthree, opacity=0.12] (1,-1.2) rectangle (2.4,1.2);
  % Region labels
  \node[branchone!80!black, scale=0.6, font=\bfseries] at (-1.7,0.95) {$w_1$};
  \node[branchtwo!80!black, scale=0.6, font=\bfseries] at (0.4,0.95) {$w_2$};
  \node[branchthree!80!black, scale=0.6, font=\bfseries] at (1.7,0.95) {$w_3$};
  % Axes
  \draw[->] (-2.3,0) -- (2.3,0) node[above left] {\scriptsize $w$};
  \draw[->] (0,-1.1) -- (0,1.1);
  % Curve
  \draw[thick, black!80, variable=\w, domain=-2.2:2.2, samples=120]
    plot ({\w}, {0.7*(1.5*\w - 0.5*\w*\w*\w)});
  % Saddle points
  \fill[red!70!black] (-1,{0.7*(-1)}) circle (2pt);
  \fill[red!70!black] (1,{0.7*(1)}) circle (2pt);
  \node[right, scale=0.65] at (0.8, 0.35 ) {$+1$};
  \node[left, scale=0.65] at (-0.8, -0.35 ) {$-1$};
\end{tikzpicture}
  \label{eq:Sw_action}
\end{equation}
This action has critical  points at $w_\pm = \pm 1$ with critical values $S_w(\pm 1) = \pm 1$.

Following the same procedure as for the zero-dimensional double well (Section~\ref{sec:0dRiemann}), we identify the Borel transform $B(t)$ as the multi-valued inverse of the action:
\begin{equation}
  t \;=\; \tfrac{3}{2}B - \tfrac{1}{2}B^3 \,.
  \label{eq:airy_Bt}
\end{equation}
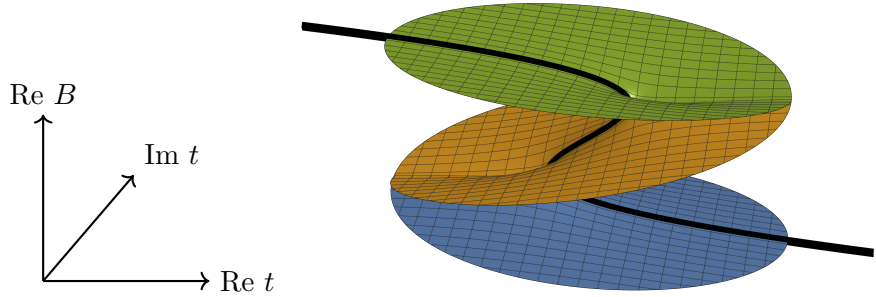
\begin{figure}[t]
  \centering
  \vspace{-5pt}
  \begin{tikzpicture}
    \node[anchor=south west,inner sep=0] (img) at (1,0)
      {\includegraphics[width=0.65\textwidth,trim=0 60 0 30,clip]{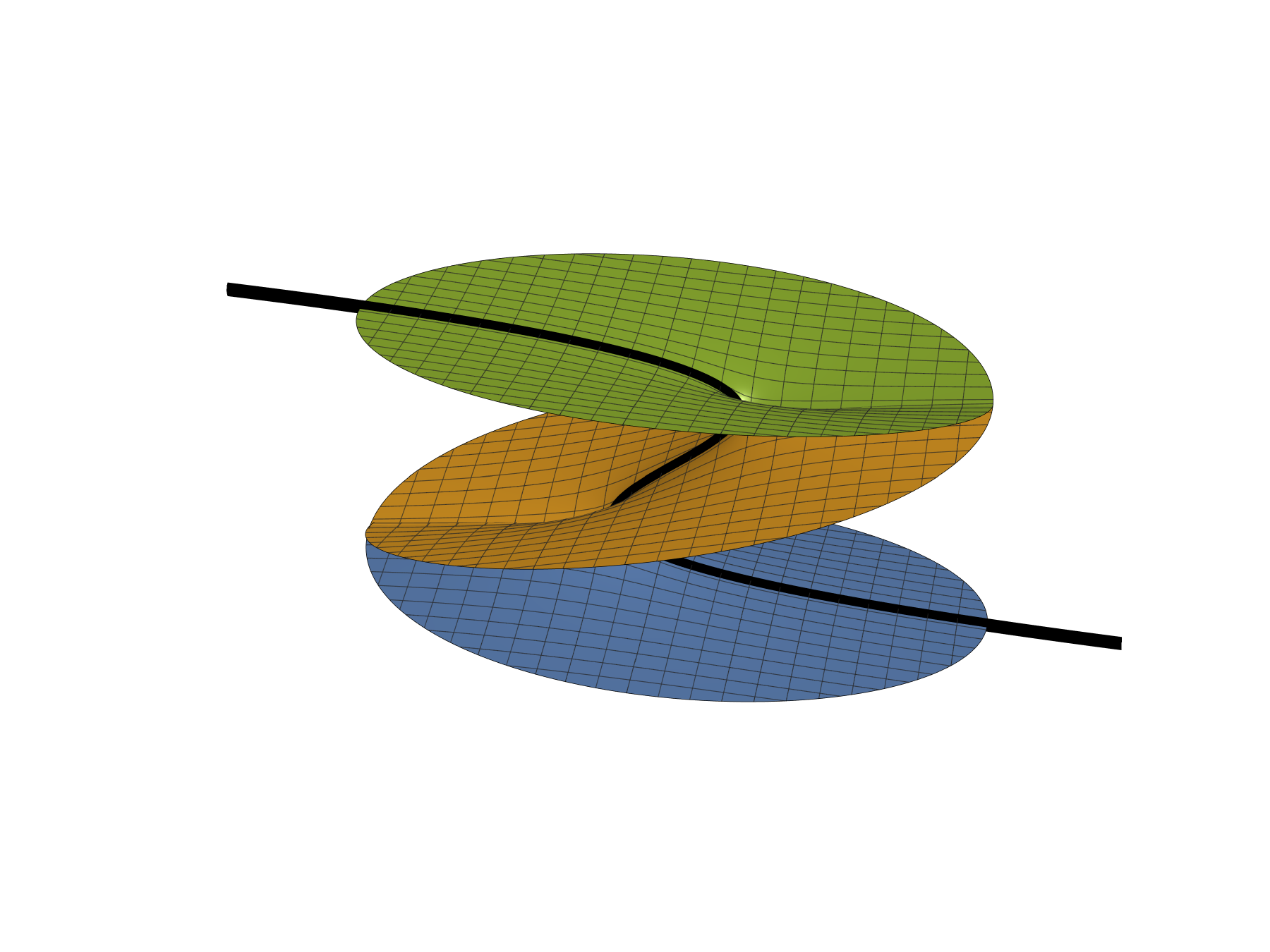}};
    % Axis origin in lower-left, clear of the surfaces
    \coordinate (O) at (-0.5,1.5);
    \draw[->,thick] (O) -- ++(2.2,0) node[right] {$\re\, t$};
    \draw[->,thick] (O) -- ++(1.2,1.4) node[above right] {$\im\, t$};
    \draw[->,thick] (O) -- ++(0,2.2) node[above] {$\re\, B$};
  \end{tikzpicture}
  \caption{Three-sheeted Riemann surface for the Airy function, showing $\re\, B$ as a function of complex $t$. The relation $t = \tfrac{3}{2}B - \tfrac{1}{2}B^3$ defines three branches $B_j(t)$, which coalesce pairwise at the branch points $t = \pm 1$. The black curves show the real branches on the real $t$-axis.}
  \label{fig:airy_riemann_surface}
\end{figure}
This equation describes a three-sheeted Riemann surface over the $t$-plane, as shown in Fig.~\ref{fig:airy_riemann_surface}. The branches are
\begin{equation}
  B_j(t) \;=\; 2\sin\!\left(\frac{1}{3}\arcsin t -\frac{4\pi}{3} + \frac{2\pi j}{3}\right)
  \;\;=\;\;
\begin{tikzpicture}[baseline=0ex, scale=0.85]
  \clip (-3.3,-1.8) rectangle (3.5,2.1);
  % axes
  \draw[->] (-2.7,0) -- (2.8,0) node[right,scale=0.7] {$t{=}S$};
  \draw[->] (0,-1.6) -- (0,1.7) node[above,scale=0.7] {$B$};
  % dashed lines at branch points
  \draw[gray, thin, dashed] (-1,-1.8) -- (-1,2.1);
  \draw[gray, thin, dashed] (1,-1.8) -- (1,2.1);
  %% SOLID Re(B): -1 < t < 1 (all three real, Im=0)
  \draw[thick, branchone, variable=\a, domain=-28.5:28.5, samples=100]
    plot ({sin(3*\a)}, {0.5*2*sin(\a-120)});
  \draw[thick, branchtwo, variable=\a, domain=-28.5:28.5, samples=100]
    plot ({sin(3*\a)}, {0.5*2*sin(\a)});
  \draw[thick, branchthree, variable=\a, domain=-28.5:28.5, samples=100]
    plot ({sin(3*\a)}, {0.5*2*sin(\a+120)});
  %% SOLID Re(B): t > 1 — B_1 real, B_2,B_3 complex (Re overlap, displaced +/-0.04)
  \draw[thick, branchone, variable=\b, domain=0:0.5, samples=50]
    plot ({(exp(3*\b)+exp(-3*\b))/2}, {-0.5*(exp(\b)+exp(-\b))});
  \draw[thick, branchtwo, variable=\b, domain=0:0.5, samples=50]
    plot ({(exp(3*\b)+exp(-3*\b))/2}, {0.25*(exp(\b)+exp(-\b)) - 0.04});
  \draw[thick, branchthree, variable=\b, domain=0:0.5, samples=50]
    plot ({(exp(3*\b)+exp(-3*\b))/2}, {0.25*(exp(\b)+exp(-\b)) + 0.04});
  %% SOLID Re(B): t < -1 — B_3 real, B_1,B_2 complex (Re overlap, displaced +/-0.04)
  \draw[thick, branchthree, variable=\b, domain=0:0.5, samples=50]
    plot ({-(exp(3*\b)+exp(-3*\b))/2}, {0.5*(exp(\b)+exp(-\b))});
  \draw[thick, branchone, variable=\b, domain=0:0.5, samples=50]
    plot ({-(exp(3*\b)+exp(-3*\b))/2}, {-0.25*(exp(\b)+exp(-\b)) - 0.04});
  \draw[thick, branchtwo, variable=\b, domain=0:0.5, samples=50]
    plot ({-(exp(3*\b)+exp(-3*\b))/2}, {-0.25*(exp(\b)+exp(-\b)) + 0.04});
  %% DOTTED Im(B): t > 1 — B_2 Im<0, B_3 Im>0 (B_1 stays real)
  \draw[thick, densely dotted, branchtwo, variable=\b, domain=0:0.5, samples=50]
    plot ({(exp(3*\b)+exp(-3*\b))/2}, {-0.433*(exp(\b)-exp(-\b))});
  \draw[thick, densely dotted, branchthree, variable=\b, domain=0:0.5, samples=50]
    plot ({(exp(3*\b)+exp(-3*\b))/2}, {0.433*(exp(\b)-exp(-\b))});
  %% DOTTED Im(B): t < -1 — B_1 Im<0, B_2 Im>0 (B_3 stays real)
  \draw[thick, densely dotted, branchone, variable=\b, domain=0:0.5, samples=50]
    plot ({-(exp(3*\b)+exp(-3*\b))/2}, {-0.433*(exp(\b)-exp(-\b))});
  \draw[thick, densely dotted, branchtwo, variable=\b, domain=0:0.5, samples=50]
    plot ({-(exp(3*\b)+exp(-3*\b))/2}, {0.433*(exp(\b)-exp(-\b))});
  % coalescence markers
  \fill[red!70!black] (-1,-0.5) circle (1.5pt);
  \fill[red!70!black] (1,0.5) circle (1.5pt);
  % Re labels (solid curves)
  \node[branchone, scale=0.55, anchor=west] at (2.45,-1.3) {$\re\, B_1$};
  \node[branchtwo, scale=0.55, anchor=south west] at (0,0.25) {$\re\, B_2$};
  \node[branchthree, scale=0.55, anchor=east] at (-1.5,1.3) {$\re\, B_3$};
  % Im label (dotted curves)
  \node[branchtwo, scale=0.55, anchor=west] at (2.45,-0.55) {$\im\, B_2$};
\end{tikzpicture}
  \label{eq:B_closed}
\end{equation}
labeled from bottom to top at $t = 0$. Beyond the branch points at $t = \pm 1$, two of the three branches become complex conjugates; their real parts (solid) are slightly displaced for clarity, and their imaginary parts are shown as dotted lines. For $t>1$ we take the continuation from $(-1,1)$ such that $\im B_2 < 0$ and $\im B_3 > 0$; for $t<-1$ the roles of $B_1, B_2$ are analogous. The branches coalesce at the critical values of $S_w$:  $B_1 = B_2 = -1$ at $t = -1$ and $B_2 = B_3 = +1$ at $t = +1$.  The three branches are related by analytic continuation: $B_1 \leftrightarrow B_2$ around $t = -1$ and $B_2 \leftrightarrow B_3$ around $t = +1$.\footnote{The three branches are related by $\sum_j B_j = 0$ for all $t$, which follows from Vieta's formulas for the sum of the roots of a cubic polynomial.} Sewing together the different sheets as shown in Fig.~\ref{fig:airy_riemann_surface} gives a concrete realization of the action--Borel correspondence of Eq.~\eqref{eq:actionBorel}: the Borel variable $t = S_w(w)$ \emph{is} the action, and $B(t) = w(t)$ \emph{is} the field configuration. 

Now consider the thimble integrals. The integral $I_+$ is defined along the thimble $\mathcal{J}_+$ passing through $w_+ = +1$. Just as in Section~\ref{sec:0dRiemann}, the thimble has two arms meeting at the saddle. Here we see that at $t = +1$, sheets 2 and~3 coalesce, so one arm lives on sheet~2 and the other on sheet~3. Changing variables from $w$ to $t = S_w(w)$, the measure on each arm is $dw = B'_j(t)\,dt$, and the two arms contribute with opposite orientations. So we can write for real positive $\zeta$ 
\begin{equation}
  I_+(\zeta) \;=\; K\!\int_{\mathcal{J}_+}\! dw\;e^{-\zeta\, S_w(w)}
  \;=\; K\!\int_{1}^{\infty}\!dt\, e^{-\zeta t}\,\bigl[B'_{2}(t) - B'_3(t)\bigr]\,.
\end{equation}
Integrating by parts, the boundary terms vanish. Then we get
\begin{align}
  I_+(\zeta) &= K\, \zeta\int_{1}^{\infty}\!dt\, e^{-\zeta t}\,\bigl[B_{2}(t) - B_3(t)\bigr]
  \\
  \label{eq:Borel_Jacobian}
  &=K\, \zeta\, e^{-\zeta}\int_0^{\infty}\!dt\, e^{-\zeta t}\,\left[B_2(t {+} 1) - B_3(t{+} 1)\right]\,.
\end{align}
where we shifted $t\to t+1$ in the second step. We can then read off the Borel transform: $\mathcal{B}[I_+](t) = B_2(t{+}1) - B_3(t{+}1)$. Note that $B_2 - B_3$ is the discontinuity across the branch cut of the Riemann surface $t = +1$ where sheets 2 and 3 coalesce. Since the integrand is single-valued for $t > 1$ there is no ambiguity and the integrand is regular at $t=1$. The Borel transform for $I_+$ does have a singularity inherited from the \emph{other} branch point of~$B(t)$: the point $t = -1$ where sheets 1 and 2 meet, which maps to $t = -2$ in the shifted variable.  This singularity lies behind the contour (on the negative real axis), so  $I_+$ is Borel summable and no tilting of the integration contour is required. Indeed for positive $\zeta$ we can verify that the series defined in Eq.~(\ref{Iplusdef}) is factorially growing, but sign-alternating, and hence Borel summable.

A similar computation applies for $\mathcal{J}_-$, where sheets 1 and 2 coalesce at $t_\star = -1$.  The integrand is now the discontinuity $B_2 - B_1$ across the cut at $t = -1$:
\begin{equation}
  \cS_\pm[I_-](\zeta) \;=\; K\, \zeta\, e^{\zeta}\int_0^{e^{\pm i\varepsilon}\infty}\! dt\, e^{-\zeta t}\, \bigl[B_{2}(t{-}1) - B_1(t{-}1)\bigr]\,.
  \label{eq:thimble_minus_t}
\end{equation}
We can again read off the Borel transform: $\mathcal{B}[I_-](t) = B_2(t{-}1) - B_1(t{-}1)$. In this case, the integrand is single-valued along its own cut ($t=0$ in these variables), but now the \emph{other} branch point ---where sheets 2 and 3 meet, or $t = +2$ in the shifted variable---sits directly on the positive real integration ray. For this reason we had to deform the integration contour, tilting  off the real axis, to go to $e^{i\varepsilon}\infty$. Going above or below the branch point gives different analytic continuations, so the integral is ambiguous: the $e^{i\varepsilon}$ tilt passes above the cut, giving the lateral Borel resummation $\cS_+[I_-]$, while $e^{-i\varepsilon}$ passes below, giving $\cS_-[I_-]$. This ambiguity in the lateral Borel resummation is reflected in the fact that the series of Eq.~(\ref{Iminusdef}) is factorially growing and non sign-alternating. The monodromy around the obstructing branch point at $t = +2$ swaps $B_2 \leftrightarrow B_3$ while leaving $B_1$ unchanged, so the jump of the integrand $B_2 - B_1$ across the cut is $(B_3 - B_1) - (B_2 - B_1) = B_3 - B_2$: minus the combination that defines $I_+$.  The two lateral Borel sums therefore differ by a contour integral encircling this branch point, giving the Stokes relation
\begin{equation}
  \boxed{(\cS_+ - \cS_-)\, I_-(\zeta) \;=\; -I_+(\zeta)}
  \label{eq:Borel_Stokes}
\end{equation}
with the Stokes constant $\fSp_{21} = 1$ (which equals the $(2,1)$ entry of the connection matrix $\fSp$) determined by the monodromy exponent at the square-root branch point.

This is the simplest instance of the {\textbf{Delabaere--Dillinger--Pham (DDP) formula~}}\cite{DelabaereDillingerPham1993,DelabaerePham1999}: the discontinuity of one Borel-resummed quantity across a Stokes line is controlled by another.  This structure generalizes: whenever a problem has resurgent Borel singularities, the discontinuity across a Stokes line at distance $A$ is proportional to $e^{-A\zeta}$ times another quantity.  We will see the same pattern for the double-well in Section~\ref{sec:DW}.

The DDP formula is often repackaged into a
\textbf{Stokes automorphism} operator
\begin{equation}
  \SA \;\equiv\; \cS_-^{-1}\cS_+\,,
  \label{eq:StokesAuto_def}
\end{equation}
acting on trans-series. This operator encodes the monodromy of a formal series around the Borel branch point: $\cS_+$ integrates above the singularity, $\cS_-$ below, so $\cS_-^{-1}\cS_+$ is the analytic continuation around it.  Using that $I_+$ is Borel summable, so $\cS_+ I_+ = \cS_- I_+$, and Eq.~\eqref{eq:Borel_Stokes}, we then have

\begin{equation}
  \begin{pmatrix} \SA\, I_+ \\ \SA\, I_- \end{pmatrix}
  = \begin{pmatrix} I_+ \\ I_- - I_+ \end{pmatrix}
  = \begin{pmatrix} 1 & 0 \\ -1 & 1 \end{pmatrix}
  \begin{pmatrix} I_+ \\ I_- \end{pmatrix}
  \equiv
  (\fSp)^{-1}
  \begin{pmatrix} I_+ \\ I_- \end{pmatrix}\,.
  \label{eq:StokesAuto_Airy}
\end{equation}
Both $\SA$ and $\fSp$ are determined by the same Stokes constant, so it's natural that they are related. But it's important to emphasize that they act on different objects: $\SA$ acts on the formal asymptotic series $I_\pm$ themselves, while $\fSp$ acts on the coefficient vectors $\vec{c} = (c_+, c_-)$ that specify how much of each series appears in a physical solution. In quantum mechanics, $\fSp$ will remain a matrix but $\SA$ as an operator will act on the much larger space of trans-series.\footnote{For the Airy function (a one-dimensional integral with two saddle points), $\SA$ is completely captured by a $2\times 2$ matrix and carries the same information as $\fSp$.  In quantum mechanics the situation is richer: the quantum-mechanical trans-series contains infinitely many multi-instanton sectors, and $\SA$ acts on this entire infinite-dimensional graded space via {\'E}calle's formula $\SA = \exp\bigl(\sum_{n\geq 1} e^{-nS/\hbar}\Delta_n\bigr)$ (Eq.~\eqref{eq:Ecalle}).  The connection matrix $\fSp$ remains finite-dimensional, acting on the two WKB solutions with entries that become functions of the Voros symbols (Eq.~\eqref{matrixform}).  Thus $\SA$ resolves the sector-by-sector resurgent structure, while $\fSp$ summarizes only the net Stokes jump of the wave function.}

\begin{figure}[t]
  \centering
  \includegraphics[width=\textwidth]{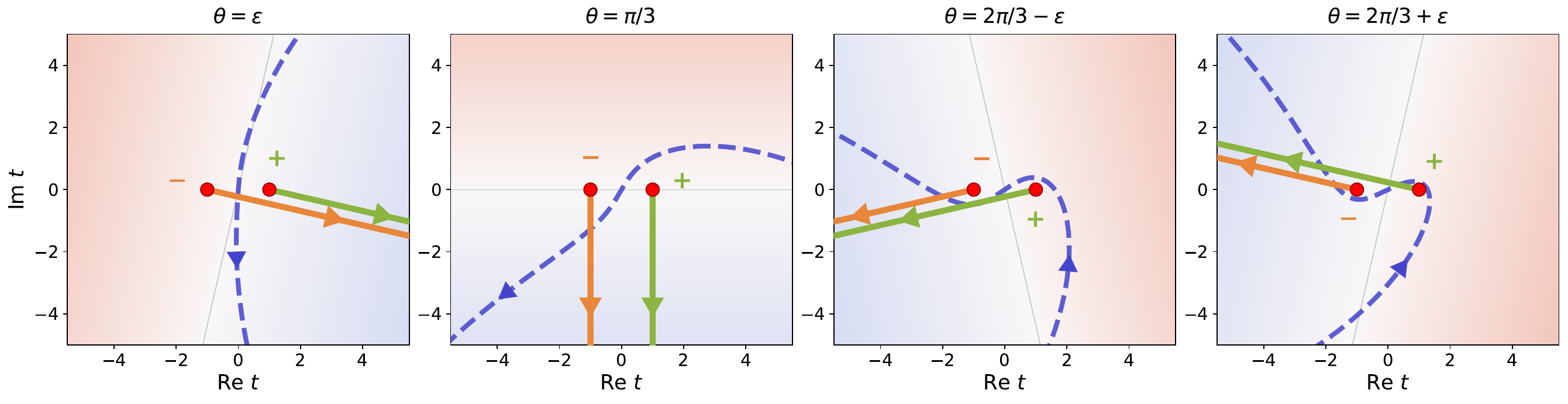}
  \caption{Image of the Airy contour (dashed blue) and the thimble contours $\mathcal{J}_+$ (green) and $\mathcal{J}_-$ (orange) mapped from $u$-space to the Borel $t$-plane for four values of $\theta = \arg z$.  The red dots mark the branch points at $t = \pm 1$. The shading indicates $\re(\zeta t)$ values, with red as the divergent region, blue at the convergent region and the gray line indicating the boundary $\re(\zeta t)=0$. Each thimble's two arms live on different sheets of the Riemann surface but project to the same curve in $t$, so the thimble images trace out natural branch cuts from $t = \pm 1$.  These cuts rotate clockwise at rate $\tfrac{3}{2}\theta$, the same rate as the Borel integration direction.  For $\theta = \varepsilon$ the Airy contour is close to the imaginary axis, passing between the two rightward-pointing cuts. As $\theta$ increases toward $2\pi/3$, the cuts rotate leftward. The Stokes phenomenon occurs when the thimble contours cross.}
  \label{fig:airy_contour_tplane}
\end{figure}

Now $I_\pm(\zeta)$ are defined as thimble integrals in $u$ space. In $t$-space these saddles originate at the branch points at $t_\star =\pm 1$ and move in steepest ascent trajectories of the action where $\re (t \zeta)$ is maximized, i.e. along straight lines defined by $\arg t = - \arg \zeta$. The lowest real-action point along the trajectories is at the saddle, so these thimbles have the asymptotic expansion $I_\pm(\zeta) \sim \exp(-t_\star \zeta)$ for any $\zeta$. For real positive $\zeta$, $I_-$ has lower action (the 1-2 branch point on the Riemann surface is lower) and the curve moves upward until it hits the $I_+$ branch point. That is why $I_-$ has a discontinuity and $I_+$ does not. This is general: the dominant saddle (smaller action at the saddle point) has an ambiguity given by directing its contour above or below the second saddle it hits. As $\zeta$ is continued, the lines associated with the thimbles rotate around. At the anti-Stokes line (where the real parts of the two asymptotic exponents are equal), $\arg \zeta =\pi/2$ ($\arg z = \pi/3$), the two contributions have equal magnitude and at the Stokes line $\arg \zeta = \pi$ ($\arg z = 2\pi/3$) they have switched places. These contours are shown in Fig.~\ref{fig:airy_contour_tplane} for various choices of $\theta = \arg z = (2/3)  \arg \zeta $.  Each thimble has two arms meeting at the saddle; these two arms live on different sheets of the Riemann surface but project onto the same curve in the $t$-plane.  A curve emanating from a branch point along which two sheets coalesce is precisely a branch cut. Thus the thimble images define a natural, $\theta$-dependent choice of branch cuts emanating from $t = \pm 1$.

In contrast to $I_\pm$ which are pinned to saddles, the contour associated with the Airy function is a fixed contour in $u$ space going between convergent regions $A$ and $B$. Since $\zeta\,t = S(u)$, the Borel variable is related to the original  integration variable $u$ by
\begin{equation}
  t \;=\; -\frac{i}{2}\,z^{-3/2}\,u\,(u^2 + 3z)\,.
  \label{eq:zeta_t}
\end{equation}
For $\arg z = \epsilon$ the Airy contour in $u$ space can be taken as a slight deformation of the real line, as seen in the first panel of Fig.~\ref{fig:contours}. From Eq.~\eqref{eq:zeta_t} we can see that this will map in $t$ space into a contour that is a slight deformation of the imaginary axis. The Borel integral is convergent for $\re (\zeta t)>0$, so we can deform this contour to a Hankel contour encircling the $t=1$ branch point, where  sheets 2 and 3 meet. With the chosen orientation, this picks up the discontinuity $B_2 - B_3$ across the cut, which is exactly the integrand of $I_+$.  This gives $\Ai = I_+$ for real positive $\zeta$, recovering the standard result. As we rotate $z = e^{i\theta}$ clockwise, the Airy contour (real $u$) deforms according to Eq.~\eqref{eq:zeta_t}. Its image in the $t$ plane is shown in Fig.~\ref{fig:airy_contour_tplane} for some values of $\theta$. We see that it always avoids the branch points, as it must since $\text{Ai}(z)$ is analytic.  The Stokes phenomenon occurs precisely when a thimble contour passes the subleading saddle, which discontinuously changes the decomposition of the Airy contour's image into thimbles.

The Borel transforms we have derived also determine the asymptotic behavior of $I_\pm$.  Writing
\begin{equation}
  I_+(\zeta) \;\sim\; e^{-\zeta}\sum_{n=0}^{\infty} \frac{a_n}{\zeta^{n}}\,,
  \qquad
  I_-(\zeta) \;\sim\; e^{+\zeta}\sum_{n=0}^{\infty} \frac{b_n}{\zeta^{n}}\,,
\end{equation}
the coefficients $a_n$ and $b_n$ are determined by the Taylor expansion of the respective Borel transforms around $t=0$.  For $I_+$, the Borel transform is $B_2(t{+}1) - B_3(t{+}1)$.  This combination passes smoothly between sheets 2 and 3 at their coalescence point $t=0$ (i.e.\ $t=+1$ before shifting), so it is \emph{not} singular there.  The only singularity of this Borel transform is at $t = -2$, the other branch point where sheets 1 and 2 meet.  Its Taylor coefficients therefore scale as $(-2)^{-n}$, giving $a_n \sim (-\frac{1}{2})^n\, n!$.  For $I_-$, the Borel transform $B_2(t{-}1) - B_1(t{-}1)$ is similarly regular at its own branch point $t=0$; its only singularity is at $t = +2$ (where sheets 2 and 3 meet), giving $b_n \sim (\frac{1}{2})^n\, n!$ with constant sign:
\begin{equation}
  a_n \;\sim\; \left(-\frac{1}{2}\right)^n n!\,,
  \qquad
  b_n \;\sim\; \left(\frac{1}{2}\right)^n n!\,.
  \label{eq:Airy_large_order}
\end{equation}
Recalling $\zeta = \frac{2}{3}z^{3/2}$ we have 
\begin{equation}
  I_+(z) \;\sim\;  e^{-\frac{2}{3}z^{3/2}}
  \sum_{n=0}^{\infty} \left(-\frac{3}{4z^{3/2}}\right)^n n!\,,
  \qquad
  I_-(z) \;\sim\; e^{+\frac{2}{3}z^{3/2}}
  \sum_{n=0}^{\infty} \left(\frac{3}{4z^{3/2}}\right)^n n!\,,
  \label{eq:Ipm_asymp_Borel}
\end{equation}
reproducing the asymptotic form of the series in Eqs.~\eqref{Iplusdef}--\eqref{Iminusdef}.  The alternating signs for $I_+$ and constant signs for $I_-$ reflect the Borel singularity lying on the negative and positive real axes respectively.

Since the Borel singularity of $I_-$ sits on the positive real integration ray, $I_-$ is not Borel summable: the lateral resummations $\cS_\pm I_-$ differ by the discontinuity across the cut at $t = +2$. But this is exactly the content of the DDP relation, Eq.~\eqref{eq:Borel_Stokes}: $(\cS_+ - \cS_-)\, I_- = -I_+$.  Since $I_-$ is purely imaginary for real $z>0$, the two lateral resummations have the same imaginary part but different real parts:
\begin{equation}
  \operatorname{Re}\,\cS_\pm\, I_-(\zeta) \;=\; \mp\frac{1}{2}\, I_+(\zeta)\,.
  \label{eq:ImBorel_minus}
\end{equation}
The real ambiguity of the dominant saddle ($I_-$, with its non-alternating series) is exactly $I_+(\zeta) \sim e^{-\zeta}$.  Since $I_- \sim i\,e^{+\zeta}$, the ambiguity is suppressed relative to $|I_-|$ by $e^{-2\zeta}$, the exponential of the difference of the two saddle-point actions. 

\subsection{Exact WKB \label{subsec:exactWKB}}
In this section and onward, we use natural units throughout where the potential $V(x)$ is taken to be a function, the mass is set to unity $m=1$  and the perturbation expansion is done in terms of $\hbar$.

The WKB expansion begins by approximating a wavefunction as the exponential of the action: $\psi (x, t) = e^{\frac{i}{\hbar} S (x, t)}$.  Plugging this WKB Ansatz into the Schr{\"o}dinger equation $i \hbar \partial_t \psi = H \psi$ with $H = \frac{p^2}{2} + V$ gives
\begin{equation}
  \partial_t S + \frac{1}{2} (\partial_x S)^2 - \frac{1}{2} i \hbar
  \partial_x^2 S + V (x) = 0 \,.
\end{equation}
In the limit $\hbar = 0$ this is the Hamilton-Jacobi equation satisfied by the classical action $S_0$. The $\hbar$ corrections to the phase correspond to including non-classical paths in the path integral. In the Hamilton-Jacobi equation one normally identifies $\frac{\partial S}{\partial q}$ with the classical momentum $p$. In the treatments of Exact WKB it is conventional to include a factor of $i$, so $P = i \partial_x S$.  Either way, the action is (up to a factor of $i$) an integral over $P$. Introducing the energy $E$ as an integration constant, the WKB Ansatz is then
\begin{equation}
  \psi (x, t) = \exp \left( - \frac{i}{\hbar} E t + \frac{1}{\hbar}
  \int_{x_0}^x d x' P (x') \right)
\end{equation}
and the Schr{\"o}dinger equation becomes
\begin{equation}
  \hbar P' + P^2 + 2E - 2 V (x) = 0 \,. \label{eq:Riccati}
\end{equation}
This nonlinear differential equation is an example of a \textbf{Riccati equation}. It can be solved iteratively in perturbation theory.

At $\hbar = 0$ the solution to Eq.~\eqref{eq:Riccati} is
\begin{equation}
    P_0(x) = \pm \sqrt{2 V (x) - 2E}
    \label{P0def}
\end{equation}
which is $i$ times the classical momentum of a particle with energy $E$ in a potential $V (x)$. Thus, $P_0$ is real in forbidden regions and imaginary in allowed regions. For the corrections, we write $P =\sum \hbar^n P_n$ and solve perturbatively in $\hbar$. The perturbation expansion can be written as a recursion relation
\begin{equation}
  P_{n + 1} (x) = -\frac{1}{2 P_0 (x)} \left[ P_n' (x) + \sum_{k = 1}^n P_k (x) P_{n + 1 - k} (x) \right] \,.
\end{equation}
In particular, once a sign is chosen for $P_0$, all of the higher-order terms are completely fixed. For example, the next few terms are
\begin{align}
P_1 &= -\frac{P_0'}{2P_0} \\
P_2 &= \frac{-3{P_0'}^2 + 2P_0 P_0''}{8P_0^3} \label{P2form}\\
P_3 &=-\frac{6{P_0'}^3 - 6P_0 P_0' P_0'' + P_0^2 P_0'''}{8P_0^5}
\end{align}
and so on.

For polynomial potentials, the spectral curve $P_0^2 = 2V(x) - 2E$ is algebraic, and each $P_n$ is a rational function of $x$ and $P_0$. For the quartic double well, this curve is elliptic (genus 1), and $P_0\,dx$ and $\partial_E P_0\,dx = -dx/P_0$ span its first cohomology. The differential $P_n\,dx$ can therefore be decomposed as
\begin{equation}
P_n\,dx = a_n(E)\, P_0\,dx + b_n(E)\, \partial_E P_0\,dx + d f_n(x) \label{eq:Pndecomp}
\end{equation}
 where $a_n$ and $b_n$ depend only on $E$ (and the potential), and $f_n(x)$ is single-valued so that $d f_n$ integrates to zero around any closed cycle. For other potentials, the spectral curve can have higher genus and additional basis differentials would be needed. In practice, the decomposition is constructed by writing $P_n = R_n(x)/P_0^{k}$ where $R_n$ is a polynomial and $k=3n-1$ is an integer, making an ansatz $f_n = N_n(x)/P_0^{k-2}$ with $N_n$ an undetermined polynomial, and matching coefficients to solve for $a_n$, $b_n$, and $N_n$. For example, $P_1 = -\frac{1}{2}\partial_x \ln P_0$ is already a total derivative, so $a_1 = b_1 = 0$ and $f_1 = -\frac{1}{2}\ln P_0$. The coefficients $a_2$ and $b_2$ are potential-dependent and will be computed explicitly for the double well in Section~\ref{sec:periodintegrals}, while additional details are left to Appendix~\ref{appendix:PicardFuchs}.

One simplifying feature of the recursion is that only the terms with even $n$ depend on the sign choice for $P_0$. To see this, we write
\begin{equation}
  P_{\pm} = \pm P_{\operatorname{even}} + P_{\operatorname{odd}}
\end{equation}
where
\begin{align}
  P_{\text{even}} &= P_0 + \hbar^2 P_2 (x) + \hbar^4 P_4 (x) + \cdots \\
  P_{\text{odd}} &= \hbar P_1 (x) + \hbar^3 P_3 (x) + \cdots
\end{align}
Then the terms with odd powers of $\hbar$ in the Riccati equation are
\begin{equation}
  \pm \hbar \partial_x P_{\text{even}} (x) \pm 2 P_{\text{even}} (x)
  P_{\text{odd}} = 0
\end{equation}
from which we deduce that $P_{\text{odd}} = -\frac{\hbar}{2} \partial_x \ln
P_{\text{even}}$ independent of the sign. Since $P_{\text{odd}} $ is a derivative its contribution to $\psi$ can be integrated allowing us to write the wavefunction as
\begin{equation}
  \psi_{\pm} (x) = \frac{1}{\sqrt{P_{\operatorname{even}} (x)}} \exp \left( \pm
  \frac{1}{\hbar} \int_{x_0}^x d x' P_{\operatorname{even}} (x') \right)
\end{equation}
up to normalization.  In summary, we can determine the WKB wavefunction perturbatively using the even terms, which can be computed recursively.

The leading WKB expansion takes $P_{\text{even}} = P_0$. At the turning points, $P_0=0$ and  $\psi_\pm(x)$ blows up. The Exact WKB approach analytically continues around this singular region. To proceed, consider taking $x$ close to a turning point at $x = x_t$. Then $P_0 \approx \sqrt{2 V' (x_t) (x - x_t)}$ and
\begin{equation}
  \psi_{\pm} (x) \approx \mathcal{N}_\pm(x_0) \, \frac{(2 V' (x_t))^{-1/4}}{(x - x_t)^{1/4}} \exp \left[ \pm \frac{2}{3\hbar} \sqrt{2 V' (x_t)} (x - x_t)^{3 / 2} \right] \,. \label{eq:psipmclose}
\end{equation}
Here we recognize the asymptotic expansion of the Airy function, Eq.~\eqref{eq:Aiasymp}, upon identifying $z = \hbar^{-2/3}(2 V'(x_t))^{1/3}(x - x_t)$. The $x$-independent prefactors $\mathcal{N}_\pm(x_0)$ only affect the normalization, but are different for the $\pm$ solutions. Rather than keep track of these normalization factors, it is simpler to absorb them into the integral by adjusting the (arbitrary) basepoint of integration from $x_0$ to the turning point $x_t$. That is, we define
\begin{equation}
  \psi_{\pm}^t (x) = \frac{1}{\sqrt{P_{\operatorname{even}} (x)}} \exp \left[ \pm
  \frac{1}{\hbar} \int_{x_t}^x d x' P_{\operatorname{even}} (x') \right]
\end{equation}
as the two branches associated with turning point $t$, located at $x = x_t$.  To track the normalization change between two turning points $x_1$ and $x_2$, we note that
\begin{equation}
  \psi_+^1 = \frac{\psi_+^2}{\sqrt{\mathcal{V}_{12}}}, \quad \psi_-^1 =
  \psi_-^2 \sqrt{\mathcal{V}_{12}} \label{Vabove}
\end{equation}
where
\begin{equation}
  \mathcal{V}_{12} = \exp \left( - \frac{2}{\hbar} \int_{x_1}^{x_2} dx \,
  P_{\operatorname{even}} (x) \right) 
\end{equation}
is known as a \textbf{Voros symbol}. We can then account for the change in normalization moving from $x_1$ to $x_2$ with a connection matrix $S_{12}$, acting on the coefficient vector $\vec{c} = (c_-, c_+)$ in $\psi = c_+ \psi_+ + c_- \psi_-$
\begin{equation}
  S_{i j} = \left(\begin{array}{cc}
    \sqrt{\mathcal{V}_{i j}} & \\
    & \frac{1}{\sqrt{\mathcal{V}_{i j}}}
  \end{array}\right) \label{Sij} \,.
\end{equation}

\begin{figure}[t]
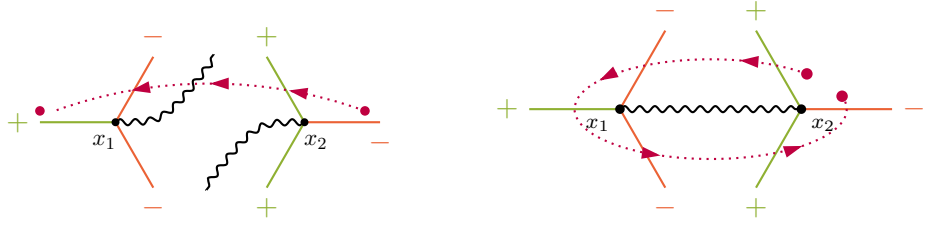

\centering
\SHOPathC
\hspace{1cm}
\SHOMonodromyF
\caption{Stokes diagrams for the simple harmonic oscillator. Left: branch cuts directed to infinity. The purple dotted curve shows the path of analytic continuation used to get the quantization condition.
Right: A different convention where the branch cut is between the turning points.
One must be cautious with this convention as the monodromy around the cut as shown is nontrivial.}
\label{fig:SHOmonodromy}
\end{figure}

The Voros symbol involves a period integral between two turning points where $P_{\text{even}}$ has algebraic branch points. For example, the harmonic oscillator potential is $V (x) = \frac{1}{2} x^2$. For this potential, $P_0 = \sqrt{x^2 - 2E} = \sqrt{(x - \sqrt{2E})(x + \sqrt{2E})}$ which has two branch points at $x = \pm\sqrt{2E}$. To get the expected oscillatory behavior in the classically allowed region where $E>V(x)$, between the turning points, we would like $P_0$ to be imaginary there. Analytic continuation from positive to negative real $x$ goes counterclockwise through the upper half of the complex $x$-plane, avoiding the branch points. This then lets us write the Voros symbol as a monodromy around the branch points:
\begin{equation}
  \mathcal{V}_{12} = \exp \left( \frac{1}{\hbar} \oint_{x_1, x_2} d x
  P_{\operatorname{even}} (x) \right) \,.
\end{equation}
The integral is along a closed  contour encircling counterclockwise the
two branch points surrounding a classically allowed region. It is convenient to also introduce $S_{i j}^{- 1}$ which we denote by $S_{j i}$. So, $S_{12}$ in Eq.~\eqref{Sij} is to be used when going from $x_1$ to $x_2$ above the branch cut if $x_1 < x_2$, but $S_{2 1} = S_{12}^{- 1}$ is used when going below the branch cut (or above the branch cut from $x_2 \rightarrow x_1$).  For two turning points surrounding a forbidden region, the integrand is real and so $S_{12}$ is used (there is no branch cut). An alternative convention is to have the branch cuts go from each turning point off to infinity, as shown on the left in Fig.~\ref{fig:SHOmonodromy}. Then one uses the branch-cut Stokes automorphism, Eq.~\eqref{Sbr}, in the analytic continuation when crossing each branch cut.

\subsection{Warm up: simple harmonic oscillator}

As a warm-up computation with Exact WKB, we consider the harmonic oscillator with $V (x) = \frac{1}{2} x^2$. The turning points are $x _{1,2}= \pm \sqrt{2 E}$:
\begin{equation}
  \SHOStokesWithPotential
  \label{eq:SHOstokesPlot}
\end{equation}
The goal is then to analytically continue from the far right to the far left along the arc as shown.
We will do this by using connection matrices where we can quantize the energies by demanding exponential decay as $|x|\to \infty$. To construct the Stokes diagram for this potential, we start on the right close to the turning point where the potential grows linearly like the Airy potential. On the left, the potential also grows, and this can be represented with its own Airy Stokes diagram. To connect them, we can either join the branch cuts or point them in complex conjugate directions. The result is shown in Fig.~\ref{fig:SHOmonodromy}, where the $+$ and $-$ labels indicate the Stokes lines where the functions $\psi_+$ and $\psi_-$ from Eq.~\eqref{eq:psipmclose} are respectively subdominant, consistent with the branch cut choice.

We begin at large real $x$ where we want $\psi$ to be exponentially decaying. Writing $\psi(x) = c_- \psi_-(x) + c_+ \psi_+(x)$ with $\psi_\pm$ defined by Eq.~\eqref{eq:psipmclose}, the decaying contribution at large positive $x$ is $\psi_-$, so the boundary condition selects $c_+ = 0$. The connection matrices $\fSp, \fSm, \fSbr$ from Section~\ref{sec:stokes_auto} were constructed with the decaying-mode coefficient in the first slot of $\vec{c}$; to reuse them here we order the vector as $\vec{c} = (c_-, c_+)$, and the starting point is $\vec{c} = (1, 0)$. Note that, as with the Airy function, it doesn't matter if we start at large $x$ above or below the Stokes line, since $\fSm \cdot \vec{c} = \vec{c}$. Moving counterclockwise, we first cross a $+$ Stokes line and get $\fSp \vec{c} = (1,1)$. With both components turned on the wavefunction is oscillatory, as expected between the turning points. To get to the next turning point, we must use a Voros connection to adjust the normalization. Then we cross a branch cut from the left region and finally another Stokes line to get to the asymptotically $x < 0$ region. The result is that on the far left
\begin{equation}
  \vec{c}{}^{\,\,\prime} = \fSm \cdot\fSbr \cdot S_{2 1} \cdot \fSp \cdot \vec{c} = \left(
  \frac{1+ \mathcal{V}_{12} }{\sqrt{\mathcal{V}_{12}}},
  \frac{-1}{\sqrt{\mathcal{V}_{12}}}  \right) \,.
\end{equation}
For $x\to -\infty$, we need the wavefunction to decay, which means the exponentially growing component must be absent. This requires $\vec{c}^{\, \, T}\cdot \fSm \fSbr S_{2 1} \fSp \cdot \vec{c} =0$ which means the first entry must vanish and therefore
\begin{equation}
     \mathcal{V}_{12}  = -1 \,.
\end{equation}
As a cross-check that we got the right condition, note that the first entry of $\vec{c}$ doesn't change as we cross the $+$ Stokes line at negative $x$, since $\fSp$ leaves the first entry invariant. Therefore our quantization condition for the harmonic oscillator is independent of our location above or below the Stokes line at negative real $x$ and reads
\begin{equation}
  \mathcal{V}_{12} = \exp \left(  \frac{1}{\hbar} \oint_{x_1, x_2} d x P_{\operatorname{even}} (x) \right) = - 1 = e^{2 \pi i \left( N + \frac{1}{2}  \right)}\,,\quad
  N =0,1,2,\ldots
\end{equation}
To solve this for $E$ we first compute the even terms in the momentum expansion:
\begin{equation}
  P_0 = \sqrt{x^2 - 2 E}, \quad P_2 = - \frac{3 x^2 + 4 E}{8 (x^2 - 2 E)^{5 /2}}, \quad P_4 = -\frac{304 E^2 + 1464 E x^2 + 297 x^4}{128 (x^2 - 2 E)^{11 / 2}}\,.
\end{equation}
In general, $P_n \sim x^{1 - 2 n}$ at large $x$. This means that if we deform the contour of integration in the Voros symbol to be large, then all the $P_n$ with $n>0$ will not contribute. The contour integral over $P_0$ does not vanish; it gives $\mathcal{V}_{12} = e^{2 i \pi E/\hbar}$. Therefore  $E = \hbar(N + \frac{1}{2})$. These are exactly the correct energies for the simple harmonic oscillator.

As an additional cross check, we can consider continuing our wavefunction all the way around a counterclockwise circle back to the positive real region. If we use the configuration on the left in  Fig.~\ref{fig:SHOmonodromy},  where the branch points go to infinity, the connection formula using the appropriate branch cuts and continuations is 
\begin{equation}
  \fSm \fSp \fSbr S_{12} \fSm  \fSp \fSm \fSbr S_{2 1} \fSp = 1 \,.
\end{equation}
This monodromy check verifies the consistency of the sign choices and connection matrices.  If instead we use the configuration on the right 
in  Fig.~\ref{fig:SHOmonodromy}, where the branch points are merged, we get the appropriate connection formula as
\begin{equation}
  \fSm \fSp S_{21} \fSm \fSp \fSm S_{2 1} \fSp = -1\,.
\end{equation}
To understand why this gives $-1$ instead of $1$, consider the contour integral $\oint \frac{dz}{\sqrt{z^2-1}}$ around a path encircling branch points at $z = \pm 1$. With a branch cut between the turning points, we can deform this to an integral along the cut, giving $\oint \frac{dz}{\sqrt{z^2-1}} = 2i \int_{-1}^{1}\frac{dz}{\sqrt{1-z^2}} = 2\pi i$, not zero. The non-zero result reflects the fact that the Stokes diagram with a shared branch cut does not fully account for all phase accumulation. This point is clearly articulated in~\cite{BucciottiReisSerone}. Additional examples can also be found in~\cite{Sueishi2020}.

\subsection{Double well}
\label{sec:DW}
For the double well, the basic procedure is the same as the harmonic oscillator, but a few complications arise. We use a double-well potential normalized as
\begin{equation}
  V (x) = \frac{1}{8} (x^2 - 1)^2 \,.
\end{equation}
This potential has minima at $x = \pm 1$ and a local maximum at $x = 0$. We choose $\frac{1}{8}$ for the normalization so that $\omega = 1$ for the harmonic oscillator at each minimum. For energy $E$ the turning points are at
\begin{equation}
  x_1 = - \sqrt{1 + \sqrt{8 E}}, \quad x_2 = - \sqrt{1 - \sqrt{8 E}}, \quad x_3 = \sqrt{1 - \sqrt{8 E}}, \quad x_4 = \sqrt{1 + \sqrt{8 E}} \,.
\end{equation}
We are primarily interested in the low-energy spectrum where $0 < E < \frac{1}{8}$ so that tunneling is relevant. States with $E > \frac{1}{8}$ are above the barrier.

\begin{figure}[t]
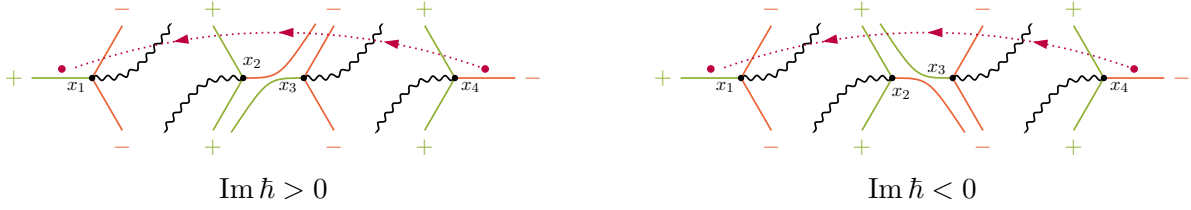

  \centering
  \begin{minipage}[b]{0.48\textwidth}
    \centering
    \scalebox{0.8}{\StokesGraphDoubleWellMisumi}\\
    $\operatorname{Im}\hbar > 0$
  \end{minipage}
  \hfill
  \begin{minipage}[b]{0.48\textwidth}
    \centering
    \scalebox{0.8}{\StokesGraphDoubleWellMisumiB}\\
    $\operatorname{Im}\hbar < 0$
  \end{minipage}
  \caption{Stokes graphs for the double well potential. The Stokes lines between $x_2$ and $x_3$ which overlap for real $\hbar$ deform into the complex plane when $\hbar$ becomes complex. The two panels show the two sign choices for $\operatorname{Im}\hbar$. The analytic continuation path for the wavefunction is shown as the purple dotted curve. Either choice of deformation leads to the same quantization condition. The branch-cut directions are a separate convention, independent of the Stokes-line deformation: either branch-cut choice can be paired with either sign of $\operatorname{Im}\hbar$.}
  \label{fig:deformed}
\end{figure}

The potential and associated Stokes graph are
\begin{equation}
  \centering
  \DWpotential
  \hspace{1cm}
  \StokesGraphDoubleWellA
  \label{overlappingStokes}
 \end{equation}
In this case, in addition to the branch cuts intersecting, the Stokes lines intersect. One of the Stokes lines from $x_2$ appears on top of one of the Stokes lines from $x_3$.  These intersections of Stokes lines give a  Stokes phenomenon similar to what we saw in the $D=0$ double well in Section~\ref{sec:zerodim} where the thimbles intersect.  It occurs because there is a non-alternating series leading to a singularity on the positive Borel axis.  The ambiguity can be resolved by giving a small imaginary piece to $\hbar$. Depending on the sign of this deformation the Stokes lines will deform upwards or downwards. We also deform the branch cuts to resolve their intersection, as we did in Fig.~\ref{fig:SHOmonodromy}. The resulting Stokes graph for one sign choice is shown in Fig.~\ref{fig:deformed}. As a consistency check, extending the path in  Fig.~\ref{fig:deformed} to a full monodromy gives
\begin{equation}
\fSm \fSp \fSbr S_{34} \fSm \fSp S_{23} \fSp \fSbr S_{12} \fSm \fSp \fSm \fSbr S_{21} \fSp \fSm S_{32} \fSm \fSbr S_{43} \fSp = 1
\end{equation}
as expected.

To impose the asymptotic boundary conditions, we proceed as with the  simple harmonic oscillator example. We begin with $\vec{c} = (1, 0)$ in the $x > x_4$ region.  The combined connection formula to the negative $x <x_1$ region is
\begin{equation}
\fSC=
  \fSm \fSbr S_{21}
  \fSp \fSm S_{32} \fSm \fSbr S_{43} \fSp
  = \frac{1}{\sqrt{\VN} \VP}
  \begin{pmatrix}
    \VN+ (1+\VP)^2 & \VP (1+\VP) \\[6pt]
    -(1+\VP +  \VN) & -\VP
  \end{pmatrix}
  \label{matrixform}
\end{equation}
where the perturbative and non-perturbative Voros symbols are
\begin{equation}
  \VP = \exp \left(\frac{1}{\hbar} \oint_{\gamma_P} d x P_{\operatorname{even}} (x) \right), \quad \VN = \exp \left( \frac{1}{\hbar} \oint_{\gamma_N} d x P_{\operatorname{even}} (x) \right) \,.
\end{equation}
The contour $\gamma_P$ goes counterclockwise around $x_1$ and $x_2$ (or around $x_3$ and $x_4$), while $\gamma_N$ goes counterclockwise around $x_2$ and $x_3$. Although the $\gamma_N$ contour passes through the branch cuts, the discontinuities cancel and the exponent in $\VN$ equals minus twice the integral of $P_{\text{even}}$ between the two turning points divided by $\hbar$. The Voros symbol $\VN$ is the Exact WKB generalization of the WKB tunneling amplitude.

The quantization condition we need is $\vec{c}^{\, \, T}  \cdot \fSC \cdot \vec{c} = 0$ where $\vec{c} = (1,0)$ as in the SHO example. This implies
\begin{equation}
 \boxed{
   (1 + \VP)^2 =- \VN  }
\label{quantPN}
\end{equation}
which is the quantization condition for the double well.

Before solving this quantization condition, we note that the double well is invariant  under $x \rightarrow - x$, so the energy eigenstates will be parity eigenstates. The parity odd states will have $\psi(0)=0$ and the parity even ones will have $\psi' (0) = 0$. If a wavefunction is even or odd, then the boundary conditions at $x = -\infty$ are determined by those at $x = +\infty$. This observation allows us to replace the boundary condition at $x = - \infty$ with one imposing $\psi (0) = 0$ or $\psi' (0) = 0$. To evaluate $\psi (0)$ we can continue the function from the asymptotic $x > 0$ region to the middle region between the $x_2$ and $x_3$ Stokes lines using $ \fSm \fSbr  S_{43} \fSp$ . Then the wavefunction in this region is
\begin{equation}
  \psi (x) =  \frac{1}{\sqrt{P_{\text{even}} (x)}} \left[c_+ \exp \left(  \frac{1}{\hbar} \int_{x_3}^x d x' P_{\text{even}} (x') \right) 
  +c_- \exp \left( - \frac{1}{\hbar} \int_{x_3}^x d x' P_{\text{even}} (x') \right) \right] \,.
\end{equation}
Noting that $\psi(0)$ involves an integral from  $x_3$ to $0$ which is  $1/4$ of the integral in the exponent of $\VN$, and accounting for the continuation of the square-root prefactor of the $\psi_-$ branch around the turning point, we can write, up to an overall normalization,
\begin{equation}
\psi(0) \propto c_+ \left(\VN^{-1/4}\right)+ c_- \left( i \VN^{1/4} \right) \,.
\end{equation}
This allows us to evaluate $\psi(0)$ by starting from $\vec{c} = (1,0)$ in the large $x$ region, continuing to the central region, and contracting with  $(\VN^{-1 / 4}, i\,\VN^{1 / 4})$:
\begin{equation}
  (\VN^{-1 / 4}, \, i \VN^{1 / 4})
  \cdot \fSm \fSbr S_{43} \fSp \cdot \vec{c}= 0
\end{equation}
which leads to
\begin{equation}
  1 + \VP = i  \sqrt{\VN} \,,\qquad \text{odd wavefunctions, $\im \hbar>0$\,.}  \label{odddown}
\end{equation}
Imposing $\psi' (0) = 0$ for the even states rather than $\psi(0)=0$ adds a minus sign to the lower component so that,
\begin{equation}
  1 + \VP =  - i  \sqrt{\VN} \,,\qquad \text{even wavefunctions, $\im \hbar>0$\,.} \label{evendown}
\end{equation}
These two quantization conditions for the even and odd parity wavefunctions are the two roots of Eq.~\eqref{quantPN}.

If we had chosen $\im \hbar <0$ instead, the line from $x_3$ would tilt upward and be crossed by the continuation path, while the line from $x_2$ would tilt downward and miss it, as shown in right panel of Fig.~\ref{fig:deformed}. This adds an extra $\fSp$ to the chain to get to the middle region where $x=0$ lives. The odd-parity condition $\psi(0)=0$ then reads
\begin{equation}
  (\VN^{-1/4},\, i\,\VN^{1/4})\,\cdot\, \fSp\,\fSm\,\fSbr\, S_{43}\, \fSp\cdot\vec{c} = 0\,,
\end{equation}
which gives
\begin{equation}
  1 + (\VP)^{-1} = -i\sqrt{\VN}\,,\qquad \text{odd wavefunctions, $\im \hbar<0$} \label{oddup}
\end{equation}
and the even condition $\psi'(0)=0$ gives
\begin{equation}
  1 + (\VP)^{-1} = +i\sqrt{\VN}\,,\qquad \text{even wavefunctions, $\im \hbar<0$.}\label{evenup}
\end{equation}
Since $P_\text{even}$ is imaginary in the allowed region (for real $\hbar$),  $\VP$ is a pure phase so that $\VP^{-1} = \VP^*$. Thus Eqs.~\eqref{oddup} and \eqref{odddown} are complex conjugates of each other (as are Eqs.~\eqref{evendown} and~\eqref{evenup}) and give the same quantization condition for the real energies order by order in real $\hbar$.

Although the exponent in $\VP$ is imaginary for real $\hbar$ at each order in perturbation theory, the resulting series at fixed $E$ is asymptotic. The overlapping Stokes lines between $x_2$ and $x_3$ for real $\hbar$, shown in Eq.~\eqref{overlappingStokes}, imply that $\VP$ is a non-alternating series and not Borel summable along the real axis; the two sign choices for $\operatorname{Im}\hbar$ correspond to two lateral Borel resummations $\cS_\pm(\VP)$ which differ non-perturbatively. By contrast, $\VN$'s Stokes lines do not intersect and $\VN$ is Borel summable: both lateral sums agree, $\cS_+(\VN) = \cS_-(\VN)$, and we simply write this common value as $\VN$. Then Eqs.~\eqref{oddup} and \eqref{odddown} become
\begin{align}
  1 + \cS_+ (\VP) &= i \sqrt{\VN} \label{SpVP} \\
   1 + [\cS_-(\VP)]^{-1} &= -i \sqrt{\VN} \,. \label{SmVP}
\end{align}
Multiplying Eq.~\eqref{SmVP} through by $\cS_-(\VP)$ gives
\begin{equation}
1+\cS_-(\VP) = -i\,\cS_-(\VP)\,\sqrt{\VN}\,.
\label{eq:SmVP_mult}
\end{equation}
Taking the difference between Eq.~\eqref{SpVP} and Eq.~\eqref{eq:SmVP_mult} gives
\begin{equation}
[\cS_+ -\cS_-](\VP) = i\,(1+\cS_-(\VP))\sqrt{\VN} = \cS_-(\VP)\cdot\VN
\,.
\label{eq:DDP_pre}
\end{equation}
where Eq.~\eqref{eq:SmVP_mult} was used again in the second step. Acting with $\cS_-^{-1}$ on both sides then gives the DDP relation for the double well\footnote{ To see that $\cS_-^{-1}[\cS_-(\VP)\cdot\VN] = \VP \VN$ we use a foundational property of resurgence theory: the lateral Borel resummations $\cS_\pm$ are \emph{algebra morphisms}, meaning they preserve products, $\cS_\pm(fg) = \cS_\pm(f)\,\cS_\pm(g)$. This follows because the Borel transform $\mathcal{B}$ converts the ordinary product of formal series into the convolution product~\cite{Sauzin2007}.
}
\begin{equation}
\boxed{
  \SA\,\VP \;=\; \VP\,(1+\VN)
}\,,
  \label{eq:DDP_DW_SA}
\end{equation}
where $\SA = \cS_-^{-1}\cS_+$ is the Stokes automorphism defined in Eq.~\eqref{eq:StokesAuto_def}. Note that $\VN$ appears in Eq.~\eqref{eq:DDP_DW_SA}, not $\sqrt{\VN}$: $\VP$ and $\VN$ are on the same footing as analytic objects on the Riemann surface of the Borel transform, and the $\sqrt{\VN}$ is generated only by the parity quantization condition. In contrast to $\VP$, the Stokes lines of $\VN$ lie at $\arg\hbar = \pm\pi/2$ (set by $S_P^0 = \pi i E$ being purely imaginary), away from the real $\hbar$ axis, which is why $\VN$ is Borel summable:
\begin{equation}
  \SA\,\VN \;=\; \VN\,,
  \label{eq:DDP_DW_VN}
\end{equation}
i.e.\ $(\cS_+ - \cS_-)\,\VN = 0$.

For completeness, we note that with $\im \hbar < 0$ the connection matrix becomes
\begin{equation}
  \fSC' = \fSm\, \fSbr\, S_{21}\, \fSp\, S_{32}\, \fSp\, \fSm\, \fSbr\, S_{43}\, \fSp
\end{equation}
and the quantization condition from $\fSC'=0$ reads
\begin{equation}
  [1 + \cS_-(\VP)]^2 = -\VN\,[\cS_-(\VP)]^2\,,
  \label{quantPN_opposite}
\end{equation}
Combining this equation with Eq.~\eqref{quantPN} with $\VP \to \cS_+(\VP)$ also leads to Eq.~\eqref{eq:DDP_DW_SA}.

\subsection{Period integrals \label{sec:periodintegrals}}
In order to solve the quantization condition, we need to be able to compute the period integrals $\oint dx\,  P_n(E,x)$. At leading order $P_0 = \sqrt{2V(x) - 2E}$ and the integrals can be done in closed form. Let us write
\begin{equation}
    \VP = \exp\left( -\frac{2}{\hbar} S_P \right), \quad
    \VN = \exp\left( -\frac{2}{\hbar} S_N \right)
    \label{VfromS}
\end{equation}
with
\begin{equation}
    S_P = \sum_{n~\text{even}} \hbar^n S_n^P,\quad
    S_N = \sum_{n~\text{even}} \hbar^n S_n^N \,.
\end{equation}
Following convention, we call these ``actions" even though they differ from the classical action by a factor of $i$ stemming from our convention $P=i p$ introduced in Section~\ref{sec:exactWKB}. The leading-order perturbative action is
\begin{equation}
  S_P^0 (E) \equiv -\frac{1}{2}\oint_{\gamma_P} d x P_0(x)=\int_{x_1}^{x_2} d x P_0(x)=
  (\pi i E)\,\,
   {}_2 F_1\!\left( \frac{1}{4},
  \frac{3}{4}, 2 ; 8 E \right)
  \label{SP0}
\end{equation}
where $x_1 = -\sqrt{1+\sqrt{8E}}$ and $x_2 = -\sqrt{1-\sqrt{8E}}$. This hypergeometric function has a convergent expansion at small $E<1/8$, given by
\begin{equation} 
    S_P^0(E) = i \sum_{k=1}^\infty c_k E^k = \pi i \left( E + \frac{3}{4} E^2 + \frac{35}{16} E^3 + \cdots \right) \,,
\end{equation} 
where 
\begin{equation}
    c_k =  \frac{\sqrt{\pi } 2^{k-1} \Gamma \left(2 k-\frac{3}{2}\right)}{\Gamma (k) \Gamma (k+1)}  \, \overset{k \rightarrow \infty}{\sim} \, \frac{8^{k-1}}{\sqrt{2} k^2} \,.
\end{equation}
The leading-order non-perturbative action is
\begin{align}\label{SN0a}
  S_N^0 &\equiv -\frac{1}{2}\oint_{\gamma_N} dx P_0(x) = \int_{x_2}^{x_3} d x P_0(x) = \frac{\pi (1 - 8 E)}{4\sqrt{2}} \, {}_2 F_1\!\left( \frac{1}{4}, \frac{3}{4}, 2 ; 1 - 8 E \right) \\ &= \frac{2}{3} + E \left( \ln \frac{E}{8} - 1 \right) + E^2 \left( \frac{3}{4} \ln \frac{E}{8} + \frac{17}{8} \right) + \cdots
\label{SN0}
\end{align}
where $x_3 = \sqrt{1-\sqrt{8E}}$.

The next period integrals we need involve $P_2$ which is computed using  Eq.~\eqref{P2form};
\begin{equation}
  P_2 (x) = \frac{- 2 x^6 + 3 x^4 - 24 E x^2 + 8 E - 1}{[(x^2 - 1)^2 - 8 E]^{5/ 2}}\,.
\end{equation}
Trying to integrate this directly is more challenging than integrating $P_0$. Even if we could do it, the subsequent integrals get harder and harder. Luckily there is a better way. After some algebra the differential $P_2(x)\,dx$ can be written as
\begin{equation}\label{eq:p2decomp}
  P_2(x)\,dx = \frac{12 E - 1}{16 E (1 - 8 E)} P_0(x)\,dx + \frac{1}{8} \partial_E P_0(x)\,dx +
   d f_2(x)
\end{equation}
where
\begin{equation}
  f_2(x) = x\frac{768 E^3 - (x^2 - 3) (x^2 - 1)^3 - 64 E^2 (18 - 19 x^2 + 3 x^4) + 12 E (13 - 22 x^2 + 14 x^4 - 6 x^6 + x^8)}{768 E (8 E - 1) P_0 (x)^3}
\end{equation}
Integrated over the closed cycle $\gamma_P$ (or $\gamma_N$), the exact form $d f_2$ drops out. So we have simply
\begin{equation}
  S_P^2 = -\tfrac{1}{2}\oint_{\gamma_P} d x\, P_2 (x) = \frac{12 E - 1}{16 E (1 - 8 E)}
  S_P^0 + \frac{1}{8} \partial_E S_P^0
\end{equation}
and
\begin{equation}
  S_N^2 = -\tfrac{1}{2}\oint_{\gamma_N} d x\, P_2 (x) = \frac{12 E - 1}{16 E (1 - 8 E)}
  S_N^0 + \frac{1}{8} \partial_E S_N^0 \,.
\end{equation}
In this way we get the next order without doing any additional elliptic integrals. 

In fact, the same procedure works for all $P_n$. Any $P_n$ can be written as
\begin{equation}
  P_n(x)\,dx = a_n (E) P_0(x)\,dx + b_n (E) \partial_E P_0(x)\,dx + d f_n(x) \,.
\end{equation}
Then
\begin{equation}
    S_P =\sum_{n~\text{even}}\hbar^n\left[ a_n S_P^0 + b_n \partial_E S_P^0\right],\quad
    S_N = \sum_{n~\text{even}}\hbar^n\left[ a_n S_N^0 + b_n \partial_E S_N^0\right]
\end{equation}
with the Voros symbols following from Eq.~\eqref{VfromS}. The reason this is possible is because the elliptic curve is genus one and its cohomology is two dimensional. This was discussed in Section~\ref{subsec:exactWKB} and a proof and algorithm to compute the $a_n$ and $b_n$ are given in Appendix~\ref{appendix:PicardFuchs}. We emphasize that to compute the $a_n$ and $b_n$ and hence to compute $S_P^n$ and $S_N^n$ requires no elliptic integrals, just some algebra. Explicitly,
\begin{align}
  a_0 &= 1, \quad b_0 = 0\\
  a_2 &= \frac{12 E - 1}{16 E (1 - 8 E)}, \quad b_2 = \frac{1}{8} \\
  a_4 &= \frac{-11280 E^3 + 5370 E^2 - 663 E + 28}{7680 E^3 (1 - 8 E)^3}, \quad
  b_4 = \frac{- 7 + 115 E - 360 E^2}{1920 E^2 (1 - 8 E)^2}
\end{align}
and so on.

Inserting the $P_n$ expansions and expanding at small $E$ (anticipating that $E\sim \hbar$) leads to 
\begin{multline}
    S_P = \pi i \Bigg[ \left(E+ \frac{3 }{4}E^2+\frac{35}{16} E^3+\cdots\right)
    +\hbar^2 \left(\frac{1}{16}+\frac{25 }{64}E+\frac{735 E^2}{256}+\cdots\right)\\
    +
    \hbar^4 \left(\frac{175}{2048}+\frac{31185 }{16384}E+\frac{1924923 }{65536}E^2+\cdots\right) \Bigg] \,.
    \label{SPexpanded}
\end{multline}
This is a perfectly sensible expansion where, at each order in $\hbar$, the individual series in $E$ converge for $E<1/8$. Since $E\sim \hbar(N+\frac{1}{2})$ the expansion reduces to a series expansion in $\hbar$. For the non-perturbative action, we find something different. The WKB expansion at small $E$ at each order in $\hbar$ is
\begin{multline}
  S_N = \frac{2}{3} + E \left( \ln \frac{E}{8} - 1 \right) + E^2 \left(
  \frac{3}{4} \ln \frac{E}{8} + \frac{17}{8} \right) + \cdots \\
  + \hbar^2 \left( -\frac{1}{24 E} + \frac{11}{48} + \frac{1}{16} \ln \frac{E}{8} + E \frac{605 + 150 \ln(E/8)}{384} + \cdots \right) \\
   + \hbar^4 \left( \frac{7}{2880 E^3} - \frac{11}{3840 E^2} +
  \frac{101}{15360 E} + \frac{17473 + 3150 \ln(E/8)}{36864} + \cdots
  \right) + \cdots
  \label{SNexpanded}
\end{multline}
The inverse powers of $E$ are problematic. Since $E\sim \hbar$ there are now an infinite number of terms scaling like $\hbar$. This different behavior of $S_P$ and $S_N$ as $E\to 0$ has a direct physical origin.  As $E\to 0$, the two pairs of turning points collide. In this limit the perturbative cycle contracts to a point: this guarantees it has a smooth limit, vanishing as $E\to 0$. There is no such guarantee for $S_N$, which simply goes to a constant.

To handle the anomalous scaling, we note that the singular behavior of the $a_n$ coefficients is
\begin{equation}
  a_{n}(E) \overset{E \rightarrow 0}{\sim}  -\frac{3}{2}\,r_{n} E^{1-n}, \qquad r_{n} = \frac{1-2^{1-n}}{n(n-1)}\,B_{n},
  \label{eq:an_leading}
\end{equation}
where $B_{n}$ are Bernoulli numbers. The $b_n$ are similar and in fact $b_n \sim - a_n E$ at leading order (for $n>2$; for $n=2$, $b_2 \sim -2 a_2 E$). These leading terms can be summed. Indeed, the $r_n$ are precisely the coefficients in the asymptotic expansion of $-\hbar\ln\Gamma(E/\hbar + \frac{1}{2})$ at small $\hbar$. This leads to
 \begin{equation}
  S_N = \frac{2}{3} + \hbar
  \ln\left(\frac{\Gamma(E/\hbar + \tfrac{1}{2})}{\sqrt{2\pi}}\right) - E\ln\frac{8}{\hbar}
  + \cO(\hbar^2).
  \label{eq:logGamma_resum}
\end{equation}
Note that the $\log\Gamma$ simultaneously resums the non-analytic $E\ln E$ piece from the $b_n \partial_E S_N^0$ and all the leading poles in $E$ to every WKB order. The resummed formula can also be derived directly~\cite{BucciottiReisSerone}: in the $E\to 0$ limit the turning points collide and the local Schr\"odinger equation reduces to the Weber (parabolic cylinder) equation, whose WKB Borel transforms are ${}_2F_1$ hypergeometric functions.  The Stokes discontinuities, computed via the hypergeometric connection formula, produce exact $\Gamma$-function prefactors, in particular the $\Gamma(E/\hbar+\frac{1}{2})$ appearing in Eq.~\eqref{eq:logGamma_resum}.  This alternative derivation is sketched in Appendix~\ref{appendix:Weber}.

After resumming the leading terms, we have a consistent expansion of $S_N$ to $\mathcal{O}(\hbar^2)$. At order $\hbar^2$ we again must contend with all $P_n$ contributing at the same order. Unlike the leading-order resummation, which follows from the Weber equation (a local model near the double turning point), the subleading resummation requires the global Picard--Fuchs coefficients $a_{2m}(E)$ and $b_{2m}(E)$.  The subleading pole tower $\tilde s_2(\kappa) = \sum_{m=1}^\infty c_{m+1}\,\kappa^{-2m}$, where $\kappa = N+\tfrac{1}{2}$ is the quantum number (so that $E = \hbar\kappa$ at leading order), collects the subleading Laurent coefficient of $S_N^{(2m)}$ from every WKB order, and has coefficients with a Bernoulli-number structure that can be identified with the asymptotic expansion of the digamma function $\psi(\kappa+\frac{1}{2}) - \ln\kappa$.  The tower resums to
\begin{equation}
  \tilde s_2(\kappa) = \frac{12\kappa^2+1}{16}\left[\psi\!\left(\kappa+\tfrac{1}{2}\right) - \ln\kappa\right] - \frac{1}{32}\,.
  \label{eq:h2_tower_wkb}
\end{equation}
The remainder $r_2$, from expanding $S_N^{(0)}(E(\hbar))$ with the energy correction $E = \hbar\kappa + \hbar^2(-\frac{3}{4}\kappa^2 - \frac{1}{16}) + \cdots$ and extracting the finite part of $\hbar^2\,S_N^{(2)}$, contains the same digamma combination with the opposite sign.  Adding $\tilde s_2 + r_2$, the $\psi - \ln\kappa$ terms cancel algebraically and the full $\hbar^2$ coefficient is simply a polynomial:
\begin{equation}
        S_N = \frac{2}{3} + \hbar \ln \left(\frac{\Gamma(\kappa + \frac{1}{2})}{\sqrt{2\pi}}\right)
    - \hbar \kappa \ln \frac{8}{\hbar}
    + \hbar^2 \left[\frac{17}{8} \kappa^2 + \frac{19}{96} \right]
    + \hbar^3 \left[\frac{125}{32} \kappa^3 + \frac{153}{128}\kappa \right]
    + \mathcal{O}(\hbar^4) \,.
    \label{SNh2}
\end{equation}
At $\cO(\hbar^3)$ the tower involves both digamma and trigamma functions, but after including the energy correction from $E_P\neq\hbar\kappa$, all special functions cancel and the full $\cO(\hbar^3)$ coefficient is again a pure polynomial in $\kappa$ (see Appendix~\ref{appendix:Weber} for the derivation).

%%%%%%%%%%%%%%%%%%%%

\subsection{Energy spectrum}
\label{sec:pertexp}
Finally, we are ready to solve the quantization condition $1 + \VP = i  \sqrt{\VN}$, for the odd modes as in Eq.~\eqref{odddown}, or $1 + \VP = -i  \sqrt{\VN}$ for the even modes, as in Eq.~\eqref{evendown}.  From Eq.~\eqref{SNexpanded} we see that $S_N = \frac{2}{3}$ at $\hbar=0$, so that $\sqrt{\VN} = \exp(-\frac{2}{3\hbar})$. This is the WKB tunneling factor which is exponentially small. So let us define
\begin{equation}
    \lambda
    = e^{-S_I/\hbar} = \exp\left(-\frac{2}{3\hbar}\right)
\end{equation}
where $S_I= \frac{2}{3}$ is the 1-instanton action in our normalization, and work order-by-order in $\lambda$. 

To order $\lambda^0$ we set $\VN=0$. Then the quantization condition reduces to $\VP +1=0$. Note that the $\pm$ plays no role, so there is no splitting between even and odd modes. That is, in the absence of tunneling the even and odd states are exactly degenerate. This is the first non-trivial result about the double-well
spectrum. 

\subsubsection*{Perturbative spectrum}
For the perturbative spectrum we need to solve $\VP = -1$ where $\VP = \exp(-\frac{2}{\hbar} S_P)$ with $S_P$ in Eq.~\eqref{SPexpanded}. The leading terms at small $\hbar$ are
\begin{equation}
  \VP =\exp\left[-\frac{2\pi i}{\hbar}
 \left(E+ \frac{3 }{4}E^2+\frac{35}{16} E^3+\cdots\right)\right] \,.
\end{equation}
To solve $\VP = - 1 = \exp[2 \pi i (N+\frac{1}{2})]$ we must have $E\sim \hbar$. The leading order result is then $E = \hbar \kappa$ with $\kappa = N +\frac{1}{2}$ which is, not surprisingly,  exactly the spectrum of the SHO associated with either minimum. To get higher order corrections in $\hbar$ to the perturbative spectrum, we expand systematically as
\begin{equation}
  E = \hbar \kappa + \hbar^2 e_2 + \hbar^3 e_3 + \cdots \,.
\end{equation}
So we want
\begin{equation}
  \exp \left[ \frac{2 \pi i}{\hbar} \left( -\hbar \kappa -\hbar^2 e_2+\cdots -\frac{3}{4} (\hbar \kappa + \hbar^2 e_2 + \cdots)^2  + \cdots \right) - 2 \pi i
  \frac{\hbar}{16} \left( 1 + \frac{25 }{4} \hbar \kappa + \cdots \right) \right] = - 1
\end{equation}
which means
\begin{equation}
  -\hbar e_2+\cdots -\frac{3}{4\hbar} (\hbar \kappa + \hbar^2 e_2 + \cdots)^2  + \cdots
  - \frac{\hbar}{16} \left( 1 + \frac{25  }{4}\hbar \kappa + \cdots \right) = 0\,.
\end{equation}
Solving order by order leads to 
\begin{equation}
  E_P = \hbar \kappa - \hbar^2 \frac{1 + 12 \kappa^2}{16} - \hbar^3 \frac{19
  \kappa + 68 \kappa^3}{64} - \hbar^4 \frac{131 + 3672 \kappa^2 + 6000
  \kappa^4}{2048} + \cdots \,.
\end{equation}
This is in perfect agreement with the same expansion computed using Rayleigh-Schr{\"o}dinger perturbation theory or the Bender-Wu approach, as discussed in Appendix~\ref{appendix:benderWu} .

\subsubsection*{1-instanton sector}
Next, we want to include non-perturbative corrections which depend on the tunneling factor. This is done by expanding perturbatively in $\lambda = \exp(-S_I/\hbar)$. To do so we write $E = E_P + \lambda E_1 + \lambda^2 E_2 + \cdots$  and solve $1+\VP(E) = \pm i \sqrt{\VN(E)}$ order by order in $\lambda$. To order $\lambda^0$ $\VN(E)=0$ and  $1 + \VP(E) = 0$ giving $E=E_P$.
 
To first order in $\lambda$ we can Taylor expand $\VP$ around $E=E_P$ so that
\begin{align}
  \label{VPEform}
1+\VP(E) &= \lambda \VP'(E_P) E_1 +\cdots=\lambda E_1 \frac{2}{\hbar} S_P'(E_P)+ \cdots\\
&= \lambda E_1 \frac{2\pi i}{\hbar}\left[1+\frac{3 \hbar \kappa }{2}+\frac{1}{64} \hbar^2 \left(348 \kappa ^2+19\right)+\frac{3}{128} \hbar^3 \kappa  \left(1052 \kappa ^2+191\right)+O\left(\hbar^4\right)\right] \,. \nonumber
\end{align}
For $\sqrt{\VN}$, since it starts at order $\lambda$, we can evaluate it at $E=E_P$. Using Eq.~\eqref{eq:logGamma_resum} this gives
	 \begin{equation} 
	   \pm i \sqrt{\VN}
	= \pm i e^{-\frac{1}{\hbar}S_N}
	   =
	   \pm \frac{i \lambda
	   \sqrt{2\pi}}{\Gamma\left(\frac{E_P}{\hbar} + \tfrac{1}{2}\right)}
	   \left(\frac{8}{\hbar}\right)^{E_P/\hbar}\left(1+ \mathcal{O}(\hbar)\right) \label{VNformgamma} \,.
	 \end{equation}
With these expansions, $1+\VP(E_P + \lambda E_1^\pm) = \pm i \sqrt{\VN(E_P )}$ lets us solve for $E_1^\pm$:
	\begin{align}
	  E_1^\pm &=\pm \frac{i \hbar }{2S_P' (E_P)} \frac{\sqrt{2\pi}}{\Gamma\!\left(\frac{E_P}{\hbar} + \tfrac{1}{2}\right)}
	   \left(\frac{8}{\hbar}\right)^{E_P/\hbar}\left(1+ \mathcal{O}(\hbar)\right) \label{E1formula}
	    \\
	   &=\pm \left(
	   \frac{\hbar}{\sqrt{2\pi}}-\frac{3 \hbar^2\kappa }{2\sqrt{2\pi}} +\cdots\right)
	   \frac{1}{\Gamma\!\left(\kappa + \tfrac{1}{2}
	   -\hbar\tfrac{1+12\kappa^2}{16}+\cdots\right)}
	   \left(\frac{8}{\hbar}\right)^{ \kappa -\hbar\tfrac{1+12\kappa^2}{16}+\cdots}
	   \left(1+ \mathcal{O}(\hbar)\right)\,. \nonumber
	\end{align}
    
To go to next order in $\hbar$ still at order $\lambda$, it is cleaner to keep $\sqrt{\VN}$ in exponential form. Writing $\sqrt{\VN(E_P)} = e^{-S_N(E_P)/\hbar} = \lambda\,e^{-(S_N - S_I)/\hbar}$ and equating with Eq.~\eqref{VPEform}, the $\lambda$ factors cancel to give
\begin{equation}
  E_1^{\pm} =\pm  \frac{i \hbar}{2 S_P'(E_P)} e^{-\frac{1}{\hbar} (S_N- S_I) } \,,
\end{equation}
which is exact in $S_N$ (no resummation applied). Relying on Eq.~(\ref{SNh2}) to expand $S_N - S_I$, we have
   \begin{multline}
       E_1^\pm = \pm \frac{\hbar}{\sqrt{2\pi}\Gamma(\kappa+\tfrac{1}{2})} \left(\frac{8}{\hbar}\right)^{\kappa}
       \Bigg[1 - \left(\frac{3\kappa}{2} + \frac{17\kappa^2}{8} + \frac{19}{96}\right)\hbar \\
       + \left(\frac{289}{128}\kappa^4 - \frac{23}{32}\kappa^3 - \frac{2125}{768}\kappa^2 - \frac{115}{128}\kappa - \frac{5111}{18432}\right)\hbar^2 + O(\hbar^3)\Bigg]
   \end{multline}
where $-\tfrac{3}{2}\kappa$ arises from expanding $i\hbar/(2S_P'(E_P))$ and $-\tfrac{17}{8}\kappa^2 - \tfrac{19}{96}$ comes from Eq.~(\ref{SNh2}); the $\hbar^2$ polynomial is obtained from the same expansion using $S_P$ to $\cO(\hbar^2)$ in Eq.~\eqref{SPexpanded} and $S_N$ to $\cO(\hbar^3)$ in Eq.~\eqref{SNh2}, together with the perturbative energy correction $E_P = \hbar\kappa - \hbar^2(\tfrac{3\kappa^2}{4} + \tfrac{1}{16}) + \cdots$. Evaluating at $\kappa=\tfrac{1}{2}$ gives the even/odd ground-state splitting
     \begin{equation}
	    \Delta E_0 = \lambda(E_1^+ - E_1^-)_{\kappa=\frac{1}{2}}
	    = e^{-\frac{2}{3\hbar}}\sqrt{\frac{16\hbar}{\pi}}\left[1-\frac{71 \hbar}{48} - \frac{6299}{4608}\hbar^2 + \mathcal{O}(\hbar^3)\right] \,, \label{eq:splitE0inst}
	\end{equation}
	while for the first excited states we have
	\begin{equation}
	    \Delta E_1 = \lambda(E_1^+ - E_1^-)_{\kappa=\frac{3}{2}}
	    = e^{-\frac{2}{3\hbar}}\frac{32}{\sqrt{\hbar\pi}}\left[1-\frac{347\hbar}{48} + \frac{5317}{4608}\hbar^2 + \mathcal{O}(\hbar^3)\right] \,. \label{eq:splitE1inst}
	\end{equation}

\subsubsection*{2-instanton sector}
The 1-instanton result determines the splitting between even and odd states. To find the common energy shift at $O(\lambda^2)$ we return to the squared quantization condition, Eq.~\eqref{quantPN}:
\begin{equation}
  (1+\VP)^2 + \VN = 0
\end{equation}
and expand $E = E_P + \lambda E_1^\pm + \lambda^2 E_2 + \cdots$. After this shift, $1+\VP(E) = \lambda \VP'(E_P) E^{\pm}_1 + O(\lambda^2)$, and $\VN(E_P) = O(\lambda^2)$. At $O(\lambda^2)$ we recover
\begin{equation}
  (\VP'(E_P))^2\, (E_1^\pm)^2 = -\frac{\VN(E_P)}{\lambda^2}
  \label{eq:E1squared}
\end{equation}
which determines $(E_1^\pm)^2$ and reproduces the 1-instanton result above.

At $O(\lambda^3)$, expanding to the next order and dividing by $E_1^\pm$ gives
\begin{equation}
  2(\VP')^2 E_2 + \VP' \VP'' (E_1^\pm)^2 = -\frac{\VN'(E_P)}{\lambda^2}
  \label{eq:E2eq}
\end{equation}
where primes denote $\partial_E$ and all quantities are evaluated at $E=E_P$. Using Eq.~\eqref{eq:E1squared} and solving for $E_2$ gives:
\begin{equation}
  E_2 = \frac{1}{2}\,(E_1^\pm)^2 \left[\frac{\VN'}{\VN} - \frac{\VP''}{\VP'}\right] \,.
  \label{eq:E2result}
\end{equation}
Since this depends on $(E_1^\pm)^2$,  $E_2$ is the \emph{same for even and odd states}: the 2-instanton sector contributes a common energy shift, not a splitting. More generally, odd instanton sectors ($\lambda^1, \lambda^3, \ldots$) produce splittings while even instanton sectors ($\lambda^2, \lambda^4, \ldots$) produce common shifts.

To evaluate Eq.~\eqref{eq:E2result} at leading order, we use
$\VN'/\VN = -2S_N'/\hbar$ and $\VP''/\VP' = S_P''/S_P' - 2S_P'/\hbar$, giving
\begin{equation}\label{eq:E2fullexpr}
  E_2 = -\frac{(E_1^\pm)^2}{2}\left[\frac{2(S_N' - S_P')}{\hbar} + \frac{S_P''}{S_P'}\right] \,.
\end{equation}
The evaluation of $S'_N$ to leading order is obtained with Eq.~\eqref{eq:logGamma_resum} which gives $S'_N  = \psi(\kappa+\tfrac{1}{2}) - \ln(8/\hbar)+\cdots$. From Eq.~\eqref{SPexpanded}, $S_P'(E_P) = \pi i(1 + \tfrac{3}{2}\hbar\kappa + \cdots)$ and $S_P''(E_P) = \pi i (\tfrac{3}{2}+\tfrac{105}{8}\hbar\kappa + \cdots)$. We then get
\begin{equation}
  E_2 = -\frac{(E_1^\pm)^2}{\hbar}\left[\Sigma + O(\hbar)\right]
  \label{eq:E2leading}
\end{equation}
where
\begin{equation}
  \Sigma \equiv -\ln\frac{8}{\hbar} + \psi\!\left(\kappa + \tfrac{1}{2}\right) - \pi i
  \label{eq:Sigmadef} \,.
\end{equation}
For the ground state ($\kappa = \frac{1}{2}$), using $\psi(1) = -\gamma_E$:
\begin{equation}
  E_2\big|_{\kappa=1/2} = \frac{(E_1^\pm)^2}{\hbar}\left[\ln\frac{8}{\hbar} + \gamma_E + \pi i + O(\hbar)\right] \,.
\end{equation}
Since $E_1$ is real, the term $-\pi i$ in $\Sigma$ gives $E_2$ a positive imaginary part: $\operatorname{Im}(E_2) = +\pi(E_1^\pm)^2/\hbar$. This cancels against the imaginary ambiguity in the lateral Borel resummation of $E_P$, ensuring that the physical energy is real.

\subsubsection*{Higher instanton sectors}
The expansion $E = E_P + \lambda E_1^\pm + \lambda^2 E_2 + \lambda^3 E_3 + \lambda^4 E_4 + \cdots$ can be continued to arbitrary order in $\lambda$ by straightforward Taylor expansion of the quantization condition. At each order $O(\lambda^{n+1})$, the equation is linear in $E_n$ and can be solved algebraically once $E_1, \ldots, E_{n-1}$ are known. Define the log-derivative ratios
\begin{equation}
  R_N \equiv \frac{\VN'}{\VN},  \qquad
  R_P \equiv \frac{\VP''}{\VP'},
  \label{eq:RNRP}
\end{equation}
with all quantities evaluated at $E = E_P$. For example at $O(\lambda^3)$ we have $E_2 = \frac{1}{2}(E_1^\pm)^2\,(R_N - R_P)$ as derived above. At $O(\lambda^4)$, the quantization condition gives $E_3$ as proportional to $(E_1^\pm)^3$:
\begin{equation}
  E_3 = \frac{(E_1^\pm)^3}{24}\left[9R_N^2 - 18\,R_N R_P + 8R_P^2 + 6R_N' - 4R_P'\right] \,,
  \label{eq:E3result}
\end{equation}
where $R_N' = \partial_E R_N$ and $R_P' = \partial_E R_P$. Evaluating at leading order using $R_N = -2S_N'/\hbar$, $R_P = S_P''/S_P' - 2S_P'/\hbar$, $R_N' = -2S_N''/\hbar$, $R_P' = -2S_P''/\hbar + (S_P'''S_P' - S_P''^2)/S_P'^2$, the 3-instanton correction becomes
\begin{equation}
  E_3 = \frac{(E_1^\pm)^3}{6\hbar^2}\left[9\Sigma^2 + \pi^2 - 3\psi'\!\left(\kappa + \tfrac{1}{2}\right) + O(\hbar)\right] \,.
  \label{eq:E3leading}
\end{equation}
Since $E_3$ depends on the sign of $E_1^\pm$, the 3-instanton sector contributes to the even/odd splitting as anticipated.

We can continue the expansion to higher orders. For instance, at $O(\lambda^5)$, one finds $E_4$ as a function of $(E_1^\pm)^2$ and $(E_1^\pm)^4$ only:
\begin{align}
  E_4 &= \frac{(E_1^\pm)^4}{24}\Big[8R_N^3 - 24\,R_N^2 R_P + 22\,R_N R_P^2 - 6R_P^3 \notag\\
  &\qquad + 12\,R_N R_N' - 8\,R_N R_P' - 12\,R_N' R_P + 7\,R_P R_P' + 2R_N'' - R_P''\Big]
  \label{eq:E4result}
\end{align}
where $R_N'' = \partial_E^2 R_N$ and $R_P'' = \partial_E^2 R_P$. At leading order we have:
\begin{equation}
  E_4 = -\frac{(E_1^\pm)^4}{6\hbar^3}\left[16\Sigma^3 + 4\Sigma\pi^2 - 12\Sigma\,\psi'\!\left(\kappa + \tfrac{1}{2}\right) + \psi''\!\left(\kappa + \tfrac{1}{2}\right) + O(\hbar)\right] \,.
  \label{eq:E4leading}
\end{equation}
Since $E_4$ depends only on even powers of $E_1^\pm$, it is independent of the sign and contributes a common energy shift. This confirms the general pattern: odd instanton sectors ($\lambda^1, \lambda^3, \ldots$) produce splittings while even instanton sectors ($\lambda^2, \lambda^4, \ldots$) produce common shifts. 

We emphasize that computing higher instanton corrections (higher powers of $\lambda$) at leading order in $\hbar$ requires only Taylor expansion of the quantization condition followed by elementary algebra. In contrast, going to higher orders in $\hbar$ within a single instanton sector is much harder: because of the $1/E^{2m-1}$ poles in the Picard--Fuchs coefficients $a_{2m}, b_{2m}$, every WKB order $P_{2m}$ contributes at the same $\hbar$ order when $E \sim \hbar$, and one must sum an entire series.

\subsection{Alien calculus constraints on the trans-series}
\label{sec:alien}

In the preceding sections we derived the trans-series $E = E_P + \lambda\, E_1^\pm + \lambda^2\, E_2 + \cdots$ by Taylor-expanding the quantization condition at each instanton order --- the approach taken historically by Zinn-Justin~\cite{ZinnJustin1981} and refined by Zinn-Justin and Jentschura~\cite{ZinnJustinJentschura1,ZinnJustinJentschura2} and independently by \'Alvarez~\cite{Alvarez2004} for higher orders. Such calculations treat each instanton sector independently. However, the different sectors are related. The resurgent structure that connects perturbative and non-perturbative sectors and the $(1/2S_I)^n\,n!$ large-order growth of the perturbative coefficients, was made manifest in this context by Dunne and \"Unsal~\cite{DunneUnsal2014WKB}. A powerful tool for extracting the inter-sector relations systematically is the alien calculus of {\'E}calle~\cite{ecalle1981fonctions1, ecalle1981fonctions2, ecalle1985fonctions3}. Alien calculus converts the DDP relation in Eq.~\eqref{eq:DDP_DW_SA},  $\SA\,\VP = \VP(1+\VN)$, into an infinite tower of constraints relating the different instanton sectors and locating all Borel singularities in a unified framework.

The central objects in alien calculus are the \emph{alien derivatives} $\Delta_n$, introduced in Section~\ref{sec:alien_calc}, which extract the Borel singularity at the $n$-instanton location $t = nS_I$.  While $\Delta_n$ extracts the monodromy around a single singularity, $\SA - 1$ gives the total monodromy around the positive real axis.  The relationship between the two is 
rather elegant~\cite{Dorigoni2014}:
\begin{equation}
 \boxed{
  \SA = \exp \left(\sum_{n \geq 1} \lambda^n 
  % e^{-nS_I/\hbar}
\Delta_n\right)\,.
  \label{eq:Ecalle}
  }
\end{equation}
To appreciate this formula, note that $\Delta_n$ satisfies the Leibniz rule $\Delta_n(fg) = (\Delta_n f)\,g + f\,(\Delta_n g)$, so the alien derivatives are derivations, while $\SA$ is a multiplicative automorphism which is the exponential of a derivation. So this is a Lie group relation: the group is the (infinite-dimensional) group of algebra automorphisms of the resurgent algebra, and its Lie algebra is the space of derivations, spanned by the $\Delta_n$ with weights $\lambda^n = e^{-nS_I/\hbar}$ fixed by the Borel singularity locations. The exponential map takes the derivation $\sum_n \lambda^n \Delta_n$ in the Lie algebra to the Stokes automorphism $\SA$ in the group, and Eq.~\eqref{eq:Ecalle} is simply this exponential relation made explicit.

Expanding the exponential generates iterated alien derivatives. For example, the total discontinuity at $\cO(\lambda^2)$ receives contributions from both $\Delta_2$ (intrinsic double-instanton singularity) and $\tfrac{1}{2}\Delta_1^2$ (two successive single-instanton singularities).  For the Airy function, we only had a single singularity at $t = 2$, so $\Delta_1^2 = 0$ and the exponential truncates: $\SA = 1 + e^{-2\zeta}\Delta_1$. For the double well, we can plug {\'E}calle's formula into the DDP relation $\SA\,\VP = \VP(1+\VN)$ from Eq.~\eqref{eq:DDP_DW_SA} giving
\begin{equation}
\left(1+\lambda \Delta_1 + \frac{1}{2} \lambda^2 \Delta_1^2 + \lambda^2 \Delta_2 + \cdots\right) \VP = 
\VP + \VP \VN \,.
\label{bothsides}
\end{equation}
Matching terms at each order in $\lambda$ will then give nontrivial constraints. The bulk of this section is occupied with computing the alien derivatives $\Delta_n E_m$ sector by sector. Once this is done, we will be able to read off various resurgent relations and verify to a few non-trivial orders that the Borel-resummed spectrum is indeed real.

\subsubsection*{First alien derivative: $\Delta_1 E_n$}

To compute the alien derivatives explicitly, since the object appearing in the quantization condition is $\sqrt{\VN} = e^{-S_N/\hbar}\propto \lambda$ it is helpful to define
\begin{equation}
  \fv(E) \;\equiv\; \frac{\sqrt{\VN(E)}}{\lambda} \;=\; e^{-(S_N(E) - S_I)/\hbar}
  \;=\; \frac{\sqrt{2\pi}}{\Gamma\!\left(\frac{E}{\hbar} + \tfrac{1}{2}\right)}\left(\frac{8}{\hbar}\right)^{\!E/\hbar} + \cdots
  \label{eq:vdef}
\end{equation}
Since $\SA\,\VN = \VN$, the function $\fv(E)$ at \emph{fixed} $E$ is Borel summable along the positive real axis:
\begin{equation}
  \Delta_n[\fv(E)] = 0 \qquad \text{for all } n \geq 1\,.
  \label{eq:VN_Borel_summable}
\end{equation}
However, when $\fv$ is evaluated at $E = E_P$, the chain rule reintroduces a nontrivial alien derivative through the shift in $E_P$:
\begin{equation}
  \Delta_n\bigl[\fv(E_P)\bigr] = \frac{\partial \fv}{\partial E}\bigg|_{E_P}  \Delta_n(E_P)\,.
  \label{eq:chain_rule_v}
\end{equation}
This distinction between fixed-$E$ and on-shell ($E = E_P$) alien derivatives is crucial: any function of $E$ that is Borel summable at fixed~$E$ acquires a nontrivial alien derivative when evaluated at $E_P$, because $E_P$ itself has a Stokes jump.

Since $\VN = \lambda^2 \fv^2$  there is no $\cO(\lambda)$ term on the right-hand side of Eq.~\eqref{bothsides}, so matching at $\cO(\lambda)$ gives
\begin{equation}
 \Delta_1( \VP) = 0\,.
  \label{eq:Delta1_VP}
\end{equation}
Since $\VP = e^{-2S_P/\hbar}$ and $\Delta_1$ is a derivation, we get immediately that $(\Delta_1 S_P)(E) = 0$ as a function of $E$. This is a statement about the period integrals in $S_P(E)$ as an asymptotic series, independent of quantization.

The quantization condition at the perturbative level is $S_P(E_P) = \pi i\kappa\hbar$ with $\kappa = 1/2 + N$. The right-hand side is an exact polynomial in $\hbar$, so it has no Borel singularities and therefore neither does the left side: $\Delta_1[S_P(E_P)] = 0$. To determine $\Delta_1(E_P)$ we can use the chain rule:
\begin{equation}
    \Delta_1[S_P(E_P)] =\frac{\partial S_P}{\partial E}\bigg|_{E_P} \,\Delta_1(E_P) + (\Delta_1 S_P)(E_P)\,,
\end{equation}
where $(\Delta_1 S_P)(E_P)$ denotes the alien derivative $\Delta_1 S_P$ (a function of $E$, computed at fixed $E$) evaluated at $E_P$. The left-hand side and the second term on the right vanish, and since generically $\partial_E S_P \neq 0$ we conclude $\Delta_1(E_P) = 0$. 

This result propagates to all sectors. To see the pattern, consider $E_1$: matching the quantization condition $1+\VP(E) = \pm i\sqrt{\VN(E)}$ at $\cO(\lambda)$ (Eq.~\eqref{VPEform}) gives
\begin{equation}
  E_1^\pm = \pm\frac{i\,\fv(E_P)}{\VP'(E_P)}\,,
\end{equation}
where $\VP'(E_P) = \partial_E\VP|_{E_P}$ arises from Taylor-expanding $\VP$ around $E_P$. Since $\Delta_1$ commutes with $\partial_E$ and annihilates $E_P$, $\VP(E)$, and $\fv(E)$ at fixed $E$, the chain rule gives $\Delta_1[\fv(E_P)] = \Delta_1[\VP'(E_P)] = 0$, and Leibniz then gives $\Delta_1 E_1 = 0$. The same argument iterates at every order: matching at $\cO(\lambda^n)$ expresses $E_n$ algebraically in $E_P$, $\fv$, $\VP$, their $E$-derivatives evaluated at $E_P$, and lower $E_{m<n}$, and $\Delta_1$ annihilates each of these. Thus
\begin{equation}
  \boxed{\Delta_1\, E_n = 0 \quad\text{for all }n\,.}
  \label{eq:Delta1_En}
\end{equation}
The nearest Borel singularity in every sector is therefore at $2S_I = \frac{4}{3}$ (the instanton--anti-instanton action, not the single-instanton action $S_I$), so the asymptotic coefficients of every sector grow at worst as $(1/2S_I)^n\, n!=(3/4)^n\, n!$. This is the first non-trivial result about the double well from the DDP formula and alien calculus.

\subsubsection*{Second alien derivative, perturbative: $\Delta_2 E_P$}
Continuing the matching in Eq.~\eqref{bothsides}, at order $\lambda^2$, since $\Delta_1 \VP=0$ is already established, we get
\begin{equation}
    \Delta_2\VP = \VP \fv^2
      \label{eq:Delta2_VP} \,.
\end{equation}
It then follows that, at fixed $E$, $(\Delta_2 S_P)(E) = -(\hbar/2)\fv^2(E)$. Separately, the quantization condition $S_P(E_P) = \pi i\kappa\hbar$ is an exact polynomial in $\hbar$ (linear in $\hbar$ and in $\kappa$), so $\Delta_2[S_P(E_P)] = 0$. The chain rule for $\Delta_2$ on the composite $S_P(E_P)$ then gives
\begin{equation}
 0 = \Delta_2[S_P(E_P)] = S_P'(E_P)\,\Delta_2(E_P) + (\Delta_2 S_P)(E_P)\,,
\end{equation}
and solving for $\Delta_2(E_P)$ gives
\begin{equation}
  \Delta_2(E_P) = \frac{\hbar\,\fv^2}{2\,S_P'(E_P)}
  \label{eq:Delta2_EP}
\end{equation}
At leading order $S_P'(E_P) = \pi i + \cO(\hbar)$, so
\begin{equation}
    \Delta_2(E_P) = \frac{\hbar}{2\pi i}  \fv^2 + \cO(\hbar^2) \,.
\end{equation}
This tells us that the leading Borel singularity of $E_P$ is at $t = 2S_I$ with residue proportional to $\fv^2$, and correspondingly that the perturbative coefficients of $E_P$ grow at leading order as $(1/2S_I)^n\,n!$.

\subsubsection*{Second alien derivative, non-perturbative: $\Delta_2 E_1$}
The next quantity we want to compute is $\Delta_2(E_1)$. For this, it is useful to introduce a shorthand at this point. Differentiating the quantization condition $S_P(E_P(\kappa,\hbar),\hbar) = \pi i\kappa\hbar$ with respect to $\kappa$ and using the chain rule gives $S_P'(E_P)\cdot\partial_\kappa E_P = \pi i\hbar$. Defining
\begin{equation}
 F \equiv \partial_\kappa E_P = \frac{\pi i\hbar}{S_P'(E_P)}
 \label{Fdef}
\end{equation}
(which at leading order is $F = \hbar + \cO(\hbar^2)$), Eq.~\eqref{eq:Delta2_EP} takes the compact form
\begin{equation}
 \Delta_2(E_P) = \frac{F\fv^2}{2\pi i}\,.
\end{equation}
From Eq.~\eqref{E1formula}, the 1-instanton coefficient is $E_1 = \pm F\fv/(2\pi)$, so the Leibniz rule reduces $\Delta_2(E_1)$ to $\Delta_2(F)$ and $\Delta_2(\fv)$. For $\Delta_2(F)$: since the Borel singularity at $t = 2S_I = \tfrac{4}{3}$ is fixed (independent of the quantum number), $\Delta_2$ and $\partial_\kappa$ commute, $[\Delta_2,\partial_\kappa] = 0$. Combined with Eq.~\eqref{eq:Delta2_EP},
\begin{equation}
  \Delta_2(F) = \Delta_2(\partial_\kappa E_P) = \partial_\kappa\,\Delta_2(E_P) = \frac{1}{2\pi i}\,\partial_\kappa(F\fv^2)\,.
  \label{eq:Delta2_F}
\end{equation}
For $\Delta_2(\fv)$ we note that since $\fv$ is Borel summable at fixed $E$, the on-shell alien derivative is given entirely by the chain rule, Eq.~\eqref{eq:chain_rule_v} at $n=2$, $\Delta_2(\fv) = \partial_E\fv|_{E_P}\,\Delta_2(E_P)$. At leading order, differentiating $\fv = e^{-(S_N - S_I)/\hbar}$ gives $\partial_E\fv = \fv\,[\ln(8/\hbar) - \psi(\kappa+\tfrac12)]/\hbar + \cdots$, so
\begin{equation}\label{eq:Delta2_FV_EP}
  \Delta_2(\fv) = \frac{F\fv^3}{2\pi i\,\hbar}\left[\ln\tfrac{8}{\hbar} - \psi(\kappa+\tfrac12)\right] + \cdots\,.
\end{equation}

With the building blocks $\Delta_2(F)$ and $\Delta_2(\fv)$ in hand, computing $\Delta_2(E_1)$ is a direct application of the Leibniz rule to $E_1 = \pm F\fv/(2\pi)$:
\begin{equation}
  \Delta_2(E_1) = \pm\frac{1}{2\pi}\bigl[\Delta_2(F)\,\fv + F\,\Delta_2(\fv)\bigr].
\end{equation}
Substituting Eqs.~\eqref{eq:Delta2_F} and \eqref{eq:Delta2_FV_EP}:
\begin{align}
  \Delta_2( E_1) &=  \pm\frac{\fv}{4\pi^2 i}\,\partial_\kappa(F\fv^2) \pm  \frac{F^2 \fv^3}{4\pi^2 i \hbar} \left[\log\left(\frac{8}{\hbar}\right) -\psi\left(\kappa+\frac{1}{2}\right)\right]  + \cdots \notag\\
  &= \pm\frac{\fv}{4\pi^2 i}\bigl[(\partial_\kappa F)\,\fv^2 + 2F\fv\,\partial_\kappa\fv\bigr] \pm  \frac{F^2 \fv^3}{4\pi^2 i \hbar} \left[\log\left(\frac{8}{\hbar}\right) -\psi\left(\kappa+\frac{1}{2}\right)\right]  + \cdots\,.
  \label{eq:Delta2_E1}
\end{align}
From Eq.~\eqref{eq:vdef}, $\partial_\kappa\ln\fv = -\psi(\kappa+\tfrac{1}{2}) + \ln(8/\hbar) + \cO(\hbar)$.  Since $F = \hbar + \cO(\kappa \hbar^2)$ and $\partial_\kappa F = \cO(\hbar^2)$, the first term in the bracket is subleading. The leading-order result is then simply
\begin{equation}
  \Delta_2(E_1) = \pm\frac{3\hbar\,\fv^3(-\psi(\kappa+\frac{1}{2}) + \ln\frac{8}{\hbar})}{4\pi^2 i} + \cO(\hbar^2)\,.
  \label{eq:Delta2_E1_LO}
\end{equation}

\subsubsection*{Second alien derivative, higher non-perturbative: $\Delta_2 E_2$}
Next we compute $\Delta_2(E_2)$. From Eq.~\eqref{eq:E2fullexpr}, $E_2 = -(E_1^\pm)^2[2(S_N'-S_P')/\hbar + S_P''/S_P']/2$ with all derivatives evaluated at $E_P$. Applying the Leibniz rule requires, besides the already-computed $\Delta_2(E_1)$, three more inputs: $\Delta_2[S_N'(E_P)]$, $\Delta_2[S_P'(E_P)]$, and $\Delta_2[S_P''(E_P)]$. We compute each via the chain rule.

For $\Delta_2[S_N'(E_P)]$: since $\fv$ is Borel summable at fixed $E$, so is $S_N'$; hence $\Delta_2 S_N'|_E = 0$ and only the chain-rule term survives,
\begin{equation}
    \Delta_2[S_N'(E_P)] = S_N''(E_P)\,\Delta_2(E_P) = \frac{\fv^2}{2\pi i}\,\psi'(\kappa+\tfrac12) + \cdots,
\end{equation}
using $S_N''(E_P) = \psi'(\kappa+\tfrac12)/\hbar + \cO(1)$ at leading order.

For $\Delta_2[S_P'(E_P)]$: both chain-rule terms contribute,
\begin{equation}
    \Delta_2[S_P'(E_P)] = S_P''(E_P)\,\Delta_2(E_P) + \Delta_2 S_P'|_E = -\fv^2\left[\ln\tfrac{8}{\hbar} - \psi(\kappa+\tfrac12)\right] + \cdots,
\end{equation}
using $S_P''(E_P) = 3\pi i/2 + \cO(\hbar)$ and $\Delta_2 S_P'|_E = -\partial_E(\hbar\fv^2/2)$ (consistent with $\Delta_2 S_P = -\hbar\fv^2/2$); the $S_P''\,\Delta_2 E_P$ term is dropped as subleading in $\hbar$.

For $\Delta_2[S_P''(E_P)]$: a similar chain-rule computation gives
\begin{equation}
    \Delta_2[S_P''(E_P)] = -\frac{\fv^2}{\hbar}\left(2\left[\ln\tfrac{8}{\hbar} - \psi(\kappa+\tfrac12)\right]^2 - \psi'(\kappa+\tfrac12)\right) + \cdots.
\end{equation}

With these inputs (and $\Delta_2(E_1^2) = 2E_1\Delta_2(E_1)$), the Leibniz rule applied to Eq.~\eqref{eq:E2fullexpr} gives
\begin{equation}
    \Delta_2(E_2) = - \frac{2 E_1 \Delta_2(E_1) \Sigma}{\hbar } - \frac{E_1^2}{\hbar} \left(\Delta_2 S_N' - \Delta_2 S_P' \right) - \frac{E_1^2}{2} \left(\frac{S_P' \Delta_2 S_P'' - S_P'' \Delta_2 S_P'}{(S_P')^2} \right) \,,
\end{equation}
where we used $E_2 = -(E_1)^2\Sigma/\hbar$ in the first term (Eq.~\eqref{eq:E2leading}). Each term is $\cO(\hbar)$ except $S_P''\Delta_2 S_P'$, which is $\cO(\hbar^2)$. Substituting the leading-order expressions, using the definition of $\Sigma$ from Eq.~\eqref{eq:Sigmadef}, and simplifying:
\begin{equation}
    \Delta_2(E_2) = \frac{i \hbar \fv^4}{4 \pi^3} \left[ -4\left(\log\tfrac{8}{\hbar} -\psi\left(\kappa+\tfrac{1}{2}\right)\right)^2  - 2i \pi \left(\log \tfrac{8}{\hbar} -\psi\left(\kappa+\tfrac{1}{2}\right)\right) + \psi'\left(\kappa+\tfrac{1}{2}\right)\right] + \cO(\hbar^2) \,. \label{eq:Delta2_E2}
\end{equation}

\subsubsection*{Higher alien derivatives}
We now turn to the 4-instanton Borel singularity at $t = 4S_I$, which is the \emph{subleading} Borel singularity of the perturbative series. To locate it we compute $\Delta_4(E_P)$.

Since $\VN = \lambda^2 \fv^2$ is $\cO(\lambda^2)$, the right-hand side of the DDP formula~\eqref{eq:DDP_DW_SA} has no contribution at $\cO(\lambda^4)$. Expanding the left-hand side via {\'E}calle's formula~\eqref{eq:Ecalle} (with $\Delta_1 = 0$, only even alien derivatives contribute), the $\cO(\lambda^4)$ terms are the direct $\Delta_4$ and the iterated $\tfrac{1}{2}(\Delta_2)^2$ from the quadratic term of the exponential:
\begin{equation}
  \Delta_4\,\VP + \tfrac{1}{2}\,(\Delta_2)^2\,\VP = 0\,.
  \label{eq:Delta4_DDP}
\end{equation}
So $\Delta_4 V_P$ is determined by $(\Delta_2)^2 V_P$. Using $\Delta_2\,\VP = \VP\,\fv^2$ (Eq.~\eqref{eq:Delta2_VP}) and $\Delta_2 \fv|_E = 0$ (Eq.~\eqref{eq:VN_Borel_summable}):
\begin{equation}
    (\Delta_2)^2\,\VP = \Delta_2(\VP\,\fv^2)= (\Delta_2\VP)\,\fv^2 =  \VP\,\fv^4\,,
\end{equation}
giving $\Delta_4\,\VP = -\tfrac{1}{2}\,\VP\,\fv^4$ at fixed $E$. Converting to $S_P$ via $\VP = e^{-2S_P/\hbar}$ gives $\Delta_4 S_P|_E = \hbar\fv^4/4$, and the chain rule with $\Delta_4[S_P(E_P)] = 0$ (the quantization condition has no Borel singularity) gives
\begin{equation}
  \Delta_4(E_P) = -\frac{\Delta_4 S_P|_E}{S_P'(E_P)} = \frac{i \hbar \,\fv^4}{4\pi} +\mathcal{O}(\hbar^2)\,.
  \label{eq:Delta4_EP}
\end{equation}
The nonzero (and purely imaginary) $\Delta_4(E_P)$ locates the subleading Borel singularity of $E_P$ at $t = 4S_I = \tfrac{8}{3}$, giving corrections that fall off as $(3/8)^n$ relative to the leading $(3/4)^n$ from $\Delta_2(E_P)$.

For the imaginary-part cancellation at $\cO(\lambda^4)$ below, we will also need the \emph{on-shell} value of $(\Delta_2)^2(E_P)$, which is the square-bracket version when $\Delta_2$ acts with the chain rule rather than at fixed $E$. Applying $\Delta_2$ to the on-shell expression $\Delta_2(E_P) = F\fv^2/(2\pi i)$ via Leibniz, and using Eqs.~\eqref{eq:Delta2_F}, \eqref{eq:Delta2_FV_EP}:
\begin{equation}
  (\Delta_2)^2(E_P) = -\frac{\fv^2\,\partial_\kappa(F\fv^2)}{4\pi^2} - \frac{F^2\fv^4}{2\pi^2\hbar}\left[\log\left(\frac{8}{\hbar}\right) -\psi\left(\kappa+\frac{1}{2}\right)\right]\,.
  \label{eq:Delta2sq_EP_corrected}
\end{equation}
In contrast to $\Delta_4(E_P)$, this is purely real. Also note that $\Delta_2(E_2)$ (Eq.~\eqref{eq:Delta2_E2}) locates the \emph{leading} singularity of $E_2$ at $t = 2S_I$, confirming that $E_2$'s coefficients diverge at the same $(3/4)^n\,n!$ rate as $E_P$'s.

\subsubsection*{Cancellation of imaginary parts}
Reality of the physical energy requires the imaginary parts to cancel at each order in $\lambda$.  At $\cO(\lambda^2)$, two terms contribute, from Eq.~\eqref{eq:Delta2_EP} and Eq.~\eqref{eq:E2leading}:
\begin{align}
 \tfrac{1}{2}\operatorname{Im}\bigl[\Delta_2(E_P)\bigr]
    &= -\frac{F \fv^2}{4\pi} = -\frac{\hbar \fv^2}{4\pi} + \cdots\,,
  \notag\\[3pt]
  \operatorname{Im}\bigl[E_2\bigr]
    &= \frac{E_1^2}{\hbar} \pi  = \frac{\hbar \fv^2}{4\pi} + \cdots \,,
  \label{eq:ImSum_lambda2}
\end{align}
where we used $F = \hbar + O(\hbar^2)$ and $E_1 = F \fv /2\pi$. The imaginary part of $E_2$ indeed cancels against the imaginary part incurred in the Borel resummation of the leading singularity in $E_P$.

At $\cO(\lambda^4)$, three terms contribute, from Eq.~\eqref{eq:Delta4_EP}, Eq.~\eqref{eq:Delta2_E2} and Eq.~\eqref{eq:E4leading}.  Writing $\Sigma_R \equiv \psi(\kappa+\tfrac{1}{2}) - \ln(8/\hbar)$ (the real part of $\Sigma$), they are 
\begin{align}
  \tfrac{1}{2}\operatorname{Im}\bigl[\Delta_4(E_P)\bigr]
    &= \frac{\hbar \fv^4}{8\pi^3}\pi^2+\cdots\,,
  \notag\\[3pt]
  \tfrac{1}{2}\operatorname{Im}\bigl[\Delta_2(E_2)\bigr]
    &= \frac{\hbar \fv^4}{8\pi^3}\left(\psi' - 4 \Sigma_R^2 \right)+\cdots \,,
  \notag\\[3pt]
  \operatorname{Im}\bigl[E_4\bigr]
    &= \frac{\hbar \fv^4}{8\pi^3}\bigl(4\Sigma_R^2 - \pi^2 - \psi'\bigr) +\cdots\,.
  \label{eq:ImSum_lambda4}
\end{align}
  Their sum vanishes identically, confirming reality of $E$ within the formal Borel framework. The same cancellation will occur at all higher orders in $\lambda$.

\subsubsection*{All-orders structure}
The results above generalize to all $\Delta_{2n}$ and $E_n$. First, the bridge pattern: $\Delta_2$ maps each $E_n$ to something proportional to $\fv^{n+2}$, connecting it to the $(n+2)$-instanton level:
\begin{equation}
  \Delta_2\, E_n \;\propto\; \fv^{n+2} \;\longleftrightarrow\; E_{n+2}\,,\qquad \Delta_1\, E_n = 0 \quad\text{for all }n\,.
  \label{eq:bridge}
\end{equation}
The even sublattice ($E_P, E_2, E_4, \ldots$) and odd sublattice ($E_1, E_3, E_5, \ldots$) are completely disconnected: no sequence of alien derivatives bridges them. This is the resurgent explanation for why the level splitting, which lives entirely in the odd sublattice, is invisible to the large-order behavior of perturbation theory.

Second, the general formula for $\Delta_{2n} V_P$. For $n \geq 2$, expanding the DDP formula to $\cO(\lambda^{2n})$ gives the recursion
\begin{equation}
  \Delta_{2n}\,\VP + \frac{1}{2}\sum_{\substack{j+k = n \\ j,k \geq 1}} \Delta_{2j}\,\Delta_{2k}\,\VP + \cdots = 0\,.
\end{equation}
Since $\Delta_{2n}(\fv) = 0$ at fixed $E$ (Eq.~\eqref{eq:VN_Borel_summable}) and $\Delta_2 \VP= \VP \fv^2$, repeated application of the Leibniz rule gives $(\Delta_2)^n\,\VP = \VP\, \fv^{2n}$, and the combinatorics of the exponential reduce to
\begin{equation}
  \Delta_{2n}\,\VP = \frac{(-1)^{n+1}}{n}\,\VP\, \fv^{2n}\,,
  \label{eq:Delta2n_VP}
\end{equation}
which for $n = 1$ reproduces $\Delta_2\VP = \VP\,\fv^2$ and for $n = 2$ reproduces $\Delta_4\VP = -\tfrac{1}{2}\VP\,\fv^4$.  Converting to $E_P$ using $V_P= e^{-2S_P/\hbar}$ and the chain rule from $\Delta_{2n}[S_P(E_P)]=0$:
\begin{equation}
  \Delta_{2n}(E_P) = \frac{(-1)^{n+1}}{2\pi i\, n}\,F\, \fv^{2n}\,.
  \label{eq:Delta2n_EP}
\end{equation}
This result exhibits the rich structure within the perturbative series. All even alien derivatives act non-trivially on $E_P$, giving Borel poles at every $t= 2n S_I$. Since $[\Delta_{2n},\,\partial_\kappa] = 0$, the analog of Eq.~\eqref{eq:Delta2_F} is $\Delta_{2n}(F) = (-1)^{n+1}\partial_\kappa(F\fv^{2n})/(2\pi i\, n)$, and Eq.~\eqref{eq:Delta2n_EP} resums into a remarkably compact form:
\begin{equation}
  \boxed{\;\sum_{n=1}^{\infty} \lambda^{2n}\,\Delta_{2n}(E_P) \;=\; \frac{F}{2\pi i}\,\ln(1 + \VN)\,.\;}
  \label{eq:alien_resum}
\end{equation}
This is arguably the central result of the section. It packages \emph{every} Borel singularity of $E_P$, residues at all $t = 2nS_I$, $n \geq 1$, into a single logarithm of the local object $\VN(E) = e^{-2S_N(E)/\hbar}$. 

One might hope to exponentiate Eq.~\eqref{eq:alien_resum} via \'Ecalle's bridge $\SA = \exp(\sum_n\lambda^n\Delta_n)$ and obtain a closed form for the total Stokes discontinuity $(\SA-1)E_P$, the ambiguity between the lateral Borel resummations $\cS_\pm E_P$. But $D = \sum_n\lambda^n\Delta_n$ is a derivation, not a function: expanding $\SA = 1 + D + \tfrac12 D^2 + \cdots$ gives $(\SA-1)E_P = D E_P + \tfrac12 D^2 E_P + \cdots$, where $D^k E_P$ means applying $D$ iteratively, not powers of $D E_P$. The resulting terms from iterated derivations drag in iterated alien derivatives such as $\Delta_{2m}\Delta_{2n}E_P$, carrying data ($\partial_\kappa F$, $\partial_E\fv$, $\psi(\kappa+\tfrac12)$, $\ln\hbar$) that does not repackage as a function of $\VN$ alone. Only at leading order in $\VN$ (equivalently $\cO(\lambda^2)$) does the iteration drop out, leaving
\begin{equation}
  (\SA - 1)\,E_P \;=\; \frac{\hbar\,\VN}{2\,S_P'(E_P)} + \cO(\VN^2)\,,
  \label{eq:ImEP_alien}
\end{equation}
which is the imaginary ambiguity of the resummed perturbative series at leading instanton order.

\subsection{Summary of the double well trans-series}
\label{sec:WKBsummary}
The Exact WKB quantization condition determines the energy levels of the double-well potential as a trans-series in the tunneling factor $\lambda = e^{-S_I/\hbar}$ where $S_I=2/3$:
\begin{equation}
  E = E_P + \lambda\, E_1^\pm + \lambda^2\, E_2 + \lambda^3\, E_3^\pm + \lambda^4\, E_4 + \cdots
  \label{eq:transseries_summary}
\end{equation}
where $\kappa = N + \frac{1}{2}$ for the $N$-th energy level, and the $\pm$ distinguishes even and odd parity states. Odd instanton sectors ($E_1, E_3, \ldots$) are proportional to odd powers of $E_1^\pm$ and produce splittings between even and odd states. Even instanton sectors ($E_2, E_4, \ldots$) depend on even powers of $E_1^\pm$ and produce common energy shifts. The first few sectors are:
\begin{align}
  E_P &= \hbar \kappa - \hbar^2 \frac{1 + 12 \kappa^2}{16} - \hbar^3 \frac{19
  \kappa + 68 \kappa^3}{64} - \hbar^4 \frac{131 + 3672 \kappa^2 + 6000
  \kappa^4}{2048} + \cdots
  \label{eq:EP_summary}\\[6pt]
  E_1^\pm &= \pm \frac{\hbar\,\fv}{2\pi}\,\bigl(1 + O(\hbar)\bigr)
  \label{eq:E1_summary}\\[6pt]
  E_2 &= -\frac{\hbar\,\fv^2}{4\pi^2}\,\Sigma\,\bigl(1 + O(\hbar)\bigr)
  \label{eq:E2_summary}\\[6pt]
  E_3^\pm &= \pm \frac{\hbar\,\fv^3}{48\pi^3}\left[9\Sigma^2 + \pi^2 - 3\psi'\!\left(\kappa + \tfrac{1}{2}\right)\right]\bigl(1 + O(\hbar)\bigr)
  \label{eq:E3_summary}\\[6pt]
  E_4 &= -\frac{\hbar\,\fv^4}{96\pi^4}\left[16\Sigma^3 + 4\Sigma\pi^2 - 12\Sigma\,\psi'\!\left(\kappa + \tfrac{1}{2}\right) + \psi''\!\left(\kappa + \tfrac{1}{2}\right)\right]\bigl(1 + O(\hbar)\bigr)
  \label{eq:E4_summary}
\end{align}
where 
\begin{equation}
  \fv(E_P) \;=\; \frac{\sqrt{\VN(E_P)}}{\lambda} \;=\; \frac{\sqrt{2\pi}}{\Gamma\!\left(\kappa+\tfrac{1}{2}\right)}\left(\frac{8}{\hbar}\right)^{\!\kappa}\bigl(1 + O(\hbar)\bigr)\,,
  \label{eq:fv_summary}
\end{equation}
is the perturbative fluctuation factor evaluated at $E = E_P$, and
\begin{equation}
  \Sigma \;\equiv\; \Sigma_R - i \pi\,, 
  \quad
  \Sigma_R = -\ln\left(\frac{8}{\hbar}\right) + \psi\left(\kappa+\frac{1}{2}\right)  
\end{equation}
with $\psi(z) = \Gamma'(z)/\Gamma(z)$ the digamma function.

Alien calculus (Section~\ref{sec:alien}) reveals the resurgent structure underlying these series. The alien derivative $\Delta_n$ probes the Borel singularity at $t = nS_I$, extracting the connection between sectors. A key result is
\begin{equation}
  \Delta_m\, E_n = 0 \quad\text{for}~m~\text{odd and all}~n\,.
\end{equation}
This tells us that the even sublattice ($E_P, E_2, E_4, \ldots$) and odd sublattice ($E_1, E_3, E_5, \ldots$) are completely disconnected: no sequence of alien derivatives bridges them. Every sector can have its nearest Borel singularity at most at $2S_I = \frac{4}{3}$ (the instanton--anti-instanton action), so all sectors diverge at most as $(3/4)^n\,n!$.

The perturbative series has even alien derivatives
\begin{equation}
  \Delta_{2n}(E_P) = \frac{(-1)^{n+1}}{2\pi i\, n}\,F\, \fv^{2n}\,.
\end{equation}
where $F \equiv \partial_\kappa E_P$. This means it has singularities in the Borel plane at $2nS_I$ for every $n \in {\mathbb Z}^+$. The first nonzero alien derivative $\Delta_2 E_P = \frac{1}{2\pi i} F \fv^2$ (Eq.~\eqref{eq:Delta2_EP}) gives the leading imaginary part of the Borel resummed $E_P$; this is the imaginary part of $E_2$ in the trans-series. Alien derivatives of the series of the non-perturbative corrections to the energy can also be computed systematically:
\begin{align}
  \Delta_2(E_1) &= \mp \frac{3\hbar\,\fv^3  \Sigma_R}{4\pi^2 i}+ \cdots\,,
  \\[4pt]
  \Delta_2(E_2) &= \frac{i \hbar \fv^4}{4 \pi^3} \left[ -4  \Sigma_R^2  + 2i \pi \Sigma_R + \psi'\left(\kappa+\tfrac{1}{2}\right)\right]  + \cdots
  \\[4pt]
  (\Delta_2)^2(E_P) &= -\frac{\fv^2\,\partial_\kappa(F\fv^2)}{4\pi^2} + \frac{\fv^2 F \,  \Sigma_R}{2\pi^2} + \cdots\,,
  \\[4pt]
  \Delta_4(E_P) &= -\frac{F\,\fv^4}{4\pi i} + \cdots\,.
\end{align}
Non-trivial checks of the formalism are the cancellations of the imaginary parts of the energy at each order in $\lambda$, which guarantee that the physical spectrum is real. At $\cO(\lambda^2)$, the imaginary part of $E_2$ cancels against the imaginary ambiguity of the Borel resummation of $E_P$:
\begin{equation}
  \operatorname{Im}[E_2] \;+\; \tfrac{1}{2}\operatorname{Im}[\Delta_2 E_P] \;=\; \frac{\hbar\,\fv^2}{4\pi} \;-\; \frac{\hbar\,\fv^2}{4\pi} \;=\; 0\,.
\end{equation}
At $\cO(\lambda^4)$, three contributions must conspire to cancel (with $(\Delta_2)^2 E_P$ purely real at leading order and hence not contributing):
\begin{equation}
  \operatorname{Im}[E_4] \;+\; \tfrac{1}{2}\operatorname{Im}[\Delta_2 E_2] \;+\; \tfrac{1}{2}\operatorname{Im}[\Delta_4 E_P] \;=\; \frac{\hbar\,\fv^4}{8\pi^3}\bigl[(4\Sigma_R^2 - \pi^2 - \psi') + (\psi' - 4\Sigma_R^2) + \pi^2\bigr] \;=\; 0\,.
\end{equation}

%% file: sections/path_integral.tex
% \!TEX root = ../DoubleDoubleMain.tex

%%%%%%%%%%%%%%%%%%%%%%%%%%%%%%%%%%%%%%%%%%%%%%%%%%%%%%%%%%%%%%%%%%%%%%%%%%%%%%%
%%%%%%%%%%%%%%%%%%%%%.    PATH INTEGRAL APPROACH.    %%%%%%%%%%%%%%%%%%
%%%%%%%%%%%%%%%%%%%%%%%%%%%%%%%%%%%%%%%%%%%%%%%%%%%%%%%%%%%%%%%%%%%%%%%%%%%%%%%

\section{Euclidean Path Integral}
\label{sec:pathintegral}  
    
The Exact WKB approach is a systematic procedure to compute the energy levels of the double well.  The quantization condition is encoded in the relation $1+\VP = \pm i\sqrt{\VN}$ where $\VP$ and $\VN$ are the Voros symbols given by period integrals over the quantum action. Solving this quantization condition perturbatively leads to a trans-series for the energies. Although Exact WKB can in principle be used to compute the energies to arbitrary order, the quantization condition makes it hard to see the origin of the resurgent structure. An alternative approach to computing the same trans-series involves the Euclidean Feynman path integral. With the path integral, each term in the trans-series for the partition function $Z$ comes from a perturbative calculation around a different saddle. One can then see the resurgent structure in the partition function directly, and independent of the energy quantization step. Although the resurgent structure is easier to see, a drawback of the path integral approach is that one must work at finite $T$ where elliptic functions abound. The path integral thereby gives a complementary view of the double well compared with Exact WKB.

\subsection{Partition function basics}
The energy levels in quantum mechanics are encoded in the partition function
\begin{equation} 
  Z  =  \sum_N e^{- T E^{(N)}(\hbar)/\hbar}
  = \sum_N \langle N | e^{- H T/\hbar} | N \rangle = \Tr (e^{- H T/\hbar}) \,.
  \label{ZENdef}
\end{equation}
For example, the ground state energy can be extracted with
\begin{equation}
  E^{(0)} = -\hbar \lim_{T \rightarrow \infty} \frac{1}{T} \ln Z \label{E0Z} \,.
\end{equation}
Since the symmetric double well is invariant under parity $\cP: x\to -x$, the energy eigenstates are parity eigenstates. They split into even and odd parity pairs $E^{(N)}_\pm$. To extract the splitting it is helpful to introduce the twisted partition function
\begin{equation}
  \widetilde Z = \Tr \left( \mathcal{P} e^{- H T/\hbar} \right)
  = \sum_N \left( e^{-E^{(N)}_+ T/\hbar} - e^{-E^{(N)}_- T/\hbar} \right)\,.
\end{equation}
Then, at small energy differences,
\begin{equation}
   \widetilde Z = - \frac{1}{\hbar} \sum_N e^{-E^{(N)}_+ T/\hbar}(E^{(N)}_+-E^{(N)}_-)T + \cdots
\end{equation}
which allows us to extract the splittings. To compute $Z$ and $\widetilde Z$ with the path integral we begin with the Euclidean propagator
\begin{equation}
    \langle b | e^{-H T/\hbar} |a \rangle = \mathcal{N} \int_{x (0) = a}^{x(T) = b} \mathcal{D} x e^{- S [x]/\hbar}\,.
\end{equation}
Taking the trace in position space gives
\begin{equation}
  Z = \int d a \langle a | e^{- H T/\hbar} | a \rangle
  = \cN \int_{x (0) = x (T)} \!\!\!\!\!\!\mathcal{D} x e^{- S [x]/\hbar}\,,
\end{equation}
while the twisted trace gives
\begin{equation}
  \widetilde Z
  = \int d a \langle -a | e^{- H T/\hbar} | a \rangle
  = \cN \int_{x (T) = -x (0)}\!\!\!\!\!\! \mathcal{D} x e^{- S [x]/\hbar}\,.
\end{equation}
So $Z$ sums periodic paths and $\widetilde Z$ sums anti-periodic paths in Euclidean time.

\subsubsection*{From $E$ trans-series to $Z$ trans-series}
\label{sec:Transseries_E_to_Z}
Both the partition functions and the energies are trans-series in $\hbar$. Using $\lambda = e^{-S_I/\hbar}$ with $S_I=2/3$ as in Exact WKB, we can write these trans-series as
\begin{equation}
E^{(N)}(\hbar)=\sum_{k\ge0}\lambda^k E_k^{(N)}(\hbar),
\qquad
  Z(T,\hbar)=\sum_{k\ge0}\lambda^k Z_k(T,\hbar)\,.
\end{equation}
Each $E_k^{(N)}(\hbar)$ is an asymptotic series in $\hbar$ and $\ln \hbar$. Throughout the path-integral analysis we use $E_k^{(N)}$ with the subscript $k$ denoting the instanton sector (matching the convention of Section~\ref{sec:exactWKB}) and the parenthesized superscript $(N)$ the state index. We drop the state superscript when working with the ground state, so $E_k\equiv E_k^{(0)}$. Expanding the exponential in the small $\lambda$ regime, using Eq.~\eqref{ZENdef} and matching powers of $\lambda$ gives
\begin{align}   \label{eq:Zk_from_Ek}
  Z_0 &= \sum_N e^{-T E_0^{(N)}/\hbar}, \\
  Z_1 &= \sum_N e^{-T E_0^{(N)}/\hbar}\!\left[-\frac{T}{\hbar} E_1^{(N)}\right], \\  \label{eq:Zk_from_Ek2}
  Z_2 &= \sum_N e^{-T E_0^{(N)}/\hbar}\!\left[-\frac{T}{\hbar} E_2^{(N)}+\frac{T^2}{2\hbar^2}\bigl(E_1^{(N)}\bigr)^2\right],
  \\ \label{eq:Zk_from_Ek3}
  Z_3 &= \sum_N e^{-T E_0^{(N)}/\hbar}\!\left[-\frac{T}{\hbar} E_3^{(N)}+\frac{T^2}{\hbar^2}E_1^{(N)}E_2^{(N)}-\frac{T^3}{6\hbar^3}\bigl(E_1^{(N)}\bigr)^3\right]
  \\ \label{eq:Zk_from_Ek4}
  Z_4 &= \sum_N e^{-T E_0^{(N)}/\hbar}\!\left[-\frac{T}{\hbar} E_4^{(N)}+\frac{T^2}{\hbar^2}\left(E_1^{(N)}E_3^{(N)} + \frac{1}{2} \bigl(E_2^{(N)}\bigr)^2\right)-\frac{T^3}{2\hbar^3}\bigl(E_1^{(N)}\bigr)^2 E_2^{(N)} + \frac{T^4}{24\hbar^4}\bigl(E_1^{(N)}\bigr)^4\right]\,.
\end{align}
These formulas are misleadingly simple. Because $E_0^{(N)}$ is a series in $\hbar$, the series for each $Z_k(T,\hbar)$ will actually mix different perturbative orders in the energies as well as different elements of the trans-series. In addition, due to parity invariance, $Z_{2m+1}=0$ (in particular $Z_1=0$), consistent with the absence of odd-instanton periodic saddles. This does \emph{not} imply vanishing one-instanton energy shifts in each parity sector, but it does imply that one-instanton energy shifts come with opposite signs for pairs of even and odd energy states. Indeed $E_1^{(N)}$ is not zero, and in fact its square shows up in the $T^2$ term of $Z_2$. As we will see, $E_1^{(N)}$ is easier to extract from the twisted partition function $\widetilde Z$ where it appears at order $\lambda$ without mixing with other orders.

Another method for extracting the energies from the partition function is through its Laplace transform, called the {\bf resolvent}:
\begin{equation} \label{eq:resolvent_def}
  G (E) = \frac{1}{\hbar}\int_0^{\infty} d T\, e^{E T/\hbar} Z = \sum_N \frac{1}{E^{(N)} - E} = \Tr
  \frac{1}{H - E} = - \frac{\partial}{\partial E} \ln D (E)
\end{equation}
where the Fredholm (spectral) determinant is $D (E) = \det (H - E)$. The energies $E^{(N)}$
are then the poles of $G (E)$ or zeros of $D (E)$.

\subsubsection*{Path integral basis and the SHO}
The normalization $\cN$ of the path integral is formally infinite. But it is a property of the measure, independent of the action, and we can therefore fix it in the free theory and apply it everywhere. Fourier modes provide a convenient basis for periodic paths:
\begin{equation}
  x(t) =c_0 + \sum_{n\in\mathbb{Z}\setminus\{0\}} c_n \psi_n,\quad \psi_n(t) = \frac{1}{\sqrt{T}} e^{i \omega_n t},\qquad
  %  \psi_n(t)=\frac{1}{\sqrt{T}}\,e^{i\omega_n t},
  %  \quad
   \omega_n=\frac{2\pi n}{T}
  \label{normalmodes} 
\end{equation}
Reality means $c_0\in\mathbb{R}$ and $c_{-n}=c_n^*$ for $n\ge 1$. These modes are eigenfunctions of $-\partial_t^2$ with eigenvalues $\omega_n^2$, so the free action is diagonal:
\begin{equation}
  S_0[x]=\int_0^T\! dt\,\frac{1}{2}\dot x^2 =\frac{1}{2}\sum_{n\in\mathbb{Z}\setminus\{0\}}\omega_n^2\,|c_n|^2\,.
\end{equation}
The constant offset is a zero mode with $\omega_0=0$ and the action does not depend on $c_0$. So the free path integral is
\begin{equation}
  Z_\text{free} = \cN \int_{x(0)=x(T)}\!\!\!\!\!\! \mathcal{D}x\,e^{-S_0[x]/\hbar}
  = \cN \int_0^L d c_0 \prod_{n=1}^\infty \int_{-\infty}^\infty d c_n d c_n^* e^{-\frac{1}{\hbar}\omega_n^2\,|c_n|^2}
  = \cN\, L\prod_{n=1}^\infty \frac{\pi\hbar}{\omega_n^2}\,,
\end{equation}
where the limits of the offset $c_0$ are fixed by the allowed range of $x$. Comparing to the spectral computation of the free-particle partition function
\begin{equation}
  Z_\text{free} = \Tr(e^{-H_0 T/\hbar})= \int_0^L dx \langle x|e^{-H_0 T/\hbar}| x\rangle
  =\int_0^L dx \int\frac{dp}{2\pi\hbar}\langle x|e^{-\frac{p^2}{2}T/\hbar}|p \rangle \langle p| x\rangle
  = L\sqrt{\frac{1}{2\pi\hbar T}}\,, \label{Zfreeqm}
\end{equation}
fixes the normalization
\begin{equation}
  \cN = \frac{1}{\sqrt{2\pi\hbar T}}\prod_{n=1}^{\infty}\frac{\omega_n^2}{\pi\hbar}\,.
  \label{PInorm}
\end{equation}
This normalization can now be used for any action.

For the SHO with $V=\frac{1}{2}x^2$ the same Fourier modes are eigenmodes of $-\partial_t^2+V''(0)=-\partial_t^2+1$ with eigenvalues $\lambda_n=\omega_n^2+1$. In the basis of Eq.~\eqref{normalmodes} the action is $S_{\text{SHO}} = \frac{T}{2}c_0^2 + \sum_{n\ge 1}\lambda_n\,|c_n|^2$, where only the zero mode picks up an explicit $T$ from $\int_0^T x^2\,dt$. Since $\lambda_0=1$, the zero mode is lifted and the $c_0$ integral becomes Gaussian, giving $\sqrt{2\pi\hbar/T}$. Each $n\ge 1$ pair contributes $\pi\hbar/\lambda_n$, so we get
\begin{equation}
  Z_{\text{SHO}} =\cN \int_{x(0)=x(T)}\!\!\!\!\!\! \mathcal{D}x\,e^{-S_{\text{SHO}}[x]/\hbar}= \frac{1}{T}\prod_{n=1}^{\infty}\frac{\omega_n^2}{\omega_n^2+1}
  = \frac{1}{T}\prod_{n=1}^{\infty}\frac{1}{1+\frac{T^2}{4\pi^2 n^2}}
  = \frac{1}{2\sinh\frac{T}{2}}\,, \label{ZSHOpi}
\end{equation}
Comparing to the spectral computation, with $E^{(N)}=\hbar(N+\frac{1}{2})$,
\begin{equation}
  Z_{\text{SHO}} = \sum_N e^{-(N+\frac{1}{2})T} = \frac{1}{2\sinh\frac{T}{2}} \label{Zform}
\end{equation}
in agreement with Eq.~\eqref{ZSHOpi}.

\subsection{Complex instantons for the double well}
Now we turn to the double well with potential $V (x) = \frac{1}{8} (x^2 - 1)^2$. This potential has two classical minima and the system has two classical static solutions: $x(t)=\pm 1$. We call these the perturbative saddles. Expanding around them gives the zero-instanton sector of the trans-series, with no $\exp(-S_I/\hbar)$ suppression. The rest of the trans-series comes from expanding around other non-trivial complex classical solutions to the equations of motion.  Our next task is to find all such solutions. After that, we need to understand which solutions contribute to the partition function, and how.

\subsubsection{Elliptic Curves}
The partition function is a path integral in Euclidean time and the corresponding Lagrangian $L = \frac{1}{2} \dot x^2 + V(x)$ is Euclidean. The saddle points are then solutions to
\begin{equation}
  \ddot{x} + U' (x) = 0
\end{equation}
where
\begin{equation}
    U (x) = - V (x) = - \frac{1}{8} (x^2 - 1)^2
\end{equation} 
is the inverted potential. Using $U(x)$ instead of $V(x)$ gives us physical intuition for the (real) solutions: they are ordinary trajectories of a particle moving in the potential $U(x)$. Since the potential is time-independent there is a conserved energy and the solutions all satisfy
\begin{equation}
  \frac{1}{2} \dot{x}^2 + U (x) = \varepsilon \label{Eeq}\,.
\end{equation}
Here $\varepsilon$ is the energy of the classical Euclidean solutions which is not the same as a quantum energy $E$ of a double-well energy eigenstate. We can solve Eq.~\eqref{Eeq} by integration.  There are two roots $\dot{x} = \pm \sqrt{2 \varepsilon - 2 U (x)}$. Integrating them gives an implicit solution for $x(t)$
\begin{equation}
  t = \pm \int^{x(t)}_{x(0)} \frac{d x'}{\sqrt{2 \varepsilon - 2 U (x')}} \label{tform} \,.
\end{equation}
These are elliptic integrals. 

Rather than attempt to integrate~\eqref{tform} directly, we employ a precise change of variables. We first find the 4 roots $x_j$ of $U(x)=\varepsilon$. These are turning points of the classical motion in the inverted well. For any fixed $x_j$, we can then change variables from $x(t)$ to $\wp(t)$ as 
\begin{equation}
  x (t) = x_j \left[ 1 + \frac{6 (x_j^2 - 1)}{1 - 3 x_j^2 + 24 \wp (t-t_0)}
  \right] \,.
  \label{Ptox}
\end{equation}
Then \eqref{Eeq} becomes the \textbf{Weierstrass equation}:
\begin{equation}
  \dot{\wp}^2 = 4 \wp^3 - g_2 \wp - g_3, \label{zWeier}
\end{equation}
with
\begin{equation}
  g_2 = \frac{6 \varepsilon + 1}{12}, \quad g_3 = - \frac{9 \varepsilon +
  1}{216} \label{gforms} \,.
\end{equation}
The solutions are the \textbf{Weierstrass functions} $\wp(t)$ defined implicitly from
\begin{equation}
  t= \int \frac{d t}{d \wp} d \wp = \int_{\infty}^{\wp} \frac{d x'}{\sqrt{4 x'^3 -
  g_2 x' - g_3}} \,.
  \label{WeierInt}
\end{equation}
 Although these functions do not depend on which turning point $x_j$ we chose, any $\wp(t)$ will generate four solutions $x(t)$ from Eq.~\eqref{Ptox}. Note also that the Weierstrass functions are conventionally defined with this definite integral; the integration constant is the global time-translation modulus encoded as $t_0$ in Eq.~\eqref{Ptox}.

\begin{figure}[t!]
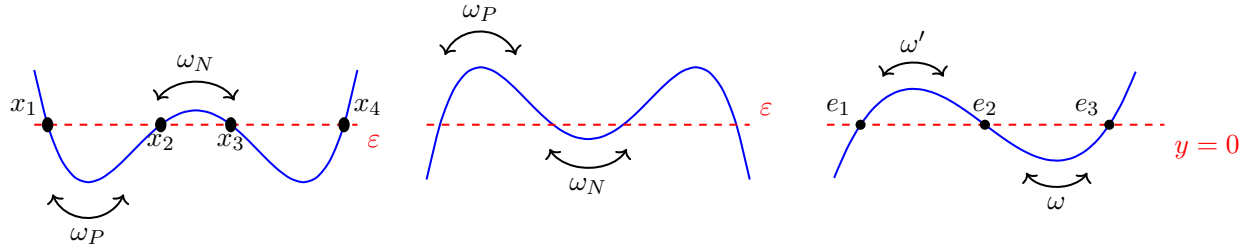

  \centering
  \scalebox{0.95}{\DoubleWellPeriods}
  \hfill
  \scalebox{0.95}{\InvertedDoubleWellPeriods}
  \hfill
  \scalebox{0.95}{\CubicPotentialPeriods}
  \caption{Left: oscillation of a particle with energy $\varepsilon$ in a double well potential has one real ``perturbative'' half-period $\omega_P$ and one imaginary non-perturbative half-period $\omega_N$. Center: in the Euclidean path integral, the potential is inverted and $\omega_P$ and $\omega_N$ are swapped, so $\omega_P$ is imaginary and $\omega_N$ is real. Right: the double well maps to an elliptic curve (a cubic potential) with half-periods $\omega$ and $\omega'$.}
  \label{fig:ec}
\end{figure}

Weierstrass functions are central objects in the study of elliptic curves. An \textbf{elliptic curve} is defined as the locus of solutions to $y^2 = 4 x^3 - g_2 x - g_3$ over some field. Over $\mathbb{R}$, the curve describes motion along a cubic potential. For quantum mechanics, we are interested in elliptic curves over $\mathbb{C}$ which have a remarkable property: the motion is doubly periodic. For a cubic potential, one period is the harmonic oscillation around the minimum. The other period is the harmonic oscillation around the maximum (equivalently the minimum of $V(x)$). This double periodicity makes elliptic curves have the topology of a torus, i.e. they are genus 1. Even though the double well is quartic, because it is symmetric it has only two independent periods and is also genus 1. Alternatively, the map in Eq.~\eqref{Ptox} maps the quartic equation for the double well directly to a cubic.  The periods for the double well are shown in Fig.~\ref{fig:ec}.

A starting point for building intuition for elliptic curves is to start with a simpler case of a genus 0 curve: the circle. The circle comprises solutions to $x^2 + y^2 = 1$, namely $y = \pm \sqrt{1 -x^2}$. We can integrate $dx/y$ by writing $x$ and $y$ as functions of a parameter $t$ along the curve. More precisely, we want to find a parameterization such that $y(t) = \partial_t x(t)$. Then the circle equation maps to the differential equation $\dot{x}^2 + x^2=1$ which is the genus-0 analog of the Weierstrass equation. We can find the parameterization by integration

\begin{equation}
    t = \int dt = \int \frac{dt}{dx} dx = \int \frac{dx}{y}  = \int^x \frac{d x}{\sqrt{1 - x^2}} =
  \sin^{-1} x \,.
  \label{thetaeq}
\end{equation}
This leads to $x=\sin t$ and $y=\dot x = \cos t$, as one might have guessed. Eq.~\eqref{thetaeq} is the analog of the period integral in Eqs.~\eqref{tform} or~\eqref{WeierInt}.  If we integrate between the turning points where $y = 0$ we get
\begin{equation}
\omega_{\operatorname{circle}} = \int_{- 1}^1 \frac{d x}{\sqrt{1 - x^2}} = \pi\,,
\end{equation}
which is, as expected, the half-period of the circle.

The elliptic curve equation $y^2 = 4 x^3 - g_2 x - g_3$ is related to Eq.~\eqref{zWeier} by the parameterization $x=\wp(t)$ and $y=\dot{x}=\wp'(t)$. The turning points correspond to no kinetic energy $\dot x = y=0$. Denoting the turning points by $e_j$ these must satisfy $4 e_j^3 - g_2 e_j - g_3 = 0$. In the original double well, there are 4 turning points where $\varepsilon = U(x)$ and $\dot{x}=0$, but due to the $x\to-x$ symmetry only two are independent. The double well and its cubic image are shown in Fig.~\ref{fig:ec}. Consider the cubic case on the right. A particle can start at rest with energy $\varepsilon$ at the central turning point $e_2$, roll down to $e_3$, reverse direction, and return to $e_2$ after one period. The complementary cycle is reached by analytic continuation $t\to i t$, which exchanges the two motions. In the Euclidean conventions used below, the first cycle corresponds to $\omega \in\mathbb{R}$ and the second to $\omega' \in i\mathbb{R}$. The two half-periods are
\begin{equation}
  \omega' = \int_{e_1}^{e_2} \frac{d x}{y(x)}, \quad
  \omega = \int_{e_2}^{e_3} \frac{d x}{y(x)},\qquad
  y(x)=\sqrt{4 x^3 - g_2 x - g_3}\,,
  \label{eq:weier_half_periods}
\end{equation}
where the square-root branch is chosen consistently on the corresponding sheet of the elliptic curve. Choosing the opposite branch $y\to -y$ flips the sign of both periods but does not change the period lattice. For the double well the half-periods are exactly the integrals between the classical turning points $x_1,x_2,x_3,x_4$:
\begin{equation}
    \omega_P = \int_{x_1}^{x_2} \frac{dx}{\sqrt{2\varepsilon - 2U(x)}},\qquad
     \omega_N = \int_{x_2}^{x_3} \frac{dx}{\sqrt{2\varepsilon - 2U(x)}},
     \label{omegaPN}
\end{equation}
as in Eq.~\eqref{tform}. With these conventions, for $- \frac{1}{8} < \varepsilon < 0$, $\omega_P$ is imaginary and $\omega_N$ is real. $\omega_P$ is $i$ times the classical half-period for motion in the double well potential $V(x)$, and $\omega_N$ is the classical half-period in the inverted double well $U(x)$. 

\begin{figure}[t]
  \centering
  \begin{minipage}[c]{0.40\textwidth}
    \centering
    \InvertedDoubleWellAxes
  \end{minipage}%
  \hspace{1.5cm}%
  \begin{minipage}[c]{0.45\textwidth}
    \centering
    \includegraphics[width=\textwidth]{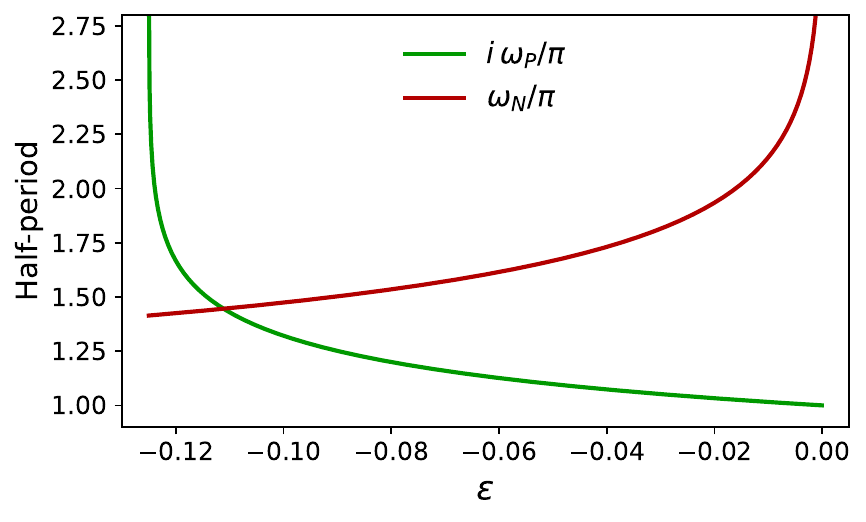}
  \end{minipage}
  \caption{Left: the inverted potential $U (x) = - \frac{1}{8} (x^2 - 1)^2$ for the double well. Choosing a real energy in the range $- \frac{1}{8} < \varepsilon < 0$, the perturbative period $\omega_P$ is imaginary and the non-perturbative period $\omega_N$ is real. $\operatorname{Im} \omega_P$ and $\operatorname{Re} \omega_N$ are shown on the right. As $\varepsilon \rightarrow 0$ we approach the instanton solution going between the two turning points so that the non-perturbative period gets large and $\omega_P \rightarrow -\pi i$.}
  \label{fig:omegas}
\end{figure}

Note that these integrals are the same period integrals that surfaced in Section~\ref{sec:periodintegrals} in the Exact WKB approach. There $P_0=\sqrt{2V-2E}$ with $V=-U$, and $E=-\varepsilon$ matching to the Euclidean energy convention. Thus $\omega_P=-\partial_E S_P^0$ evaluated at $E=-\varepsilon$. The explicit forms for these periods are
\begin{equation}
  \omega_P (\varepsilon) = -\frac{2 i}{ \sqrt{1 + \sqrt{- 8
  \varepsilon}}} K \left( \frac{4 \sqrt{-2 \varepsilon}}{1 + \sqrt{- 8
  \varepsilon}} \right), \quad \omega_N (\varepsilon) =  \frac{\sqrt{8}}{\sqrt{1+\sqrt{1+8 \varepsilon}}} K \left( \frac{1+8 \varepsilon -\sqrt{1+8 \varepsilon}}{4 \varepsilon } \right) \label{periods}
\end{equation}
with $K$ the Elliptic $K$ function. These have equivalent hypergeometric forms given by $\partial_E$ acting on Eqs.~\eqref{SP0} and~\eqref{SN0} with $E=-\varepsilon$. The dependence of these periods on $\varepsilon$ is shown in Fig.~\ref{fig:omegas}. 

The Weierstrass functions usually written as $\wp(t;\omega,\omega')$ have a number of useful properties. They are doubly periodic with half periods $\omega$ and
$\omega'$. This means that 
\begin{equation}
  \wp (t ; \omega , \omega') = \wp (t + 2 \omega ; \omega ,
  \omega') = \wp (t + 2 \omega' ; \omega , \omega') \,.
  \label{wpperiod}
\end{equation}
They are even functions $\wp (t) = \wp (- t)$, which follows from time-reversal invariance of the Weierstrass equation. They are meromorphic functions with double poles at all lattice points $t = 2k \omega + 2k'\omega'$ for any $k,k' \in {\mathbb Z}$. Near a pole $t_\star$, the local behavior is
\begin{equation}
  \wp(t)=\frac{1}{(t-t_\star)^2}+\frac{g_2}{20}(t-t_\star)^2+\frac{g_3}{28}(t-t_\star)^4+\cdots
\end{equation}
so there is no simple-pole term. At the half periods, the Weierstrass functions are finite and, for this labeling convention, give the turning points:
\begin{equation}
  \wp (\omega ; \omega , \omega') = e_1, \quad \wp (\omega' ; \omega ,
  \omega') = e_3, \quad \wp (\omega + \omega' ; \omega , \omega') = e_2\,.
\end{equation}
The functions have a scaling law
\begin{equation}
  \wp (\lambda t ; \lambda \omega, \lambda \omega' ) = \frac{1}{\lambda^2} \wp (t ; \omega, \omega' )
\end{equation}
so that 
\begin{equation}
  \omega^2 e_1 = \omega^2 \wp (\omega ; \omega , \omega') = \wp (1 ;1, \tau )\,,
\end{equation}
where $\tau = \frac{\omega'}{\omega}$ is the modular parameter of the elliptic curve.

\subsubsection{Periodic saddles}\label{sec:complex_instantons}
The generic solutions to the Euclidean equations of motion for the double well are written in Eq.~\eqref{Ptox} in terms of Weierstrass functions $\wp(t;\omega,\omega')$ and the choice of one of the four turning points $x_j$. To restrict these solutions to saddle points of the partition function, we must next impose periodic boundary conditions $x(t) = x(t+T)$. Conveniently, the solutions to the double well inherit periodicity from the Weierstrass functions, as in Eq.~\eqref{wpperiod}, with half-periods given explicitly in Eq.~\eqref{omegaPN}. So all the generic double-well solutions are periodic under
\begin{equation}
  t \rightarrow t + 2 k \omega_N + 2 k' \omega_P \,.
\end{equation}
Here $k,k'\in\mathbb{Z}$ label how many times the trajectory winds around the two primitive period directions of the Weierstrass solution, $2\omega_N$ and $2\omega_P$. In terms of period integrals, $k$ and $k'$ enumerate the homology class of the integration contour. 

For a solution to be periodic in time $T$ we must then have
\begin{equation}
  2 k \omega_N (\varepsilon) + 2 k' \omega_P (\varepsilon) = T \,,
  \label{quantcond}
\end{equation}
for some integers $k, k'$. This is only solved at quantized values of the classical energy $\varepsilon$. Since $\omega_N$ is real and $\omega_P$ imaginary for real $\varepsilon$, the quantization condition forces $k'=0$ for real $\varepsilon$; nonzero $k'$ requires complex $\varepsilon$ and the corresponding paths are complex.

We are ultimately interested in the $T \rightarrow \infty$ limit. In this limit, the solutions to Eq.~\eqref{quantcond} at finite $k, k'$ require large $\omega_N$ or $\omega_P$. As we can see in Fig.~\ref{fig:omegas}, we can achieve large $\omega_N$ with $\varepsilon \approx 0$. Such small energies correspond to starting near the top of one of the hills in the inverted potential, staying there a very long time, rolling over to the other side, staying there a long time, then rolling back. These are the famous instanton trajectories. Expanding around $\varepsilon = 0$ gives
\begin{equation}
  \omega_N = - \ln \frac{- \varepsilon }{8} + \frac{\varepsilon}{2} \left( 10 + 3 \ln
  \frac{- \varepsilon }{8} + \cdots \right)
\end{equation}
\begin{equation}
  \omega_P = - \pi i \left( 1 - \frac{3}{2} \varepsilon + \frac{105}{16}
  \varepsilon^2 + \cdots \right) \,.
\end{equation}
Then, solving Eq.~\eqref{quantcond} we get
\begin{equation} \label{eq:epsilonexpand}
  \varepsilon = - 8 e^{-i \pi \frac{k'}{k}} e^{- \frac{T}{2 k}} + \cdots \,.
\end{equation}
For $k' = 0$ these are the usual periodic $n=2k$ instanton solutions. For nonzero $k'$ the energy $\varepsilon$ has a phase rotation. This phase rotation gives $\omega_N $ an imaginary part to cancel against $\omega_P = -\pi i$. 

\begin{figure}[t]
  \centering
  \includegraphics[width=\textwidth]{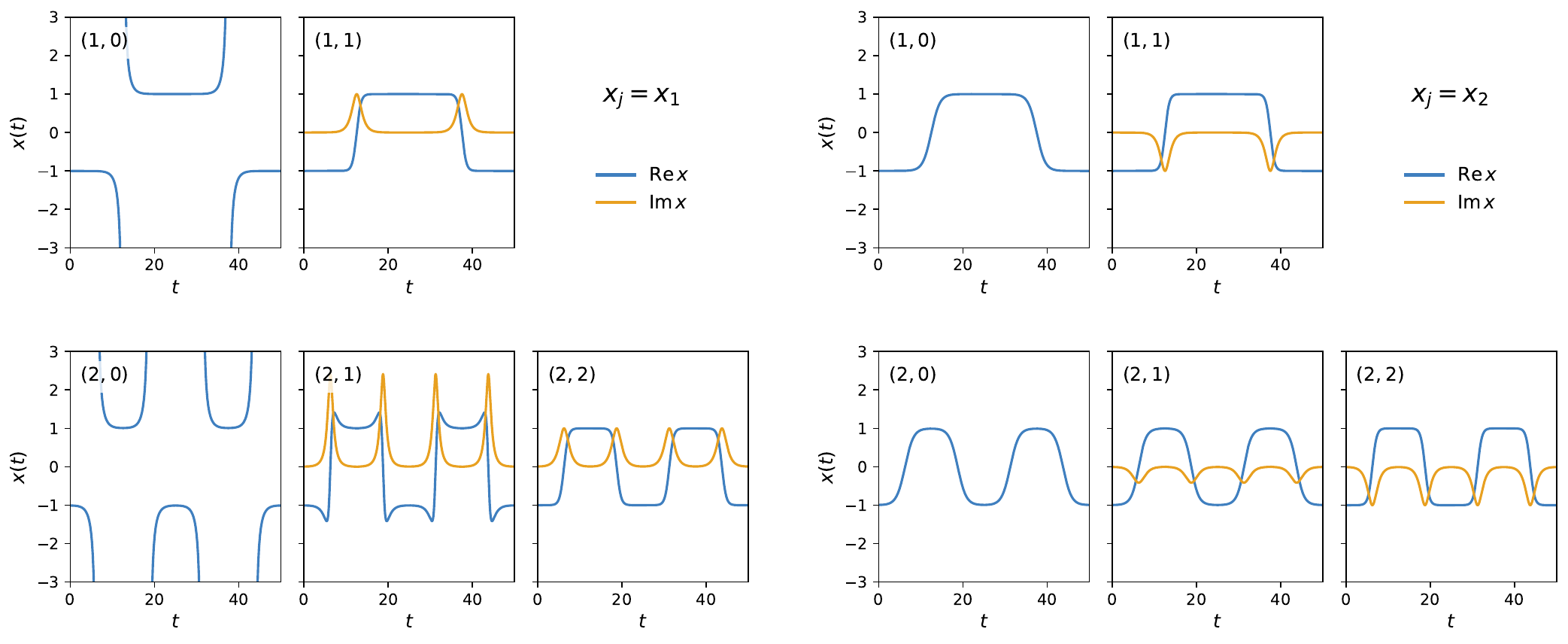}
  \caption{Exact instanton solutions from the Weierstrass $\wp$ formula~\eqref{Ptox} with $\varepsilon$ determined by the quantization condition~\eqref{quantcond}, for $k=1,2$ and $0\le k'\le k$. Left: $x_1$ branch; right: $x_2$ branch. For real $\varepsilon$ ($k'=0$), the $x_1$-branch solution diverges where $x_{(2)}(t)$ crosses zero, consistent with Eq.~\eqref{xinversion}. For complex $\varepsilon$ ($k'\neq 0$), both branches remain bounded. Real parts in blue, imaginary parts in orange.}
  \label{fig:solutions}
\end{figure}

To enumerate the solutions, we first note that the pairs $(k,k')$ and $(-k,-k')$ describe the same geometric trajectory with opposite orientation ($t\to -t$), or equivalently shifting the offset by $t_0 \to T - t_0$.  So we can take $k>0$ without loss of generality. Furthermore, taking $k' \rightarrow k' + 2 k$ leaves $\varepsilon$ unchanged. So an independent set of solutions is indexed by $-k < k' \le k$, where the solution branches $k'$ and $-k'$ are complex conjugates. We also have a choice of the turning point $x_j$ in Eq.~\eqref{Ptox}.  $x_1$ and $x_4$ give solutions related by the $x\to -x$ symmetry of $V(x)$, as do $x_2$ and $x_3$, so the independent branches are labeled by $x_1$ and $x_2$. Example solutions, using the exact $\omega_N, \omega_P$, are shown in Fig.~\ref{fig:solutions}, for different values of $(k,k')$ and different turning points $x_1, x_2$.  These are all exact classical solutions at finite $T$.

Let us denote the two solution classes by $x_{(1)}(t)$ and $x_{(2)}(t)$. These two branches are related by inversion: since $x_1^2+x_2^2=2$ it follows from Eq.~\eqref{Ptox} that 
\begin{equation}
  x_{(1)}(t)\cdot x_{(2)}(t) = x_1\, x_2 = \sqrt{1+8\varepsilon}\,.
  \label{xinversion}
\end{equation}
For real $\varepsilon$ (i.e.\ $k'=0$), $x_{(2)}(t)$ crosses zero as the instanton traverses $x=0$ between the two wells, so $x_{(1)}(t)=x_1 x_2/x_{(2)}(t)$ diverges --- the trajectory escapes to $x=\pm\infty$ and has infinite action. For complex $\varepsilon$ ($k'\neq 0$), $x_{(2)}(t)$ acquires a nonzero imaginary part throughout its orbit, so both branches remain bounded. Since the map $x\to \sqrt{1+8\varepsilon}/x$ is not a symmetry of $V(x)$, the two branches are distinct saddle points with generically different actions. Both share the leading action $2kS_I$, and for the self-conjugate case $k'=k$ their actions are exactly equal, but for $k'<k$ they differ at subleading order $\mathcal{O}(\varepsilon)$. Therefore, the real finite-action multi-instanton solutions are all in the $x_2$ branch and we will assume $x_j=x_2$ and just write $x(t)$ for these solutions from now on, unless otherwise specified.

The action for the $(k,k')$ instantons can be computed as
\begin{equation}\label{eq:actionexactsols}
  S_{k,k'} = \int_0^T d t \left[ \frac{1}{2} \dot{x}^2 - U (x) \right]
  = \oint_{\mathcal{C}_{k,k'}} P_0\,d x - \varepsilon T\,,
\end{equation}
where $\mathcal{C}_{k,k'}$ denotes the closed Euclidean trajectory winding around the elliptic curve. The momentum here is the same momentum used in Exact WKB from Eq.~\eqref{P0def}:
\begin{equation}
  P_0(x)=\dot{x}=\sqrt{2\varepsilon-2U(x)}=\sqrt{2V(x)-2E},\qquad E=-\varepsilon.
\end{equation}
We can use the same actions as derived in the Exact WKB section, rewriting Eqs.~\eqref{SP0} and \eqref{SN0} in elliptic-integral form:
\begin{align}
  S_P^0 (\varepsilon) &= \frac{i}{3}\sqrt{1+\sqrt{-8\varepsilon}}\left[
  E(m_P)-\left(1-\sqrt{-8\varepsilon}\right)K(m_P)\right],\qquad
  m_P=\frac{4\sqrt{-2\varepsilon}}{1+\sqrt{-8\varepsilon}},
\\
  S_N^0 (\varepsilon) &= \frac{\sqrt{2}}{3}\sqrt{1+\sqrt{1+8\varepsilon}}\left[
E(m_N)+\left(\sqrt{1+8\varepsilon}-1\right)K(m_N)\right],\qquad
  m_N=\frac{1+8\varepsilon-\sqrt{1+8\varepsilon}}{4\varepsilon}.
\end{align}
Expanding for small $\varepsilon$ (large $T$) gives
\begin{equation}
  S_N^0(\varepsilon)=\frac{2}{3}-\varepsilon\!\left(\ln\frac{-\varepsilon}{8}-1\right)
+\mathcal{O}\!\left(\varepsilon^2\ln\varepsilon\right),\qquad
  S_P^0(\varepsilon)=-\pi i\,\varepsilon+\mathcal{O}\!\left(\varepsilon^2\right).
\end{equation}
The classical action for the $(k,k')$ instanton is then
\begin{equation} 
  \label{eq:actionexactsolsb}
S_{k,k'}=2k\,S_N^0(\varepsilon)+2k'\,S_P^0(\varepsilon)-\varepsilon\,T\,,
\end{equation}
where $T$ and $\varepsilon$ are implicitly related by Eq.~\eqref{quantcond}. Substituting these into the action formula and using Eq.~\eqref{eq:epsilonexpand} and expanding gives
\begin{align}
  S_{k,k'} & =2kS_I
  -2k\,\varepsilon\!\left(\ln\frac{-\varepsilon}{8}-1\right)
  -2\pi i k'\varepsilon-\varepsilon T+\cO\!\left(\varepsilon^2\ln\varepsilon\right)\\
&=2kS_I+2k\,\varepsilon+\cO\!\left(\varepsilon^2\ln\varepsilon\right),
\end{align}
where $S_I=\frac{2}{3}$ and we used $T=-2k\ln\frac{-\varepsilon}{8}-2\pi i k'+\mathcal{O}(\varepsilon\ln\varepsilon)$. Note that there is no separate linear $k'$ term at this order: it cancels against the $-\varepsilon T$ piece. There is implicit $k'$ dependence though, since
\begin{equation}
  \varepsilon(T)=-8e^{-i\pi k'/k}e^{-T/(2k)}+\cdots,
\end{equation}
and the same result can alternatively be written as
\begin{equation}
  S_{k,k'}(T)=2kS_I-16k\,e^{-i\pi k'/k}e^{-T/(2k)}
  +\mathcal{O}\!\left(T\,e^{-T/k}\right).
  \label{eq:action_largeT_periodic}
\end{equation}
The instanton action contains the $2k S_I$ term, confirming that only saddles with even $2k$ instanton numbers contribute to the partition function, as expected from having imposed periodic boundary conditions.  Note that the leading correction to the action depends on $k'$ only through the phase of an exponentially suppressed correction.  

Although we have found all the bounded periodic solutions, there are also solutions that run off to infinity in $x$, such as the $k'=0$ solutions with $x_j=x_1$. It is an unusual feature of the quartic potential that it takes finite time to roll to infinity. That is
\begin{equation}
  \int_{- \infty}^{x_1} \frac{d x}{\sqrt{\frac{1}{4} (1 - x^2)^2 + 2
  \varepsilon}} = \frac{\omega_N}{2} \,.
\end{equation}
For a quadratic potential this is not true -- it takes an infinite amount of time to get to infinity. In any case, these solutions have infinite action so they never contribute to the path integral and we can ignore them.

In summary, the complete set of independent finite-action periodic complex saddles for the double well are given by
\begin{equation}
    x(t) =    x_j \left[ 1 + \frac{6 (x_j^2 - 1)}{1 - 3 x_j^2 + 24 \wp (t-t_0;\omega_P, \omega_N)}
  \right] \,,
\end{equation}
where $x_j=x_1,x_2$ is a turning point with $U(x_j) = \varepsilon$, $\wp(t;\omega,\omega')$ is the Weierstrass function, $\omega_P$ and $\omega_N$ are the half-periods in Eq.~\eqref{periods}, and the classical energy $\varepsilon$ is quantized by 
\begin{equation}
  2 k \omega_N (\varepsilon) + 2 k' \omega_P (\varepsilon) = T \,.
\end{equation}
The $k$ and $k'$ are integers and independent solutions are indexed by $k>0$ and $-k < k' \le k$ (equivalently $k' \mod 2k$), giving $2k$ saddles at instanton number $n=2k$, with $k'$ and $-k'$ forming complex-conjugate pairs. These periodic saddles have even instanton number $n=2k$, and their leading action is $S_{k,k'} = 2kS_I + \mathcal{O}(e^{-T/(2k)})$.

\subsubsection{Anti-periodic saddles}\label{sec:antiperiodic_solutions}

The saddles from Sec.~\ref{sec:complex_instantons} have even instanton number $n=2k$ due to the periodic boundary conditions. To obtain odd-instanton saddles, we consider the twisted partition function
\begin{equation}
  \widetilde{Z}(T) = \Tr\!\left(\mathcal{P}\,e^{-H T/\hbar}\right)
  = \cN \int_{x(0)=-x(T)} \!\!\!\!\!\!\!\!\!\!{\mathcal D} x\; e^{-\frac{1}{\hbar}S[x]}\,,
  \label{eq:Ztilde_def}
\end{equation}
where $\mathcal{P}$ is the parity operator $x\to -x$, and the path integral runs over anti-periodic paths with $x(t+T)=-x(t)$. The saddle points of this path integral are solutions of $\ddot x = V'(x)$ satisfying anti-periodic boundary conditions.

To find the corresponding solutions we exploit a sign-flip property of the Weierstrass solution Eq.~\eqref{Ptox} with $x_j=x_2$: a shift by the non-perturbative half-period $\omega_N$ flips the sign of the solution,
\begin{equation}
  x(t + \omega_N) = -x(t)\,.
  \label{eq:Z2_antiperiod}
\end{equation}
This holds because at the modulus $t = t_0$ the trajectory sits at $x_2$ with $\dot x = 0$, and after one half-period $\omega_N$ it reaches the opposite turning point $x_3 = -x_2$, again with $\dot x = 0$. So $\tilde{x}(t) \equiv x(t+\omega_N)$ and $\hat{x}(t)\equiv -x(t)$ both satisfy the equation of motion (since $V(x)$ is even) with the same initial data $(-x_2,0)$ at $t = t_0$; by uniqueness they agree for all $t$.

Combining Eq.~\eqref{eq:Z2_antiperiod} with the full-period symmetries $x(t+2k\omega_N) = x(t+2k'\omega_P) = x(t)$ gives $x(t+T) = -x(t)$ whenever $T = (2m+1)\omega_N + 2k'\omega_P$. The analog of Eq.~\eqref{quantcond} for anti-periodic boundary conditions is therefore
\begin{equation}
  (2m+1)\,\omega_N(\varepsilon) + 2k'\,\omega_P(\varepsilon) = T\,,
  \label{eq:AP_quantcond}
\end{equation}
with integers $m \geq 0$ and $-m \le k' \le m$ (equivalently $k' \mod (2m+1)$), giving $2m+1$ saddles at instanton number $n = 2m+1$, with $k'$ and $-k'$ forming complex-conjugate pairs. Their leading action at large $T$ is $S_{m,k'}^{\rm AP} = (2m+1)\,S_I + \mathcal{O}(e^{-T/(2m+1)})$.

The simplest case is $m=0$, $k'=0$, for which we have the single-instanton quantization condition $\omega_N(\varepsilon)=T$. Rewriting the Weierstrass solution in the Jacobi elliptic form, here shifted to be defined on $t\in [0,T]$, we have
\begin{equation}
  x_\cI(t) = \sqrt{\frac{2\sig^2}{1+\sig^2}}\;\operatorname{sn}\!\left(\frac{t-T/2}{\sqrt{2(1+\sig^2)}},\,\sig^2\right),
  \label{xCIfirst}
\end{equation}
with $\operatorname{sn}(u,\sig^2)$ the Jacobi elliptic sine function and $\sig$ the elliptic modulus. This modulus is related to the classical energy by
\begin{equation}
  \sig^2 = \frac{1 - \sqrt{-8\varepsilon}}{1 + \sqrt{-8\varepsilon}}\,.
  \label{eq:sig_from_eps}
\end{equation}
The modulus $\sig$ is the ratio of the two quarter-periods of the Jacobi elliptic functions, encoding how close the solution is to the zero-energy instanton: $\sig\to 1$ as $\varepsilon\to 0$ (recovering $\operatorname{sn}(u,1) = \tanh u$), while $\sig\to 0$ corresponds to $\varepsilon\to -1/8$ (the bottom of the inverted potential). We will make use of this solution in Sections~\ref{sec:morse} and~\ref{sec:subleading_T}.

Together with Eq.~\eqref{Ptox}, Eqs.~\eqref{quantcond} and~\eqref{eq:AP_quantcond} classify all finite-action saddle points of the double-well path integral: periodic saddles with even instanton number $n=2k$, and anti-periodic saddles with odd instanton number $n=2m+1$.

\subsubsection{Approximate solutions} \label{sec:approx_solutions}
For large $T$, the exact solutions can be well approximated by sewing together hyperbolic tangents. This picture will prove helpful in providing a natural set of coordinates on the quasi-zero mode manifold.  

The starting point is the single instanton at $T=\infty$ which is a hyperbolic tangent centered at some point $t_0$:
\begin{equation} \label{eq:single_instanton}
  x_1^s (t) = \tanh \left( \frac{t - t_0}{2} \right) \,.
\end{equation}
This is an exact solution to the Euclidean equations of motion for any $t_0$, but only satisfies the boundary conditions at $T=\infty$. For finite $T$, $\tanh((t-t_0)/2)$ only satisfies $x(T/2)=-x(-T/2)$ at $t_0=0$.  Even with $t_0=0$, the derivatives at $t=\pm T/2$ are equal rather than equal-and-opposite, so anti-periodicity ($\dot x(T/2) = -\dot x(-T/2)$) is violated and the equations of motion fail at the boundaries. In contrast, the exact $n=1$ solution built from Eq.~\eqref{xCIfirst} with the quantization condition $\omega_N(\varepsilon)=T$ satisfies the appropriate anti-periodic boundary conditions exactly and can be centered at any $t_0$. Although not exact, the $\tanh$ approximate instanton and sewn-together $\tanh$ approximate multi-instantons provide good approximations to the exact solutions, up to corrections that are exponentially small in $T$.

\begin{figure}[t]
  \centering  \includegraphics[width=0.5\textwidth]{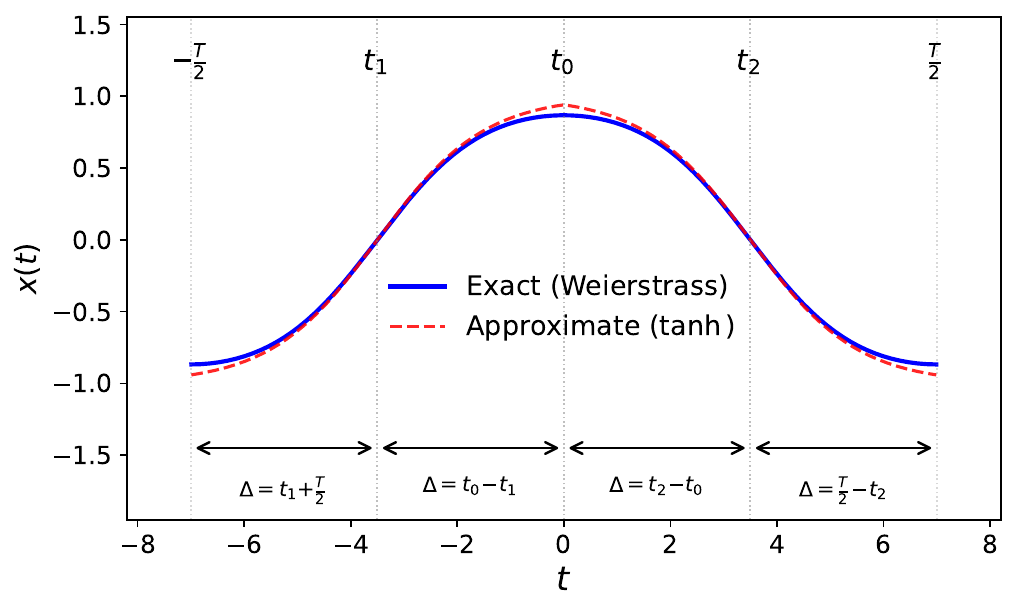}
  \caption{Exact (Weierstrass $\wp$) and approximate ($\tanh$) 2-instanton solutions for the double well with $T=14$. The instanton is centered at $t_1=-T/4$, the anti-instanton at $t_2=T/4$, with transition point $t_0=0$. The action decomposes as $S = \Shat(\Delta_1) + \Shat(\Delta_2) + \Shat(\Delta_3) + \Shat(\Delta_4)$ where each $\Delta$ is the distance from an instanton center to the nearest boundary.}
  \label{fig:instanton2}
\end{figure}
To calculate the action for the $\tanh$ instanton, it is helpful to view Eq.~\eqref{eq:single_instanton} as the approximate instanton centered at $t_0$, with two half-instanton contributions extending a distance $T/2$ from the center. On a finite interval $- \frac{T}{2} < t < \frac{T}{2}$ it has action
\begin{equation}
  S [x_1^s] = \Shat \left( t_0 + \frac{T}{2}  \right) + \Shat \left( \frac{T}{2} - t_0 \right) \,,
\end{equation}
where $\Shat(\Delta)$ gives the action contribution from a half-instanton extending a distance $\Delta$ from its center to a boundary:
\begin{equation}
  \Shat (\Delta) = \int_0^{\Delta} dt\, \frac{1}{4} \operatorname{sech}^4\!\left(\frac{t}{2}\right) = \tanh\left(\frac{\Delta}{2}\right) \left[\frac{1}{3} + \frac{1}{6} \cosh^{-2}\left(\frac{\Delta}{2}\right)\right] \,.
\end{equation} 
For large $\Delta$, 
\begin{equation}
    \Shat(\Delta) =  \frac{S_I}{2} \left(1 - 6 e^{- 2 \Delta} + \cdots \right) \label{largeDelta}
\end{equation}
so that as $T \rightarrow \infty$, $S [x_1^s] \rightarrow S_I$. The parameter $t_0$ can be complex; taking $t_0 = i \pi$ gives $x_1^s (t) = \coth  \frac{t}{2}$, a discontinuous solution that rolls to $\pm \infty$ in finite time.

For multi-instanton solutions, consider first an instanton--anti-instanton pair approximated by
\begin{equation}
  \xii (t) = \tanh \left( \frac{t - t_1}{2} \right) \theta (t_0 -
  t) - \tanh \left( \frac{t - t_2}{2} \right) \theta (t - t_0)
\end{equation}
where $t_1$ and $t_2$ are the instanton and anti-instanton centers and $t_0$ is the transition point. This approximate $n=2$ solution is shown and compared to the exact $x_{1,0}$ solution in Fig.~\ref{fig:instanton2}. Using the global shift symmetry to set $t_0=0$, periodicity forces $t_1 = - t_2$, giving the action 
\begin{equation}
  S [\xii] = 2 \Shat (t_2) + 2 \Shat ( \dfrac{T}{2} - t_2
  ) \approx 2 S_I \left( 1 -3 e^{- 2 t_2} -3 e^{- 2 \left( \frac{T}{2} -
  t_2 \right)} \right) .
  \label{eq:SII_sewn}
\end{equation}
This gives the ``sewn'' large-$T$ action for $n=2$, where $t_2$ plays the role of a quasi-collective-coordinate on which the action is only exponentially weakly dependent. Writing $t_2 = t_2^R + i \phi_2$ the saddle-point condition $\partial_{t_2} S[\xii] = 0$ becomes
\begin{equation}
  \exp\left(-2 t_2^R -2 i \phi_2\right) - \exp\left(-T+2 t_2^R + 2 i \phi_2\right) = 0,
\end{equation}
with solution $t_2^R=T/4$ and $\phi_2=0$ or $\phi_2=\pi/2$. The two distinct saddle solutions are then
\begin{equation}
  \xii (t) = \tanh \left( \frac{t + \frac{T}{4} + i\phi_2}{2} \right)
  \theta (-t) - \tanh \left( \frac{t - \frac{T}{4} - i\phi_2}{2} \right) \theta
  (t)
\end{equation}
with $\phi_2 = 0, \pi/2$, corresponding to the $(k,k')=(1,0)$ and $(1,1)$ exact solutions.

Motivated by the $n=2$ case, for $n=2k$ we include phase shifts from the start and write
\begin{equation}
  x_{2k}^s (t) = \sum_{p = 1}^{2k} (- 1)^{p+1} \tanh \left( \frac{t - t_p + i\phi_p}{2} \right) \theta (\tau_p < t < \tau_{p + 1}) .
\end{equation}
Here $t_p$ are the constituent instanton centers, $\phi_p$ their phase offsets, and $\tau_p$ for $p=1,\ldots,2k$ are the transition points between consecutive constituents on the circle (the analog of $t_0$ in the $n=2$ case), defined cyclically with $\tau_{2k+1} \equiv \tau_1 + T$.
The sewn action is
\begin{equation} \label{eq:action_centers}
  \begin{aligned}
    S [x_{2k}^s] &= \sum_{p=1}^{2k} \left[\Shat (t_p-\tau_p-i\phi_p) + \Shat (\tau_{p + 1}-t_p+i\phi_p)\right] \\
    &\simeq 2kS_I-3S_I\sum_{p=1}^{2k}\left[e^{-2(t_p-\tau_p-i\phi_p)}+e^{-2(\tau_{p+1}-t_p+i\phi_p)}\right].
  \end{aligned}
\end{equation}
Extremizing with respect to $\tau_p$ gives
\begin{equation}
  \exp\left(-2(t_p-\tau_p-i\phi_p) \right)=\exp\left(-2(\tau_p-t_{p-1}+i\phi_{p-1})\right),
\end{equation}
so each nearest-neighbor separation appears twice and we have $\tau_p = (t_p + t_{p-1} -i \phi_p -i \phi_{p-1})/2$. Defining the complex separations directly by
\begin{equation}
  \alpha_p\equiv (t_{p+1}-t_p)-i(\phi_{p+1}-\phi_p)\,,
\end{equation}
and evaluating $S[x_{2k}^s]$ on the $\tau_p$-extremum constraint above, we get the reduced sewn action
\begin{equation} \label{eq:effectiveaction_approx}
  \Seff^{(s)}
  \simeq
  2kS_I-6S_I\sum_{p=1}^{2k}e^{-\alpha_p} \,,
\end{equation} 
with periodicity constraint $\sum_{p=1}^{2k}\alpha_p=T$ (mod $2\pi i$ in the phase branch). Now extremizing all remaining quasi-collective coordinates at once using a Lagrange multiplier:
\begin{equation}
  \frac{\partial}{\partial \alpha_p}
  \left[ \Seff^{(s)} +\lambda\Bigl(\sum_q\alpha_q-T\Bigr)\right]=0
  \;\Longrightarrow\;
  6S_I\,e^{-\alpha_p}+\lambda=0,
  \qquad p=1,\ldots,2k.
\end{equation}
Because the same $\lambda$ appears for every $p$, all $e^{-\alpha_p}$ are equal. Using $\sum_p\alpha_p=T$ and the phase branch condition gives
\begin{equation}
  \alpha_p=\frac{T}{2k}-i\frac{k'\pi}{k}, \qquad k'\in\mathbb{Z} \,.
\end{equation}
Choosing the time origin so that the first instanton center is at $t=T/(4k)$,
\begin{equation}
  t_p = \left( p - \frac{1}{2} \right) \frac{T}{2 k},
    \qquad
  \phi_p = \frac{k'}{2k} (2 p - 1) \pi,
  \qquad
  \tau_p = (p-1)\frac{T}{2k}.
  \label{onshell}
\end{equation}
So the sewn construction reproduces the correct qualitative saddle structure: equally spaced constituents, constant phase increment, and a discrete branch label $k'$ (with $k'\sim k'+2k$ and $k'\leftrightarrow -k'$ giving conjugate branches), matching the large-$T$ pattern of the exact saddles.

Finally, we can make explicit the correspondence to the two turning-point branches of the exact elliptic solutions discussed around Eq.~\eqref{xinversion}. The sewn $\tanh$ configurations above naturally approximate the $x_2$-branch saddles (the trajectories that traverse between the two wells). The second branch can be obtained without repeating the elliptic-curve analysis: indeed one checks that, for any solution $x(t)$ at fixed $\varepsilon$, the inverted path
\begin{equation}
  x(t)\ \longmapsto\ \frac{\sqrt{1+8\varepsilon}}{x(t)}
\end{equation}
obeys the same equation of motion (away from zeros of $x$) and hence maps the $x_2$-branch saddle to the $x_1$-branch saddle. This map preserves periodicity provided $x(t)$ does not cross~$0$. For the real saddles ($k'=0$) the $x_2$-branch instanton chain crosses $x=0$, so the inverted branch diverges and has infinite action; for the complex saddles ($k'\neq0$) the orbit is complex and stays away from~$0$ on the real-$t$ contour, so both branches are finite-action and agree with the exact solutions (cf.\ Fig.~\ref{fig:solutions}).

Evaluating the action on the phase-locked saddles using Eq.~\eqref{eq:effectiveaction_approx} (and a $k'\rightarrow -k'$ relabeling) we get
\begin{equation}
  S_{k,k'}^{s}(T)
  =
  2kS_I-8k\,e^{-i\pi k'/k}e^{-T/(2k)}+\cdots.
\label{eq:SII_sewn_largeT}
\end{equation}
This is to be contrasted with the action of the exact saddles from Sec.~\ref{sec:complex_instantons} where we got
Eq.~\eqref{eq:action_largeT_periodic}:
\begin{equation}
   S[x_{k,k'}]
  =
  2kS_I-16k\,e^{-i\pi k'/k}e^{-T/(2k)}+\cdots\,.
  \label{eq:SII_exact_largeT}
\end{equation}
Comparing Eqs.~\eqref{eq:SII_sewn_largeT} and~\eqref{eq:SII_exact_largeT} shows a factor-of-two difference in the exponentially small coefficient. This comes from the hard sewing mismatch at the joining region: the coefficient of the $e^{-T/(2k)}$ term is sensitive to the detailed smooth crossover there. Rather than model that crossover explicitly, we keep the same nearest-neighbor functional form and simply rescale its coefficient by a factor of two. This rescaling does not affect the equations of motion, and on the equations of motion the exponentially suppressed term will double in size, agreeing with the exact answer.

To summarize, we take the effective action to be
\begin{equation}
  S_{\mathrm{eff}}
  \simeq
  2kS_I-6S_I\sum_{p=1}^{2k}\left[e^{-2(t_p-\tau_p-i\phi_p)}+e^{-2(\tau_{p+1}-t_p+i\phi_p)}\right].
\end{equation}
Setting the transition points $\tau_p$ to their on-shell values in Eq.~\eqref{onshell} the effective action simplifies to
\begin{equation}
  \boxed{
    S_{\mathrm{eff}}
    =
    2kS_I-12S_I\sum_{p=1}^{2k}e^{-\alpha_p},\qquad \sum_{p=1}^{2k}\alpha_p=T
    \label{Seffalpha} 
  } 
\end{equation}
where
\begin{equation}
    \alpha_p= (t_{p+1}-t_p)-i(\phi_{p+1}-\phi_p)
  \label{eq:Seff_largeT_calibrated} \,.
\end{equation}
Extremizing with respect to $t_p$ and $\phi_p$ reproduces Eq.~\eqref{onshell} and now the effective action agrees with the $T \rightarrow \infty$ limit of the exact action.

\subsection{Picard-Lefschetz theory and the dilute instanton gas}

Now that we have found all the complex saddles of the double well, what do we do with them? 
Historically, the way these saddles have been incorporated is through the dilute instanton gas approximation which we review next.

\subsubsection{Dilute instanton gas}
The dilute instanton gas (DIG) is the leading semiclassical approximation to the path integral at large $T$ and small $\hbar$, restricted to the region of configuration space where the path consists of well-separated (anti-)instanton events. It assumes that any path $x(t)$ contributing to the path integral can be cleanly associated with a single $n$-instanton saddle $x_n^\star$. So summing over paths amounts to summing over these different sectors. Roughly speaking, the DIG treats the real path integral as if the integration domain could be partitioned into sectors associated with different instanton numbers,
\begin{equation}
    \GR = \sum_n \Gamma_n,\qquad
    \Gamma_n = \Big\{x~\text{with}~\|x-x_n^\star\|
    < \|x-x_m^\star\|~\forall m\neq n\Big\}\,.
    \label{eq:decomp_dig_paths}
\end{equation}
The DIG assumes that the dominant configurations are dilute $n$-instanton paths, meaning that the constituent events are well separated (``dilute''), $\alpha_p\gg1$, so that this schematic decomposition is sensible.  In the dilute regime, the path integral calculation simplifies dramatically. Because the overlap of neighboring instanton tails is exponentially small, the action is then approximated by a sum of isolated instanton actions, $S[x]\approx nS_I$, with corrections of order $\cO(e^{-\alpha_p})$. For the same reason, the fluctuation operator is approximately block diagonal, so the determinant factorizes into $n$ copies of the single-instanton determinant up to corrections of the same order. Finally, if the instantons are dilute, the $n$ approximate translation modes can be treated as independent collective coordinates, so the path integral measure reduces to an approximately flat measure on the ordered instanton centers, giving the familiar $T^n/n!$ volume factor together with the standard Jacobian from the zero-mode normalization.

The instanton in the DIG has the approximate form discussed in Section~\ref{sec:approx_solutions}: $x_I = \tanh(t/2)$. Around this solution, at $T=\infty$, one can compute the fluctuation determinant exactly
\begin{equation}
  K
  \;\equiv\;
  \left(\frac{S_I}{2\pi\hbar}\right)^{\!1/2}
  \left(\frac{\det \opO_0}{\det\nolimits'\opO_I}\right)^{\!1/2}
  =
  \left(\frac{12\,S_I}{2\pi\hbar}\right)^{\!1/2},
  \label{eq:K_DIG_prefactor}
\end{equation}
where $\opO_0 = -\partial_t^2+1$ is the fluctuation operator around the perturbative saddles at $x=\pm 1$ and
\begin{equation}
    \opO_I = -\partial_t^2 - U''[x_I] =-\partial_t^2+1
    -\frac{3}{2}\operatorname{sech}^2\left(\frac{t}{2}\right)
     \label{eq:O_I_DIG}
\end{equation}
is the fluctuation operator around the instanton.
In Eq.~\eqref{eq:K_DIG_prefactor}, the $\det'$ indicates that the zero mode has been removed from the determinant and $\sqrt{S_I/(2\pi\hbar)}$ is the compensating collective-coordinate Jacobian. The factor of 12 is the determinant ratio $\det\opO_0/\det'\opO_I$ which can be computed either by the Gel'fand--Yaglom method or by noting that $\opO_I$ is a Schr\"odinger operator with a P\"oschl--Teller potential whose spectrum is known exactly. These results can be found in textbook treatments of the DIG~\cite{ColemanErice,MarinoLectures,Kleinert,ZinnJustinBook}\footnote{There is an important subtlety in the collective-coordinate change of variables which is not discussed in these textbook references: the map from $c_0$ to $t_0$ implicitly assumes a one-to-one correspondence, but in fact it is multi-valued, with the number of preimages $N_{\psi_0}[x]$ equal to the number of times the path $x(t)$ crosses the barrier at $x=0$, and the correct measure includes a factor of $1/N_{\psi_0}[x]$~\cite{CollectiveCoordinateFix}. In the double well each additional crossing costs $e^{-S_I/\hbar}$, so $N_{\psi_0}=1$ up to non-perturbative corrections and the textbook result is recovered at leading order. In the free theory there is no such suppression and ignoring this factor gives the wrong answer for the free twisted partition function by a factor that grows as $\sqrt{T}$. This collective coordinate fix will be important for the integral over quasi-zero modes in Sections~\ref{sec:n3_twisted_geometry} and~\ref{sec:n4}.}. We will generalize this calculation to finite $T$ in Section~\ref{sec:transverse_det} from which Eq.~\eqref{eq:K_DIG_prefactor} can be recovered in the large-$T$ limit.

In the DIG approximation, each instanton sector contributes a factor $K\,T\,e^{-S_I/\hbar}$ to the partition function, with $K$ the fluctuation determinant, $e^{-S_I/\hbar}$ from the instanton action, and $T$ from the integral over the collective coordinate $t_0$ for the instanton center. Then the DIG estimate for the partition function takes the schematic form
\begin{equation}
  Z
  \approx
  2\,Z_\text{SHO}(T)\sum_{k=0}^{\infty}\frac{1}{(2k)!}\left(K T\,e^{-S_I/\hbar}\right)^{2k}
  =
  2\,Z_\text{SHO}(T)\,\cosh\!\Bigl(KT\,e^{-S_I/\hbar}\Bigr),
  \label{eq:DIG_cosheff}
\end{equation}
where $Z_\text{SHO} = 1/(2\sinh\frac{T}{2})$ is the SHO partition function and only the periodic saddles (with even $n$ instantons) contribute. The factor of 2 is from allowing the instanton trajectories to start from either well and $1/(2k)!$ accounts for the overcounting from integrating the $2k$ centers over $[0,T]^{2k}$ instead of the ordered region $0<t_1<t_2<\cdots<t_{2k}<T$.

Unfortunately, the DIG approximation is not systematically improvable. It has three fundamental limitations
\begin{enumerate}
\item \emph{Merger/overlap region:} When integrating around each instanton sector, one cannot naturally prevent the integration region from entering other instanton sectors. For example, the integral over instanton separations can enter a region where an instanton-anti-instanton pair annihilates. Ignoring these transition boundaries would lead to overcounting. To properly account for the boundaries would require a very complicated partition of the integration domain into instanton sectors, making higher order corrections exceedingly difficult.
\item \emph{Non-Gaussian fluctuations:} 
In the double well, the instanton separation directions are actually quasi-zero modes, not exact zero modes at finite $T$ (as the basic DIG assumes). These quasi-zero modes have exponentially small negative eigenvalues under the multi-instanton fluctuation operator $\opO_{n}$. In the Gaussian approximation, these negative eigenvalues would lead $Z$ to be imaginary. 
\item \emph{Exponentially suppressed terms:}
If one is only interested in the leading energy splitting, the DIG gives the same answer as WKB. However, if one wants the full trans-series, including systematically all of the exponentially suppressed terms, one must properly account for all exponentially small effects, including not just the domain boundaries, quasi-zero modes, and collective-coordinate intersection numbers~\cite{CollectiveCoordinateFix}, but the full non-linear dependence of the action on the instanton separations.
\end{enumerate}
These limitations seem essentially insurmountable. However, they all follow from the insistence on dividing the integration region into real paths in Eq.~\eqref{eq:decomp_dig_paths}. As we will see, generalizing to complex paths elegantly sidesteps the challenges of the DIG and leads to a systematically improvable expansion.

\subsubsection{Thimble decomposition}\label{sec:thimble_decomposition} 
The mathematically rigorous way to incorporate the multi-instanton saddles is not to chop up the real integration measure as in Eq.~\eqref{eq:decomp_dig_paths} but to decompose it into thimbles. Recall from Section~\ref{sec:complex_instantons} that the periodic saddles are labeled by $(k,k')$ with instanton number $n=2k$ and $k'$ the branch label (the anti-periodic saddles are labeled analogously by $(m,k')$ with $n=2m+1$ in Section~\ref{sec:antiperiodic_solutions}). The thimble decomposition of the partition function can be written as
\begin{equation}
  Z = \int_{\GR} \mathcal{D} x\, e^{-\frac{1}{\hbar} S[x]} = \sum_{k,k'} \eta_{k,k'}\, Z_{k,k'},
  \qquad \GR = \sum_{k,k'} \eta_{k,k'}\, \mathcal{J}_{k,k'}
 \label{ZsumGeneral} \,,
\end{equation}
where $\GR$ is the space of real periodic paths and ${\mathcal{J}_{k,k'}}$ are the thimbles passing through the saddles $x_{k,k'}$. The objects
\begin{equation}
    Z_{k,k'} = \int_{\mathcal{J}_{k,k'}} \mathcal{D} x\, e^{-\frac{1}{\hbar} S [x] }
\end{equation}
 are the path integrals over the thimbles. The coefficients $\eta_{k,k'}$ in Eq.~\eqref{ZsumGeneral} are the intersection numbers. They are determined by the number of times, with orientation, the unstable trajectories ${\mathcal K}_{k,k'}$ (the anti-thimbles, orthogonal to the ${\mathcal J}_{k,k'}$) intersect the original integration contour over real paths. Generally $\eta_{k,k'}=0,\pm 1$. As we will see, all complex saddles (those with $k'\neq 0$) have $\eta_{k,k'}=0$ and do not contribute.

The decomposition in Eq.~\eqref{ZsumGeneral} produces the trans-series for $Z$ with each instanton sector coming from the series expansion around the appropriate saddle. An exact expression for these series is
\begin{align}
  Z_{k,k'}
  &\sim
    \int_{\Gamma_{k,k'}}  \mathcal{D} x\,
  e^{-\frac{1}{\hbar} S[x_{k,k'}] -\frac{1}{2 \hbar} S''[x_{k,k'}]\, x^2 }\, \sum_{n = 0}^{\infty} \frac{1}{n!} \frac{1}{\hbar^n}\left(- S[x] + S [x_{k,k'}] + \frac{1}{2} S''[x_{k,k'}] \, x^2\right)^n
  \\
 &=  e^{-\frac{1}{\hbar} S[x_{k,k'}]} \sum_{n,m=0}^\infty c^{(k,k')}_{n,m}\, \hbar^n \ln^m \hbar \,,
\end{align}
where the contour $\Gamma_{k,k'}$ is the affine hyperplane through the saddle in the directions of steepest descent. All saddles other than the perturbative saddles at $x=\pm 1$ (which have $k=0$) will have at least one negative eigenvalue in the Hessian $S''[x_{k,k'}]$, requiring at least one imaginary integration direction. The Borel resummation of these series will reproduce the full $Z_{k,k'}$, and when these are summed to produce $Z$, the imaginary parts will cancel. In the path integral picture, the resurgent structure is most transparent in the partition function $Z$, not the energies. How the trans-series for $Z$ translates into a trans-series for $E$ will be worked out in detail in Section~\ref{sec:pi_spectrum}.

To proceed, we need to start by determining the intersection numbers $\eta_{k,k'}$ which determine which saddles actually contribute. We will prove that only the real saddles contribute, which restricts the sum to $k'=0$, i.e.\ real paths with an even number $n=2k$ of instantons. We will find that each such saddle with $k>0$ has Morse index $\mu_{k,0} = 2k-1$, meaning there are $2k-1$ real directions in which the action decreases. The action is nearly flat in all of these directions, so that they correspond to quasi-zero modes of the fluctuation operator.

\subsubsection{Morse indices from the Lam\'e equation}
\label{sec:morse}
For a real saddle point $x_*$, the Morse index is the number of negative eigenvalues of the fluctuation operator
\begin{equation}
  \opO_\star = -\partial_t^2 -U''(x_\star) = -\partial_t^2 + \tfrac{1}{2}\!\left(3x_*^2-1\right) \label{fluctop} \,,
\end{equation}
acting on functions with the same boundary conditions as the saddle (periodic or anti-periodic with period $T$). The Morse index determines the local thimble structure: each negative eigenvalue contributes one integration direction along $i\mathbb{R}$ rather than $\mathbb{R}$. For complex saddles, the spectrum is generically complex and there is no canonical ordering by sign, so the Morse index (``number of negative eigenvalues'') is not a well-defined invariant. In any case, as we will see in Section~\ref{sec:intersection_numbers}, all complex saddles have zero intersection number and do not contribute to $Z$. So we focus on the real saddles here.

At $T=\infty$ the single-instanton solution is exactly $x_I(t) = \tanh(t/2)$, and the fluctuation operator is $\opO_I = -\partial_t^2 + 1 - \frac{3}{2}\operatorname{sech}^2(t/2)$. Changing variables to $u=t/2$ gives
\begin{equation} \label{eq:schro_opp_pt}
  \opO_I = \frac{1}{4}\left[-\partial_u^2 + 4 - 6\operatorname{sech}^2(u)\right].
\end{equation}
This is, up to an energy shift, a Schr\"odinger operator for the P\"oschl--Teller potential $V_{n_L}(u)=- n_L(n_L+1)\operatorname{sech}^2(u)$ with $n_L=2$, and its spectrum is known exactly: a zero mode ($\lambda = 0$, the translation mode), a single bound state ($\lambda = 3/4$), and a continuum starting at $\lambda = 1$. This is the exact spectrum at $T=\infty$.

At finite $T$ the $\tanh$ profile no longer satisfies the boundary conditions (as discussed in Section~\ref{sec:approx_solutions}). The exact single-instanton anti-periodic saddle for $t \in [-T/2,T/2]$ is given in Eq.~\eqref{xCIfirst}:
\begin{equation}
  x_\cI(t) = \sqrt{\frac{2\sig^2}{1+\sig^2}}\;\operatorname{sn}\!\left(\frac{t}{\sqrt{2(1+\sig^2)}},\,\sig^2\right),
\end{equation}
where $\sig$ is the elliptic modulus. The general $n=2k$ periodic saddle $x_{k,0}$ from Sec.~\ref{sec:complex_instantons} traverses the same classical orbit $k$ times, so it is given by the same function with the period condition\footnote{We can check that our formulation can be obtained from Eq.~\eqref{periods} after using an ascending Landen transformation.} $T = 2k \omega_N(\varepsilon) = 4k\,K(\sig^2)\sqrt{2(1+\sig^2)}$ (compared to $T = \omega_N(\varepsilon)= 2K(\sig^2)\sqrt{2(1+\sig^2)}$ for the single instanton). In both cases, the fluctuation operator~\eqref{fluctop} has an eigenvalue equation $\opO_\star\,\psi(t) = \lambda\,\psi(t)$ which becomes, in the variable $u = t/\sqrt{2(1+\sig^2)}$
\begin{equation}
  -\psi''(u) + 6\sig^2\,\operatorname{sn}^2(u,\sig^2)\;\psi(u) = \mathcal{E}\,\psi(u)\,,
  \qquad \mathcal{E} = (1+\sig^2)(1+2\lambda)
  \label{eq:Lame_main}\,.
\end{equation}
This is the \textbf{Lam\'e equation} with index $n_L=2$ (since $n_L(n_L+1)\sigma^2=6\sigma^2$), where the subscript distinguishes the Lam\'e index from the instanton number $n=2k$. More details of the derivation and the Lam\'e equation are given in Appendix~\ref{appendix:Lame}. The equation has the same form for all real saddles, but the elliptic modulus $\sigma$ depends on the instanton number $n$. For complex saddles the operator is non-Hermitian and the band-structure discussion below does not directly apply.

Like the P\"oschl--Teller case, the Lam\'e equation is exactly solvable~\cite{DLMF,WhittakerWatson}. Its spectrum is gapped, so the Morse index is simply the number of eigenvalues below zero. The potential $6\sig^2\operatorname{sn}^2(u)$ has fundamental period $2K(\sig^2)$ and the eigenvalues always fall within three allowed bands, separated by two forbidden gaps. These bands and the numerical eigenvalues for the $6$-instanton case are shown in Fig.~\ref{fig:lame_bands}; the band-edges are computed exactly in Appendix~\ref{appendix:Lame}. The physical interpretation of the three bands is most transparent in the $\sig\to 1$ ($T\to\infty$) limit, where the Lam\'e equation reduces to P\"oschl--Teller (Appendix~\ref{app:PT_recovery}). In this limit both gaps close: the zero mode $\lambda=0$, the bound state at $\lambda=3/4$, and the continuum threshold at $\lambda=1$ are isolated spectral points. As $\sig$ decreases from~$1$ (finite $T$), each of these points fans out into a band:
\begin{itemize}
\item \textbf{Band~I} ($\lambda < 0$): grows from the zero mode. Its eigenvalues are exponentially small and negative, corresponding to quasi-zero modes---the relative separations of instantons.
\item \textbf{Band~II} ($0 < \lambda < 1$): grows from the bound state at $\lambda = 3/4$. These are the ``shape'' modes of the instantons.
\item \textbf{Band~III} ($\lambda > 1$): grows from the continuum threshold. These are the scattering modes.
\end{itemize}
For saddles with $n=2k$ instantons on a circle there are $n$ eigenvalues in Band~I (including the zero mode, leaving $n-1$ quasi-zero modes), $n$ eigenvalues near $3/4$ in Band~II, and an unbounded discrete tower in the semi-infinite Band~III.
These results are derived in Appendices~\ref{app:Lame_spectrum} and~\ref{app:Floquet_counting}, where the band-edge spectrum and Floquet/Bloch counting of the Lam\'e equation are explained in detail.

Two features of the spectrum are important for the Morse index. First, all eigenvalues in Bands~II and~III have $\lambda > 0$. Second, there is always an exact zero mode ($\lambda=0$, eigenfunction $\operatorname{cn}(u)\operatorname{dn}(u) \propto \dot x_*$) which sits at the \emph{top} of Band~I. Therefore all negative eigenvalues lie in Band~I and the Morse index is
\begin{equation}
  \boxed{\mu_{k,0}=2k-1=n-1}\,.
  \label{eq:Morse_index}
\end{equation}
The first cases are $\mu_{1,0}=1$ (real $n=2$) and $\mu_{2,0}=3$ (real four-instanton, $n=4$), consistent with the explicit mode counting in Sec.~\ref{sec:n4}. For the perturbative saddles $x_*(t)=\pm 1$, the operator $\opO_0 = -\partial_t^2 +1$ has only positive eigenvalues, confirming $\mu=0$. For the single instanton with anti-periodic boundary conditions, only the AP band-edge $\Delta=-2$ in Band~I is allowed, giving $\lambda=0$ and no negative eigenvalues, so $\mu=0$. The band structure for the $k=3$ case is shown in Fig.~\ref{fig:lame_bands}; the same Floquet machinery is also used in Section~\ref{sec:transverse_det} to compute the determinant of the same fluctuation operator.

\begin{figure}[t]
  \centering
  \includegraphics[width=0.8\textwidth]{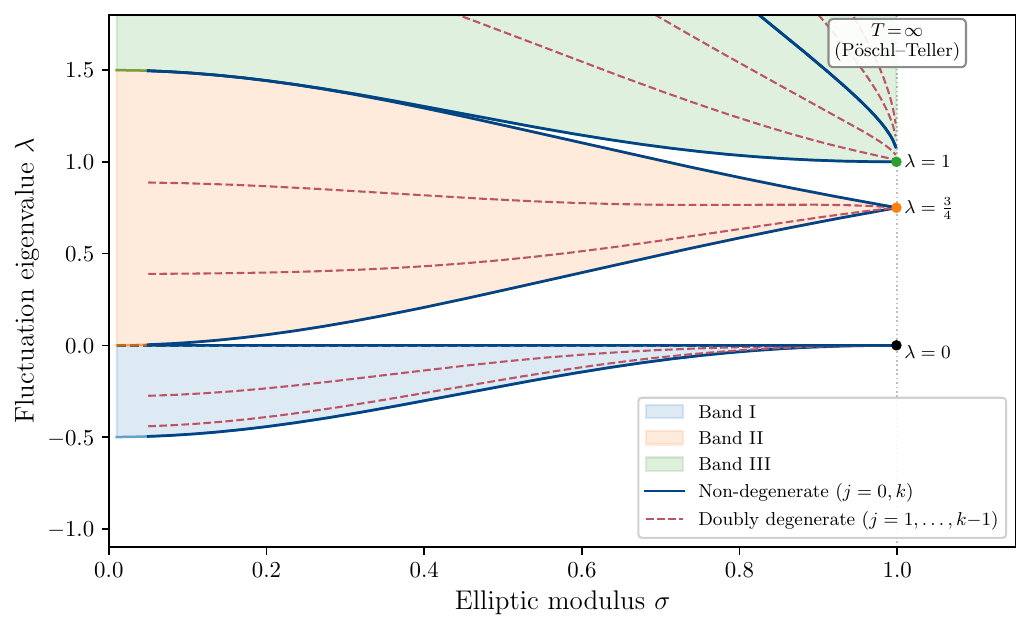}
  \caption{Band structure of the Lam\'e equation as a function of the elliptic modulus~$\sigma$, with the eigenvalues for a $k{=}3$ (6 instanton) periodic saddle overlaid. The zero mode sits at the top of Band~I ($\lambda=0$, solid). At large $T$, $\sigma \sim 1-8e^{-T/(4k)}$ ($T/12$ for the plotted $k=3$ case), so  $\sig\to 1$  as $T\to\infty$. In this limit both gaps close and the spectrum degenerates to the P\"oschl--Teller result (Appendix~\ref{app:PT_recovery}). Solid curves are non-degenerate eigenvalues ($j=0$ and $j=k$); dashed curves are doubly degenerate ($j=1,\ldots,k{-}1$). In Bands~I and~II the $j{=}0$ and $j{=}k$ eigenvalues trace the band edges, but Band~III is semi-infinite so the $j{=}k$ eigenvalue (first anti-periodic point, $\Delta=-2$) lies in the interior of the band.}
  \label{fig:lame_bands}
\end{figure}

\subsubsection{Intersection numbers of the complex saddles}
\label{sec:intersection_numbers}

Next we turn to the intersection number $\eta_{k,k'}$ which appears in the Picard--Lefschetz decomposition $\GR = \sum \eta_{k,k'} \mathcal{J}_{k,k'}$ of the path integral. The intersection number counts the (signed) intersections between the original integration cycle $\GR$ of \emph{real} periodic paths and the unstable thimble passing through the $x_{k,k'}$ saddle. Along the downward flow,
\begin{equation}
  \partial_u x(u,t)=-\overline{\frac{\delta S}{\delta x(u,t)}} ,
\end{equation}
we have decreasing $\re S$ and constant imaginary part
\begin{equation}
  \frac{d}{du}\re S[x(u)] = -\left\|\frac{\delta S}{\delta x}\right\|^2\le 0,
  \qquad
  \frac{d}{du}\im S[x(u)] = 0.
\end{equation}

\begin{lemma}\label{lem:real_thimble}
If the saddle $x_*$ is not real, its unstable thimble ${\mathcal K}_*$ cannot intersect $\Gamma_{\mathbb R}$, and hence $\eta_*=0$.
\end{lemma}
\begin{proof}
Because the action is real on $\Gamma_{\mathbb R}$, the real cycle is invariant under the flow: if $x(u_0,t)\in\Gamma_{\mathbb R}$ then $\delta S/\delta x$ is real, so $\partial_u x$ is real and the entire flow line remains in $\Gamma_{\mathbb R}$. Now suppose $x_0\in{\mathcal K}_*\cap\Gamma_{\mathbb R}$. By definition of the unstable thimble, the flow line through $x_0$ satisfies $x(u)\to x_*$ as $u\to-\infty$. But since $x_0$ is real, the flow line stays in $\Gamma_{\mathbb R}$ for all $u$, so its limit is a \emph{real} critical point, contradicting the assumption that $x_*$ is complex. 
\end{proof}

\noindent 
The proof relies on two properties: (i) $S$ is holomorphic on the complexified configuration space, so the gradient flow is well-defined and preserves $\im \, S$ ; and (ii) $S$ is real on $\Gamma_{\mathbb{R}}$ and compatible with complex conjugation so the real cycle is flow invariant. Both hold for the double-well path integral\footnote{The lemma can fail when either condition is violated; for instance, in theories with fermionic degrees of freedom, integrating out the fermions can produce contributions to the effective action that introduce branch-cuts or complex phases~\cite{Behtash:2015kna, Behtash:2015loa}.}.

Since every saddle with $k'\neq0$ is complex-valued (regardless of whether $\im S$ happens to vanish), we conclude
\begin{equation}
  \boxed{\eta_{k,k'}=0\qquad (k'\neq 0)}\,.
\end{equation}
In particular, the self-conjugate branch $k'=k$ is also excluded, even though $\im S_{k,k}=0$.

The exclusion of non-self-conjugate complex branches is easier to understand with a quick consistency check. Along the flow $\im S$ is conserved and therefore matches the large-$T$ expansion of the saddle action in Eq.~\eqref{eq:action_largeT_periodic},
\begin{equation}
  \im S_{k,k'}(T)=16k\,\sin\!\left(\frac{\pi k'}{k}\right)e^{-T/(2k)}
  +\mathcal{O}\!\left(T\,e^{-T/k}\right) \,.
\end{equation}
So for $0<k'<k$ one finds $\im S_{k,k'}\neq0$ along the flow, which cannot possibly intersect $\Gamma_\mathbb{R}$ where the action is real, consistent with $\eta_{k,k'}=0$, and the fact that only the real saddles contribute to $Z$ for real $\hbar$.

\subsection{Partition function in an instanton background}
\label{sec:exact_det}
At this point, we have shown that at finite $T$ the path integral over all real paths $\GR$ can be decomposed into a sum of thimble integrals passing through the real saddles $x_{k,0}$:
\begin{equation}
  Z = \int_{\GR} \mathcal{D} x\, e^{-\frac{1}{\hbar} S[x]} = \sum_k \eta_{k,0}\, Z_{k,0} \label{ZsumReal} \,,
\end{equation}
with $\eta_{k,0}=\pm1$ depending on the convention for the orientation of the thimble contour $\cJ_{k,0}$. The real saddles contributing to $Z$ have instanton number $n=2k$. For $k\ge 1$ they have one exact zero mode (global translation) and $n-1$ quasi-zero modes with exponentially small negative eigenvalues; the $k=0$ perturbative saddles have no zero modes. Each $Z_{k,0}$ can be computed perturbatively around the $x_{k,0}$ saddle\footnote{The path integral measure includes an overall normalization constant $\mathcal{N}$ (Eq.~\eqref{PInorm}) which combines with the Gaussian integration over all modes to produce $Z_{\text{SHO}} = 1/(2\sinh(T/2))$ for the perturbative saddle. In what follows, all fluctuation determinants are regularized in the sense that they implicitly absorb $\mathcal{N}$. We note that determinant ratios such as $\det\mathcal{O}_0/\det_\perp\mathcal{O}_{k,k'}$ are independent of this regularization.}. For $k\ge 1$ these series are all of the form
\begin{equation}
  Z_{k,0} = e^{-\frac{2k S_I}
  {\hbar}}
  \sum_{a=0}^\infty
  \sum_{b=0}^{2k-1} c_{a,b}\, \hbar^a\ln^b \hbar \,.
\end{equation}
That is, $Z_{k,0}$ isolates the $n=2k$-instanton sector in the full partition function trans-series. It is nevertheless useful to write the general factorization for $Z_{k,k'}$, even though only the real branch $k'=0$ contributes to $Z$. As we will see, logarithms and imaginary parts in the trans-series, which underlie its resurgent structure, come entirely from the integral over the finite-dimensional manifold of quasi-collective coordinates.

We factor the full infinite-dimensional thimble into a finite-dimensional integration over the $2k-1$ Quasi-Zero Mode (QZM) directions and an infinite-dimensional integration over transverse modes with eigenvalues of order one. The essential ingredient for resurgence is the finite-dimensional piece,
\begin{equation} \label{eq:quasi_collective_integral}
  \cI_{k,k'} = \int_{\cJ_{k,k'}}  \!\!\!d^{2k-1}\alpha\, e^{-\frac{1}{\hbar} S_\text{eff}(\alpha)} \,,
\end{equation}
where, with a slight abuse of notation, we now use $\cJ_{k,k'}$ for the middle-dimensional complex cycle in quasi-collective coordinate space (the finite-dimensional analog of the full thimble defined in Eq.~\eqref{ZsumGeneral}); $\alpha_p$ are the quasi-collective coordinates; and $S_\text{eff}(\alpha)$ is the effective action on the quasi-collective coordinate space, measured relative to the saddle.  The structure of $S_\text{eff}$ will be worked out in detail beginning in the next section.  For real instantons ($k'=0$, $n=2k$) the lateral thimble integral reduces to $\cI_{k,0}^{\pm} \equiv \cI_n^{\pm}$ as defined in Eq.~\eqref{eq:gen_thimble_def}, with the saddle-action $nS_I$ subtracted from $\Seff$ via Eq.~\eqref{eq:Seffn_def}.

Although all the resurgent structure is contained within $\cI_{k,k'}$, the full $Z_{k,k'}$ is needed to compute the energy spectrum and requires integrating over the transverse directions and the zero mode. It has the general form
\begin{equation}
\boxed{
  Z_{k,k'}
  =  \sqrt{\frac{\|\dot x_{k,k'}\|^2}{2\pi\hbar}}\;T\;
  e^{-\frac{1}{\hbar} S_{k,k'}}\,
(\det\nolimits_\perp \opO_{k,k'})^{-1/2}\,
\, e^{T \DV}\,
 e^{2k\DL} \, Y_{k,k'} \,
 \cI_{k,k'}
 } \,.
  \label{eq:Z_factored}
\end{equation}
Here, $\|\dot x_{k,k'}\|^2 \equiv \int_0^T \dot x_{k,k'}^2\,dt$ is the collective-coordinate Jacobian factor, computed in Eq.~\eqref{eq:xdot2_action} below, and the $T$ comes from the integral over that coordinate; $S_{k,k'}$ is the classical action of the saddle in Eq.~\eqref{eq:actionexactsolsb}; $(\det_\perp \opO_{k,k'})^{-1/2}$ comes from the Gaussian integral over the transverse modes with the zero mode and all quasi-zero modes removed (Section~\ref{sec:transverse_det_largeT});  $T\DV$ is the sum over connected perturbative vacuum bubbles (Section~\ref{sec:Feynman});  $\DL$ is the sum over corrections to those vacuum bubbles coming from the instanton background (Section~\ref{sec:higher_loop_instanton} and Appendix~\ref{appendix:loop_corrections}) and $Y_{k,k'}$ is the remaining Jacobian from the change to quasi-collective coordinate space.  A version of Eq.~\eqref{eq:Z_factored} with the factors computed and inserted can be seen in Eq.~\eqref{Zpluggedin}. The remainder of this section computes these various factors. 

The measure on the QZM and the introduction of the zero-mode deserve explicit comment. Properly isolating the modes leading to Eq.~\eqref{eq:quasi_collective_integral} requires carefully accounting for the introduction of the (quasi) collective coordinates. For this, one can use a Faddeev--Popov projection onto the $n$ exact orthonormal eigenmodes $\Psi_a(t)$ of $\opO_{k,k'}$, since the eigenmode basis is a valid global basis. Defining collective coordinates as global linear projections $c_a[x] \equiv \int_0^T \Psi_a(t)\,(x(t)-x_{k,k'}(t))\,dt$ and decomposing $x(t) = x_{k,k'}(t) + \sum_a c_a\,\Psi_a(t) + x_\perp(t)$, the path-integral measure factorizes as $\mathcal D x = \prod_a dc_a\,\mathcal D x_\perp$ and the $c_a$ range over the global QZM thimble. Crucially, as pointed out in~\cite{CollectiveCoordinateFix}, one must account for the multi-valuedness of the map from local fluctuation coordinates to the collective coordinates. In particular, the introduction of the exact collective coordinate produces an extra \textit{CC-fix factor} $1/N_{\psi_0}[x]=1/n$, where $N_{\psi_0}[x]$ counts the number of times the path crosses the origin. In Eq.~\eqref{eq:Z_factored} this residual factor is included within $Y_{k,k'}$.

The formula in Eq.~\eqref{eq:Z_factored} simplifies dramatically as $T\to\infty$. For instance, in this limit, at $(n=2k,k'=0)$ we have $Y_{k,k'} \sqrt{\|\dot x_{k, k'}\|^2/2\pi \hbar}  \to n^{-1}(S_I/2\pi \hbar)^{n/2}$, $S_{k,k'}\to n S_I$, both $\DV$ and $\DL$ become $T$-independent, $\det\nolimits_\perp \opO_{k,k'} \to 12^{-2k}$, and $\cI_{k,k'}$ reduces to a polynomial of degree $n-1$ in $T$. Even then, this formula goes beyond the dilute instanton gas.  Importantly, however, the factorized form in~\eqref{eq:Z_factored} is rigorous at finite $T$.\footnote{\label{footIQZ}The chief subtlety is defining the quasi-zero manifold globally, which as described can be done using a Faddeev--Popov projection. If instead one defines the QZM manifold using local tanh-based modes (rather than the exact $\Psi_a$), the measure factorization receives a Jacobian whose deviation from unity is set by the Gram matrix $G_{ab} = \langle\psi_a,\psi_b\rangle$ of the would-be zero modes; off-diagonal entries of $G_{ab}$ are $\cO(e^{-\alpha})$ from neighboring-instanton overlaps and are subleading to $\Seff(\alpha)$, so we drop them throughout. A worked example of the Faddeev--Popov construction beyond 1-loop is given in Appendix~\ref{appendix:loop_corrections}.} Many of the objects can even be computed exactly: the $\|\dot x_{k,k'}\|$ and $S_{k,k'}$ factors and the 1-loop functional determinant around any instanton background can be computed exactly, and $\DV$ can be computed exactly in perturbation theory at least to 3 loops. However, finite-$T$ calculation of $\DL$ and $\cI_{k,k'}$ are more challenging. Working at finite $T$ around the perturbative saddle and to leading order at large $T$ around the instanton saddles is enough to see the full perturbative energy spectrum and the leading non-perturbative corrections (Section~\ref{sec:pi_spectrum}). More importantly, it is enough to see the full resurgent structure of the partition function  (Sections~\ref{sec:n2thimble}-\ref{sec:n4}). This structure has a geometric origin in the manifold of quasi-collective coordinates where the thimbles can be visualized as steepest-descent contours of the relatively simple effective action $\Seff$ given in Eq.~\eqref{Seffalpha}.

\subsubsection{Exact 1-loop determinant}
\label{sec:transverse_det}
The 1-loop contribution to the path integral around a saddle is $(\det \opO)^{-1/2}$, where $\opO$ is the quadratic fluctuation operator. In our case,
\begin{equation}
  \opO_{k,k'}=-\partial_t^2 - U''[x_{k,k'}(t)],
  \qquad
  U''(x)=\tfrac{1}{2}-\tfrac{3}{2} x^2.
  \label{eq:fluct_op_periodic}
\end{equation}
One way to compute this determinant is to find all eigenvalues $\lambda_n$ of $\opO$ and take their (regularized) product. A slicker way is to use the Gel'fand--Yaglom method~\cite{GelfandYaglom,DunneGY} which avoids finding the $\lambda_n$; instead, one solves the homogeneous equation $\opO\,\psi = 0$ as an initial-value problem and reads off the determinant from the solution evaluated at the boundary. For Dirichlet boundary conditions on $[0,T]$, one solves $\opO\,\psi = 0$ with $\psi(0) = 0$, $\psi'(0) = 1$ and obtains $\det\opO = \psi(T)$. For periodic boundary conditions, the generalization~\cite{Forman1987,DunneGY} uses the same idea but requires propagating \emph{two} independent solutions to build the monodromy matrix (since periodicity constrains both $\psi$ and $\psi'$, not just one endpoint value). If $\opO$ has a zero mode, as is the case for our non-trivial periodic saddles ($k\ge 1$), the determinant vanishes and one needs the reduced determinant $\det'$ with the zero mode removed. This is obtained by instead solving $(\opO-\lambda)\,\psi = 0$, differentiating in $\lambda$ and evaluating at $\lambda=0$. We now carry out this program.

Since $x_{k,k'}(t)$ is periodic with period $T$, so is $U''(x_{k,k'}(t))$, and $\opO_{k,k'}$ is a Hill operator~\cite{MagnusWinkler} (i.e. a Schr\"odinger operator for a periodic 1D potential). Differentiating the equation of motion $\ddot x = -U'(x)$ with respect to $t$ shows that the translation mode $\dot x_{k,k'}(t)$ is an exact periodic zero mode:
\begin{equation}
  \opO_{k,k'}\,\dot x_{k,k'}(t)=0, \qquad \dot x_{k,k'}(t+T)=\dot x_{k,k'}(t).
\end{equation}
Therefore the full periodic determinant vanishes,
\begin{equation}
  \det \opO_{k,k'}=0 \qquad (k\ge1),
\end{equation}
and the physically relevant quantity is the reduced determinant $\det'\,\opO_{k,k'}$ with the translation zero mode removed.
 
Following the Gel'fand--Yaglom strategy, we solve $(\opO - \lambda)\,\psi = 0$ as an initial-value problem with two sets of initial conditions. Let $\psi_1(t;\lambda)$ and $\psi_2(t;\lambda)$ be the fundamental solutions with
\begin{equation} \label{eq:bcs_psi12}
  \psi_1(0,\lambda)=1,\; \psi_1'(0,\lambda)=0;\qquad \psi_2(0,\lambda)=0,\; \psi_2'(0,\lambda)=1.
\end{equation}
Propagating these across one full period gives the monodromy matrix:
\begin{equation}
  M(\lambda) = \begin{pmatrix} \psi_1(T;\lambda) & \psi_2(T;\lambda) \\ \psi_1'(T;\lambda) & \psi_2'(T;\lambda) \end{pmatrix}.
\end{equation}
Since $(\opO-\lambda)$ has no first-derivative term, $\det M = 1$ (the Wronskian is conserved). A periodic solution exists iff its Cauchy data are fixed by $M$, which requires $M$ to have an eigenvalue $1$. Coupled with $\det M =1$, this implies that the \emph{Hill discriminant} $\Delta(\lambda) \equiv \operatorname{tr} M(\lambda) = \psi_1(T;\lambda) + \psi_2'(T;\lambda)$ equals $2$ at the periodic eigenvalues. These eigenvalues of $\opO$ are therefore the roots of $\Delta(\lambda) = 2$, and the full periodic determinant is~\cite{MagnusWinkler,Forman1987}
\begin{equation}
  \det(\opO - \lambda) = \Delta(\lambda) - 2
  \label{eq:detP_discriminant}\,.
\end{equation}
For instance, around the perturbative saddle we have
\begin{equation}
  \Delta_0(\lambda) = 2\cosh(T\sqrt{1-\lambda}),
  \qquad
  \det \opO_0 = 4\sinh^2(T/2) \quad (\lambda=0),
\end{equation}
for the regularized determinant, in which the path integral measure normalization $\mathcal{N}$ is absorbed.

For the non-trivial saddles, $\lambda = 0$ is a simple periodic eigenvalue, namely the translation mode, so the reduced determinant with the zero mode removed is
\begin{equation}
  \det\nolimits'\,\opO_{k,k'}
  = -\Delta'(0).
  \label{eq:detprime_discriminant}
\end{equation}
This reduces computing $\det'$ to computing the $\lambda$-derivative of the Hill discriminant at $\lambda = 0$. To compute $\Delta'(0)$, we start by recovering two independent solutions of $\opO_{k,k'} \phi = 0$. The first solution is the translation mode $\phi_1 = \dot x_{k,k'}$, which is periodic. A second independent solution is obtained by differentiating the classical solution with respect to the conserved energy $\varepsilon$ in Eq.~\eqref{Eeq}. Since the equation of motion does not depend explicitly on $\varepsilon$,
\begin{equation}
  \opO_{k,k'}\,\partial_\varepsilon x_{k,k'}(t)=0,
\end{equation}
so $\phi_2 = \partial_\varepsilon x_{k,k'}$. Unlike $\phi_1$, it is \emph{quasi}-periodic: differentiating the periodicity condition $x_{k,k'}(t+T(\varepsilon);\varepsilon) = x_{k,k'}(t;\varepsilon)$ with respect to $\varepsilon$ gives
\begin{equation}
  \phi_2(t+T) = \phi_2(t) - T'(\varepsilon)\,\phi_1(t),
  \qquad T'(\varepsilon)\equiv \frac{dT}{d\varepsilon}\,.
  \label{eq:quasiperiodic_eps}
\end{equation}
The Wronskian can be computed from the energy conservation law $\frac{1}{2}\dot x^2 + U(x) = \varepsilon$. Differentiating with respect to $\varepsilon$:
\begin{equation}
  \dot x\,\partial_\varepsilon \dot x + U'(x)\,\partial_\varepsilon x = 1.
\end{equation}
Since $\ddot x = -U'(x)$, this recovers the Wronskian as
\begin{equation}
  W(\phi_1,\phi_2) \equiv \phi_1\,\phi_2' - \phi_1'\,\phi_2 = \dot x\,\partial_\varepsilon \dot x - \ddot x\,\partial_\varepsilon x = 1 \,.
  \label{eq:wronskian_jacobi}
\end{equation}

To find $\Delta'(0) = \partial_\lambda[\psi_1(T;\lambda)+\psi_2'(T;\lambda)]|_{\lambda=0}$, we first define $\tilde \psi_i(t) = \partial_\lambda \psi_i(t;\lambda)|_{\lambda=0}$. Recalling that $(\opO - \lambda)\,\psi_i(t,\lambda) = 0$, these functions satisfy the inhomogeneous equation
\begin{equation}
  \opO_{k,k'}\,\tilde \psi_i = \psi_i(t;0),\qquad \tilde \psi_i(0) = \tilde \psi_i'(0) = 0\,,
\end{equation}
since the initial conditions are $\lambda$-independent. We solve this equation by variation of parameters using $\phi_1, \phi_2$ with Wronskian $W=1$ and $\opO \phi=0$. For $\opO\,\tilde \psi = \psi$ with $\tilde \psi(0) = \tilde \psi'(0) = 0$, the standard formula gives
\begin{equation}
  \tilde \psi(t) = \phi_1(t)\int_0^t \phi_2(s)\,\psi(s,0)\,ds - \phi_2(t)\int_0^t \phi_1(s)\,\psi(s,0)\,ds.
  \label{eq:var_of_params}
\end{equation}
The unperturbed fundamental solutions $\psi_1(t;0)$ and $\psi_2(t;0)$ with the boundary conditions of Eq.~\eqref{eq:bcs_psi12} are linear combinations of $\phi_1$ and $\phi_2$. Writing $a = \phi_1(0)$, $b = \phi_1'(0)$, $c = \phi_2(0)$, $d = \phi_2'(0)$ with $ad - bc = 1$:
\begin{equation}
  \psi_1(t;0) = d\,\phi_1(t) - b\,\phi_2(t),\qquad \psi_2(t;0) = -c\,\phi_1(t) + a\,\phi_2(t)\,.
\end{equation}
Substituting into Eq.~\eqref{eq:var_of_params}, evaluating at $t=T$, and using the monodromy relations $\phi_1(T) = \phi_1(0) = a$, $\phi_1'(T) = b$, $\phi_2(T) = c - T'a$, $\phi_2'(T) = d - T'b$ from Eq.~\eqref{eq:quasiperiodic_eps}, the cross terms all cancel (after using $ad-bc=1$) and only one integral survives:
\begin{equation}
  \Delta'(0) = \tilde \psi_1(T) + \tilde \psi_2'(T) = T'(\varepsilon)\int_0^T dt\,\phi_1(t)^2 = T'(\varepsilon)\int_0^T dt\,\dot x_{k,k'}(t)^2.
  \label{eq:deltaprime_master}
\end{equation}

Up to this point we have not used the explicit instanton solutions: the above analysis applies to the fluctuation operator around any periodic orbit of any 1D classical Hamiltonian. Because $x_{k,k'}$ are elliptic functions, both factors in Eq.~\eqref{eq:deltaprime_master} can be expressed in terms of the elliptic curve data. The period is $T = 2k\,\omega_N(\varepsilon) + 2k'\,\omega_P(\varepsilon)$ from Eq.~\eqref{quantcond}, so
\begin{equation}
  T'(\varepsilon) = 2k\,\omega_N'(\varepsilon) + 2k'\,\omega_P'(\varepsilon).
\end{equation}
The remaining integral is the abbreviated action around the closed orbit. Using $\dot x^2 = 2\varepsilon - 2U(x)$ and Eq.~\eqref{eq:actionexactsolsb}, we directly recover
\begin{equation}
  \int_0^T dt\,\dot x_{k,k'}^2 = \oint_{\cC_{k,k'}} \dot x\, dx = 2k\,S_N^0(\varepsilon) + 2k'\,S_P^0(\varepsilon),
  \label{eq:xdot2_action}
\end{equation}
where $S_N^0$ and $S_P^0$ are the same period integrals appearing in the saddle action. Combining Eqs.~\eqref{eq:detprime_discriminant}, \eqref{eq:deltaprime_master}, and \eqref{eq:xdot2_action} gives the exact finite-$T$ reduced determinant:
\begin{equation}
  \boxed{
  \det\nolimits'\,\opO_{k,k'}(T)
  =
  -\bigl(2k\,\omega_N'(\varepsilon)+2k'\,\omega_P'(\varepsilon)\bigr)\,
  \bigl(2k\,S_N^0(\varepsilon)+2k'\,S_P^0(\varepsilon)\bigr)
  } \,,
  \label{eq:detprime_exact}
\end{equation}
where $\varepsilon = \varepsilon_{k,k'}(T)$ is determined implicitly by the quantization condition~\eqref{quantcond}. This is a closed-form expression for the reduced 1-loop determinant around any periodic saddle, valid at any finite $T$, with no dilute-gas or large-separation approximation. It depends on the saddle only through the elliptic curve data $(\omega_N, \omega_P, S_N^0, S_P^0)$ evaluated at the quantized energy.

To extract the large-$T$ behavior we send $T\to\infty$ at fixed $(k,k')$, so $\varepsilon \to 0$. Using the small-$\varepsilon$ expansions $\omega_N(\varepsilon)=-\ln(-\varepsilon/8)+\cO(\varepsilon\ln\varepsilon)$ and $S_N^0(\varepsilon)= S_I + \cO(\varepsilon\ln\varepsilon)$ with $S_I = 2/3$, the two factors become
\begin{equation}
  T'(\varepsilon) = -\frac{2k}{\varepsilon}+\cO(\ln\varepsilon),
  \qquad
  2k\,S_N^0(\varepsilon)+2k'\,S_P^0(\varepsilon)=2k\,S_I+\cO(\varepsilon\ln\varepsilon).
\end{equation}
Together with the large-$T$ form of the quantized energy, $\varepsilon_{k,k'}(T)=-8\,e^{-i\pi k'/k}e^{-T/(2k)}+\cdots$, this gives the leading behavior at large $T$:
\begin{equation}
  \det\nolimits'\,\opO_{k,k'}(T)
  \approx
  \frac{4k^2 S_I}{\varepsilon_{k,k'}(T)}
  \approx
  -\frac{k^2 S_I}{2}\,e^{i\pi k'/k}\,e^{T/(2k)}
  \label{eq:detprime_largeT}
\end{equation} 
The reduced determinant grows exponentially with $T$, as $e^{T/(2k)}$. For real saddles ($k'=0$) this is real and negative, consistent with the Morse index $\mu_{k,0}=2k-1$ being odd. The sign can be checked directly: for $\varepsilon\in(-1/8,0)$ one has $\omega_N'(\varepsilon)>0$ and $S_N^0(\varepsilon)>0$, so the product in Eq.~\eqref{eq:detprime_exact} is positive and the overall minus sign gives $\det'\,\opO_{k,0} < 0$.

For comparison, the vacuum determinant on periodic functions is $\det\nolimits\,\opO_0 = 4\sinh^2(T/2) \sim e^T$ at large $T$, so the ratio
\begin{equation} \label{eq:detprime_ratio_largeT}
  \frac{\det'\,\opO_{k,k'}}{\det\,\opO_0}
  \sim
  -\frac{k^2 S_I}{2}\,e^{i\pi k'/k}\,e^{-(2k-1)T/(2k)}
\end{equation}
is exponentially suppressed for $k\ge 1$, reflecting the $n-1$ quasi-zero eigenvalues that are exponentially small at large separation.

\subsubsection{Exact 1-loop twisted partition function for single instanton background}\label{sec:subleading_T}
The exact $\det'$ formula in Eq.~\eqref{eq:detprime_exact} was derived for periodic saddles. Adapted to anti-periodic boundary conditions, it gives the exact 1-loop fluctuation determinant around the AP single instanton. Since this saddle has only the translation zero mode (and no quasi-zero modes), the result is the exact 1-loop twisted partition function $\widetilde Z_1(T)$ at finite $T$. Expanding at large $T$ produces the full trans-series in $e^{-T}$.

Concretely, we wish to expand about the exact finite-$T$ saddle for $n=1$ from Eq.~\eqref{xCIfirst}, shifted to put the instanton at $t=0$:
\begin{equation}
  x_\cI(t) = \sqrt{\frac{2\sig^2}{1+\sig^2}}\;\operatorname{sn}\!\left(\frac{t}{\sqrt{2(1+\sig^2)}},\,\sig^2\right),
  \label{eq:exact_n1_saddle}
\end{equation}
where $\sig^2 = (1-\sqrt{-8\varepsilon})/(1+\sqrt{-8\varepsilon})$ and with the quantization condition $\omega_N(\varepsilon)=T$.

Expanding the quantization condition at small $\varepsilon$ leads to
\begin{equation}
  T  = - \ln \frac{-\varepsilon}{8} + 
  \varepsilon\left(\frac{3}{2}\ln\frac{-\varepsilon}{8} +5\right) +
  \varepsilon^2\left(- \frac{105}{16}\ln\frac{-\varepsilon}{8} -\frac{389}{16}\right) + \cdots
\end{equation}
which can be inverted perturbatively to 
\begin{equation}
  \varepsilon = -8\, e^{-T} + (320 - 96T)\,e^{-2T} + (-1728 T^2 + 9312 T - 10592)e^{-3T} + \cdots\,.
  \label{ETconversion}
\end{equation}
Just as in Eq.~\eqref{eq:actionexactsols} the corresponding action $S[x_\cI]$ which depends on $T$ is the Legendre transform of the classical period integral $S_N^0(\varepsilon)$ where we identify the energy $E$ of the quantum state with $-\varepsilon$ \footnote{The $-\varepsilon T$ term arises because on shell, energy conservation gives $\frac{1}{2}\dot{x}^2 = V(x) + \varepsilon$, so the Euclidean Lagrangian becomes $L_E = \frac{1}{2}\dot{x}^2 + V = \dot{x}^2 - \varepsilon$, and the action splits as $S_E = \int_0^T (\dot{x}^2 - \varepsilon)\,dt = \oint P_0\,dx - \varepsilon T = S_N^0(\varepsilon) - \varepsilon T$. Since $T = \partial_\varepsilon S_N^0 = \omega_N$, this is precisely the Legendre transform of the period integral $S_N^0(\varepsilon)$ from the energy variable $\varepsilon$ to the time variable $T$. The period integral $S_N^0$ is natural in the energy domain where Exact WKB and Voros symbols live; its Legendre transform $S_E = S_N^0 - \varepsilon T$ is Hamilton's principal function, natural in the time domain where the path integral lives. The two are dual descriptions of the same saddle.}. 
That is
\begin{equation}
  S[x_\cI] = S_N^0(\varepsilon) - \varepsilon T
\end{equation}
where $S_N^0(\varepsilon)$ is given explicitly in Eq.~\eqref{SN0a} after $E\rightarrow - \varepsilon$.
\begin{equation}
  S_N^0  = \int_{x_2}^{x_3} d x P_0(x) = \frac{\pi (1 + 8 \varepsilon)}{4\sqrt{2}} \, {}_2 F_1\!\left( \frac{1}{4}, \frac{3}{4}, 2 ; 1 + 8 \varepsilon \right) \,.
\end{equation} 
The action then expands as
\begin{align}
S[x_\cI] &= S_I + \varepsilon +\varepsilon^2\!\left(-\frac{3}{4}\ln\frac{-\varepsilon}{8}-\frac{23}{8}\right) + \varepsilon^3\left(\frac{271}{16} + \frac{70}{16}\ln\frac{-\varepsilon}{8}\right) + \cdots
\\
 &= \frac{2}{3} - 8\,e^{-T} + (136-48T)\,e^{-2T}
 +(-576 T^2 + 2720 T -2624) e^{-3T} + \cdots \,,
\label{eq:S1_exact_largeT} 
\end{align}
which matches the form of Eq.~\eqref{Seffalpha} with $n=1$.

The exact fluctuation determinant around $x_\cI$ follows the Gel'fand--Yaglom calculation of \S\ref{sec:transverse_det}, adapted to anti-periodic boundary conditions. The zero mode $\dot{x}_\cI$ is now anti-periodic, so it is an exact zero of $\opO_\cI = -\partial_t^2 - U''[x_\cI]$ which must be removed to get a non-zero determinant. The monodromy matrix $M$ still satisfies $\det M =1$, but the anti-periodic condition $M \psi = -\psi$ now requires $M$ to have eigenvalues $-1$. This implies that the Hill discriminant is now $\Delta(\lambda)=-2$ on the eigenvalues. The anti-periodic analog of Eq.~\eqref{eq:detP_discriminant} is $\det(\opO-\lambda) = \Delta(\lambda)+2$, so the reduced determinant is still $\det' \opO_\cI = -\Delta'(0)$. However, a key difference is that the monodromy relations for the $\phi_{1,2}$ now read $\phi_1(T) = -\phi_1(0) = -a$, $\phi_1'(T) = -b$, $\phi_2(T) = -c + T'a$, $\phi_2'(T) = -d + T'b$. Thus $\Delta'(0)=-T'(\varepsilon)\int_0^T dt\,\dot x_\cI(t)^2$ which is Eq.~\eqref{eq:deltaprime_master} with an overall sign. Therefore we have
\begin{equation} 
  \det\nolimits'\,\opO_\cI
  = \omega_N'(\varepsilon)\,S_N^0(\varepsilon)
  = -\frac{3\bigl[S_N^0(\varepsilon)\bigr]^2}{2\varepsilon(1+8\varepsilon)}\,,
  \label{eq:detprimeAP_exact}
\end{equation}
where we use the relation $\omega_N'=-3S_N^0/(2\varepsilon(1+8\varepsilon))$. This is the anti-periodic analog of Eq.~\eqref{eq:detprime_exact}, valid at any finite~$T$. It is real and positive since we have excluded the zero mode and for the single instanton background this leaves us with only positive eigenvalues.

Combining the on-shell action $S[x_\cI]=S_N^0(\varepsilon)-\varepsilon T$, the collective-coordinate integral and Jacobian $T\sqrt{S_N^0/(2\pi\hbar)}$, and the determinant~\eqref{eq:detprimeAP_exact}, the exact 1-loop twisted partition function around the true saddle is
\begin{equation}
  \widetilde{Z}_1(T)
  = \frac{T}{\sqrt{2\pi\hbar\,\omega_N'(\varepsilon)}}
  \;e^{-\frac{1}{\hbar}S[x_\cI]} \,,
\end{equation}
which can be rewritten as
\begin{equation}
  \boxed{
  \widetilde{Z}_1(T)
  = T\sqrt{\frac{-2\varepsilon(1+8\varepsilon)}{6\pi\hbar\,S_N^0(\varepsilon)}}
  \;e^{-\frac{1}{\hbar}(S_N^0(\varepsilon)-\varepsilon T)}
  } \,.
  \label{eq:Ztilde1_exact}
\end{equation}
This closed-form expression is exact at 1 loop for all~$T$. Expanding at large $T$
\begin{multline}
  \widetilde{Z}_1(T) = \frac{2T\,e^{-T/2}}{\sqrt{\pi\hbar}}\,
  \left[1+(12T-46)\,e^{-T}+(252T^2-1410T+1510)\,e^{-2T}+\cdots\right]\;\\
\times  \exp\left[-\frac{1}{\hbar}
\left(S_I - 8\,e^{-T}+(136-48T)\,e^{-2T}+\cdots\right)\right]\,.
  \label{eq:Ztilde1_exact_largeT} 
\end{multline}
Notably, the corrections to the prefactor are at integer powers of $e^{-T}$, with no $e^{-T/2}$ corrections.

\subsubsection{Transverse determinant and quasi-zero modes at large $T$}\label{sec:transverse_det_largeT}
Returning to the periodic multi-instanton saddles, the reduced determinant $\det'$ from Eq.~\eqref{eq:detprime_exact} still includes the $2k-1$ quasi-zero eigenvalues. These modes correspond to relative instanton separations, and for the real saddle their eigenvalues are exponentially small and negative. They cannot be treated in the Gaussian approximation: the Gaussian integral would give an imaginary answer for each negative eigenvalue. Instead, we remove them from the determinant and integrate over the corresponding quasi-collective coordinates separately. This requires knowing their eigenvalues so that we can divide them out of Eq.~\eqref{eq:detprime_exact}.

We know from the DIG that for $n = 2k$ well-separated instantons on the circle, there is a zero mode $\psi_a(t) = \dot x_I(t-t_a)$ associated with each instanton transition $t_a$, with $\|\psi_a\|^2 = \int \dot x_I^2\,dt = S_I$. At finite separation $\alpha = T/n$, the $n$ would-be zero modes hybridize through nearest-neighbor overlap. One linear combination, the global translation, remains an exact zero mode and is already removed in $\det'$. The remaining $n-1$ combinations are the quasi-zero modes. The quasi-zero mode eigenvalues can be computed from the Hessian of the restricted action on the quasi-collective-coordinate manifold. Taking the instanton centers $t_a$ as coordinates, the nearest-neighbor form, valid to leading order in $e^{-T/n}$, is obtained from the $k'=0$ form of Eq.~\eqref{Seffalpha}:
\begin{equation}
  S_\text{eff}(\{t_a\}) = nS_I - 12S_I\sum_{p=1}^{n} e^{-(t_{p+1}-t_p)} + \cdots,
\end{equation}
with equal separations $t_{p+1}-t_p=T/n$. The Hessian is 
\begin{equation}
  H_{ab} \equiv \frac{\partial^2 S_\text{eff}}{\partial t_a\,\partial t_b} = -12S_I\,e^{-T/n}\,L_{ab},
  \label{eq:QCC_hessian}
\end{equation}
where $L$ is the discrete Laplacian on the $n$-cycle ($L_{aa}=2$, $L_{a,a\pm 1}=-1$, others zero). The QZM subspace is spanned by the would-be zero modes $\psi_a$ defined above; projecting $\opO\psi = \lambda\psi$ onto this basis and using $\|\psi_a\|^2 = S_I$, the QZM eigenvalues come out as those of $H/S_I$. Since the eigenvalues of $L$ are $4\sin^2(\pi m/n)$, the eigenvalues of $\opO_{k,0}$ on the QZM subspace are
\begin{equation}
  \lambda_m = -48\,e^{-T/n}\sin^2\!\left(\frac{\pi m}{n}\right),
  \qquad m=0,1,\ldots,n-1.
  \label{eq:QZM_eigenvalues} 
\end{equation}
The $m=0$ mode is the exact zero mode (global translation, $\lambda_0 = 0$). The remaining $n-1$ modes are the quasi-zero modes, all negative, with eigenvalues of order $e^{-T/n}$.

We next define the transverse determinant $\det\nolimits_\perp$ from $\det'$ by removing the QZM eigenvalues $\lambda_m$ computed from $S_\text{eff}$:
\begin{equation}
  \det\nolimits_\perp\opO_{k,0}
  \equiv
  \frac{\det'\,\opO_{k,0}}{\prod_{m=1}^{n-1}\lambda_m}.
  \label{eq:detpp_def}
\end{equation}
This product can be evaluated using $\prod_{m=1}^{n-1}\sin^2(\pi m/n) = n^2/4^{n-1}$ so that
\begin{equation}
  \prod_{m=1}^{n-1}\lambda_m
  = (-1)^{n-1}\,n^2\cdot 12^{n-1}\,e^{-(n-1)T/n}.
  \label{eq:QZM_product}
\end{equation}
Using Eq.~\eqref{eq:detprime_ratio_largeT} with $k'=0$ and $n=2k$, we get
\begin{equation}
  \frac{\det_\perp \opO_{k,0}}{\det \opO_0}
  \approx
  \frac{
  \left(-\frac{k^2S_I}{2}\right)e^{-(2k-1)T/(2k)}
  }{
  \left[-4k^2\,12^{2k-1}\right]e^{-(2k-1)T/(2k)}
  }
  =
  \frac{S_I}{8\cdot 12^{2k-1}}
  =
  \frac{1}{12^{2k}},
  \label{eq:detpp_ratio_DIGnorm}
\end{equation}
where the last equality uses $S_I=2/3$. Both $\prod_m\lambda_m$ and $\det'/\det\opO_0$ scale as $e^{-(2k-1)T/(2k)}$ at leading $T$; that they match is a non-trivial check that the $\lambda_m$ from $S_\text{eff}$ correctly capture the $n-1$ quasi-zero eigenvalues encoded in the exact $\det'$ in Eq.~\eqref{eq:detprime_exact}. The $T$-dependence cancels in the ratio, leaving 
\begin{equation}
\boxed{
  \frac{\det_\perp\opO_{k,0}}{\det \opO_0}
  \sim
  \left(\frac{\det\nolimits'\opO_I}{\det \opO_0}\right)^{2k}
  =\frac{1}{12^{2k}}
  }
  \label{eq:detperp} \,.
\end{equation}
This is finite and $T$-independent and agrees with the DIG factor in Eq.~\eqref{eq:K_DIG_prefactor} raised to the $n=2k$ power.

Eq.~\eqref{eq:detperp} is valid at leading order at large $T$. The finite-$T$ form would keep $\det'\opO_{k,0}$ exact via Eq.~\eqref{eq:detprime_exact} and divide by the exact Band-I Floquet eigenvalues of the Lam\'e discriminant (see Appendix~\ref{appendix:Lame}). The resulting expression would be complicated and also not particularly useful. The challenge is that at the same level of accuracy we would need also to define and integrate along the QZM manifold (see Footnote~\ref{footIQZ}). In contrast, for $n=1$, there are no QZMs, and the expression in Eq.~\eqref{eq:Ztilde1_exact} is exact and useful: it allows us to compute non-perturbative splittings of the excited state energies (see Section~\ref{sec:NP_excited_state}).

\subsubsection{Feynman diagrams}
\label{sec:Feynman}
The functional determinant is the 1-loop or next-to-leading order (NLO) contribution to the partition function. We have been able to compute this exactly around any saddle at finite $T$, in Eq.~\eqref{eq:detprime_exact}. To go to 2 loops and higher we use Feynman diagrams. We first compute the 2-loop and 3-loop diagrams around the perturbative saddle which we can do exactly at finite $T$. Together these determine $\DV$ through order $\hbar^2$. In Section~\ref{sec:higher_loop_instanton} we compute corrections around instanton saddles.

To derive the Feynman rules, we first expand the potential around one of the perturbative saddles ($x=-1$):
\begin{equation}
  V (- 1 + x) = \frac{1}{2}
  x^2 - 3 \frac{x^3}{3!} + 3 \frac{x^4}{4!} \label{Expanded} \,.
\end{equation}
Then we treat the $x^3$ and $x^4$ terms as interactions and evaluate the path integral order-by-order in $\hbar$. That is, we write
\begin{equation}
    Z_P 
    = \sum_{n=0}^\infty {\mathcal N}\int \mathcal{D}x
    e^{ - \frac{1}{\hbar} \int d t \frac{1}{2} x(-\partial_t^2 +1 )x}
    % e^{ - \frac{1}{\hbar} \int d t [-\frac{1}{2}x^3 + \frac{1}{8} x^4]}
     \frac{1}{n!}\left[ - \frac{1}{\hbar} \int d t \left(-\frac{1}{2}x^3 + \frac{1}{8} x^4\right) \right]^n
\end{equation}
where $Z_P$ is the perturbative contribution from expanding around a single perturbative saddle (the full 0-instanton sector is $Z_0 = 2\,Z_P$ from summing both wells via the Picard--Lefschetz decomposition).  We then rescale $x \to \sqrt{\hbar} x$ so that the kinetic term is canonically normalized. The vertices of the Feynman rules then include a cubic interaction of strength $3 \sqrt{\hbar} x^3$ and a quartic interaction of strength $-3 \hbar x^4$. The propagator is the periodic Green's function of the harmonic operator,
\begin{equation}
  \left(-\partial_t^2+1\right)\prop_0(t)=\delta_T(t),\qquad \prop_0(t+T)=\prop_0(t)\,.
\end{equation}
Using the Fourier modes $\psi_n$ from Eq.~\eqref{normalmodes} with eigenvalues $\lambda_n=\omega_n^2+1$:
	\begin{equation}
	\!\!\!  \prop_0(t_1,t_2)
	  = \sum_n \frac{\psi_n(t_1)\,\psi_n^*(t_2)}{\lambda_n}
	  = \frac{1}{T}\sum_{n\in\mathbb{Z}}\frac{e^{i\omega_n (t_1-t_2)}}{\omega_n^2+1}
	   = \frac{\cosh \left(
	  \frac{ T}{2} -  | t_1-t_2 | \right)}{2  \sinh \frac{  T}{2}} 
    =\!\!\! \vcenter{\hbox{\PropagatorSHO}}  \,.
	  \label{DeltaPprop}
	\end{equation}
	The propagator is time-translation invariant due to the translation invariance of the perturbative saddle we are expanding around. It is a kind of thermal propagator, with $T$ playing the role of inverse temperature. In the limit $T\to \infty$ it reduces to the zero-temperature propagator $\prop_0(t)\to\frac{1}{2}e^{-|t|}$.

An alternative way to compute the propagator is by ``unfolding'' the circle (method of images). We start with the Green's function of $(-\partial_t^2+1)$ on the infinite line
	\begin{equation}
	  \prop_\infty(t)=\frac{1}{2}e^{-|t|}\,,
	\end{equation}
    and we enforce periodization by placing ``image sources'' at each integer multiple of $T$. Thus, the periodic Green's function is
	\begin{align}
	  \prop_0(t_1,t_2)
	  &=\sum_{m\in\mathbb{Z}}\prop_\infty(t_1-t_2+mT)
	  =\frac{1}{2}\sum_{m\in\mathbb{Z}}e^{-|t_1-t_2+mT|}
	  = \frac{e^{-|t_1-t_2|}+e^{-(T-|t_1-t_2|)}}{2(1-e^{-T})}\\
	  &= \frac{\cosh\!\left(\frac{T}{2}-|t_1-t_2|\right)}{2\sinh\frac{T}{2}},
	  \label{DeltaPimages}
	\end{align}
	which reproduces~\eqref{DeltaPprop}. The equivalence between the mode sum and image sum is an instance of the Poisson summation formula applied to $\prop_\infty(t)$. 
  At coincident points we have
	\begin{equation}
	  \prop_0(t,t)=\frac{1}{2}+\sum_{m=1}^\infty e^{-mT}=\frac{1}{2}\coth\frac{T}{2}
	  \label{DeltaPladder}
	\end{equation}
  This representation will be helpful for understanding the higher-loop structure in Section~\ref{sec:pi_spectrum}.

With these Feynman rules, we can compute the partition function as a sum over vacuum bubbles. The leading Gaussian (1-loop) contribution is the same SHO partition function we already calculated
\begin{equation}
  Z_P=  Z_{\text{SHO}}(1+\cO(\hbar)) \,.
\end{equation}
The first nontrivial corrections from the interaction vertices appear at order $\hbar$ (2 loops). There are 3 connected diagrams: the figure-8, sunset, and dumbbell:
 \begin{align}
  G_4 &= \vcenter{\hbox{\FeynmanQuarticVertex}}
 = - \hbar \frac{3}{8} \int_{-\frac{T}{2}}^{\frac{T}{2}} d t \prop_0(t,t)^2 
  = -\hbar T \frac{3 }{32 }
  \coth^2 \left( \frac{ T}{2} \right) \label{G4cont}\\
  G_{3 a} &= \vcenter{\hbox{\FeynmanCubicBubble}}
  = \hbar \frac{9}{2\times 3!} \int_{- \frac{T}{2}}^{\frac{T}{2}} d t_1
  \int_{- \frac{T}{2}}^{\frac{T}{2}} d t_2 \prop_0(t_1,t_2)^3 
  = \hbar T  \frac{3}{4}
    \left( \frac{1}{12} + \frac{1}{4} \frac{1}{\sinh^2
  \frac{ T}{2}} \right)\\
  G_{3 b} &= \vcenter{\hbox{\FeynmanCubicSunset}}
  = \hbar \frac{9}{8} \prop_0(0,0)^2 \int_{- \frac{T}{2}}^{\frac{T}{2}} d
  t_1  \int_{- \frac{T}{2}}^{\frac{T}{2}} d t_2 \prop_0(t_1,t_2) 
  = \hbar T \frac{9 }{32 } \coth^2 \left( \frac{ T}{2} \right) \,.
  \label{eq:twoloopdiagrams}
\end{align}
The exact sum at finite $T$ is
\begin{equation}
  G_4 + G_{3 a} + G_{3 b}
  = \frac{\hbar T}{4}\,\frac{\cosh T + 2}{\cosh T - 1}
  = \hbar T\!\left(\frac{1}{4} + \frac{3}{2}\,e^{-T} + \cO(e^{-2T})\right) \,.
  \label{eq:2loop_exact}
\end{equation}
Each connected vacuum-bubble contribution has an overall center-of-mass factor of $T$; the remaining relative-time integrals can still depend on $T$ through finite-size effects. Disconnected vacuum bubbles exponentiate. At 2-loop order the partition function around the $x=-1$ saddle is therefore
\begin{equation}
   Z_P= Z_\text{SHO}\,\exp\!\left[\frac{\hbar T}{4}\,\frac{\cosh T + 2}{\cosh T - 1} +\mathcal O(\hbar^2)\right] \,.
  \label{ZDWZSHO}
\end{equation}
More generally, since connected diagrams exponentiate, we can write
\begin{equation}
  Z_P = Z_{\text{SHO}}\, e^{T \DV(T,\hbar)}
  \label{DWPexpform} \,,
\end{equation}
where 
\begin{equation}
\DV(T,\hbar) = \frac{\hbar}{4}\,\frac{\cosh T + 2}{\cosh T - 1} + \mathcal{O}(\hbar^2)
  \label{DVtwoloop}
\end{equation}
is the connected vacuum-bubble sum per unit center-of-mass time. In the large-$T$ limit $\DV = \frac{1}{4}\hbar + \cdots$. As we will see, this limit will only give the ground-state energy shift and the full $T$ dependence is needed for the excited states. 

The finite-$T$ 3-loop contribution to $\DV$ can be computed with the same perturbative Feynman rules. Writing $u=e^{-T}$, direct diagram evaluation gives
\begin{equation}
\DV  =
\hbar \frac{1+4u+u^2}{4(1-u)^2} + 
  \hbar^2 \left[  \frac{3(1+u)(3u^2+28u+3)}{32(1-u)^3}
  -\frac{9u(u^2+3u+1)\ln u}{8(1-u)^4}\right] + \cO(\hbar^3) \,.
  \label{DVthreeloop}
\end{equation}
Taking the large-$T$ limit gives
\begin{equation}
  \label{DV3loop}
  \DV = \frac{1}{4}\hbar + \frac{9}{32}\hbar^2 + \mathcal{O}(\hbar^3,\hbar e^{-T},\hbar^2 T e^{-T}) \,.
\end{equation}
This agrees with the dilute-instanton-gas result computed by Escobar-Ruiz, Shuryak, and Turbiner~\cite{Escobar-Ruiz:2015nsa}, after translating conventions.
We will verify these results by comparing to the perturbative energy spectrum in Section~\ref{section:finiteT_spectrum}. In fact, there is a 1-to-1 correspondence between the Feynman diagrams and the terms in the perturbative expansion of the energy levels, so that the result for $\DV$ to any order in $\hbar$ can be determined algebraically from the Bender--Wu expansion.

\subsubsection{Higher-loop instanton corrections}
\label{sec:higher_loop_instanton}
Feynman diagram computations around the perturbative saddle are encoded in the $\DV$ factor in Eq.~\eqref{eq:Z_factored}, with the explicit result in Eq.~\eqref{DVthreeloop}. Higher-loop results around the instanton saddles lead to the $\DL$ correction in Eq.~\eqref{eq:Z_factored}. Using the dilute-instanton gas approximation, the single-instanton 2-loop calculation was first carried out by W\"ohler and Shuryak \cite{Wohler:1994pg} (correcting earlier attempts \cite{Olejnik:1989id}), with the 3-loop extension by Escobar-Ruiz, Shuryak, and Turbiner \cite{Escobar-Ruiz:2015nsa}. We provide here a summary of the method and results, with more details in Appendix~\ref{appendix:loop_corrections}.

The $\DL$ factor arises from connected Feynman diagrams in an instanton background with the perturbative background contribution subtracted. To compute the diagrams, we expand a general path as $x(t) = x_{k,k'}(t-t_0) + \xi(t-t_0)$ and then treat the $x_{k,k'}$ insertions as background fields with $\xi$ the propagating field. Taking the 1-instanton background $x_I(t)$ for concreteness, at 2 loops, the same three connected vacuum-bubble topologies are needed as in the perturbative sector (Eq.~\eqref{eq:twoloopdiagrams}):
\begin{align}
  G_4
  &= \vcenter{\hbox{\FeynmanQuarticVertexInst}}
  = -\frac{3\hbar}{8} \int dt\,\prop_I(t,t)^2\,,\\
  G_{3a}
  &= \vcenter{\hbox{\FeynmanCubicBubbleInst}}
  = \frac{3\hbar}{4} \int dt_1\,dt_2\,x_I(t_1)x_I(t_2)\,\prop_I(t_1,t_2)^3\,,\\
  G_{3b}
  &= \vcenter{\hbox{\FeynmanCubicSunsetInst}}
  = \frac{9\hbar}{8} \int dt_1\,dt_2\,x_I(t_1)x_I(t_2)\,\prop_I(t_1,t_1)\prop_I(t_1,t_2)\prop_I(t_2,t_2)\,,
\end{align}
where $\prop_I(t_1,t_2)$ is the $\xi$ propagator. 

These integrals have two key differences compared to the perturbative sector. First, the propagator depends on $t_1$ and $t_2$ \emph{separately}, not just on $t_1-t_2$, because the instanton breaks time-translation invariance. Second, the cubic vertex depends on the instanton background $x_I(t)$. These complications make the loops rather challenging. However, once the perturbative contributions to the same graphs are subtracted, the diagrams are all finite and can be computed numerically if needed.

A fourth diagram exists at 2 loops in the instanton sector that has no perturbative analog. The change of variables to the collective coordinate $t_0$ (the instanton position) introduces a Jacobian whose expansion in the transverse fluctuation $\xi(t) = x(t+t_0) - x_I(t)$ has a linear piece $-(1/S_I)\!\int\!dt\,\ddot x_I\xi$, which acts as a tadpole-source vertex. Although a true \emph{action} tadpole is forbidden by the equation of motion $\ddot x_I = V'(x_I)$ at the saddle, the source from the measure does not vanish: it leaves an additional 2-loop topology with the source contracting through one propagator into a self-loop on a cubic vertex,
\begin{equation}
  G_1
  = \vcenter{\hbox{\FeynmanCubicJacobian}}
  = \frac{9\hbar}{4} \int dt_1\,dt_2\,x_I(t_1)\,\ddot x_I(t_2)\,\prop_I(t_1,t_1)\prop_I(t_1,t_2)\,.
\end{equation}
This diagram is derived in Appendix~\ref{appendix:loop_corrections} and gives a critical contribution (without it the shift in the energies would not agree with Exact WKB; see
Section~\ref{sec:pi_spectrum}).

The result of the computation to 3 loops, at leading order in the $T\to \infty$ limit, is 
\begin{equation}
  \DL(\hbar) \;=\; -\frac{71}{48}\,\hbar \;-\; \frac{315}{128}\,\hbar^2 \;+\; \cO(\hbar^3)\,.
  \label{eq:DLn_summary}
\end{equation}
This is the single-instanton result. Around an $n$-instanton background, the result gets multiplied by $n$ since the instantons do not interact at infinite $T$. Finite-$T$ corrections to $\DL$ require using the exact-instanton (Lam\'e) propagator rather than its $T\to\infty$ approximation (P\"oschl--Teller). The exact propagator can likely be computed in closed form using the Lam\'e spectrum, but the resulting expression would be a complicated elliptic function of $t_1$ and $t_2$ and the time integrals would be formidable. These finite-$T$ corrections would be needed only at the level of precision required to extract $N$-dependent corrections to the non-perturbative parts of the excited-state energies.  On the Exact WKB side, the analogous integrals are those of the quantum action $S_N^m$, which can be done algebraically using the closure of the Picard--Fuchs equation. Thus from a practical point of view it does not seem worthwhile to attempt the finite-$T$ calculation on the path integral side.

\subsection{\texorpdfstring{$n=2$}{n=2} thimble decomposition}
\label{sec:n2thimble}
We now have almost all the ingredients in Eq.~\eqref{eq:Z_factored}: the action, the transverse fluctuation determinant with the quasi-zero modes removed, the perturbative vacuum-bubble factor $\DV$, and the instanton-background loop correction $\DL$. It remains to integrate over the quasi-zero modes. As a reminder, the full integration contour is decomposed as a sum over Lefschetz thimbles passing through saddles (Section~\ref{sec:thimble_decomposition}), and only the real instanton saddles contribute in the decomposition (see Section~\ref{sec:intersection_numbers}). The thimbles factorize into transverse and quasi-collective-coordinate pieces as described around Eq.~\eqref{eq:quasi_collective_integral}, and we are now interested in computing the finite-dimensional quasi-collective-coordinate integrals. These integrals are along complex thimbles and are in general complex. We begin here with the $n=2$ case, which is already very intricate. Our goal is to understand the origin of the imaginary parts from the geometry of the quasi-zero mode space and to see how they cancel with the Borel resummation of the $n=0$ sector asymptotic series.

Concretely, the program of this section is:
\begin{enumerate}
    \item Establish the quasi-zero-mode geometry around the $n=2$ saddle: the collective coordinate $\alpha$ for the instanton separation, the QZM Jacobian, and the complex saddle structure in the $\alpha$-plane. 
    \item Compute the thimble integral $\cI_2^\pm$ through the real $n=2$ saddle and identify the imaginary part which contributes to $Z_2$.
    \item Extract the leading-order asymptotic series of $Z_0$ from the integral over real $\alpha$ directions between the $n=0$ and $n=2$ saddles.
    \item Assemble the resurgent structure, where the lateral Borel resummation of $Z_0$ produces an imaginary ambiguity that exactly cancels the explicit imaginary part of $Z_2$.
\end{enumerate}                                                                                                                       
Following sections treating the $n>2$ decomposition will follow a similar road.

\subsubsection{Quasi-zero-mode geometry for $n=2$}
\label{sec:quasi}
\begin{figure}[t!]
  \centering
  \includegraphics[width=0.49\textwidth]{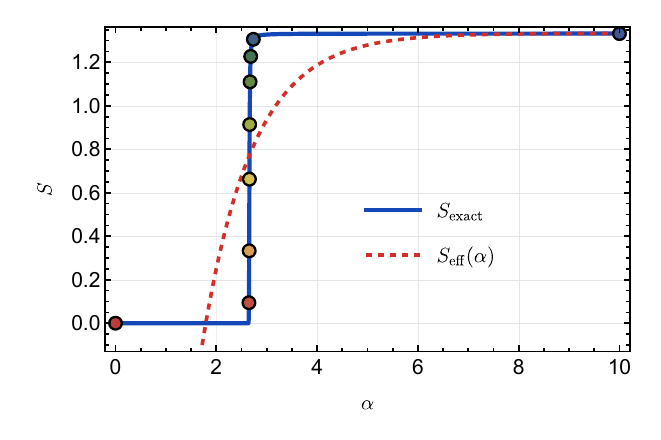}
  \includegraphics[width=0.49\textwidth]{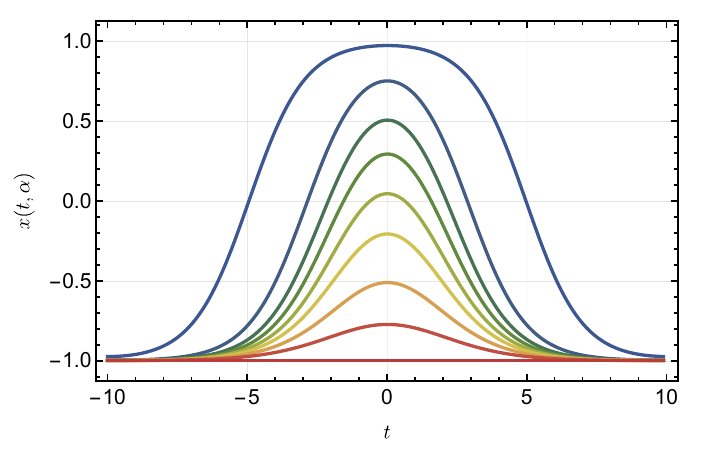}
  \caption{Starting from the exact $n=2$ saddle at finite $T$, we follow gradient flow to the $n=0$ saddle at $x=-1$. Left: action along the flow, plotted as $S(\alpha)$ with the normalization $\alpha(u{=}0)=T/2$ and $\alpha(u_{\rm end})=0$. Right: representative profiles $x(t)$ along the same trajectory, labeled by $\alpha$. The overall normalization of the flow parameter $u$ is arbitrary.}
  \label{fig:flow_action_n2}
\end{figure}

We begin the $n=2$ analysis with the effective action from Eq.~\eqref{Seffalpha}. Recall that this action is computed by approximating the instanton solutions by sewing together tanh functions, evaluating the action, and then rescaling the exponentially suppressed terms to agree with the action of the exact saddle at large $T$. The resulting effective action $S_{\mathrm{eff}}$ captures the leading large-$T$ QZM dynamics and matches the first exponentially small correction to the exact saddle action. We can check numerically that the exact action agrees with $S_{\mathrm{eff}}$ along the QZM direction near the saddle (see Fig.~\ref{fig:flow_action_n2}). With $\alpha_1 = \alpha$ and $\alpha_2 = T-\alpha$ and $n=2k=2$ the effective action is 
\begin{equation}
    S_{\mathrm{eff}}(\alpha)
    =
    2 S_I-12S_I  \left[e^{-\alpha} + e^{-(T-\alpha)}\right]
    \label{eq:Seffalpha_n2} \,.
\end{equation}
Following the convention of Eq.~\eqref{eq:Seffn_def}, we subtract the saddle action and define
\begin{equation}
    S_\text{eff}^{{(2)}}(\alpha) \equiv S_\text{eff}(\alpha) - 2S_I = -12 S_I\!\left[e^{-\alpha}+e^{-(T-\alpha)}\right]\,.
    \label{eq:Seff2_def}
\end{equation}
Here $\alpha$ is the distance between the instantons, so its natural range is $0 \le \re \alpha \le T$.  The standard interpretation of this action is that $e^{-\alpha}$  and $e^{-T+\alpha}$ represent interactions between instantons on a circle of circumference $T$. The action is invariant under $\alpha \to \alpha + 2\pi i$ and we therefore take $-\pi \le \im \alpha \le \pi$. There are then two extrema of this action in the fundamental domain, at $\alpha = T/2$ and $\alpha = T/2 + i \pi$. These correspond to the real instanton $x_{1,0}$ and the complex instanton $x_{1,1}$ shown in Fig.~\ref{fig:solutions} on the $x_2$ branch.

To integrate over $\alpha$ we need to establish the normalization of the associated mode. Around the (1,0) saddle there is one exact zero mode parameterized by a global time translation parameter $t_0$ and a quasi-zero mode parameterized by $\alpha$. We would like to parametrize this 2D space close to the saddle as
\begin{equation}
x(t)=x_{1,0}(t)+c_0\, \Psi_0(t) +c_1\,\Psi_1(t),
\label{xPsi12}
\end{equation}
where $\Psi_0$ and $\Psi_1$ are respectively the normalized zero mode and unstable mode of the exact Hessian. We need to go from the integration measure over $t_0$ and $\alpha$ to the orthonormal coordinates $c_0$ and $c_1$. We start with two instantons with centers at $t_1,t_2$. When the instantons are far separated, all the quasi-zero modes are associated with the independent motion of each instanton center. So at leading order we can use $\| \dot{x}_a\|^2=S_I$, giving a Jacobian factor $\sqrt{S_I/(2\pi \hbar)}$ for each mode in $t_1, t_2$ coordinate space. For the $n=2$ case we are considering here, $t_0 = (t_1+t_2)/2$ is the global translation coordinate and  $\alpha=t_1-t_2$ is the difference between the instanton centers. So we have
\begin{equation} \label{eq:coordinate_change_measure}
     dc_0^2 + d c_1^2 =  S_I(dt_1^2+dt_2^2)=2S_I\,dt_0^2+\frac{S_I}{2}\,d\alpha^2.
\end{equation}
Thus the canonically normalized coordinates are related by $c_0 = \sqrt{2 S_I} t_0$ and 
\begin{equation}
  \alpha=\frac{T}{2}-\sqrt{\frac{2}{S_I}}\,c_1+{\cal O}(c_1^2)\,.
  \label{eq:alpha_c1_match}
\end{equation}
So in particular, the Jacobian associated with $d\alpha$ differs by a factor of $1/\sqrt{2}$ from that of an individual center coordinate.  As a check, we can expand Eq.~\eqref{eq:Seffalpha_n2} about $\alpha=T/2$:
\begin{equation}
     S \left(\frac{T}{2}-\sqrt{\frac{2}{S_I}}\,c_1 \right) 
     = 2 S_I  - 24 S_I  e^{- \frac{T}{2}}
  - 24  e^{- \frac{T}{2}} c_1^2 + \cdots
\end{equation}
So the Hessian entry $d^2 S /dc_1^2$ is $-48 e^{-T/2}$, in exact agreement with Eq.~\eqref{eq:QZM_eigenvalues} for $n=2,\;m=1$.

\begin{figure}[t!]
  \centering
  \begin{tabular}{cc}
    \includegraphics[width=0.45\textwidth]{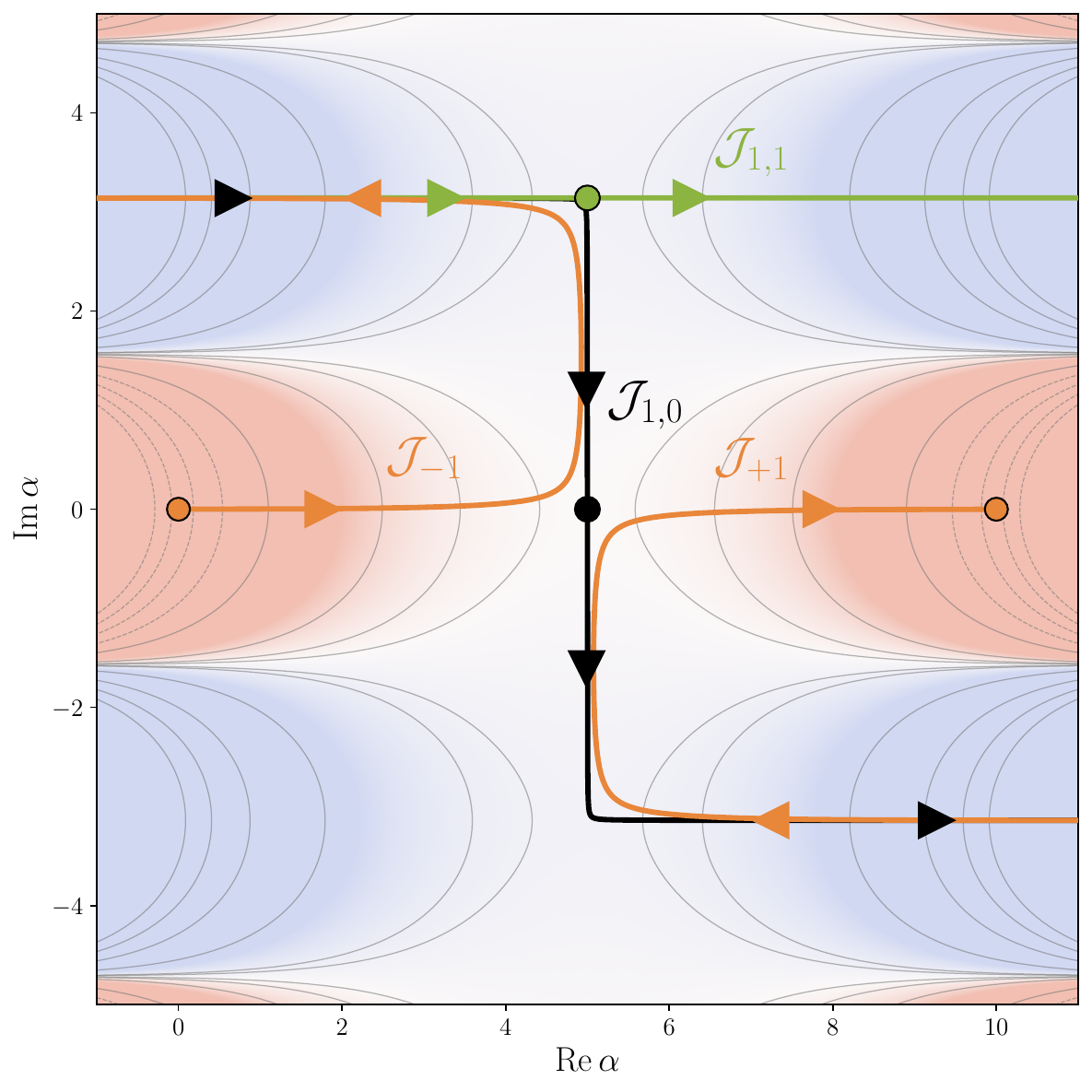} &
    \includegraphics[width=0.45\textwidth]{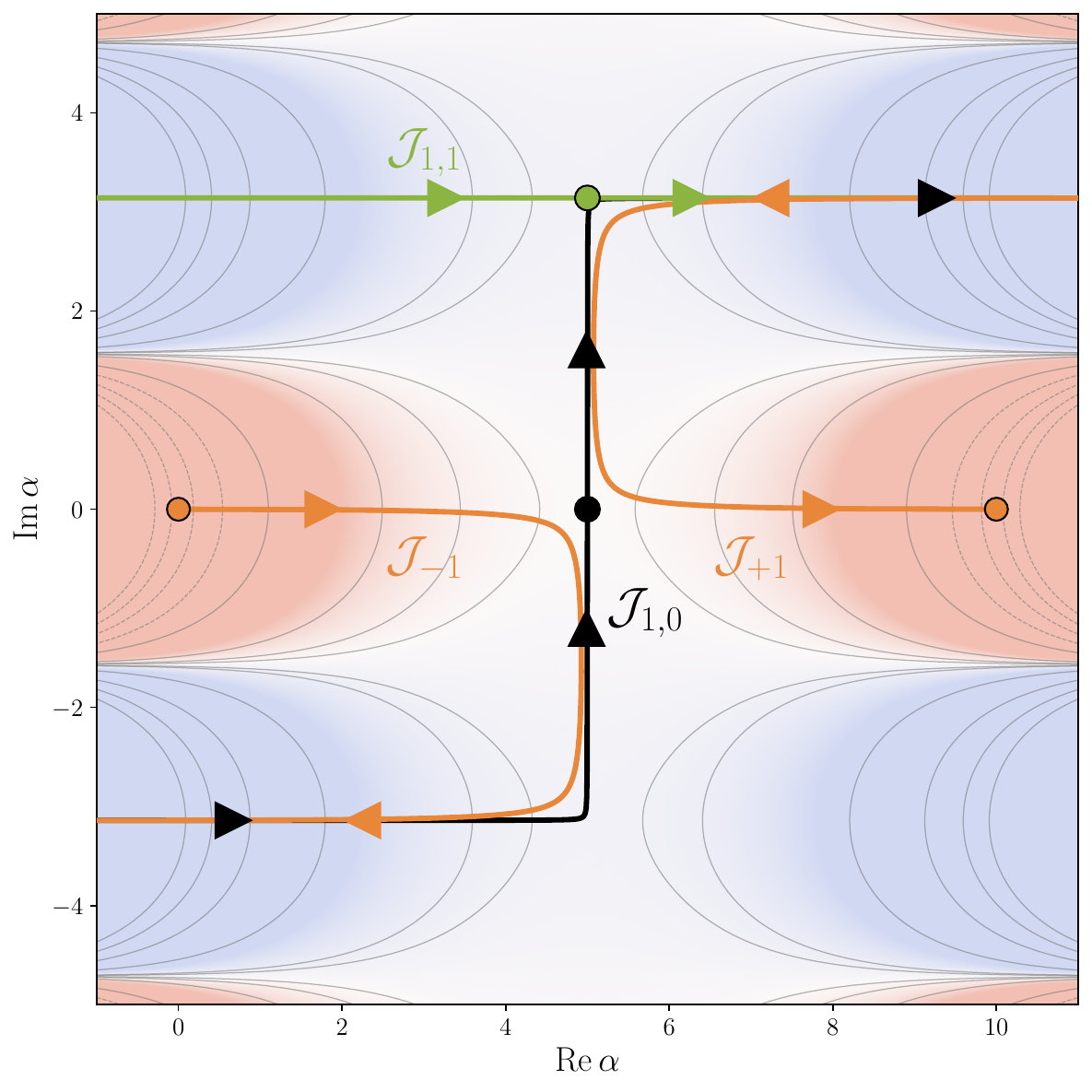} \\
    $\im \hbar > 0$ & $\im \hbar < 0$
  \end{tabular}
  \caption{Lefschetz thimbles using the $n=2$ effective action $S_\text{eff}$ from Eq.~\eqref{eq:Seffalpha_n2} with $\hbar \to \hbar e^{i\varepsilon}$. The perturbative saddles at $x=\pm 1$ are plotted at $\alpha=0,T$ and shown as the orange dots, with their thimbles in orange. The $(1,0)$ real saddle is the black dot at $\alpha=\frac{T}{2}$. Its thimble goes off in the imaginary direction then veers real at the complex Stokes point associated with the $(1,1)$ saddle. The green thimble $\cJ_{1,1}$ through the (1,1) saddle maintains an imaginary offset $\im \alpha = \pi$. The imaginary direction is compact, so the $\cJ_{1,1}$ thimble shown with $\im \alpha = \pi$ is the same as the one with $\im \alpha=-\pi$. 
  }
  \label{fig:thimbles_simple}
\end{figure}

Since $\alpha$ is the quasi-collective coordinate associated with a quasi-zero mode with eigenvalue $\sim e^{-T/2}$, we cannot integrate over it in the Gaussian approximation, and instead integrate over its full thimble. The thimbles for $S_\text{eff}$ are shown in Fig.~\ref{fig:thimbles_simple}. For these plots we deformed $\hbar \to \hbar e^{i\varepsilon}$ to resolve the Stokes ambiguity. The real part of the action for the (1,0) saddle increases in the imaginary $\alpha$ direction until it hits the (1,1) saddle. Note that the (1,1) saddle is mirrored at $\pm i \pi$ due to the invariance of the action under $\alpha \to \alpha + 2\pi i$. The saddles at $x=\pm 1$ are not extrema of $\Seff$ since the effective action is only valid at large $T$. However, they live in the directions going towards $\alpha=0$ and $\alpha=T$. These boundary points are indicated in the figure as orange dots, with associated thimbles starting on the real axis, proceeding to the $(1,0)$ Stokes point, before following along the $\cJ_{1,0}$ trajectory in the complex plane.

Although $\alpha$ is only well defined when the instantons are well separated, there is still a unique steepest descent trajectory descending from the (1,0) saddle. We can find this trajectory numerically. The exact action $S[x] = \int_0^T dt\,[\tfrac12\dot x^2 - U(x)]$ is a functional of the path $x(t)$. Starting from a path slightly perturbed from $x_{1,0}$ in the direction of the unstable mode, we evolve under gradient flow, $\partial_u x(t) = -\delta S/\delta x(t)$, computing the action integral on the discretized path at each flow step. The result is shown in Fig.~\ref{fig:flow_action_n2} where the exact action along the flow is compared to $\Seff$, both agreeing in the large-$\alpha$ regime near the $n=2$ saddle. The flow trajectory interpolates between the real $n=2$ instanton saddle and the static $n=0$ path at $x=-1$. 

\subsubsection{$n=2$ thimble integral}
We now move to compute the thimble integrals in the complexified $\alpha$ space. We begin with the $n=2$ thimble integral through $\cJ_{1,0}^\pm$, where the superscript refers to the sign of $\im \hbar$ as in Fig.~\ref{fig:thimbles_simple}, used to resolve the Stokes ambiguity. This integral can be decomposed into two equal horizontal arms coming in from infinity to the complex saddles, and two equal vertical arms, going between the complex saddles and the real ones:
\begin{equation} 
\cI_2^\pm (T) =   \int_{\cJ_{1,0}^\pm} d\alpha\,e^{-\Seff^{(2)}/\hbar}
=
  2 I^\text{arm}(T) \pm 2 I^\text{vert}(T)
\label{I2psum}
\end{equation}
The arm integral is 
\begin{equation}
    \begin{aligned} \label{eq:arms_integral}
    I^\text{arm}(T)
    &=
    \int_{ i\pi -\infty}^{i\pi } d\alpha\;
    e^{-\Seff^{(2)}/\hbar}
    =
    \int_{-\infty}^{0} dx\;
    \exp\!\left[-\frac{24S_I}{\hbar}e^{-T/2}\cosh x\right]
    \\
    &=
    K_0\!\left(\frac{24S_I}{\hbar}e^{-T/2}\right),
    \end{aligned}
\end{equation}
where we parameterized $\alpha=T/2+i\pi+x$ and used $\int_{-\infty}^{0}e^{-a\cosh x}dx=K_0(a)$. For large $T$,
\begin{equation}
     I^\text{arm}(T)
  =
  \frac{T}{2}-\gamma_E
    -\log\!\left(\frac{12S_I}{\hbar}\right)
    +\mathcal O\!\left(T e^{-T}\right)\,.
\end{equation}
The vertical integral over the imaginary contour between the two saddles gives
\begin{equation}
    \begin{aligned}
       I^\text{vert}(T)
    &=
    \int_{\frac{T}{2}+i\pi}^{\frac{T}{2}} d\alpha\;
    e^{- \Seff^{(2)}/\hbar}
    =
    -i\,\int_{0}^{\pi} dx\;
    \exp\!\left[-\frac{24S_I}{\hbar}e^{-T/2}\cos x\right]
    \\
    &=
    -i \pi \,I_0\!\left(\frac{24S_I}{\hbar}e^{-T/2}\right),
    \end{aligned}
    \label{eq:n2_vertical_segment}
\end{equation}
where now we used $\alpha=T/2+i\pi-ix$  and  $\int_{0}^{\pi}e^{-a\cos x}dx=\pi I_0(a)$.
For large $T$,
\begin{equation}
    I^\text{vert}(T)
  =- i \pi
  \left[ 1+
  % \frac{1}{4}\left(\frac{24S_I}{\hbar}e^{-T/2}\right)^2+
    \mathcal O\!\left(e^{-T}\right)
  \right].
\end{equation}
Adding these as in Eq.~\eqref{I2psum} gives
\begin{equation} 
\cI_2^\pm (T)
= 2
  \left[
    \frac{T}{2} - \gamma_E - \log\!\left(\frac{12S_I}{\hbar}\right)
    \mp \pi i
  \right] + \cO(T e^{-T}).
  \label{eq:j_10_thimble}
\end{equation}

An alternative way to compute $\cI_2^\pm(T)$ is directly along the thimble without decomposing it into parts. This can be done by continuing $\hbar \to e^{i\pi} \hbar$ so that the thimble aligns with $\RR$, and then we can use
\begin{equation}
    \cI_2^\RR(T) = \int_{-\infty}^\infty d\alpha\, e^{\Seff^{(2)}/\hbar}
    = \int_{-\infty}^\infty d\alpha \exp\left[-\frac{ 24S_I}{\hbar} e^{-T/2}\cosh(\alpha-T/2)\right] 
    =\cG_2(\ntwoScale^2)
\end{equation}
where $\cG_n$ is the all-zero-parameter Meijer $G$-function (the inverse Mellin transform of $\Gamma(s)^n$):
\begin{equation}
  \cG_n(z)
  \equiv
  G_{0,n}^{n,0}\!\left(z\,\big|\,\underbrace{0,\ldots,0}_{n}\right)
  =
  \frac{1}{2\pi i}\int^{c+i \infty}_{c-i \infty} ds\,\Gamma(s)^n z^{-s},
  \qquad c>0,
  \label{eq:cG_def}
\end{equation}
with
\begin{equation}
    \ntwoScale = \frac{12S_I}{\hbar} e^{-T/2} \,.
    \label{ntwodef}
\end{equation}
Then,
\begin{equation}
  \cI_2^\pm(T)
  =
  \cI^{\mathbb R}_2(\ntwoScale e^{\pm i\pi})
  =
  \cG_2(\ntwoScale^2 e^{\pm 2\pi i})
  = 2 \,K_0(- 2\ntwoScale \pm i \varepsilon) \,.
  \label{eq:n2_exact_thimble_G}
\end{equation}
At large $T$, small $\ntwoScale$
\begin{equation}
\cG_2(\ntwoScale^2)
  =
  -\ln \ntwoScale^2 - 2\gamma_E \,,
  \label{eq:A1_G_largeT}
\end{equation}
which when combined with Eqs.~\eqref{eq:n2_exact_thimble_G} and~\eqref{ntwodef} agrees with Eq.~\eqref{eq:j_10_thimble}. This method will be more practical for the higher instanton sectors. 

The integral $\cI_2^\pm$ over the $n=2$ thimble is a multiplicative factor in the $n=2$ sector of the partition function trans-series, as in Eq.~\eqref{eq:Z_factored}. Recall that we write $Z =\sum_n \lambda^n Z_n(T,\hbar)$ where $\lambda=e^{-S_I/\hbar}$.  The transverse determinant factor is $(\det_\perp\mathcal{O}_2)^{-1/2} = Z_{\text{SHO}} \cdot (\det\mathcal{O}_0/\det_\perp\mathcal{O}_2)^{1/2}$. The determinant ratio with the zero mode and quasi-zero mode removed is given in~\eqref{eq:detpp_ratio_DIGnorm} at large $T$: $(\det\mathcal{O}_0/\det_\perp\mathcal{O}_2)^{1/2} = 12$. The collective-coordinate Jacobian from Eq.~\eqref{eq:coordinate_change_measure} has $\sqrt{ S_I/\pi\hbar}$ from the zero mode and $\sqrt{S_I/ 4 \pi \hbar}$ from the quasi-collective coordinate, contributing $S_I/(2\pi\hbar)$ to the path-integral measure. The two wells contribute a factor of 2, which cancels against the correction factor from the change to collective coordinate, which is $1/n=1/2$, where $n$ is the number of times the origin is crossed. Combining $\cI_2^+$ from Eq.~\eqref{eq:j_10_thimble} with these factors,
\begin{align}
  \label{eq:Z2_master1}
  Z_2^\pm
  &=
  Z_\text{SHO}
  \!\left(\frac{\det\opO_0}{\det_\perp\opO_2}\right)^{\!1/2}
  \frac{S_I}{2\pi\hbar}\,\int_0^T d t_0
  \cI_2^\pm  
  \\
  &=  Z_\text{SHO} \frac{6 S_I T}{\pi \hbar} 
  \left[
    T - 2\gamma_E - 2\log\!\left(\frac{12S_I}{\hbar}\right)
    \mp 2\pi i
  \right].
  \label{eq:Z2_master}
\end{align}
This is a complex result, with the imaginary part coming from $2I^\text{vert}=-2\pi i$. Its imaginary part is
\begin{equation}
  \label{eq:imZ2}
    \im Z_2^\pm =\mp 2\pi Z_\text{SHO} \frac{6 S_I T}{\pi \hbar} \,.
\end{equation}
This is qualitatively and quantitatively different from the dilute-gas approximation which associates the $\lambda^2$ terms effectively with an integral along the real $\alpha$ direction. In the dilute gas, the equivalent of $Z_2^\pm$ is real. 

\subsubsection{$n=0$ asymptotic series}
Next we consider the implications of $\Seff$ for the perturbative series. The leading asymptotic growth of the trans-series in the $n=0$ sector is determined by the nearest saddle, which is the one at $n=2$, and can be reproduced by integrating along the valley direction $\alpha$ between the two saddle points.  This integral is a little more subtle because, as shown in Fig.~\ref{fig:flow_action_n2}, we cannot trust $\Seff$ away from the $\alpha = \frac{T}{2}$ region. However, because the asymptotic growth comes from the existence of the $n=2$ saddle, the detailed structure of the action in the $\alpha\approx 0$ region is not critical: a change in $\Seff$ in this region will only give finite corrections to the terms in the series which can be included as matching corrections to a perturbative calculation with the full action. We will extract the asymptotic series from the integral along $\alpha$ between the $n=0$ saddle, represented by an endpoint at $\alpha=\amin$ up to the $n=2$ saddle at $\alpha=T/2$. The value of $\amin$ can be fixed by location of the Borel pole at $t=2S_I$, or equivalently that the action gap is $\Seff(T/2)-\Seff(\amin)=2S_I$, which gives $\amin=\ln 6$. 

The real integral along $\alpha$ between the saddles that gives the asymptotic growth is then
\begin{equation}
  \label{I02def}
    I_{0\to2}(T) = \int_{\amin}^{\frac{T}{2}} d\alpha\, e^{-\frac{1}{\hbar} \Seff}
    =
    e^{-2S_I/\hbar}\int_{\amin}^{T/2} d\alpha\,
    \exp\!\left[+\frac{12S_I}{\hbar}\left(e^{-\alpha}+e^{-(T-\alpha)}\right)\right]\,.
\end{equation}
Since $\alpha < T/2$ we can use $\exp\!\left[\frac{12S_I}{\hbar}e^{-(T-\alpha)}\right]=1+\mathcal O(e^{-T/2})$ uniformly on the full integration range. Therefore,
\begin{align} \label{eq:asy_int_n2}
  I_{0\to2}(T)
  &=
  e^{-2S_I/\hbar}\int_{\amin}^{T/2} d\alpha\,
  \exp\!\left[\frac{12S_I}{\hbar}e^{-\alpha}\right]
  +\mathcal O\!\left(T e^{-T/2}\right)
  \\
  &=
  e^{-2S_I/\hbar}\int_{\ntwoScale}^{\umin} \frac{du}{u}\,e^u
  +\cO\!\left(T e^{-T/2}\right),
  \label{eq:asy_int_n2b}
\end{align}
where $u\equiv \frac{12S_I}{\hbar}e^{-\alpha}$ with $\umin=\frac{12S_I}{\hbar}e^{-\amin}$. This integral can be done using $\int_a^b du\, e^u/u = \Ein(-a) - \Ein(-b) + \ln(b/a)$, where
\begin{equation}\label{eq:Ein_E1_def}
  \Ein(z)\equiv\int_0^z dt\,\frac{1-e^{-t}}{t}\,.
\end{equation}
Then
\begin{equation}
\label{I02asyEE}
  I_{0\to2}(T)
  =
  e^{-2S_I/\hbar}
  \left[
    \Ein(-\ntwoScale) - \Ein(-\umin) + \ln\frac{\umin}{\ntwoScale} 
  \right]
  +\cO\!\left(T e^{-T/2}\right)\,,
\end{equation}
The function $\Ein(z)$ is an entire function on the complex plane, and $\Ein(z)$ is real for real $z$. For positive real $z$, its large-$z$ expansion on the negative axis is the real formal trans-series
\begin{equation}\label{eq:Ein_trans}
  \Ein(-z)
  \sim
  \gamma_E + \ln(-z)  - e^{z}\sum_{n=0}^{\infty} \frac{n!}{z^{n+1}}\,,
\end{equation}
Since $\ntwoScale \sim e^{-T/2}\to 0$ as $T\to \infty$, the $\Ein(-\ntwoScale)$ contribution to Eq.~\eqref{I02asyEE}  vanishes in this limit (as can be seen from the $z\to 0$ limit of Eq.~\eqref{eq:Ein_E1_def}) and we can drop it. We cannot drop the $\ntwoScale$ in the logarithm in Eq.~\eqref{I02asyEE} though since it is linear in $T$. Inserting the trans-series expansion of $\Ein$ we then get
\begin{align}
  I_{0\to2}^\text{asy}(T)
  &\sim e^{-2S_I/\hbar}\left[- \Ein(-\umin)
   + \ln\frac{\umin}{\ntwoScale} \right]
  \\
  &=  e^{-2S_I/\hbar}\left[
    \exp\!\left(\frac{12 S_I}{\hbar}e^{-\amin}\right)\sum_{n=0}^{\infty} n!\left(\frac{\hbar\,e^{\amin}}{12 S_I}\right)^{n+1}
    + \frac{T}{2} - \gamma_E - \ln\!\left(\frac{12 S_I}{\hbar}\right) \pm i \pi
  \right] \,.
\end{align}
The $\amin$ factors are obviously awkward, and we will remove them shortly with $\amin=\ln6$. However, before doing so, we observe critically that the imaginary ambiguity of the lateral Borel resummation of the perturbative series can be read directly from the factorial tail in Eq.~\eqref{eq:Ein_trans}:
\begin{equation}
    \im \cS_\pm\!\left[
    \exp\!\left(\frac{12 S_I}{\hbar}e^{-\amin}\right)
    \sum_{n=0}^{\infty} n!\left(\frac{\hbar\,e^{\amin}}{12 S_I}\right)^{n+1}
    \right]
    =
    \pm \pi \,.
\end{equation}
Therefore, the imaginary ambiguity from the resummation of the perturbative series only is
\begin{equation}
    \im \cS_\pm\big[ I_{0\to2}^\text{asy}(T)\big]
    =
    \pm \pi e^{-2S_I/\hbar}\,,
\end{equation}
independent of $\amin$. Naturally, since $I_{0\to2}$ is real by definition, within that integral the resummation ambiguity cancels against the explicit $\pm i \pi$ factor. However, for a quantity that involves only the perturbative series without the explicit additional $\pm i \pi$, like the thimble integrals $\mathcal{J}_{\pm 1}$, or the $n=0$ asymptotic series, the resulting imaginary ambiguity is not canceled. Its magnitude is exactly half of the imaginary part of $\cI_2^\pm$ in Eq.~\eqref{eq:j_10_thimble} as expected from the geometry, and is independent of the boundary value chosen for $\amin$. That being said, the real part of the trans-series does depend on $\amin$, and since the $n=0$ sector cannot be suppressed by any power of $e^{-S_I/\hbar}$, we are led to the unique choice $\amin=\ln 6$. This choice corresponds to $\Seff(\amin)=0$, so that the action difference between the boundaries is $\Seff(T/2)-\Seff(\amin)=2S_I$. With this choice, the asymptotic series becomes
\begin{equation}
  \label{Irealasy}
I_{0\to2}^\text{asy}(T)
 \sim \sum_{n=0}^{\infty} n!\left(\frac{\hbar}{2S_I}\right)^{n+1} 
   +\cO(e^{-T/2})\,.
\end{equation}
Note that terms suppressed by $e^{-2S_I/\hbar}$ are non-perturbative and not part of the formal asymptotic series expansion. With $\amin=\ln6$, so $\umin =2S_I/\hbar$, the full integral from 
Eq.~\eqref{I02asyEE} becomes in the large-$T$ limit
\begin{equation}\label{eq:Ireal_largeT}
  I_{0\to2}(T)
  =
  e^{-2S_I/\hbar}
  \left[    -\Ein\!\left(-\frac{2S_I}{\hbar}\right) +\frac{T}{2}
    -\ln 6     +\cO\!\left(T\,e^{-T/2}\right)
  \right] \,.
\end{equation}
The asymptotic expansion of $I_{0\to2}(T)$ using Eq.~\eqref{eq:Ein_trans} agrees with $I_{0\to2}^\text{asy}(T)$. 

The full thimble integral through the perturbative saddle is then the sum of the $0\to 2$ segment, one vertical segment with opposite orientation, and one arm with opposite orientation:
\begin{align}
\cI_0^{\pm} &= 
    \int_{\cJ_{\pm 1}} d\alpha\,e^{-\Seff/\hbar}
  =I_{0\to2}(T)
  -I^\text{arm}(T)  \mp I^\text{vert}(T) \\
  &=
  -e^{-2S_I/\hbar}
    E_1\!\left(-\frac{2S_I}{\hbar}\pm i\varepsilon\right)\,,
  \label{J1int}
\end{align}
which is indeed the lateral Borel resummation of Eq.~\eqref{Irealasy}. Here we have expressed the answer in terms of another exponential integral function
\begin{equation}
  E_1(z\pm i\varepsilon) = \Gamma(0,z\pm i\varepsilon) = \int_{z\pm i\varepsilon}^{\infty} dt\,\frac{e^{-t}}{t}\,,
\end{equation}
where $E_1$ is related to $\text{Ein}$ by
\begin{equation}\label{eq:Ein_E1_relation}
  \operatorname{Ein}(z) = E_1(z\pm i\varepsilon) + \ln(z\,e^{\gamma_E} \pm i\varepsilon)\,.
\end{equation}
Unlike Ein, which is entire, $E_1(z)$ has a logarithmic branch cut for $z<0$.

With all the relevant thimble integrals in hand for the Picard--Lefschetz decomposition of the real $\alpha$ contour, we can combine them together.  The range of that original integral is $0 \leq \re \alpha \leq T$, which gets decomposed as  
\begin{equation}
  \int_0^T d\alpha\,e^{-\Seff/\hbar}
  =\left[ \int_{\cJ_{-1}} +\int_{\cJ_{1,0}}+ \int_{\cJ_1} \right]d\alpha\,e^{-\Seff/\hbar} =
  e^{-2S_I/\hbar}
  \left[
    -2\Ein\!\left(-\frac{2S_I}{\hbar}\right)
    +T - 2\ln 6
  \right] \,.
\end{equation}
This sum is manifestly real as expected. Indeed, this is nothing but $2 I_{0\to2}(T)$, as can be seen from the geometry.  The $\ln 6$ is an artifact of using $\Seff$ away from the region near the $n=2$ saddle where it is derived, and is subleading at large $T$; we only include it for completeness.

\subsubsection{Resurgent structure at order $\lambda^2$}
\label{sec:E0_E2_transseries}
Next, we construct the trans-series for $Z_0$ and $Z_2$, then convert to the energies using Eq.~\eqref{eq:Zk_from_Ek}. We will verify the cancellation of the Borel ambiguity at order $\lambda^2$. We will focus on the ground state energy only in this section and leave the discussion of the full spectrum to Section~\ref{sec:pi_spectrum}.

First, we consider the $Z_0$ series. This represents the perturbative expansion around the perturbative saddles at $x=\pm 1$. More precisely, it is the sum of the two expansions, since both the intersection numbers are $\eta_\pm = 1$. The saddle point approximation around each individual saddle was computed with Feynman diagrams in Section~\ref{sec:Feynman}. Keeping the leading large-$T$ vacuum-bubble correction from Eq.~\eqref{DVthreeloop},
\begin{equation}
  Z_{0}
  =
  2\,Z_\text{SHO}\,\exp\!\left[T\!\left(\frac{\hbar}{4}+\mathcal O(\hbar^2)\right)\right]
   =
   2\,Z_\text{SHO}\left( 1 + \frac{T}{4}\hbar + \cdots \right) \,.
\end{equation}
The exponent represents the sum of connected graphs, each of which has a factor of $T$ from time-translation invariance.  Higher order terms are in principle computable from Feynman diagrams as in Eq.~\eqref{DVthreeloop}, although the diagrammatic approach is not very efficient, as compared to say, Bender--Wu (see Appendix~\ref{appendix:benderWu}). 

The leading asymptotic behavior of the series in $Z_0$ is determined by the path integral in the direction towards the nearest Stokes point, at the $n=2$ instanton saddle. The relevant integral is $I_{0\to2}^\text{asy}$ in the space of quasi-zero-modes around the $n=2$ saddle. Its asymptotic expansion is given in Eq.~\eqref{Irealasy}. The coordinates used for this integral involve the zero mode $t_0$ for the instanton-pair center and the quasi-zero mode $\alpha$ for the instanton-pair separation. The remaining modes are spectators to the asymptotic growth and can be treated as Gaussian. Thus the prefactor is the same as in Eq.~\eqref{eq:Z2_master1} and we have
\begin{equation}
Z_0 \sim 2 Z_\text{SHO}
  \!\left(\frac{\det\opO_0}{\det_\perp\opO_2}\right)^{\!1/2}
  \frac{S_I}{2\pi\hbar}\,\int_0^T d t_0\,
  \cI_{0\to2}^\text{asy}  
\end{equation}
where the overall factor of 2 comes from summing the series around the two perturbative saddles. Plugging in the factors as in Eqs.~\eqref{eq:Z2_master} and~\eqref{Irealasy} and combining with the leading perturbative vacuum-bubble correction gives
\begin{equation}
  \label{Z0form}
    Z_0
    =2\,Z_\text{SHO}\left( 1 + \frac{T}{4}\hbar+ \cdots + T  \frac{6S_I}{\pi\hbar}  \sum_{n=2}^{\infty}\left(\frac{\hbar}{2S_I}\right)^{n+1} n!  \right)\,,
\end{equation}
 with the $\cdots$ terms containing additional analytic $\hbar$ dependent functions as well as terms with subleading asymptotic growth (e.g. $(4S_I)^{-n}n!$). 

 The lateral Borel resummation of the $Z_0$ trans-series restores the full integral over the $\cJ_{\pm 1}$ thimbles, as in Eq.~\eqref{J1int}. So
\begin{equation}
    \cS[Z_0]
    = 2 Z_\text{SHO}\left[
      1 + \frac{T}{4}\hbar
      - T \frac{6S_I}{\pi\hbar} e^{-2S_I/\hbar}
      E_1\!\left(-\frac{2S_I}{\hbar}+i\varepsilon\right)
      +\cdots
    \right]\,,
\end{equation}
so that using $\im E_1(-g+i\varepsilon) = -\pi$ we find
\begin{equation}
    \im \cS[Z_0] = 2\pi \, Z_\text{SHO}\, T \frac{6S_I}{\pi\hbar} e^{-2S_I/\hbar} + \cdots \,.
    \label{imZ0}
\end{equation}
Combining with Eq.~\eqref{eq:imZ2}  we then have for the full partition function
\begin{align}
  \im \cS_\pm [Z] &= \im\cS_\pm[Z_0] + \lambda^2 \im Z_2^\pm + \cdots \\
  &= 
  2\pi \, Z_\text{SHO}\, T \frac{6S_I}{\pi\hbar} e^{-2S_I/\hbar} -2\pi \, Z_\text{SHO}\, T \frac{6S_I}{\pi\hbar} e^{-2S_I/\hbar} =0 \,.
\end{align}
The leading Borel ambiguity at order $\lambda^2$ exactly cancels.

Finally, we translate from the partition function to the energy trans-series using Eq.~\eqref{eq:Zk_from_Ek}. Recall from Section~\ref{sec:Transseries_E_to_Z} the convention $E^{(N)}(\hbar) = \sum_k \lambda^k E_k^{(N)}(\hbar)$, where the subscript $k$ is the instanton sector and $(N)$ labels the energy state ($N=0$ ground, $N=1$ first excited, \ldots); we drop the state superscript when working with the ground state, so $E_k\equiv E_k^{(0)}$. At the order needed here, the perturbative energy follows from the leading large-$T$ term in Eq.~\eqref{DVthreeloop}:
\begin{equation}
  E_0(\hbar)=\frac{\hbar}{2}-\frac{\hbar^2}{4}+\cO(\hbar^3)\,.
\end{equation}
Since $Z_1=0$ in the untwisted trace, the one-instanton splitting is not read off directly; it is either calculated by using the twisted partition function or it appears through $(E_1)^2$ in $Z_2$. At large $T$, the ground-state doublet dominates $Z_2$. Both parity states have the same perturbative energy and the same $(E_1)^2$, while $E_2$ is parity-independent. Using Eq.~\eqref{eq:Zk_from_Ek2}, the large-$T$ form is
\begin{equation}
  Z_2
 =
  2\,Z_\text{SHO}
  \left[
    \frac{T^2}{2\hbar^2}\bigl(E_1\bigr)^2
    -\frac{T}{\hbar}E_2
  \right] \,.
  \label{eq:Z2_from_energies}
\end{equation}
Rewriting the path-integral result~\eqref{eq:Z2_master} in powers of $T$ we have
\begin{equation}
  Z_2
  =
  2Z_\text{SHO}\frac{3S_I}{\pi\hbar}
  \left[
    T^2-2T\!\left(\gamma_E+\log\!\frac{12S_I}{\hbar}+\pi i\right)
  \right]
  \label{Z2_largeT}
\end{equation}
for the $\im\hbar>0$ lateral prescription. Matching powers of $T$ gives the path-integral energy trans-series for the ground-state doublet ($S_I=2/3$):
\begin{align}
  E_1 &= \pm 2\sqrt{\frac{\hbar}{\pi}}\,,
  \label{eq:E1_result}
  \\[4pt]
  E_2 &=
  \frac{4}{\pi}
  \left[\gamma_E+\ln\frac{8}{\hbar}\pm\pi i\right].
  \label{eq:E2_result}
\end{align}
The $\pm$ in $E_1$ distinguishes even and odd parity while the $\pm\pi i$ in $E_2$ reflects the choice of lateral Borel resummation. These results are in exact agreement with what we found with Exact WKB in Section~\ref{sec:exactWKB}: Eqs.~\eqref{eq:E1_summary} and~\eqref{eq:E2_summary} with $\kappa=1/2$ for the ground state. In particular, both the one-instanton splitting and two-instanton corrections agree exactly at leading order in $\hbar$ and we recover the coefficient of $\ln\hbar$, the imaginary part~$\pm\pi i$ and the finite real constant~$\gamma_E$. Interestingly, in the path-integral result $\gamma_E$ enters through the small-argument expansion $K_0(a)\sim -\gamma_E-\ln(a/2)$ of the arm integral, while in the exact-WKB result it enters as $\psi(1)=-\gamma_E$ from the Weber equation.

Excited-state energies ($E_0^{(2)}$, $E_0^{(3)}$, \ldots) are not accessible from the present large-$T$ matching, as they contribute to $Z_2$ only through exponentially suppressed terms $\cO(e^{-3T/2})$. For the excited states, we need to keep the full $T$ dependence, which can be done in the path integral and is the subject of Section~\ref{sec:pi_spectrum}.

\subsection{\texorpdfstring{$n=3$}{n=3} thimble decomposition}
\label{sec:n3_twisted_geometry}
The next simplest case is $n=3$. For an odd number of instantons, we need to use the twisted partition function. We use the same convention for the trans-series grading as with $Z$:
\begin{equation}
  \widetilde Z
  =
  \lambda\,\widetilde Z_1
  +\lambda^3\,\widetilde Z_3+\cdots,
  \qquad
  \lambda=e^{-S_I/\hbar}\,.
\end{equation}
The lateral Borel resummation of the asymptotic series in the $n=1$ coefficient has an ambiguity of order $\lambda^2$ and should be canceled by the thimble integral in the $n=3$ sector. We will verify this cancellation geometrically and, at leading large $T$, explicitly by computing the relevant parts of the two trans-series sectors.

\subsubsection{Quasi-zero-mode geometry for $n=3$}
In the $n=3$ case, the effective action from Eq.~\eqref{Seffalpha} has the form
\begin{equation}
   \Seff
    =
   3S_I-12S_I
  \left(e^{-\alpha_1}+e^{-\alpha_2}+e^{-\alpha_3}\right),
    \qquad \alpha_1+\alpha_2+\alpha_3=T \,.
\end{equation}
Here $\alpha_p$ are the distances between successive instanton centers.  This action has an extremum with $\alpha_p = T/3$ for all $p$. This is the real $n=3$ saddle around which $\Seff$ is locally valid. It is useful to introduce two coordinates $u$ and $v$ via
\begin{equation}
  \alpha_1=\frac{T}{3}-u,
  \qquad
  \alpha_2=\frac{T}{3}-v,
  \qquad
  \alpha_3=\frac{T}{3}+u+v\, .
\end{equation}
The $u$ and $v$ coordinates correspond to the collapse of the first or second instanton--anti-instanton pairs:
\begin{equation}
\begin{tikzpicture}[xscale=0.13, yscale=0.78, baseline=(current bounding box.center),
  declare function={
    mytanh(\s) = (exp(\s)-exp(-\s))/(exp(\s)+exp(-\s));
  }]
  % A three-event twisted chain plus the image of the first event.
  \draw[gray, dashed, thin] (-8,1) -- (54,1) node[right, black] {\footnotesize $+1$};
  \draw[gray, dashed, thin] (-8,-1) -- (54,-1) node[right, black] {\footnotesize $-1$};
  \draw[thick] (-8,-1) -- (-5,-1);
  \draw[thick, domain=-5:5, samples=60] plot (\x, {mytanh(0.5*\x)});
  \draw[thick] (5,1) -- (15,1);
  \draw[thick, domain=15:25, samples=60] plot (\x, {-mytanh(0.5*(\x-20))});
  \draw[thick] (25,-1) -- (35,-1);
  \draw[thick, domain=35:45, samples=60] plot (\x, {mytanh(0.5*(\x-40))});
  \draw[thick] (45,1) -- (54,1);
  \draw[-{Stealth[length=4pt]}, gray!80] (-8, -1.6) -- node[below] {\footnotesize $\alpha_3$} (0, -1.6);
  \draw[<->, gray!80] (0, -1.6) -- node[below] {\footnotesize $\alpha_1$} (20, -1.6);
  \draw[<->, gray!80] (20, -1.6) -- node[below] {\footnotesize $\alpha_2$} (40, -1.6);
  \draw[{Stealth[length=4pt]}-, gray!80] (40, -1.6) -- node[below] {\footnotesize $\alpha_3$} (54, -1.6);
  \definecolor{modeu}{rgb}{0.78, 0.24, 0.20}
  \definecolor{modev}{rgb}{0.20, 0.50, 0.78}
  % u: collapse the first +1 plateau, alpha_1.
  \draw[-{Stealth[length=5pt]}, thick, modeu] (2, 0.55) -- (7, 0.55)
    node[below, midway] {\scriptsize $u$};
  \draw[-{Stealth[length=5pt]}, thick, modeu] (18, 0.55) -- (13, 0.55)
    node[below, midway] {\scriptsize $u$};
  % v: collapse the intermediate -1 plateau, alpha_2.
  \draw[-{Stealth[length=5pt]}, thick, modev] (22, -0.55) -- (27, -0.55)
    node[above, midway] {\scriptsize $v$};
  \draw[-{Stealth[length=5pt]}, thick, modev] (38, -0.55) -- (33, -0.55)
    node[above, midway] {\scriptsize $v$};
\end{tikzpicture}
\end{equation}
Collapsing the third pair corresponds to $\alpha_3\to 0$, i.e.\ $u+v\to -T/3$, which is not an independent coordinate. In the $(u,v)$ coordinates, the effective action is 
\begin{equation}
  \label{Seffuv}
  \Seff(u,v)=
   3S_I-12S_I e^{-T/3}
  \left(e^u+e^v+e^{-u-v}\right) \,.
\end{equation}
Following the convention of Eq.~\eqref{eq:Seffn_def}, the deviation from the saddle action is
\begin{equation}
  \Seff^{(3)}(u,v) \equiv \Seff(u,v) - 3S_I = -12 S_I e^{-T/3}\left(e^u + e^v + e^{-u-v}\right) \,.
  \label{eq:Seff3_def}
\end{equation}
This action is shown in Figure~\ref{fig:n3_action_surface}. The $n=3$ saddle is the zenith of the plateau at $\Seff(u,v)\approx 3S_I$. The triangular structure of the simplex $\{\alpha_p>0;~\sum \alpha_p=T\}$ is evident in the symmetry of the action surface. When any of the three possible adjacent instanton pairs collapses, the same $n=1$ saddle results. Although $\Seff$ is not accurate near the $n=1$ saddle (just as the $n=2$ version of $\Seff$ did not describe the $n=0$ saddle in Section~\ref{sec:n2thimble}), the saddle's action is $S_I$, shown as a horizontal plane in the figure. The boundary edges of the simplex, where one $\alpha_p$ saturates $\ln 6$, lie on this plane and are drawn as jagged lines in the right panel.

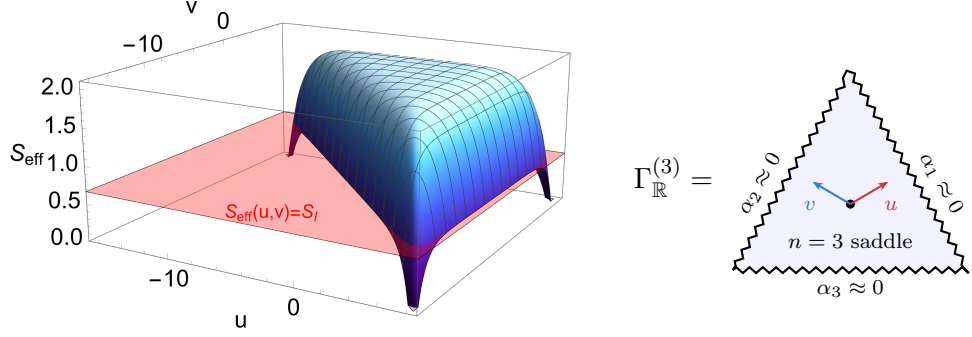
\begin{figure}[t]
\centering
\begin{tabular}{@{}c@{\hspace{0.05\linewidth}}c@{}}
  \raisebox{-0.6\height}{\includegraphics[width=0.45\linewidth]{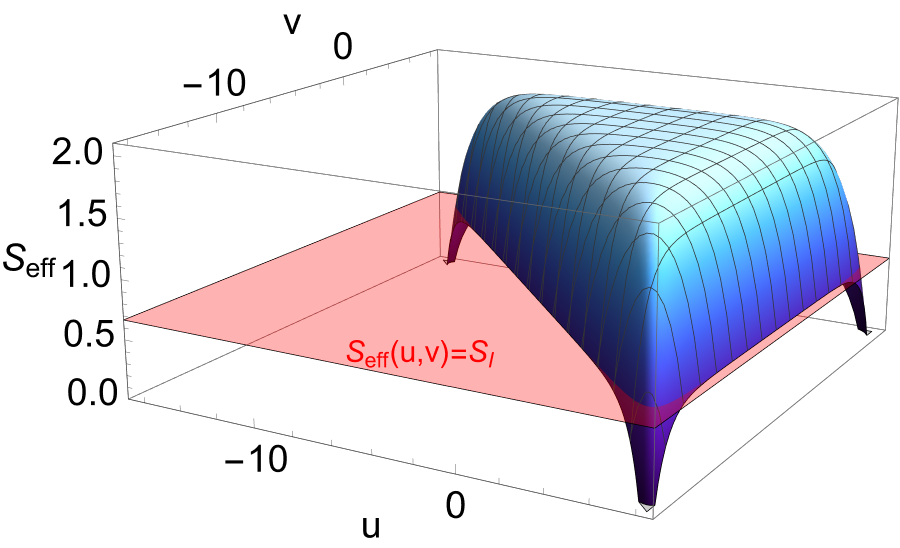}}
  &
  \raisebox{-0.5\height}{%
  $\Gamma_{\mathbb R}^{(3)}=$
  \begin{tikzpicture}[scale=0.74, baseline=(current bounding box.center)]
    \coordinate (A) at (0,0);
    \coordinate (B) at (4.2,0);
    \coordinate (C) at (2.1,3.55);
    \coordinate (O) at (2.1,1.18);
    \fill[blue!5] (A)--(B)--(C)--cycle;
    \draw[thick, decorate,
      decoration={zigzag, segment length=5pt, amplitude=1.1pt}]
      (A)--(B)--(C)--cycle;
    \node[below] at (2.1,0) {\scriptsize $\alpha_3 \approx 0$};
    \node[rotate=60] at (0.48,1.55) {\scriptsize $\alpha_2\approx 0$};
    \node[rotate=-60] at (3.72,1.55) {\scriptsize $\alpha_1 \approx 0$};
    \fill[black] (O) circle (2.3pt);
    \node[below, yshift=-7pt] at (O) {\scriptsize $n=3$ saddle};
    \definecolor{modeu}{rgb}{0.78, 0.24, 0.20}
    \definecolor{modev}{rgb}{0.20, 0.50, 0.78}
    \draw[-{Stealth[length=4pt]}, thick, modeu] (O) -- (2.78,1.58)
      node[pos=0.62, below right] {\scriptsize $u$};
    \draw[-{Stealth[length=4pt]}, thick, modev] (O) -- (1.42,1.58)
      node[pos=0.62, below left] {\scriptsize $v$};
  \end{tikzpicture}}
\end{tabular}
\caption{The $n=3$ collective-coordinate geometry.  On the left
the effective action $\Seff(u,v)$ is shown. This action is valid near the $n=3$ saddle, where $\alpha_p=T/3$ and $u=v=0$, and descends to the $n=1$ saddle where $S=S_I$, approximated by the red plane. The right panel shows the real $u,v$ domain with the $n=3$ saddle at the center. The triangle edges at $\alpha_p\approx \ln 6$ represent the $n=1$ instanton. The jagged lines indicate that the $n=1$ saddle of the full action is not precisely described by $\Seff$ on these edges.
}
\label{fig:n3_action_surface}
\end{figure}

The Picard--Lefschetz decomposition of $\widetilde{Z}$ involves the thimble $\cJ_1$ through the $n=1$ saddle which contributes to $\widetilde Z_1$ and the thimble $\cJ_3$ through the $n=3$ saddle, which contributes to $\widetilde Z_3$.  The $\cJ_1$ thimble departs from the $n=1$ saddle and hits the Stokes point at the $n=3$ saddle where it becomes complex and lines up with $\cJ_3$ but with opposite orientation. The imaginary parts of the $\cJ_1$ and $\cJ_3$ thimbles cancel because they come from an identical cycle with opposite orientation. The geometric picture makes the cancellation trivial, as it did in $n=2$. To determine the corresponding parts of the trans-series we have to do the relevant integrals.

\subsubsection{$n=3$ thimble integral}

We begin by computing the integral $\cI_3^\pm$ along $\cJ_3^\pm$, where $\pm$ refer to the sign of the imaginary part in the lateral deformation:
\begin{equation}
  \mathcal I_3^\pm(T)
  =
  \int_{\cJ_3^\pm} du\,dv\,
  e^{-\Seff^{(3)}/\hbar}
  =
  \int_{\cJ_3^\pm} du\,dv\,
  \exp\!\left[
    \nthreeScale\left(e^u+e^v+e^{-u-v}\right)
  \right],
  \label{eq:n3_A2_thimble_integral}
\end{equation}
where
\begin{equation}
  \nthreeScale\equiv\frac{12S_I}{\hbar}e^{-T/3}.
\end{equation}
The contour $\cJ_3$ is a middle-dimensional cycle in ${\mathbb C}^2$ attached to the real saddle at $u=v=0$ and ending in convergent asymptotic regions.  To evaluate the integral, it is useful to start from same integral over $\RR^2$ with the sign in the exponent flipped
\begin{equation}
  \cI^{\mathbb R}_3(\mu)
  =
  \int_{\mathbb R^2} dx\,dy\,
  \exp\!\left[-\mu\left(e^x+e^y+e^{-x-y}\right)\right],
  \qquad
  \re\mu>0,
\end{equation}
which gives
\begin{equation}
  \cI^{\mathbb R}_3(\mu)
  =
  \cG_3(\mu^3),
  \label{eq:A2_Macdonald}
\end{equation}
with $\cG_3$ a Meijer G-function defined in Eq.~\eqref{eq:cG_def}.  The physical saddle has the opposite sign in the exponent, so it is obtained by continuing the real cycle to the negative $\mu$ axis.  During this continuation the real cycle deforms into the thimble $\cJ_3^\pm$, giving
\begin{equation}
  \cI_3^\pm(T,\hbar)
  =
  \cI^{\mathbb R}_3(\nthreeScale e^{\pm i\pi})
  =
  \cG_3(\nthreeScale^3 e^{\pm 3\pi i}).
  \label{eq:n3_exact_thimble_G}
\end{equation}
Here $\nthreeScale^3e^{\pm3\pi i}$ records the sheet reached by the lateral continuation of the real convergent integral.  This is the direct two-dimensional replacement of the factor $\cG_2(\ntwoScale^2 e^{\pm 2\pi i}) = 2K_0(2\ntwoScale e^{\pm\pi i})$ that appeared in the $n=2$ thimble integral, Eq.~\eqref{eq:n2_exact_thimble_G}.

To expand at large $T$, where $\nthreeScale\to0$, we can use
\begin{equation}
\cG_3(z)
  =
  \frac{1}{2}\left(\log z+3\gamma_E\right)^2
  +\frac{\pi^2}{4}
  +\cO\!\left(z\log^2 z\right).
  \label{eq:A2_G_largeT}
\end{equation}
We then get
\begin{equation}
  \cI_3^{\pm}
  =
    \frac{9}{2}\LT^2
    -\frac{17\pi^2}{4}
    \pm 9\pi i\,\LT
    +\cO\!\left(T^3 e^{-T}\right)
  \label{eq:n3_I3_largeT}
\end{equation}
where
\begin{equation} 
  \LT
  =\log\nthreeScale+\gamma_E
  =-\frac{T}{3}+
  \log\!\left(\frac{12S_I}{\hbar}\right)
  +\gamma_E \,. \label{eq:l3_def}
\end{equation}
This combination $\LT$ appearing in $\cI_3^{\pm}$ is nearly identical to the combination appearing in $\cI_2^+$ in Eq.~\eqref{eq:j_10_thimble}: the $T$ comes from the length of the plateau and the $\ln \hbar$ from its exponentially small curvature. 
The imaginary part of the $n=3$ thimble integral is then
\begin{equation}
  \im\,\cI_3^{\pm}
  =
  \pm\,9\pi\,\LT \,.
  \label{eq:n3_exact_imag_largeT}
\end{equation}
 Combining with the collective-coordinate measure and the transverse determinant in Eq.~\eqref{eq:detperp}, the trans-series coefficient is
\begin{align}
  \widetilde Z_3^{\pm}
  &=
  \frac{2T}{3}\, Z_\text{SHO}
  \left(\frac{S_I}{2\pi\hbar}\right)^{3/2}
  \left(\frac{\det\opO_0}{\det_\perp\opO_3}\right)^{1/2}
\cI_3^{\pm}\\ 
  &=
  \frac{2T}{3}\,Z_\text{SHO}
  \left(\frac{12 S_I}{2\pi\hbar}\right)^{3/2}
\left[
    \frac{9}{2}\LT^2
    -\frac{17\pi^2}{4}
    \pm 9\pi i\,\LT
  \right]
  \label{Z3I3}
\end{align}
The factor of $2$ comes from the two parity-related three-instanton configurations. The factor of $1/3$ multiplying $\widetilde Z_3^\pm$ is the CC-fix factor, the correction to the standard collective-coordinate measure. We emphasize that the change of variables from local fluctuation coordinates to the collective coordinate $t_0$ is in fact multi-valued, with the number of preimages $N_{\psi_0}[x]$ equal to the number of barrier crossings of the path. The correct measure contains a compensating factor $1/N_{\psi_0}[x]$~\cite{CollectiveCoordinateFix}. For an $n$-instanton path $N_{\psi_0}=n$ at leading order, so the CC-fix factor is $1/n$. For $n=2$ in Eq.~\eqref{eq:Z2_master1} the CC-fix factor canceled in $Z_2$ against the factor of 2 coming from the two-well multiplicity.

\subsubsection{$n=1$ asymptotic series}
\label{sec:n1asy}
Around the $n=1$ saddle, the 1-loop path integral and its large-$T$ behavior were given in Eqs.~\eqref{eq:Ztilde1_exact} and~\eqref{eq:Ztilde1_exact_largeT}. Stripping the $e^{-S_I/\hbar}$ factor in accordance with the trans-series definition $\widetilde Z = \lambda \widetilde Z_1 + \lambda^3 \widetilde Z_3 + \cdots$ and including both orientations,
\begin{equation}
  \widetilde Z_1
  = 2 Z_\text{SHO}\,
   T\left(\frac{12S_I}{2\pi\hbar}\right)^{1/2}
  + \cdots,
   \label{eq:Ztilde1_leading}
\end{equation}
which is real at all orders in the perturbative expansion. The leading asymptotic growth of the $n=1$ coefficient is governed by the nearest twisted saddle at $n=3$. As in the $n=2$ analysis of $I_{0\to2}^{\text{asy}}$, this growth is generated by the integral over quasi-collective coordinates. We first write the integral over the real simplex as
\begin{equation}
  \mathcal I_{1\to3}
  =
  \int_{\Gamma_\triangle}du\,dv\,
  e^{-\Seff/\hbar}
  =
  e^{-3S_I/\hbar}
  \int_{\Gamma_\triangle}du\,dv\,
  \exp\!\left[\nthreeScale\left(e^u+e^v+e^{-u-v}\right)\right] \,.
  \label{eq:n3_lower_real_integral}
\end{equation}
This is the $n=3$ analog of the integral $I_{0\to2}$ over $\alpha$ in the $n=2$ case (Eq.~\eqref{eq:asy_int_n2}), with the simplex $\Gamma_\triangle$ playing the role of the real segment $[\amin,T/2]$. The three sides of the triangle correspond to the three adjacent-pair collapse channels and are equivalent under cyclic relabeling, so we can pick the sector $\alpha_1\le\alpha_2,\alpha_3$ (i.e.\ $-2u\le v\le u$) and multiply by~$3$. Substituting $w=v+u/2$ symmetrizes the inner range to $w\in[-3u/2,3u/2]$ and brings the action into $\cosh$ form,
\begin{equation}
    \cI_{1\to3} = 3 e^{-3S_I/\hbar}\int_0^{\umin} du\,e^{\nthreeScale e^{u}}
    \int_{-3u/2}^{3u/2} dw \, e^{2\nthreeScale e^{-u/2}\cosh w} \,.
     \label{eq:n3_one_edge_integral_full}
\end{equation}
It is natural to take $\umin = \frac{T}{3}-\log 6$
in direct analogy with $\amin=\log 6$ in Eq.~\eqref{eq:Ireal_largeT}. With this choice $\Seff$ spans the full $2S_I$ action gap between the $n=1$ and $n=3$ saddles at leading order in $T$ ($\Seff \to S_I$ at $u=\umin$ and $\Seff\to 3S_I$ at $u=0$), matching the range of the full action between those saddles. As in the $n=2$ case, this fixes the location of the Borel singularity, leaving only matching constants from the small-$\alpha$ region which do not affect the leading asymptotic behavior.

The integral in Eq.~\eqref{eq:n3_one_edge_integral_full} is over the whole triangle in Fig.~\ref{fig:n3_action_surface}. It contains the directions of descent into the $n=1$ saddle in the 3 directions (the triangle edges). However, it also includes regions where a second separation becomes small (the triangle corners). These corners are problematic: they are well outside of the validity of $\Seff$, with action well below $S_I$, and are not relevant to the asymptotic growth of $\cI_{1\to3}$. To isolate the asymptotic growth of the $n=1$ series we should sum the 3 collapse channels independently, where in each channel the third instanton is a spectator.  The action in Eq.~\eqref{Seffuv} has three terms, $e^u$, $e^v$ and $e^{-u-v}$, corresponding to the three collapse channels, so to isolate one channel we can keep one of these terms at a time, treating the other two as constant. Focusing on the $u$ channel, we keep $e^u$ and drop the other two terms: $e^v \to 0$ and $e^{-u-v}\to 0$. This amounts to setting $e^{2\nthreeScale e^{-u/2}\cosh w}\to 1$ in Eq.~\eqref{eq:n3_one_edge_integral_full}. Equivalently, we drop the $w$ dependence so that $w$ becomes a collective coordinate. We still need the volume of $w$, with the $\pm 3u/2$ integration boundaries, like we would need the volume of an honest collective coordinate such as $t_0$.   Thus we expect the leading asymptotic growth to be given by
\begin{equation} \label{eq:n3_w_block}
  \cI_{1\to3}
  =
  3e^{-3S_I/\hbar}\int_0^{\umin}du\,e^{\nthreeScale e^u}
  \int_{-3u/2}^{3u/2}dw
  +\cdots \,,
\end{equation}
where the $\cdots$ represent terms not relevant to the leading asymptotic growth of $\widetilde{Z}_1$. 

Changing variables to $\xi=\nthreeScale e^u$ gives
\begin{equation}
  \cI_{1\to3}
  =
  9e^{-3S_I/\hbar}\int_{\nthreeScale}^{\ximin}
  \frac{d\xi}{\xi}\,e^\xi\log(\xi/\nthreeScale)
  +\cdots ,
  \qquad
  \ximin=\nthreeScale e^{\umin}=\frac{2S_I}{\hbar} \,.
  \label{eq:n3_open_face_projection}
\end{equation}
This integral is dominated by its upper endpoint region, near the $n=1$ saddle at $\xi = \ximin$ where the integrand is exponentially large. Near the upper endpoint we split
\begin{equation}
  \label{twoterms}
  \log(\xi/\nthreeScale)
  =
  \log(\ximin/\nthreeScale)+\log(\xi/\ximin) \,.
\end{equation}
 The first term is
\begin{equation}
  \log(\ximin/\nthreeScale)
  =
  -\LT+\bigl(\log \ximin+\gamma_E\bigr) \,,
\end{equation}
which gives the dominant behavior of the integral, growing linearly with $T$. The second term in Eq.~\eqref{twoterms} vanishes at the upper endpoint and can only contribute at order $T^0$ or lower. Retaining the part proportional to $\LT$, and noting that the resulting integral is the same as in the $n=2$ case, Eq.~\eqref{eq:asy_int_n2b}, we then get
\begin{equation} \label{eq:n3_residue}
  \cI_{1\to3}
  =
  - 9\LT \, e^{-3S_I/\hbar}
  \int_{\nthreeScale}^{\ximin}\frac{d\xi}{\xi}\,e^\xi + \cdots
  =
  +9 \LT \, e^{-3S_I/\hbar}\Ein(-\ximin)+\cdots \,,
\end{equation}
where in the last step we used the identity from Eq.~\eqref{eq:asy_int_n2b}, $\int_a^b du\, e^u/u = \Ein(-a) - \Ein(-b) + \ln(b/a)$, with $\Ein(-\nthreeScale)\to 0$ as $\nthreeScale\to 0$ and the resulting $\ln(\ximin/\nthreeScale)$ piece absorbed into the $\cdots$. Since this factorial tail expression is obtained from integrating over the quasi-zero modes about the $n=3$ saddle, we must also account for an extra $1/3$ factor from the CC-fix\footnote{Note that in Eq.\eqref{Irealasy} the CC-fix factor from the $n=2$ quasi-zero mode sector was implicit. There we had integrated along the single descent direction from $\alpha \in [\amin, T/2]$, which corresponds to integrating along a single face of the relevant simplex. Therefore the CC-fix factor of $1/2$ canceled against the multiplicative factor of 2 associated with the 2 distinct collapse channels (or faces of the 1-simplex).}. Therefore, using $\ximin=2S_I/\hbar$, and the trans-series expansion of $\Ein$ in Eq.~\eqref{eq:Ein_trans} we arrive at
\begin{equation} \label{eq:I13_factorial}
  \cI_{1\to3}^{\rm asy}
=
  -3\LT\,e^{-S_I/\hbar}
  \sum_{n=0}^{\infty}n!\left(\frac{\hbar}{2S_I}\right)^{n+1} + \cdots \,.
\end{equation}
The omitted $\cdots$ terms are not relevant to the leading asymptotic growth. 

\subsubsection{Resurgent structure at order $\lambda^3$}

To convert $\cI_{1\to 3}^\text{asy}$ into the twisted partition function trans-series, we need to add the transverse determinant. Since $\cI_{1\to 3}^\text{asy}$ was computed using the quasi-zero mode coordinates from around the $n=3$ saddle, the relevant transverse determinant is over coordinates transverse to this space, $(\det\opO_0/\det_\perp\opO_3)^{1/2}=12^{3/2}$. Combining the leading $\widetilde Z_1$ from Eq.~\eqref{eq:Ztilde1_leading} with the factorial tail in Eq.~\eqref{eq:I13_factorial}, weighted by the determinant ratio, and accounting for the $\lambda$ factor in the definition of $\widetilde Z_1$ we get:
\begin{equation}
   \widetilde Z_1
   = 2T\,Z_\text{SHO}\!\left(\frac{12S_I}{2\pi\hbar}\right)^{1/2}\!\left[
     1 + \cdots
     \,-\,3\LT\,\frac{12S_I}{2\pi\hbar}\sum_{n=0}^{\infty} n!\!\left(\frac{\hbar}{2S_I}\right)^{n+1}
   \right]\,.
   \label{eq:Ztilde1_form}
\end{equation}
The lateral Borel resummation of this series gives
\begin{equation}
   \cS_\pm \bigl[\widetilde Z_1 \bigr]
   = 6 T \LT \,Z_\text{SHO}\!\left(\frac{12S_I}{2\pi\hbar}\right)^{3/2}\!e^{-2S_I/\hbar}
     E_1\!\left(-\frac{2S_I}{\hbar}\pm i\varepsilon\right) \,.
\end{equation}
With $\im E_1(-2S_I/\hbar \pm i\varepsilon)=\mp \pi$, the imaginary part is
\begin{equation}
   \im\,\cS_\pm \bigl[\widetilde Z_1 \bigr]
=
   \mp\,6 \pi T \LT \,Z_\text{SHO}\!\left(\frac{12S_I}{2\pi\hbar}\right)^{3/2}\!e^{-2S_I/\hbar}\,.
   \label{eq:n3_cancellation}
\end{equation}
This exactly cancels the imaginary part from the $n=3$ sector of the trans-series in Eq.~\eqref{Z3I3}. More precisely
\begin{eqnarray}
    \im \cS_\pm [\widetilde{Z} ] = \im \cS_\pm [e^{-S_I/\hbar}\widetilde{Z}_1 ] + \im  [e^{-3S_I/\hbar}\widetilde{Z}_3 ] +\cdots = 0
\end{eqnarray}
establishing the reality of the Borel-resummed twisted partition function to order $\lambda^3$. This is the $n=3$ counterpart of the $\lambda^2$ cancellation in $Z$ in Eq.~\eqref{imZ0} which we have verified in this subsection to leading order in $\LT$.

%%%%%%%%%%%%%%%%%%% N=4 %%%%%%%%%%%%%%%%%%%

\subsection{\texorpdfstring{$n = 4$}{n=4} thimble decomposition}
\label{sec:n4}
The geometric thimble picture of the quasi-zero-mode space is similar for any $n$. At $n=4$ there is a new feature: the imaginary part of the integral along the $n=4$ thimble is not completely canceled by the imaginary parts from thimbles associated with the nearest lower saddle at $n=2$. Instead, a thimble passing through $n=0$ is also needed. Since the effective action $\Seff(\alpha_p)$ we have been using is derived at large $T$, it will not describe these distant saddles well enough to establish the complete cancellation. We can nevertheless see the cancellation geometrically and verify the cancellation of the leading component of the imaginary part at large $T$. We next discuss this new feature of the $n=4$ case and then briefly summarize the resurgent structure for general $n$. 

\subsubsection{Quasi-zero-mode geometry for $n=4$}

At $n=4$ we are back to the untwisted partition function. This case illustrates a new feature: there are 3 saddles, with $n=0,2,4$, all of which are relevant to the Picard--Lefschetz decomposition and the resurgent structure at order $\lambda^4$. For $n=4$, the effective action from Eq.~\eqref{Seffalpha} has the form
\begin{equation}
   \Seff
    =
    4S_I-12 S_I\left[e^{-\alpha_1}+e^{-\alpha_2}+e^{-\alpha_3}+e^{-\alpha_4}\right],
    \qquad \alpha_1+\alpha_2+\alpha_3+\alpha_4=T \,.
\end{equation}
Here $\alpha_p$ are the distances between successive instanton centers. The simplex $\Gamma_\RR^4 = \{\alpha_p>0;~\sum \alpha_p=T\}$ is a tetrahedron in $\mathbb R^3$, generalizing the triangle from Figure~\ref{fig:n3_action_surface}.  Imposing the constraint $\sum \alpha_p = T$  with a Lagrange multiplier and extremizing the action leads to $\alpha_p^\star = T/4$ for all $p$. This is the real $n=4$ saddle around whose local neighborhood $\Seff$ is valid.

We can diagonalize the Hessian with normal coordinates $t_0, x_1, x_2, y$ as
\begin{equation}
  \alpha_1 = \frac{T}{4}- x_1 - y,\quad
  \alpha_2 = \frac{T}{4} - x_2 + y,\quad
  \alpha_3 = \frac{T}{4} + x_1 - y,\quad
  \alpha_4 = \frac{T}{4} + x_2 + y\,.
  \label{eq:xmodes_n4}
\end{equation}
so that
\begin{align}\label{eq:seff_n4}
     \Seff &= 4S_I - 12 S_I e^{-\frac{T}{4}} \left[ e^{x_1 +y} + e^{x_2-y} + e^{-x_1+y} + e^{-x_2-y} \right] \\
     &= 4 S_I - 48 S_I e^{-\frac{T}{4}} - 12 S_I e^{-\frac{T}{4}}(x_1^2 + x_2^2 + 2y^2) +\cdots
\end{align}
 There is no dependence on the exact zero mode coordinate $t_0$. 
The three normal modes act on the instanton separations as follows:
\begin{equation}
\label{eq:n4_mode_diagram}
\begin{tikzpicture}[xscale=0.13, yscale=0.7, baseline=(current bounding box.center),
  declare function={
    mytanh(\s) = (exp(\s)-exp(-\s))/(exp(\s)+exp(-\s));
  }]
  % Centers at 0, 20, 40, 60 with separation T/4 = 20.
  % Dashed lines at x = pm 1
  \draw[gray, dashed, thin] (-7,1) -- (67,1) node[right, black] {\footnotesize $+1$};
  \draw[gray, dashed, thin] (-7,-1) -- (67,-1) node[right, black] {\footnotesize $-1$};
  % Piecewise curve: flat segments and tanh transitions
  \draw[thick] (-7,-1) -- (-5,-1);
  \draw[thick, domain=-5:5, samples=60] plot (\x, {mytanh(0.5*\x)});
  \draw[thick] (5,1) -- (15,1);
  \draw[thick, domain=15:25, samples=60] plot (\x, {-mytanh(0.5*(\x-20))});
  \draw[thick] (25,-1) -- (35,-1);
  \draw[thick, domain=35:45, samples=60] plot (\x, {mytanh(0.5*(\x-40))});
  \draw[thick] (45,1) -- (55,1);
  \draw[thick, domain=55:65, samples=60] plot (\x, {-mytanh(0.5*(\x-60))});
  \draw[thick] (65,-1) -- (67,-1);
  % Instanton centers (orange) and anti-instanton centers (blue)
  \fill[orange] (0, 0) circle (4pt);
  \fill[blue!70] (20, 0) circle (4pt);
  \fill[orange] (40, 0) circle (4pt);
  \fill[blue!70] (60, 0) circle (4pt);
  % Separation labels below
  \draw[<->, gray!80] (0, -1.6) -- node[below] {\footnotesize $\alpha_1$} (20, -1.6);
  \draw[<->, gray!80] (20, -1.6) -- node[below] {\footnotesize $\alpha_2$} (40, -1.6);
  \draw[<->, gray!80] (40, -1.6) -- node[below] {\footnotesize $\alpha_3$} (60, -1.6);
  %
  % Define colors to match Figure~\ref{fig:n4_mode_flows}
  \definecolor{modex1}{rgb}{0.78, 0.24, 0.20}
  \definecolor{modex2}{rgb}{0.18, 0.42, 0.80}
  \definecolor{modey}{rgb}{0.20, 0.58, 0.30}
  % x1 mode: alpha_1 shrinks (I1 and Ibar2 approach)
  \draw[-{Stealth[length=5pt]}, thick, modex1] (1, 0.2) -- (6, 0.2)
    node[above, midway] {\footnotesize $x_1$};
  \draw[{Stealth[length=5pt]}-, thick, modex1] (14, 0.2) -- (19, 0.2)
    node[above, midway] {\footnotesize $x_1$};
  %
  % x2 mode: alpha_2 shrinks (Ibar2 and I3 approach)
  \draw[-{Stealth[length=5pt]}, thick, modex2] (21, 0.2) -- (26, 0.2)
    node[above, midway] {\footnotesize $x_2$};
  \draw[{Stealth[length=5pt]}-, thick, modex2] (34, 0.2) -- (39, 0.2)
    node[above, midway] {\footnotesize $x_2$};
  %
  % y mode: alpha_1,3 shrink and alpha_2,4 grow (breathing)
  \draw[-{Stealth[length=5pt]}, thick, modey] (1, -0.2) -- (6, -0.2)
    node[below, midway] {\footnotesize $y$};
  \draw[{Stealth[length=5pt]}-, thick, modey] (14, -0.2) -- (19, -0.2)
    node[below, midway] {\footnotesize $y$};
  \draw[-{Stealth[length=5pt]}, thick, modey] (41, -0.2) -- (46, -0.2)
    node[below, midway] {\footnotesize $y$};
  \draw[{Stealth[length=5pt]}-, thick, modey] (54, -0.2) -- (59, -0.2)
    node[below, midway] {\footnotesize $y$};
\end{tikzpicture}
\end{equation}
 The $x_1$ normal mode compresses $\alpha_1$ and stretches $\alpha_3$, while $x_2$ compresses $\alpha_2$ and stretches $\alpha_4$. Each corresponds to a single instanton--anti-instanton pair collapsing. For example, as $x_1\to T/4$, $\alpha_1\to 0$ so that the first and second instantons annihilate. The surviving pair then sits at separation $\alpha_3 = T/2$, which is the $n=2$ saddle of the reduced two-instanton problem on the same circle. The $y$ mode is a breathing mode that compresses $\alpha_1,\alpha_3$ while stretching $\alpha_2,\alpha_4$. As $y\to T/4$ both $\alpha_1,\alpha_3\to 0$ and the configuration degenerates to $n=0$. 
 
\begin{figure}[t]
\centering
    \includegraphics[width=\textwidth]{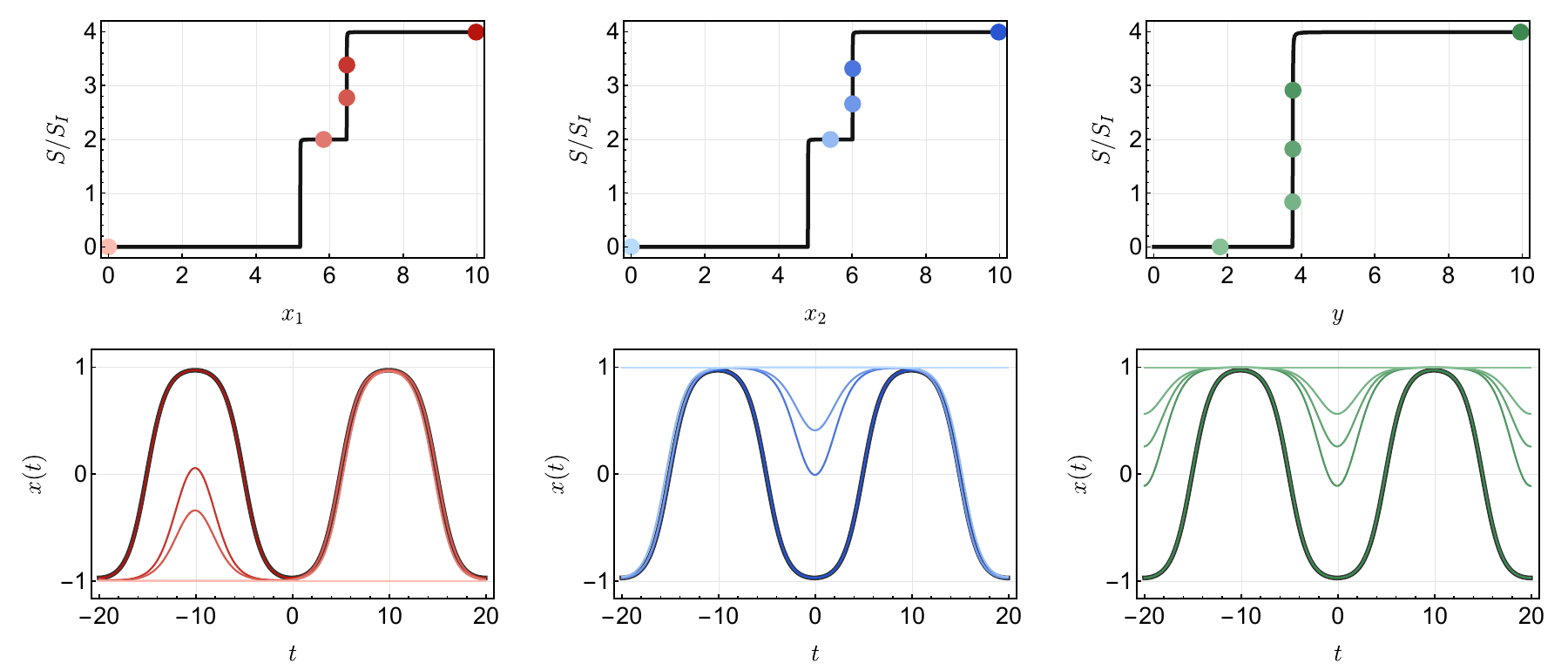}
\caption{Downward-flow trajectories from the $n=4$ saddle along the three unstable normal-mode directions $x_1$, $x_2$, and $y$ at $T=40$. The top row shows the normalized action $S/S_I$ along the flow coordinate, and the bottom row shows representative profiles $x(t)$ (darker curves are earlier in the flow). The $x_1$ and $x_2$ modes each annihilate a single instanton--anti-instanton pair, reducing the action from $4S_I$ to $\approx 2S_I$; the surviving pair then contracts and annihilates, reaching $S\approx 0$. The action plateaus near $2S_I$ because any well-separated kink pair has $S\approx 2S_I$ regardless of separation. That the kink separation at the plateau appears closer to $T/4$ than $T/2$ is an artifact of the flow integration: the QZM eigenvalues scale as $e^{-T/4}$ (so $\sim e^{-10}$ at $T=40$), making the time to escape the saddle exponentially sensitive to the size of the initial perturbation along $\Psi_1$.
}
\label{fig:n4_mode_flows}
\end{figure}

 The action $\Seff(x_1,x_2,y)$ is only a local expression near the $n=4$ saddle, which is where $x_1,x_2$ and $y$ are sensible normal modes.  The corresponding directions can nevertheless be followed in the full path space by gradient flow, as shown in Fig.~\ref{fig:n4_mode_flows}.  The $x_1$ and $x_2$ directions first approach the $n=2$ boundary sector, while the $y$ direction directly approaches the perturbative $n=0$ sector.  The plateau near $2S_I$ is therefore not a feature of the local $\Seff$ chart; it is a feature of the full action along the flow. The natural QZM manifold (including the zero mode) has dimension $n$ near the $n$-instanton saddle, 
\begin{equation}
  n=0:\ 0,\qquad
  n=2:\ (t_0,\alpha),\qquad
  n=4:\ (t_0,x_1,x_2,y).
\end{equation}
Thus the lower sectors should be viewed as boundary levels of the compactified quasi-zero-mode space, not as ordinary interior critical points of the same local $n=4$ coordinate action. 

For the $n=4$ thimble integral, we follow the same strategy as in the $n=3$ analysis: rather than work in the diagonal coordinates $(x_1,x_2,y)$ of Eq.~\eqref{eq:xmodes_n4}, which are local normal modes of the Hessian but mix the four collapse channels, we use simplex coordinates that resolve each adjacent-pair collapse on its own face. Following the general-$n$ setup of Section~\ref{sec:gen_n_thimble}, we set
\begin{equation}
  \alpha_p = \frac{T}{4} - u_p \quad (p=1,2,3),\qquad
  \alpha_4 = \frac{T}{4} + u_1+u_2+u_3,
  \label{eq:n4_simplex_coords}
\end{equation}
in which the effective action reads
\begin{equation}
  \Seff(u) = 4 S_I - 12 S_I e^{-T/4}\Bigl[e^{u_1}+e^{u_2}+e^{u_3}+e^{-u_1-u_2-u_3}\Bigr]\,,
  \label{eq:n4_seff_simplex}
\end{equation}
with the deviation from the saddle action
\begin{equation}
  \Seff^{(4)}(u) \equiv \Seff(u) - 4S_I = -12 S_I e^{-T/4}\Bigl[e^{u_1}+e^{u_2}+e^{u_3}+e^{-u_1-u_2-u_3}\Bigr]\,.
  \label{eq:Seff4_def}
\end{equation}
The collective-coordinate domain $\Gamma_\triangle^4$ is a tetrahedral 3-simplex, with four codimension-1 faces $\alpha_p\to 0$ each describing one adjacent pair collapsing down to a $n=2$ configuration. The Picard--Lefschetz decomposition pairs the saddle thimble $\cJ_4$ through the real $n=4$ saddle at $u=0$ with boundary thimbles attached to the lower-$n$ saddles, with imaginary parts canceling channel by channel as in the $n=3$ case.

\subsubsection{$n=4$ thimble integral}
\label{sec:n4_thimble_integral}
We first compute the integral along the thimble through the real $n=4$ saddle. This integral is almost identical 
to Eq.~\eqref{eq:n3_A2_thimble_integral} and is evaluated the same way:
\begin{align}
  \cI^\pm_4(T)
  &= \int_{\cJ_4^\pm} du_1\,du_2\,du_3\,e^{-\Seff^{(4)}/\hbar}
  = \int_{\cJ_4^\pm} du_1\,du_2\,du_3\,
  \exp\!\left[
	\nfourScale\Bigl(e^{u_1}+e^{u_2}+e^{u_3}+e^{-u_1-u_2-u_3}\Bigr)
  \right] \nonumber \\
  &= \cG_4(\nfourScale^4 e^{\pm 4\pi i}) \,,
  \label{eq:n4_thimble_def}
\end{align}
where $\nfourScale\equiv\frac{12S_I}{\hbar}e^{-T/4}$. The result is the $n=4$ analog of Eq.~\eqref{eq:n3_exact_thimble_G}. At large $T$, $\nfourScale\to0$, and $\cI_4^\pm$ expands to a polynomial in $T$ (cf. Eq.~\eqref{eq:gen_smallz} below):
\begin{equation}
  \cI_4^\pm
  =
    -\frac{32}{3}\LF^3 + \frac{92\pi^2}{3}\LF - \frac{4\zeta(3)}{3}
    \pm i\!\left(-32\pi \LF^2 + \frac{28\pi^3}{3}\right)
    + \cO \bigl(T^4 e^{-T}\bigr) \,,
  \label{eq:n4_I4_largeT}
\end{equation}
where
\begin{equation}
  \LF \equiv \log\nfourScale + \gamma_E
  = \log\!\left(\frac{12 S_I}{\hbar}\right) - \frac{T}{4} + \gamma_E\,.
  \label{eq:l4_def}
\end{equation}
Therefore
\begin{equation}
  \im\,\cI_4^\pm
  =
  \mp 32\pi\, \LF^2 \pm \frac{28\pi^3}{3}
  + \cO(T^4 e^{-T}) \,.
  \label{eq:n4_exact_imag_largeT}
\end{equation}
Compared to $\im\,\cI_3^\pm=\pm9\pi \LT$, the leading logarithmic degree is now quadratic instead of linear. Combining $\cI_4^\pm$ with the collective-coordinate measure, the transverse determinant from Eq.~\eqref{eq:detperp}, the two-well multiplicity factor of $2$, and the CC-fix factor $1/4$ for the 4-instanton paths, gives
\begin{align}
  Z_4^\pm
  &= \frac{2}{4}\,T\,Z_\text{SHO}
  \left(\frac{S_I}{2\pi\hbar}\right)^{2}
  \left(\frac{\det\opO_0}{\det_\perp\opO_4}\right)^{1/2}\cI_4^\pm \\ \label{eq:z4n4thimble}
  &= \frac{T}{2}\,Z_\text{SHO}
  \left(\frac{12 S_I}{2\pi\hbar}\right)^{2}
  \left[
   -\frac{32}{3}\LF^3 + \frac{92\pi^2}{3}\LF - \frac{4\zeta(3)}{3}
    \pm i\!\left(-32\pi \LF^2 + \frac{28\pi^3}{3}\right)\right]\,.
\end{align}
Here the prefactor $2/4=2/N_{\psi_0}$ combines the two-well multiplicity factor with the CC-fix factor $1/N_{\psi_0}$.

\subsubsection{$n=2$ asymptotic series}
Next, we compute the leading asymptotic series around the $n=2$ saddle, following Section~\ref{sec:n1asy} closely. The full real-cycle integral in the $n=4$ quasi-zero-mode space is
\begin{equation}
  \mathcal I_{\triangle}^4
  = e^{-4S_I/\hbar}
  \int_{\Gamma_\triangle^4} du_1\,du_2\,du_3\,
\exp\!\left[\nfourScale\Bigl(e^{u_1}+e^{u_2}+e^{u_3}+e^{-u_1-u_2-u_3}\Bigr)\right]\,.
  \label{eq:n4_lower_real_integral}
\end{equation}
This is the analog of $\cI_{1\to3}$ in Eq.~\eqref{eq:n3_lower_real_integral} but now the integration cycle $\Gamma_\triangle^4$ is over a tetrahedron instead of a triangle. The center of the tetrahedron is the $n=4$ saddle, and the directions from the center to a face along each $u_p$ correspond to the annihilation of an adjacent instanton-anti-instanton pair. Analogous to the $n=3$ case, the dominant asymptotic growth of the $n=2$ trans-series comes from integrations along a single quasi-zero mode coordinate $u_p$ at a time with the others treated as static and well separated. So, labeling $u$ the coordinate to be integrated, and $w_1, w_2$ the static ones, we have
\begin{equation}
\mathcal I_{2\to4}^{\text{asy}}
= 4 e^{-4S_I/\hbar}\int_0^{\umin} du\,e^{\nfourScale e^{u}}
    \int_{\Gamma_2(u)} d w_1 d w_2 
    +\cdots ,
     \label{eq:n4_one_face_cluster}
\end{equation}
where $\umin=T/4-\log 6$ and, for the face where $\alpha_1$ collapses,
\begin{equation}
  \Gamma_2(u)
  =
  \{(w_1,w_2):\; w_1\le u,\; w_2\le u,\; w_1+w_2\ge -2u\}.
\end{equation}
The area of this triangle is $8u^2$. Therefore
\begin{equation}
  \mathcal I_{2\to4}^{\text{asy}}
  =
  32e^{-4S_I/\hbar}\int_0^{\umin}du\,u^2 e^{\nfourScale e^u}
  +\cdots
  =
  32e^{-4S_I/\hbar}\int_{\nfourScale}^{\ximin}
  \frac{d\xi}{\xi}\,e^\xi\log^2(\xi/\nfourScale)
  +\cdots ,
\end{equation}
with $\ximin=\nfourScale e^{\umin}=2S_I/\hbar$. Expanding around the upper endpoint as in the $n=3$ case and accounting for the CC-fix factor of $1/4$ gives the relevant asymptotic expansion as 
\begin{equation}
  \mathcal I_{2\to4}^{\text{asy}}
=
  8\,\LF^2\,e^{-2S_I/\hbar}
  \sum_{n=0}^{\infty}n!\left(\frac{\hbar}{2S_I}\right)^{n+1} + \cdots\,.
  \label{I2asy}
\end{equation}
This is the piece whose lateral Borel ambiguity cancels the leading $\LF^2$ piece of $\im Z_4^\pm$. The omitted terms are of order $\mathcal{O}(\LF)$ and are sensitive to how the boundary of the simplex is handled.

Combining Eq.~\eqref{I2asy} with the leading $n=2$ sector, weighted by the determinant ratio, including the parity factor of $2$ from the two-well multiplicity (just as for $Z_4$ above), and removing the $e^{-2S_I/\hbar}$ factor in accordance with the trans-series definition $Z = Z_0 + \lambda^2 Z_2 + \lambda^4 Z_4 + \cdots$, gives
\begin{equation}
  Z_2
  =
  T Z_{\text{SHO}}
  \left(\frac{12S_I}{2\pi\hbar}\right)
  \left[
    1+\cdots
    +16\,\LF^2
    \frac{12S_I}{2\pi\hbar}
    \sum_{n=0}^{\infty}n!\left(\frac{\hbar}{2S_I}\right)^{n+1}
  \right].
\end{equation}
The factorial-tail coefficient is $16 = 8\times 2$, with the $2$ from the two-well multiplicity applied when forming the $Z_2$ trans-series coefficient. The lateral resummation of this tail gives
\begin{equation}
  \cS_\pm[Z_2]
  =
  T Z_{\text{SHO}}
  \left(\frac{12S_I}{2\pi\hbar}\right)
  \left[
    1+\cdots
    -16\,\LF^2
    \frac{12S_I}{2\pi\hbar}
    e^{-2S_I/\hbar}E_1\!\left(-\frac{2S_I}{\hbar}\pm i\varepsilon\right)
  \right],
\end{equation}
so that
\begin{equation}
  \im\,\cS_\pm[Z_2]
  =
  \pm\,T Z_{\text{SHO}}
  \left(\frac{12S_I}{2\pi\hbar}\right)^2
  e^{-2S_I/\hbar}\,16\pi \LF^2 \,.
  \label{eq:n4_Z2_imag}
\end{equation}
In the full trans-series, the order-$\lambda^4$ imaginary contribution is $\lambda^2\,\im\cS_\pm[Z_2] + \lambda^4\,\im Z_4 = e^{-2S_I/\hbar}\,\im\cS_\pm[Z_2] + e^{-4S_I/\hbar}\,\im Z_4$.  Both pieces carry the same $e^{-4S_I/\hbar}$ prefactor, and comparing with Eq.~\eqref{eq:z4n4thimble} we see that the leading $\LF^2$ ambiguity of $\lambda^2\,\cS_\pm[Z_2]$ cancels that of $\lambda^4 Z_4$.

This is the partition-function analog of the alien-calculus cancellation at $\cO(\lambda^4)$ derived in Eq.~\eqref{eq:ImSum_lambda4}:
\begin{equation}
  \im[E_4] + \tfrac{1}{2}\im[\Delta_2 E_2] + \tfrac{1}{2}\im[\Delta_4 E_P] = 0\,.
\end{equation}
The partition-function version reads
\begin{equation}
  \underbrace{\im[Z_4]}_{T\LF^2 + T + \cdots} \;+\; \tfrac{1}{2}\underbrace{\im[\Delta_2 Z_2]}_{T\LF^2 + T + \cdots} \;+\; \tfrac{1}{2}\underbrace{\im[\Delta_4 Z_0]}_{T + \cdots} \;=\; 0\,,
  \label{eq:n4_alien_cancel}
\end{equation}
where the underbraces show the leading $\LF$-degrees in falling powers. At the highest power $T\LF^2$, only $\im[Z_4]$ and $\im[\Delta_2 Z_2]$ contribute --- these are precisely what we have computed (the $\LF^2$ pieces of $\cI_4^\pm$ and the leading collapse-face asymptotic of $\cI_{2\to 4}$, respectively), and they cancel as discussed. The third piece $\im[\Delta_4 Z_0]$ first enters at the subleading $\LF$ level: it comes from the codimension-2 collapse face of the $n=4$ simplex, where two instanton--anti-instanton pairs collapse simultaneously, and we have not computed it. The leftover $\pm 28\pi^3/3$ in $\im\cI_4^\pm$ (Eq.~\eqref{eq:n4_exact_imag_largeT}) is the $\LF^0$ piece of $\im[Z_4]$ that awaits cancellation against $\im[\Delta_4 Z_0]$ together with subleading-$\LF^0$ pieces of $\im[\Delta_2 Z_2]$.

Note that the full imaginary part is not canceled with what we have computed so far. The remaining part of order $\LF^0$ must be canceled by a combination of the subleading terms in the $n=2$ sector and the $n=0$ contribution.  This cancellation is guaranteed by the Picard--Lefschetz decomposition of $\Seff$, and can in principle be verified. That is, the $\LF^0$ pieces computed from $\Seff$ must close the cancellation among $\Seff$-based quantities ($\cI_4^\pm$, $\cI_{2\to4}$, $\cI_{0\to2}$) by construction, since the integral over the tetrahedron is real. The real issue is that since $\Seff$ is not approximating the full action well in the boundary-sensitive regions generating these $\LF^0$ pieces, the resulting trans-series terms should not be expected to agree with terms in the actual trans-series of $Z$. The leading-$\LF$ cancellation is the part that is robust to the $\Seff$ approximation. In other words, closing the imaginary cancellation to order $\LF^0$ would just be testing $\Seff$ against itself and would not tell us anything new about the double well.

\subsection{General \texorpdfstring{$n$}{n}}
\label{sec:gen_n_thimble}

The structure exposed by the $n=2$, $n=3$, and $n=4$ analyses generalizes cleanly to all $n$. At any instanton number $n$, the effective action from Eq.~\eqref{Seffalpha} reads
\begin{equation}
  \Seff = n S_I - 12 S_I \sum_{p=1}^{n} e^{-\alpha_p},
  \qquad \sum_{p=1}^{n} \alpha_p = T \,,
\end{equation}
with a real saddle at $\alpha_p = T/n$ for every $p$. We can pick $n-1$ local coordinates $u_1,\ldots,u_{n-1}$ that resolve the $n-1$ adjacent-pair collapse channels, with a constraint that determines the last separation:
\begin{equation}
  \alpha_p = \frac{T}{n} - u_p \quad (p=1,\ldots,n-1),
  \qquad
  \alpha_n = \frac{T}{n} + \sum_{p=1}^{n-1} u_p \,.
\end{equation}
Each $u_p$ collapses an adjacent pair when $u_p \to T/n$, while $\alpha_n \to 0$ corresponds to $\sum_p u_p \to -T/n$ and is not an independent direction. In these coordinates the action is
\begin{equation}
  \Seff(u) = n S_I - 12 S_I e^{-T/n} \Bigl[\sum_{p=1}^{n-1} e^{u_p} + e^{-\sum_p u_p}\Bigr]\,.
\end{equation}
It is convenient to subtract the saddle action $nS_I$ from $\Seff$, defining the deviation
\begin{equation}
  \Seff^{(n)}(u) \equiv \Seff(u) - nS_I = -12 S_I e^{-T/n}\Bigl[\sum_{p=1}^{n-1} e^{u_p} + e^{-\sum_p u_p}\Bigr]\,,
  \label{eq:Seffn_def}
\end{equation}
so that $\Seff^{(n)}=0$ at the saddle and the $e^{-nS_I/\hbar}$ factor does not need to be carried through every formula in the QZM analysis.
The real integration domain $\Gamma_{\mathbb R}^{(n)} = \{u \in \mathbb R^{n-1}:\ \alpha_p \ge 0 \ \forall p\}$ is the $(n-1)$-simplex, generalizing the triangle of the $n=3$ case (Fig.~\ref{fig:n3_action_surface}). Each of its $n$ codimension-one faces $\alpha_p \to 0$ is an adjacent-pair collapse channel down to the $n-2$ saddle in the same parity sector, on which the effective action $\Seff$ is no longer trustworthy and matching to the full action is required.

\subsubsection{General-$n$ thimble integral}

The integral over $\cJ_n^\pm$ is
\begin{equation}
  \cI^\pm_n(T)
  = \int_{\cJ_n^\pm} d^{n-1}u\,e^{-\Seff^{(n)}/\hbar}
  = \int_{\cJ_n^\pm} d^{n-1}u\,
  \exp\!\left[
    \nnScale \Bigl(\sum_{p=1}^{n-1} e^{u_p} + e^{-\sum_p u_p}\Bigr)
  \right],
  \label{eq:gen_thimble_def}
\end{equation}
with
\begin{equation}
  \nnScale \equiv \frac{12 S_I}{\hbar} e^{-T/n} \,.
\end{equation}
and $\cJ_n^\pm$ the middle-dimensional Lefschetz thimble through the saddle at $u_p=0$. To recover the thimble integral it is easier to first consider the real-cycle integral
\begin{equation}
  \cI_n^{\mathbb R}(\mu)
  = \int_{\mathbb R^{n-1}} d^{n-1}\xi\,
  \exp\!\left[-\mu \sum_{i=1}^{n} e^{a_i\cdot \xi}\right],
  \qquad \re \mu > 0 \,, \label{eq:real_cycle_int_n}
\end{equation}
where the $n$ exponent vectors $a_p = \mathbf{e}_p$ for $p=1,\ldots,n-1$ and $a_n = -\sum_{p=1}^{n-1}\mathbf{e}_p$ obey the linear relation $\sum_i a_i = 0$. The collapse to a single-variable Meijer $G$-function comes from a Mellin--Barnes computation. First, we use the Cahen–Mellin integral representation
\begin{equation}
    e^{-y} = \frac{1}{2\pi i}\int_C ds\,\Gamma(s) y^{-s} \,,
\end{equation}
where $C$ is the vertical line in the complex plane passing through $c>0$, to rewrite the integrand of Eq.~\eqref{eq:real_cycle_int_n} as 
\begin{equation}
  e^{-\mu \sum_i e^{a_i\cdot\xi}}
  = \frac{1}{(2\pi i)^n} \int_{C^n} \prod_{i=1}^n ds_i\,\Gamma(s_i)\,\mu^{-\sum_{i=1}^n s_i}\,
  \exp\!\left[-\left(\sum_{i=1}^n s_i a_i\right)\!\cdot\xi\right]\,.
\end{equation}
Choosing each $s_i$ on the same vertical contour $\re s_i = c$ as $s_i= c+i t_i$, the linear relation $\sum_i a_i = 0$ removes the real part of $\sum_i s_i a_i$ and the $\xi$ integral collapses to a Dirac delta:
\begin{equation}
  \int_{\mathbb R^{n-1}} d^{n-1}\xi\,e^{-i(\sum_{i=1}^n t_i a_i)\cdot\xi}
  = (2\pi)^{n-1}\delta^{(n-1)}\!\left(\sum_{i=1}^n t_i a_i\right) \,.
\end{equation}
The constraint $\sum_{i=1}^n t_i a_i = 0$ in $\mathbb R^{n-1}$ pins down $t_1 = t_2 = \cdots = t_n \equiv t$, so that after taking the integrals over $s_1, \cdots, s_{n-1}$, we are left with a single contour integral
\begin{equation}
  \cI_n^{\mathbb R}(\mu)
  = \frac{1}{2\pi i}\int_C ds\,\Gamma(s)^n \mu^{-ns}
  = \cG_n(\mu^n) \,.
  \label{eq:gen_real_to_G}
\end{equation}
This is exactly the Mellin--Barnes representation of the all-zero-parameter Meijer $G$-function in Eq.~\eqref{eq:cG_def}, appearing in Eq.~\eqref{eq:n2_exact_thimble_G}, Eq.~\eqref{eq:A2_Macdonald} and Eq.~\eqref{eq:n4_thimble_def} for $n=2,3,4$ respectively. The physical thimble integral~\eqref{eq:gen_thimble_def} has the opposite sign in the exponent and is obtained by analytic continuation $\mu \to \nnScale e^{\pm i\pi}$. Continuous tracking of the rotation gives $\mu^n \to \nnScale^n e^{\pm n\pi i}$ and the thimble integral reads
\begin{equation}
  \cI_n^{\pm}(T)
  = \cI_n^{\mathbb R}\!\bigl(\nnScale e^{\pm i\pi}\bigr)
  = \cG_n\!\bigl(\nnScale^n e^{\pm n\pi i}\bigr) \,.
  \label{eq:gen_thimble_G}
\end{equation}
which is the generalization of the cases for $n=2,3,4$.

To evaluate~\eqref{eq:gen_thimble_G} at large $T$, we close the contour in the Mellin--Barnes integrand
\begin{equation}
  \cI_n^{\pm}(T)
  = \frac{1}{2\pi i}\int_C ds\,\Gamma(s)^n\,\nnScale^{-ns}\,e^{\mp i n\pi s}
\end{equation}
to the left.  The poles of $\Gamma(s)^n$ are at $s=0,-1,-2,\ldots$\,; the residue at $s=-k$ carries a factor $\nnScale^{nk}\propto e^{-kT}$ and is exponentially suppressed at large $T$ relative to the $s=0$ residue.  Keeping only the leading $s=0$ residue,
\begin{equation}
  \cI_n^{\pm}(T)
  = \mathop{\rm Res}_{s=0}\!\left[\Gamma(s)^n\,\nnScale^{-ns}\,e^{\mp i n\pi s}\right]
  + \cO(e^{-T})\,,
\end{equation}
where the factors of $i$ from $1/(2\pi i)$ and from the closed contour have cancelled.  To extract the order-$n$ residue, we write $\Gamma(s)^n = \Gamma(1+s)^n/s^n$ and Taylor-expand using $\log\Gamma(1+s) = -\gamma_E s + \sum_{k\ge 2}(-s)^k\zeta(k)/k$:
\begin{equation}
  \label{Pnexpand}
  \Gamma(1+s)^n\,\nnScale^{-ns}
  = \exp\!\left[-n\Ln\,s + \sum_{k=2}^{\infty}\frac{n(-1)^k\zeta(k)}{k}\,s^k\right]
  = \sum_{m\ge 0} P_m^{(n)}(n\Ln)\,s^m\,,
\end{equation}
where
\begin{equation}
  \Ln \equiv \log\nnScale + \gamma_E
  =  - \frac{T}{n} + \log\!\left(\frac{12 S_I}{\hbar}\right) + \gamma_E\,,
\end{equation}
The polynomials $P_m^{(n)}(u)$ are defined by Eq.~\eqref{Pnexpand} and can be written explicitly as a sum over integer partitions of $m$:
\begin{equation}
  P_m^{(n)}(u)
  =
  \sum_{
  q_1+2q_2+\cdots+m q_m=m}
  \frac{(-u)^{q_1}}{q_1!}
  \prod_{r=2}^{m}\frac{1}{q_r!}
  \left[\frac{n(-1)^r\zeta(r)}{r}\right]^{q_r}.
  \label{eq:Pmn_partition}
\end{equation}
Each $P_m^{(n)}(u)$ is a polynomial of degree $m$ in $u$, with $n$-dependent coefficients. For example,
\begin{align}
  P_0^{(n)}(u) &= 1, & P_1^{(n)}(u) &= -u, \\
  P_2^{(n)}(u) &= \frac12 u^2 + \frac{n\pi^2}{12}, & P_3^{(n)}(u) &= -\frac16 u^3 - \frac{n\pi^2}{12}u - \frac{n\zeta(3)}{3} \,.
\end{align}
Combining with $e^{\mp i n\pi s} = \sum_j (\mp i n\pi)^j s^j/j!$, the residue at $s=0$ is the coefficient of $s^{n-1}$ in the product, which is then picked out by $m+j = n-1$. So,
\begin{equation}
\boxed{
 \cI_n^{\pm}(T)
  =
  \cG_n\!\left[ \nnScale^n e^{\pm n\pi i}\right]=
  \sum_{j=0}^{n-1}
  P_{n-1-j}^{(n)}(n\Ln)\,
  \frac{(\mp i n\pi)^j}{j!}
  +\cO(e^{-T})\,.
}
  \label{eq:gen_smallz}
\end{equation}
This is the full $\cI_n^{\pm}$ at large $T$, with both real and imaginary parts. 

\subsubsection{Resurgent structure at general $n$}
The polynomials $P_m^{(n)}(n\Ln)$ are real, so the imaginary part comes from the odd-$j$ terms, where $(\mp i n\pi)^j = \mp i\,(-1)^{(j-1)/2}(n\pi)^j$:
\begin{equation}
  \im\,\cI_n^{\pm}(T)
  =
  \mp\!\!\sum_{\substack{j=1\\ j\,\text{odd}}}^{n-1}
  \frac{(n\pi)^j}{j!}\,(-1)^{(j-1)/2}\;
  P_{n-1-j}^{(n)}\!\bigl(n\Ln\bigr)
  \;+\;\cO\!\bigl(\nnScale^n \Ln^{n-2}\bigr)\,.
  \label{eq:gen_imag_largeT}
\end{equation}
This produces imaginary contributions at every degree $\Ln^{n_1} \pi^{n_2} \prod_{k\geq 3}\zeta^{n_k}(k)$ where $n_2 \geq 1$ is odd, $k\geq 3$ is odd and $n_1 + n_2 + \sum_{k \geq 3} k n_k =n-1$. Evaluating~\eqref{eq:gen_imag_largeT} explicitly for the first few $n$,
\begin{align}
  \im\,\cI_2^{\pm} &= \mp\,2\pi,\\
  \im\,\cI_3^{\pm} &= \pm\,9\pi\,\LT,\\
  \im\,\cI_4^{\pm} &= \mp\,32\pi\,\LF^2 \pm \frac{28\pi^3}{3},\\
  \im\,\cI_5^{\pm} &= \pm\,\frac{625\pi}{6}\,T_5^3 \mp \frac{375\pi^3}{4}\,T_5 \pm \frac{25\pi\zeta(3)}{3},
\end{align}
recovering the results of the previous sections. Notably, the leading large $T$ imaginary part is obtained by evaluating Eq.~\eqref{eq:gen_imag_largeT} at $j=1$. Making use of $P_m^{(n)}(n T_n) = (-n T_n)^m/m! + \mathcal{O}(T_n^{m-2})$ we have $\im\,\cI_n^{\pm} = \mp n \pi T^{n-2}/(n-2)! + \cdots$. Similarly, the leading large $T$ real part is obtained from Eq.~\eqref{eq:gen_smallz} by taking $m=n-1$ and $j=0$ to give $\re\,\cI_n^{\pm} =(-n T_n)^{n-1}/(n-1)! + \cdots = T^{n-1}/(n-1)! - n\bigl(\log(12S_I/\hbar)+\gamma_E\bigr)T^{n-2}/(n-2)! + \cdots$.

This leading imaginary part is canceled by the resummation of the $n-2$ asymptotic series.  In the $n$-instanton quasi-zero-mode space we calculate the real-cycle integral over the simplex $\Gamma_\Delta^{(n)}$ as
\begin{equation}
  \mathcal I_{\Delta}^{(n)}
  =
  \int_{\Gamma_\Delta^{(n)}} d^{n-1}u\,
  \exp\!\left[
    \kappa_n \left(\sum_{p=1}^{n-1}e^{u_p}
    +e^{-\sum_{p=1}^{n-1}u_p}\right)
  \right],
  \qquad
  \kappa_n=\frac{12S_I}{\hbar}e^{-T/n}\,.
\end{equation}
The leading large-order growth of the \((n-2)\)-instanton sector is controlled by the codimension-one faces of this simplex, where one single instanton--anti-instanton pair collapses.  Generically picking the face where $\alpha_1$ is the smallest separation and writing $u\equiv u_1$, $w_a\equiv u_{a+1}$, $a=1,\ldots,n-2$,  the sector condition $\alpha_1\leq\alpha_p$ gives
\begin{equation}
  \Gamma_{n-2}(u)
  =
  \left\{
    w_a\leq u,\quad
    \sum_{a=1}^{n-2}w_a\geq -2u
  \right\} \,,
\end{equation}
which has volume
\begin{equation}
    \operatorname{Vol}\Gamma_{n-2}(u) = \frac{(nu)^{n-2}}{(n-2)!} \,.
\end{equation}
Summing over the $n$ equivalent collapse faces and including the CC-fix factor $1/n$, the leading face finite part is therefore
\begin{equation}
  \mathcal I_{n-2\to n}^{\rm asy}
  =
  \frac{n^{n-2}}{(n-2)!}
  \int_0^{u_{\max}}du\,u^{n-2}e^{\kappa_ne^u}
  +\cdots ,
  \qquad
  u_{\max}=\frac{T}{n}-\log 6 \,.
\end{equation}
After changing variables to $\xi=\kappa_ne^u$ and retaining the part of the integral near the upper endpoint region to get the leading order in $\Ln$, we get the asymptotic series
\begin{equation}
  \mathcal I_{n-2\to n}^{\rm asy}
  =
  \frac{n^{n-2}}{(n-2)!} (-\Ln)^{n-2}\,e^{2 S_I/\hbar} \sum_{\ell=0}^{\infty}
  \ell!\left(\frac{\hbar}{2S_I}\right)^{\ell+1}
 +\cdots \,.
\end{equation}
This is the piece whose lateral Borel resummation cancels the leading $\Ln^{n-2}$ imaginary part of $\cI_n^{\pm}$ in Eq.~\eqref{eq:gen_imag_largeT} and demonstrates that for any instanton sector $n$, the asymptotic series of the $n-2$ instanton sector is required to have a real and unambiguous partition function.

\subsubsection{Partition function at general $n$}
Combining $\cI_n^{\pm}$ with the collective-coordinate measure, the transverse determinant from Eq.~\eqref{eq:detperp}, the two-well multiplicity factor of $2$, and the CC-fix factor $1/n$, the $n$-instanton partition function at NLO at large $T$ becomes
\begin{equation}
  Z_n^{\pm}
  = \frac{2}{n}\,Z_{\text{SHO}}\,T\,
  \left(\frac{12 S_I}{2\pi\hbar}\right)^{n/2}\!\!
  \cI_n^{\pm}, 
\end{equation}
where the prefactor $2/n$ combines the two-well multiplicity and CC-fix factors. The lateral imaginary part of this $n$-instanton trans-series is~\eqref{eq:gen_imag_largeT} dressed by the same prefactors.

Equivalently, the leading contribution from the $n$-instanton thimble and the leading asymptotic tail generated by the neighboring $(n+2)$-instanton saddle combine into the schematic large-$T$ prediction
\begin{equation}
\begin{aligned}
  Z_n^\pm
  &=
  Z_{\text{SHO}}\,T
  \left(\frac{6S_I}{\pi\hbar}\right)^{n/2}
  \left[
    \frac{2\,T^{n-1}}{n!}
    -\frac{2\,T^{n-2}}{(n-2)!}\Big(\log\!\frac{12S_I}{\hbar}+\gamma_E\pm\pi i\Big)
    +\cdots
  \right.\\
  &\hspace{7.2cm}\left.
    +\frac{12S_I}{\pi\hbar}\,\frac{T^n}{n!}
    \sum_{\ell=0}^{\infty}\ell!\left(\frac{\hbar}{2S_I}\right)^{\ell+1} + \cdots
  \right].
\end{aligned}
\label{eq:Zn_box}
\end{equation}
Summing over even $n$ and using $Z_{\text{SHO}}=1/(2\sinh(T/2))$ leads to
\begin{equation}
  Z^{\pm}
  = \frac{\cosh(e^{-S_I/\hbar} K T)}{\sinh(T/2)}\left[
    1 - T K^2e^{-2S_I/\hbar} \Big(\log\frac{12S_I}{\hbar}+\gamma_E \pm i\pi\Big)
    + T K^2 \sum_{\ell=0}^{\infty}\ell!\left(\frac{\hbar}{2S_I}\right)^{\ell+1}
    + \cdots
  \right]
  \label{eq:Z_resummed_cosh}
\end{equation}
where $K=\sqrt{6S_I/(\pi\hbar)}$. The twisted partition function sums over odd $n$ leading to an identical result with the replacement $\cosh\to\sinh$.  The single $T$ inside the bracket is from the center of the instanton--anti-instanton pair whose collapse leads to the subleading factors shown. The $\cosh/\sinh$ prefactor is the result from the dilute instanton gas. The $\log\hbar$, imaginary part, and asymptotic series are all beyond the DIG.

The expression in Eq.~\eqref{eq:Z_resummed_cosh} when expanded gives the trans-series for the partition function. Note however that not every term in this expansion comes from the saddle-point approximation: the imaginary part and $\ln \hbar$ factors come from the full integral along the thimble in the quasi-zero-mode subspace. This was necessary because the quasi-zero modes have exponentially small (in $T$) eigenvalues so the saddle-point approximation in those directions is ill-defined as $T\to\infty$. Using Eq.~\eqref{eq:Ein_trans}, the imaginary parts cancel and the resummed result is manifestly real:
\begin{equation}
  \cS_\pm[Z_\pm]
  = \frac{\cosh(e^{-S_I/\hbar} K T)}{\sinh(T/2)}\left[
    1 - T K^2e^{-2S_I/\hbar} \Big(\Ein(-\tfrac{2S_I}{\hbar})+ \log 6\Big) + \cdots
  \right]\,,
  \label{eq:Z_resummed_real}
\end{equation}
where the constant $\log 6$ arises from combining $\log(12 S_I/\hbar)+\gamma_E$ in Eq.~\eqref{eq:Z_resummed_cosh} with the $\gamma_E + \log(2 S_I/\hbar)$ generated when the factorial tail is rewritten via Eq.~\eqref{eq:Ein_trans}. The cancellation is a rigorous large-$T$ path-integral result, as illustrated for $\cO(\lambda^4)$ by Eq.~\eqref{eq:n4_alien_cancel}. Subleading-$T$ cancellations involve higher alien derivatives ($\Delta_4, \Delta_6, \ldots$) acting on lower sectors and correspondingly higher-codimension collapse faces, neither of which we have computed; they are absorbed into the trailing $\cdots$.

Finally, we collect together everything we have computed so far. Specializing to real instantons ($k'=0$, $n=2k$) and inserting into the master factorized expression for the partition function in Eq.~\eqref{eq:Z_factored} the formula for $\|\dot x_{k,0}\|^2 = n\,S_N^0(\varepsilon)$ from Eq.~\eqref{eq:xdot2_action}, the saddle action $S_{k,0} = n\,S_N^0(\varepsilon)-\varepsilon T$ from Eq.~\eqref{eq:actionexactsolsb}, the combined collective-coordinate Jacobian and transverse determinant factor $(12 S_I/(2\pi\hbar))^{n/2}/(2\sinh(T/2))$ at large $T$ (with two-well multiplicity and CC-fix factor $2/n$, see Eq.~\eqref{eq:detperp}), and $\cI_n^\pm$ in $\cG_n$ form from Eq.~\eqref{eq:gen_thimble_G}, we get
\begin{equation}
  \label{Zpluggedin}
\boxed{
  Z_n^\pm
  = \frac{2}{n}\sqrt{\frac{S_N^0(\varepsilon)}{S_I}}\,
    \frac{T}{2\sinh(T/2)}\,
    \left(\frac{12 S_I}{2\pi\hbar}\right)^{n/2}\,
    e^{-\bigl(n\,S_N^0(\varepsilon)-\varepsilon T\bigr)/\hbar}\,
    e^{T\DV}\,e^{n\DL}\,
    \cG_n\left[\left(\frac{12S_I}{\hbar}\right)^{n}e^{-T\pm n\pi i}\right]
}\,.
\end{equation}
The saddle action is exact at finite $T$; the prefactor $\sqrt{S_N^0(\varepsilon)/S_I}$ is the finite-$T$ correction to the zero-mode Jacobian and equals~$1$ at large $T$ where $S_N^0\to S_I$; the combined Jacobian and transverse-determinant factor $(12 S_I/(2\pi\hbar))^{n/2}/(2\sinh(T/2))$ is the standard large-$T$ form, as discussed below Eq.~\eqref{eq:detperp}. The expression is exact at finite $T$ for the sectors with no QZMs ($n=0,1$, including via the $n=1$ result Eq.~\eqref{eq:Ztilde1_exact}). For $n\ge 2$ where the QZMs are present it drops terms exponentially suppressed in $T$. Although the QZM thimble integral factor is written in its Meijer $G$-function form, this should be understood only as its large-$T$ expansion which reduces $\cG_n[\cdots]$ to the complex polynomial of degree $n-1$ in Eq.~\eqref{eq:gen_smallz}.  The perturbative vacuum-bubble $\DV$ and instanton-loop $\DL$ factors are in Eqs.~\eqref{DVthreeloop} and~\eqref{eq:DLn_summary}. We next proceed to extract the energy spectrum and trans-series from this partition function data.

% %%%%%%%%%%%%%%%%%%% ENERGY SPECTRUM %%%%%%%%%%%%%%%%%%%

\subsection{Energy spectrum from the path integral}
\label{sec:pi_spectrum}
In the preceding sections we computed the partition function trans-series: the perturbative sector  $\DV$ and $Z_0$ at finite $T$ from vacuum diagrams around a perturbative saddle (Section~\ref{sec:Feynman}), and the non-perturbative sectors $n\DL$ and $Z_n$ at large $T$ from instanton-background vacuum diagrams (Section~\ref{sec:higher_loop_instanton}) and thimble integrals over quasi-zero modes (Sections~\ref{sec:n2thimble}--\ref{sec:n4}). We now extract the energy spectrum from this data.

Our starting point is the partition function which encodes the energy levels through
\begin{equation}
  Z(T) = \sum_{E}  e^{-E T/\hbar}  \,.\label{eq:Z_from_Es}
\end{equation}
To invert this and extract the energies from the partition function, we can use the \emph{resolvent} $G(E)$ defined in Eq.~\eqref{eq:resolvent_def} and the \emph{spectral determinant} $D(E) = \det(H - E)$. The energies are the poles of $G(E)$, equivalently the zeros of $D(E)$. This is the path-integral formulation of quantization: in place of the Schr\"odinger equation boundary condition, one has $D(E) = 0$. For the SHO, the resolvent is
\begin{equation}
  G_\text{SHO}(E) = \frac{1}{\hbar}\int_0^\infty dT\,e^{\frac{ET}{\hbar}}\,\frac{1}{2\sinh\frac{T}{2}}
  = -\frac{1}{\hbar}\psi\!\left(\frac{1}{2} - \frac{E}{\hbar}\right) + \text{const},\,
\end{equation}
where $\psi = \Gamma'/\Gamma$ is the digamma function. Up to regular terms, the SHO spectral determinant is $D_\text{SHO}(E) = 1/\Gamma\!\left(\frac{1}{2} - \frac{E}{\hbar}\right)$, whose zeros are at $E = \hbar \kappa$ with $\kappa= N+\frac{1}{2}$, reproducing the harmonic spectrum.

\smallskip

  Making use of the resolvent and the form of the partition function, in this section we will:                                  
  \begin{enumerate}                                                             
    \item Reproduce the Bender--Wu corrections to the ground-state energy by taking a large-$T$ expansion and looking at the $\DV$ contributions to the perturbative partition function $Z_P$.
    \item Extend the analysis by reproducing the perturbative corrections to the excited-state energies via the $\DV$ contributions to the finite $T$ formulation of $Z_P$.                                      
    \item Compute the leading (in $\hbar$) non-perturbative ground-state splitting $E^{(1)} - E^{(0)} \propto e^{-S_I/\hbar}$ from $\widetilde Z_1$ and extract the common shift $\propto e^{-2S_I/\hbar}$ from $Z_2$  in the large $T$ limit.
    \item Extend the analysis to compute the leading non-perturbative splitting for all excited-states $E_+^{(N)} - E_-^{(N)}\propto e^{-S_I/\hbar}$ via the finite $T$ formulation of the exact 1-loop $\widetilde Z_1$. 
    \item Incorporate higher-loop corrections (in $\hbar$) to the ground-state splitting $E^{(1)} - E^{(0)} \propto e^{-S_I/\hbar}$ using the single instanton loop correction $\DL$ in $\widetilde Z_1$ at large $T$.
    \item Sketch the structural form of higher-loop corrections for excited-state splittings which requires the exact finite $T$ formulation.       
  \end{enumerate}

\subsubsection{Perturbative corrections to the ground state energy ($\DV$ large $T$) \label{sec:ground_state_energy}}
At large $T$, the factorized formula for the partition function in Eq.~\eqref{eq:Z_factored} reduces to
\begin{equation}
  Z = 2\,Z_\text{SHO}\,e^{\DV(\hbar) T}\!\left[1+\sum_{n=2,4,\ldots} \frac{1}{n}\, T\, K^n \lambda^n\, e^{n\,\DL(\hbar)}\,\cI_n^\pm(T,\hbar)\right]\,,
  \label{eq:Z_structure_recap}
\end{equation}
where $K=\sqrt{6S_I/(\pi\hbar)}$, $\lambda=e^{-S_I/\hbar}$, $\DL(\hbar)$ is the single-instanton loop correction, and $\cI_n^\pm(T,\hbar)$ is the thimble integral over the quasi-zero modes.  At leading order at large $T$, each $\cI_n^\pm(T,\hbar)$ is a \emph{polynomial} in $T$, as can be seen in Eq.~\eqref{eq:gen_smallz}, and both $\DV(\hbar)$ and $\DL(\hbar)$ are $T$-independent. 

Taking a Laplace transform, each power of $T$ converts to a derivative,
\begin{equation}
  \frac{1}{\hbar}\int_0^\infty dT\,e^{\frac{ET}{\hbar}}\,T^j\,e^{\DV T}\,Z_\text{SHO}(T)
  = (\hbar\partial_E)^j\,G_\text{SHO}(E + \hbar\DV) \,.
  \label{eq:Tj_resolvent}
\end{equation}
This leads to a large-$T$ formula for the spectral determinant:
\begin{equation}
  \ln D(E) =  2\left[1 + \sum_{n=2,4,\ldots} \frac{1}{n}\,K^n\,\lambda^n\,(\hbar\partial_E)\, e^{n\,\DL(\hbar)}\,\cI_n^\pm(\hbar\partial_E,\hbar)\right]
 \ln\!\left[\frac{1}{2} - \frac{E}{\hbar} - \DV(\hbar)\right]
   \,.
  \label{eq:lnD_master}
\end{equation}
Since the large-$T$ form allows us to compute only the ground state energy, we replaced
 $D_\text{SHO}(E) = 1/\Gamma\!\left(\frac{1}{2} - \frac{E}{\hbar}\right) \to \frac{1}{2} - \frac{E}{\hbar}$ with only the single relevant zero.

Eq.~\eqref{eq:lnD_master} implies that the vacuum bubbles $\DV$ shift the argument of the logarithm. This translates to perturbative corrections to the ground state energy
\begin{eqnarray}
  E^{(0)} = \frac{1}{2}\hbar - \hbar \DV(\hbar) \,.
\end{eqnarray}
 The 3-loop result in Eq.~\eqref{DV3loop} at large $T$ gives $\DV = \frac{1}{4}\hbar + \frac{9}{32}\hbar^2 + \cO(\hbar^3)$ in agreement with Bender--Wu. At higher orders, the Bender--Wu recursion as reviewed in Appendix~\ref{appendix:benderWu} gives the ground state energy to as high order in $\hbar$ as desired, and therefore gives $\DV$ to high order. Computing $\DV$ from Bender--Wu is a much easier calculation than the direct Feynman-diagram approach, but the direct 2-loop and 3-loop calculations that we have worked out explicitly provide nontrivial checks.

\subsubsection{Perturbative corrections to the excited state energies ($\DV$ finite $T$)}
\label{section:finiteT_spectrum}
The partition function in Eq.~\eqref{eq:Z_from_Es} encodes all the energy levels and not just the ground state. Therefore, information from the excited states must be in the path integral computation. To extract it, we just need to go beyond the leading large-$T$ limit. We have already done this at the 1-loop level in the zero-instanton sector where $Z_P = Z_\text{SHO}= 1/(2\sinh\frac{T}{2})$ leads to $G_\text{SHO}(E) = - \hbar^{-1}\psi\!\left(\frac{1}{2} - \frac{E}{\hbar}\right)$, giving the full spectrum. If we had taken the large-$T$ limit, $Z_\text{SHO}(T) \approx e^{-T/2}$, then we would have found $G_\text{SHO}(E) \approx 1/(E - \frac{1}{2}\hbar)$ giving only the ground state energy.

Keeping the order-$\hbar$ part of $\DV$ from Eq.~\eqref{DVthreeloop} and expanding gives for the zero-instanton partition function:
\begin{equation}
  Z_P = \frac{1}{2\sinh\frac{T}{2}}e^{T \DV }
  = \frac{1}{2\sinh\frac{T}{2}}\left(1+ \hbar\frac{T}{4}\frac{\cosh T + 2}{\cosh T - 1}  + \cdots\right) 
  \label{eq:ZDWZSHO_recap}
\end{equation}
This order-$\hbar$ term can be written suggestively as
\begin{equation}
  \frac{\cosh T + 2}{\cosh T - 1} = 1 + \frac{3}{2\sinh^2\frac{T}{2}} = 1+ 6\sum_{m=1}^\infty m\, e^{-mT}
  \label{eq:coshexpand}
\end{equation} 
then
\begin{align}
  Z_P &= \frac{1}{2\sinh\frac{T}{2}}\left[1 + \frac{\hbar T}{4}\left(1 + 6\sum_{m=1}^\infty m\, e^{-mT}\right) + \cdots\right]\nonumber\\
  &= \sum_{N=0}^\infty e^{-(N+\frac{1}{2})T}\left[1 + \frac{\hbar T}{4}\left(1 + 6\sum_{m=1}^\infty m\, e^{-mT}\right) + \cdots\right] \,.
  \label{eq:Z0_expanded}
\end{align}
Now each factor $e^{-mT}$ appearing in the exponential correction shifts $N \to N+m$. Thus, collecting all contributions to the coefficient of $e^{-(N+\frac{1}{2})T}$:
\begin{equation}
  Z_P = \sum_{N=0}^\infty e^{-(N+\frac{1}{2})T}\left[1 + \frac{\hbar T}{4}\left(1 + 6\sum_{m=1}^{N} m\right) + \cO(\hbar^2)\right] \,,
  \label{eq:Z0_Ncoeff}
\end{equation}
where the upper limit $m \le N$ arises because the term $e^{-(N+\frac{1}{2})T}\cdot e^{-mT}$ contributes to the coefficient of $e^{-(N'+\frac{1}{2})T}$ with $N' = N+m$. Relabeling the final index $N'\to N$, the original index must be non-negative, which requires $m \le N$. Using $\sum_{m=1}^N m = \frac{N(N+1)}{2}$, the coefficient of $e^{-(N+\frac{1}{2})T}$ is
\begin{equation}
  1 + \frac{\hbar T}{4}\Big[1 + 3N(N+1)\Big] + \cO(\hbar^2) \,.
\end{equation}
We identify this with $e^{-E^{(N)}T/\hbar}$ expanded to first order in $\hbar$. Writing $E^{(N)} = \hbar(N+\frac{1}{2}) + \hbar^2 a_N + \cO(\hbar^3)$, we have $e^{-E^{(N)}T/\hbar} = e^{-(N+\frac{1}{2})T}(1 - \hbar\, a_N T + \cdots)$, so matching gives
\begin{align}
  E^{(N)} &= \hbar\!\left(N + \frac{1}{2}\right) - \hbar^2\!\left[\frac{1}{4} + \frac{3}{4}N(N+1)\right] + \cO(\hbar^3)\\
  &= \hbar \kappa + \hbar^2\!\left[-\frac{1}{16} - \frac{3}{4}\kappa^2\right] + \cO(\hbar^3) + \cdots
  \label{eq:EN_2loop} 
\end{align}
in exact agreement with the Bender--Wu result~\cite{BenderWu} and with the perturbative Exact WKB quantization condition in Section~\ref{sec:pertexp}.

The 3-loop expression in Eq.~\eqref{DVthreeloop} gives the next perturbative coefficient by the same extraction. Writing $\DV=\hbar\DV^{(1)}+\hbar^2\DV^{(2)}+\cO(\hbar^3)$, the partition function contains
\begin{equation}
  Z_P
  =
  \frac{1}{2\sinh\frac{T}{2}}
  \left[1+\hbar T\DV^{(1)}+\hbar^2\left(T\DV^{(2)}+\frac{T^2}{2}\bigl(\DV^{(1)}\bigr)^2\right)+\cO(\hbar^3)\right].
\end{equation}
Matching the coefficient of $e^{-(N+\frac{1}{2})T}$ to
$E^{(N)}=\hbar e_1(N)+\hbar^2 e_2(N)+\hbar^3 e_3(N)+\cdots$ gives
\begin{equation}
  e_3(N)
  =
  -\frac{1}{32}(2N+1)(17N^2+17N+9)
  =
  -\frac{17}{16}\kappa^3-\frac{19}{64}\kappa\,,
\end{equation}
again in agreement with Bender--Wu. The terms proportional to $\ln u$ in Eq.~\eqref{DVthreeloop} are essential in this check: after using $T=-\ln u$, they contribute to the $T^2$ part required by exponentiation of the 2-loop energy shift.

Several features of this calculation are worth emphasizing. First, retaining the full $T$-dependence of the vacuum diagrams is essential. If we had dropped the $e^{-T}$ terms in Eq.~\eqref{eq:2loop_exact}, keeping only $\frac{\hbar T}{4}$, we would obtain only the $N$-independent constant $-\frac{\hbar^2}{4}$. This is the shift $\DV = \frac{\hbar}{4}$ of the ground state, but the $N$-dependent pieces that distinguish excited states would be lost. The dilute instanton gas, which is only sensible at large $T$, is therefore only capable of describing the ground state. Computing the full perturbative spectrum requires the finite-$T$ corrections which are necessarily beyond the DIG. 

Second, the partition function is a single object that encodes all energy levels simultaneously. The structure $Z = \sum_N e^{-E^{(N)}T/\hbar}$ means that different powers of $e^{-T}$ talk to different energy levels. The $e^{-mT}$ corrections in the vacuum diagrams are not negligible; they are the mechanism by which the path integral knows about excited states. This is the same principle that will recur in the non-perturbative sectors: the partition function trans-series is structurally simpler than the energy trans-series, and all level-dependent complexity arises from the extraction step.

Third, this direct matching of $e^{-(N+\frac{1}{2})T}$ coefficients is the simplest way to go from $Z$ to the spectrum at this order. The resolvent or spectral determinant provides a more invariant packaging of the same information and is useful for comparison with exact WKB, but for the finite-$T$ checks below, the coefficient-extraction method is sufficient and more transparent.

\subsubsection{Non-perturbative corrections to the ground state energy ($S_N^0=S_I$, large $T$)}
\label{sec:NP_ground_state} 
The first non-perturbative effect, which we have already discussed in Section~\ref{sec:E0_E2_transseries}, is the splitting between even and odd parity states at the one-instanton level. The parity splitting is encoded in the twisted partition function
\begin{align}
  \widetilde{Z}(T) &= \Tr(\cP \,e^{-HT/\hbar})
  =  \sum_E \pm e^{-E T/\hbar}
  = e^{-E^{(0)} T/\hbar}- e^{-E^{(1)} T/\hbar} + \cdots \\ \label{eq:Z1twisted_expand}
  &\approx e^{-E^{(0)} T/\hbar} \frac{(E^{(1)}-E^{(0)})T}{\hbar}\, + \cdots \,.
\end{align} 
The leading contribution to $\widetilde{Z}$ comes from the one-instanton sector where we found in Eq.~\eqref{eq:Ztilde1_exact_largeT} that at large $T$
\begin{equation}
  \widetilde{Z}_1(T) = Z_\text{SHO}(T)\,
  \frac{2T}{\sqrt{\pi\hbar}}\, e^{-\frac{S_I}{\hbar}}\,
  \approx \, e^{-T/2}\,\frac{2T} {\sqrt{\pi\hbar}}\,e^{-\frac{S_I}{\hbar}} \,.
  \label{eq:Z1_largeT}
\end{equation}
This must be multiplied by 2 to include both the instanton and anti-instanton saddle point contributions. Matching to Eq.~\eqref{eq:Z1twisted_expand} leads immediately to
\begin{eqnarray}
  E^{(1)} - E^{(0)} = 4\sqrt{\frac{\hbar}{\pi}}\,e^{-\frac{S_I}{\hbar}} + \cdots
  \label{E1minusE0}
\end{eqnarray}
which is the well-known one-instanton energy splitting of the ground state doublet. This is consistent with both the WKB result and the path integral derivation of Section~\ref{sec:E0_E2_transseries}. We recall that from the path integral we were also able to extract the leading correction in $e^{-2S_I /\hbar}$ and that this derivation proceeded by using at large $T$ 
\begin{equation}
    Z_2 =  2\,e^{-\frac{T E_0}{\hbar}}
    \!\left[\frac{T^2}{2\hbar^2}\bigl(E_1\bigr)^2
    -\frac{T}{\hbar} E_2
    \right],
\end{equation}
and matching to 
\begin{equation}
  Z_2
  =
  e^{-\frac{T}{2}}\frac{4}{\pi\hbar}
  \left[
    T^2-2T\!\left(\gamma_E+\log\!\frac{12S_I}{\hbar}+\pi i\right)
  \right] \,.
\end{equation}
The exponent and $T^2$ factors agree with the leading perturbative energy and one-instanton splitting already found above, and the $T$ term fixes the common two-instanton correction to the lowest doublet. Equivalently, the two lowest energies have the lateral trans-series
\begin{align}
  E^{(0)}
  &=
  E_P^{(0)}(\hbar)
  -2\sqrt{\frac{\hbar}{\pi}}\,e^{-\frac{S_I}{\hbar}}
  +\frac{4}{\pi}\!\left[\gamma_E+\ln\frac{8}{\hbar}\pm i\pi\right]e^{-\frac{2S_I}{\hbar}}
  +\cdots, \\
  E^{(1)}
  &=
  E_P^{(0)}(\hbar)
  +2\sqrt{\frac{\hbar}{\pi}}\,e^{-\frac{S_I}{\hbar}}
  +\frac{4}{\pi}\!\left[\gamma_E+\ln\frac{8}{\hbar}\pm i\pi\right]e^{-\frac{2S_I}{\hbar}}
  +\cdots.
  \label{eq:lowest_doublet_PI}
\end{align}
The two-instanton term is common to both members of the doublet. Its imaginary part depends on the lateral Borel prescription and cancels the corresponding ambiguity in the resummation of the perturbative series $E_P^{(0)}(\hbar)$, leaving the real median shift proportional to $\gamma_E+\ln(8/\hbar)$.

The story can be pushed to three and four-instanton order. Writing the two lowest energies as $E_\pm = E_0 \pm \lambda E_1+\lambda^2E_2\pm\lambda^3E_3+\lambda^4E_4+\cdots$, one can use the twisted version of Eq.~\eqref{eq:Zk_from_Ek3} and the untwisted Eq.~\eqref{eq:Zk_from_Ek4}, matching those partition functions to the large $T$ results obtained in Eq.~\eqref{Z3I3} and Eq.~\eqref{eq:z4n4thimble} respectively.  For instance, one can check that the $T^3$ coefficient matching allows for the extraction of $(E_1)^3$, giving an expression consistent with Eq.~\eqref{E1minusE0}. The $T^2$ coefficient gives the same $E_2$ expression as Eq.~\eqref{eq:lowest_doublet_PI}, while the order $T$ matching gives $E_3$ and
\begin{eqnarray}
  E^{(1)} - E^{(0)} = 4\sqrt{\frac{\hbar}{\pi}}\,e^{-\frac{S_I}{\hbar}} + \sqrt{\frac{64}{9\pi^3 \hbar}} \,e^{-\frac{3S_I}{\hbar}} \left[9 \left( \gamma_E + \ln \frac{8}{\hbar}\pm i \pi \right)^2+\frac{\pi^2}{2} \right] +\cdots \,.
  \label{E1minusE0order3}
\end{eqnarray}
This expression matches with the Exact WKB result of Eq.~\eqref{eq:E3_summary}. Similarly, one can check that the expressions for $E_1, E_2, E_3$, extracted from the matching between Eq.~\eqref{eq:Zk_from_Ek4} and Eq.~\eqref{eq:z4n4thimble}, are all consistent with previous instanton results and Exact WKB. In that case the order $T$ coefficient allows for the extraction of $E_4$, with a result that once again perfectly reproduces the Exact WKB formula of Eq.~\eqref{eq:E4_summary}. Under the sector-matching prescription discussed in Section~\ref{sec:n4}, the leading large-$T$ ambiguities in $\cS[Z_0]$, $\cS[Z_2]$, and $Z_4$ at $\cO(\lambda^4)$ cancel. Leveraging the general $n$-instanton thimble expression of Eq.~\eqref{eq:gen_smallz}, accounting for the CC-fix factor, the transverse determinant and using the representation of Eq.~\eqref{Pnexpand}, we can isolate the linear term in $T$ in $Z_n$ (or $\widetilde Z_n$). This gives a closed form formula for the leading term in $E_n$ as 
\begin{equation} \label{eq:Enclosedform}
    E_n =  - \frac{(-1)^n}{n!} \sqrt{\frac{4^n}{\pi^n \hbar^{n-2}}}  \frac{d^{n-1}}{ds^{n-1}} \left.\left[ \Gamma(1+s)^n \left(\frac{8}{\hbar}\right)^{-ns} e^{\mp i n \pi s} \right] \right|_{s=0}  + \mathcal{O}(\hbar^{2-n/2})\,.
\end{equation}
We emphasize that at every order the large-$T$ limit of the partition function gives only the lowest-doublet corrections, solely applicable for the ground state. To obtain the excited-state spectrum, we need the finite-$T$ corrections, which we turn to next.

\subsubsection{Non-perturbative corrections to the excited state energies ($S_N^0$ finite $T$)}
\label{sec:NP_excited_state}
Next, we proceed to study the non-perturbative corrections to the excited states. For this computation we need the full $T$ dependence. The exact twisted partition function has the spectral representation
\begin{equation}
  \widetilde{Z} =\sum_N\left[  e^{-E^{(N)}_+ T/\hbar} - e^{-E^{(N)}_- T/\hbar}\right]
  \,.
  \label{eq:Ztwisted_spectral_exact}
\end{equation}
Writing the one-instanton energy splitting as $E^{(N)}_\pm = E_0^{(N)} \pm \lambda\,\Delta_N + \cdots$ with $\lambda = e^{-S_I/\hbar}$, this becomes
\begin{equation}
  \widetilde{Z}
  = \sum_N -2\,e^{-E_0^{(N)}T/\hbar}\,\sinh\!\!\left(\frac{\lambda\,\Delta_N\, T}{\hbar}\right) + \cdots.
  \label{eq:ZDelta}
\end{equation}
With this convention $E_+^{(N)}$ denotes the even-parity state in the twisted trace, so $\Delta_N$ is negative; the positive splitting is $E_-^{(N)}-E_+^{(N)}=-2\lambda\Delta_N$. Using the harmonic leading order result $E_0^{(N)} = (N+\tfrac{1}{2})\hbar + \cdots$ and expanding to first order in $\Delta_N$ we then get
\begin{equation}
  \left.\widetilde{Z}\right|_{\lambda}
  = -e^{-\frac{S_I}{\hbar}} \frac{2T}{\hbar} \sum_N \Delta_N\,
	   e^{-(N+\frac{1}{2})T} \,.
	  \label{eq:Ztilde1_spectral_excited}
\end{equation} 
We then pull out the $\exp(-T/2)$ factor and change to $u=e^{-T}$ as the expansion variable giving
\begin{equation}
  \left.\widetilde{Z}\right|_{\lambda}
  = -e^{-\frac{S_I}{\hbar}} \,\frac{2T}{\hbar}\,e^{-T/2}\,g(u)\,,
  \quad
  g(u)=\sum_{N=0}^\infty \Delta_N u^N \,.
  \label{Ztog}
\end{equation}
We now compute the same order-$\lambda$ object from the one-instanton saddle. Including both saddle orientations (instanton and anti-instanton, related by parity), the exact 1-loop result in Eq.~\eqref{eq:Ztilde1_exact} gives
\begin{align}
  \widetilde{Z}_\text{NLO}
  &= 2\,\widetilde{Z}_1(T)
  = 2T\sqrt{\frac{-2\varepsilon(1+8\varepsilon)}{6\pi\hbar\,S_N^0(\varepsilon)}}
	e^{-\frac{1}{\hbar}(S_N^0(\varepsilon)-\varepsilon T)}
	    \label{eq:Ztilde1_full}
	  \\
	  & = \frac{4T\,e^{-T/2}}{\sqrt{\pi\hbar}}\,
	  \bigl[1+(12T-46)\,e^{-T}+\cdots\bigr]\;
	  e^{-\frac{1}{\hbar}
\left(S_I - 8\,e^{-T}+(136-48T)\,e^{-2T}+\cdots\right)} \,.
	  \label{eq:Ztilde1_expanded}
\end{align}
Matching $\widetilde{Z}_\text{NLO}$ to $\left.\widetilde{Z}\right|_{\lambda}$ lets us read off the one-instanton splittings $\Delta_N$ from the coefficient of $u^N$. At NLO,
\begin{align} 
g(u) &= - \hbar  e^{T/2}
\sqrt{\frac{-2\varepsilon(1+8\varepsilon)}{6\pi\hbar\,S_N^0(\varepsilon)}}	e^{-\frac{1}{\hbar}(S_N^0(\varepsilon)-S_I -\varepsilon T)}  \\
&=-2\sqrt{\frac{\hbar}{\pi}} \big[1-46 u - 12 u \ln u + \cdots\big] e^{\frac{1}{\hbar}\left(8u- 136 u^2 - 48 u^2 \ln u + \cdots\right)} \,.
 \label{eq:gu_expanded} 
\end{align}
We want to expand this as a series in $u$. As $\hbar \to 0$ the exponent dominates and as we expand for small $u$ the leading term in the exponent dominates. Including this term only to start, we get
\begin{equation}
g(u)\sim -2\sqrt{\frac{\hbar}{\pi}}  e^{8u/\hbar} =
-2\sqrt{\frac{\hbar}{\pi}}
\sum_{N=0}^\infty \frac{1}{N!}\left(\frac{8}{\hbar}\right)^N u^N
\end{equation}
This leads to
\begin{equation}
  \Delta_N = -2\sqrt{\frac{\hbar}{\pi}}\,\frac{(8/\hbar)^N}{N!}
  = -\frac{\hbar}{\sqrt{2 \pi}}\,\frac{1}{\Gamma(\kappa+\frac{1}{2})}\left(\frac{8}{\hbar}\right)^\kappa\,,
  \label{eq:DeltaN_action}
\end{equation}
where we have used the WKB $\kappa = N + \frac{1}{2}$ to match previous notation. This is exactly the leading non-perturbative splitting we computed using Exact WKB. Recall that with Exact WKB, this result required a resummation of the leading terms in $\VN$ to all orders in the Riccati equation. Here, it comes from simply taking the first subleading term at large $T$ in the classical action around the exact $n=1$ instanton saddle.

Now consider the additional terms in $g(u)$ in Eq.~\eqref{eq:gu_expanded}. The leading effect of these terms is
\begin{align}
g(u) &=
-2\sqrt{\frac{\hbar}{\pi}}
\sum_{N=0}^\infty \frac{1}{N!}\left(\frac{8}{\hbar}\right)^N u^N \left( 1-46 u - 12 u \ln u  - 136 \frac{u^2}{\hbar}-48 \frac{u^2}{\hbar}\ln u + \cdots\right) \\
&=
-2\sqrt{\frac{\hbar}{\pi}}\left\{ 1+\sum_{N=1}^\infty \frac{1}{N!}\left(\frac{8}{\hbar}\right)^N u^N\left[1
- \hbar \frac{23N}{4}
- \hbar \frac{3N}{2} \ln u
- \hbar \frac{17N(N-1)}{8}
-\hbar \frac{3N(N-1)}{4}  \ln u
\right]
\right\} \,.
\label{logmuterms}
\end{align} 
Thus the non-logarithmic factors give contributions to $\Delta_N$ that are subleading in $\hbar$. Although they are 1-loop in the path integral, they contribute only higher order in $\hbar$ terms to the splittings. Note also that these shift $N$ upwards so they do not affect the ground state.

The $\ln u$ terms require an explanation. Since $g(u) = \sum_N \Delta_N u^N$ is a power series with constant coefficients, terms proportional to $u^N \ln u$ cannot be matched. However, since these terms are suppressed by $\hbar$ they are NNLO corrections to the energies. At this order, the perturbative energies themselves get corrections, so we need to compute the non-perturbative splittings on top of those corrections. In Eq.~\eqref{eq:ZDelta} we used the leading order result $E_0^{(N)} = \hbar(N+\frac{1}{2})$,  but we know from Eq.~\eqref{eq:EN_2loop} that $E_0^{(N)} = \hbar(N+\tfrac{1}{2})-\hbar^2(\tfrac14 + \tfrac34 N(N+1)) + \cdots$ and therefore
\begin{equation}
  e^{-E_0^{(N)}T/\hbar} = e^{-(N+\frac{1}{2})T}\left[1 + \hbar\left(\frac{1}{4} + \frac{3N(N+1)}{4}\right) T + \cdots\right].
\end{equation}
The first new term $\hbar/4$ is an $N$-independent shift, correcting the $E = \hbar/2$ zero point energy of the SHO. Thus we can pull this term out in the definition of the full sum, like we did with the leading $\exp(-T/2)$ piece in Eq.~\eqref{Ztog}. The $N$-dependent part, however, differentiates between levels and provides a nontrivial constraint. Using $T=-\ln u$, the spectral sum becomes
\begin{equation}
    \widetilde{Z} = -e^{-\frac{S_I}{\hbar}} \,\frac{2T}{\hbar}\,
    e^{-T/2 + \hbar(T/4)}
    \sum_{N=0}^\infty \Delta_N u^N\left[1 - \hbar\,\frac{3N(N+1)}{4} \ln u + \cdots\right]\,,
\end{equation}
so that the expected form of $g(u)$ in Eq.~\eqref{Ztog} to match to is updated to 
\begin{equation}
    g(u) =
    \sum_{N=0}^\infty \Delta_N u^N\left[1 -\hbar \frac{3N}{2}\ln u - \hbar\,\frac{3N(N-1)}{4} \ln u + \cdots\right]\,.
\end{equation}
where we used $\tfrac{3N(N+1)}{4} = \tfrac{3N}{2} + \tfrac{3N(N-1)}{4}$. This form exactly matches the two $\ln u$ terms in Eq.~\eqref{logmuterms}. 

The cancellation of the $\ln u$ terms in the matching is a highly nontrivial check on the consistency of the path integral calculation mixing three different orders. The $\ln u$ terms in $g(u)$ arise from the 1-loop determinant and the classical action around the instanton saddle, but they must cancel against the $\ln u$ terms that arise from the perturbative corrections to the energies. The check validates agreement between the 2-loop vacuum bubbles in $\DV$, which determine the perturbative spectrum, the classical action $S[x_\cI]$ which gave the $48u^2 \ln u$ term and the 1-loop functional determinant $\det{}'(\cO_\cI)$, which gave the $12 u \ln u$ term.  

\subsubsection{Higher-order corrections to the ground-state splitting  ($\DL$ large $T$)}
\label{sec:DL_corrections}
So far we have extracted non-perturbative information at two distinct levels of approximation around the instanton. In Section~\ref{sec:NP_ground_state} we used the large-$T$ limit of the 1-loop one-instanton partition function $\widetilde Z_1(T)$, which is enough to fix the ground-state doublet splitting at leading non-perturbative order. In Section~\ref{sec:NP_excited_state} we used the full $T$-dependence of the 1-loop one-instanton, which extends this to the leading splittings $\Delta_N$ of all excited-state pairs through the generating function $g(u)$. Two further refinements remain. The first is 2-loop and higher corrections around the instanton in the large $T$ limit, captured by $\DL$ at large $T$. These give an $N$-independent multiplicative dressing of every $\Delta_N$ that we discuss now. The second is 2-loop and higher corrections at \emph{finite} $T$, which give $N$-dependent corrections to the splittings $\Delta_N$ for excited states; this is discussed in Section~\ref{sec:spectral_structure}.

The leading large-$T$ corrections on the instanton background are collected in $\DL$, given to 3 loops in Eq.~\eqref{eq:DLn_summary}: $\DL =-\frac{71}{48}\,\hbar-\frac{315}{128} \hbar^2 + \cdots$. These corrections exponentiate into the factor $e^{n\DL(\hbar)}$ in $Z_n$. For the one-instanton sector, this means the twisted partition function gets corrected to
\begin{equation}
  \widetilde{Z}_1(T)  = e^{-T/2}\,\frac{2T} {\sqrt{\pi\hbar}}\,e^{-\frac{S_I}{\hbar}+ \DV T +\DL} \,.
  \label{eq:Z1_largeT_corr}
\end{equation}
The $\DV$ gets absorbed into a universal shift of all the energies, and $\DL$ affects the energy splittings by correcting 
Eq.~\eqref{E1minusE0} to
\begin{equation}
  E^{(1)} - E^{(0)} = 4\sqrt{\frac{\hbar}{\pi}}\,e^{-\frac{S_I}{\hbar} + \DL}+ \cdots
  = 4\sqrt{\frac{\hbar}{\pi}}\,e^{-S_I/\hbar}\!\left(1 - \frac{71}{48}\,\hbar - \frac{6299}{4608}\,\hbar^2 + \cO(\hbar^3)\right) + \cdots
  \label{E1minusE0b}
\end{equation}
The trans-series for the lowest two energies in Eq.~\eqref{eq:lowest_doublet_PI} picks up half these corrections each. Both the 2-loop $-\tfrac{71}{48}\hbar$ and the 3-loop $-\tfrac{6299}{4608}\hbar^2$ coefficients in Eq.~\eqref{E1minusE0b} agree with the Exact WKB ground-state splitting in Eq.~\eqref{eq:splitE0inst}. The path-integral $\hbar^2$ coefficient combines the bare 3-loop $\DL$ piece $-\tfrac{315}{128}$ with the $\tfrac12\DL^2$ exponentiation cross-term $+\tfrac12(\tfrac{71}{48})^2 = +\tfrac{5041}{4608}$ to give $-\tfrac{6299}{4608}$. On the Exact WKB side the same coefficient drops out of the closed form Eq.~\eqref{E1formula} using $S_P$ to $\cO(\hbar^2)$ and $S_N$ to $\cO(\hbar^3)$ from Eqs.~\eqref{SPexpanded} and \eqref{SNh2}.

\subsubsection{Higher-order corrections to the excited-state splittings ($\DL$ finite $T$)}
\label{sec:spectral_structure}
At large $T$, $\DL$ is a constant in both $T$ and level number so it multiplies every $u^N$ in the spectral sum by the same factor, and the ratios $\Delta_N/\Delta_0$ are functions only of $\hbar$. To see the level-dependence from the path integral, one needs the $\cO(e^{-T/n})$ refinements to $\DL$ which break the simple $n\DL$ proportionality. Computing these corrections with the path integral would require evaluating Feynman diagrams using the full Lam\'e Green's function and elliptic function vertex profiles, which is a substantial undertaking. We will not do this here, but we can understand the structural form of the corrections and how they generate the transcendental functions that appear in the Exact WKB quantization condition. 

To go from infinite $T$ to finite $T$, one can use the image expansion fruitfully employed in Section~\ref{section:finiteT_spectrum}.  Around the perturbative saddle, the $N$-dependent energy shift came from terms in Eq.~\eqref{eq:Z0_expanded} of the form
\begin{equation}
    Z_P=
  \cdots + \hbar \frac{3T}{2} \sum_{N=0}^\infty e^{-(N+\frac{1}{2})T}\sum_{m=1}^\infty m\, e^{-mT}
  =\cdots +  \hbar \frac{3T}{2}  e^{-T/2}\sum_{N=0}^\infty u^N \frac{u}{(1-u)^2} \,.
\end{equation}
The $m e^{-mT}$ structure comes from the winding-number decomposition of the periodic propagator in Eq.~\eqref{DeltaPimages}, $\Delta_P(\tau)=\sum_m e^{-|\tau+mT|}$. A 2-loop vacuum bubble has two independent windings $m_1,m_2$, and the vertex integration contributes one additional unit of winding. Thus the total winding $m=m_1+m_2+1$ has degeneracy $m$, giving
\begin{equation}
\sum_{m_1=0}^{\infty}\sum_{m_2=0}^{\infty} u^{m_1+m_2+1} = \frac{u}{(1-u)^2} = \sum_{m=1}^{\infty} m\,u^m.
\end{equation}
The correction factor has a double pole at $u=1$. After multiplying by the SHO tower $\sum_N u^N=1/(1-u)$, coefficient extraction gives the cumulative sum $\sum_{m=1}^N m=\frac12N(N+1)$, hence the polynomial dependence on the level number. More generally, rational poles in $u$ give polynomial dependence on $N$.

In the instanton sectors, the image decomposition still works, but the weights of the images are no longer fixed by translation invariance. The instanton-background propagator $\Delta_I(t_1,t_2)$ depends on $t_1$ and $t_2$ separately, not only on $t_1-t_2$, and its instanton-dependent part is localized near the core. Vertex integrations are therefore weighted by the core profile. As a toy model, suppose the broken-translation-invariant part of one propagator gives an extra image weight $1/(m_1+1)$ while a second propagator still carries an ordinary winding $m_2$. Then the perturbative counting sum is replaced by
\begin{equation}
  \sum_{m_1,m_2\ge0}\frac{u^{m_1+m_2+1}}{m_1+1}
  =
  -\,\frac{\ln(1-u)}{1-u}\,.
\end{equation}
This example is only schematic, but it illustrates the mechanism: once the core profile makes different windings contribute with non-polynomial weights, the rational functions generated by translation-invariant counting are replaced by logarithms. When we extract the energy shift from  $-\ln(1-u)/(1-u)$ we would use
\begin{equation}
  -\frac{\ln(1-u)}{1-u}=\sum_{N\ge1}H_N\,u^N,
\end{equation}
where $H_N=\psi(N+1)+\gamma_E$ is the $N$-th harmonic number. Under $N\to\kappa-\tfrac12$ this becomes $\psi(\kappa+\tfrac12)+\gamma_E$, the digamma structure of the Exact WKB pole tower at $\cO(\hbar^2)$ (see Eq.~\eqref{eq:h2_tower_wkb}). Similarly, logarithmic structures such as
\begin{equation}
  \frac{\ln^2(1-u)}{(1-u)^k}
  \label{eq:log2_from_DL}
\end{equation}
generate combinations of harmonic sums whose analytic continuation contains $\psi'(\kappa+\frac{1}{2})$, matching the trigamma structures that enter at $\cO(\hbar^3)$ (see Appendix~\ref{appendix:Weber}).

%% file: sections/conclusions.tex
% \!TEX root = ../DoubleDoubleMain.tex
\section{Conclusions}
\label{sec:conclusions}

The symmetric double-well potential is one of the oldest and most instructive problems in quantum mechanics. It sits at the intersection of perturbation theory and non-perturbative physics, of algebraic geometry and analysis, of path integrals and spectral theory. It is also physical: double-well physics arises throughout nature, such as in atomic systems like $H_2^+$. The double-well energy spectrum is described by a trans-series comprising asymptotic series with coefficients in $\hbar$ growing as $n!$, non-perturbative factors of the form $e^{-n S_I/\hbar}$ with $S_I=2/3$ and logarithmic factors of the form $\ln^n\hbar$.  In the double-well, resurgence, asymptotic series, Stokes phenomena, Borel resummation, ambiguity cancellation, Lefschetz thimbles, elliptic curves, the Picard--Fuchs, Weber and Lam\'e equations all appear in a model that remains explicitly computable. The zero-dimensional warmup in Section~\ref{sec:zerodim} introduced these tools (Borel transforms, alien calculus, thimble decomposition, the action--Borel correspondence, and the Riemann-surface picture of the Borel plane) in a setting where everything can be checked by hand. Exact WKB and the Euclidean path integral then expose different parts of the same structure in the full quantum-mechanical problem. Exact WKB gives a sharp spectral and algebraic description through periods, Voros symbols, and Stokes automorphisms, while the path integral gives a geometric description in terms of saddles, thimbles, quasi-zero modes, and fluctuation determinants.  This paper has presented both approaches from first principles, in a common notation with explicit bookkeeping, showing how they fit together into a single, tightly constrained framework.

\subsubsection*{Exact WKB summary}
On the Exact WKB side, the starting point is the Schr\"odinger equation and the requirement of normalizable wavefunctions. The typical failure mode of the WKB approximation (its breakdown at turning points where the WKB ansatz explodes) is elegantly resolved through analytic continuation. Near turning points, the Schr\"odinger equation linearizes to  Airy's equation, whose solutions have well-known asymptotics and Stokes phenomena.  By carefully tracking the analytic continuation across Stokes lines emanating from the turning points, one obtains an exact quantization condition for the energy levels. This quantization condition is expressed as
\begin{equation}
  (1 + \VP)^2 = -\VN\,,
  \label{eq:conclusion_quantization}
\end{equation}
where $\VP$ and $\VN$ are Voros symbols, defined as exponentials of period integrals over the WKB momentum along cycles of an elliptic curve.

The non-perturbative Voros symbol $\VN$ is Borel summable but the perturbative Voros symbol $\VP$ is not. The resurgent structure arises because the Stokes discontinuity of $\VP$ is controlled by $\VN$ through the Delabaere-Dillinger-Pham formula $\SA\,\VP = \VP(1+\VN)$, where $\SA=\cS_+ \cS_-^{-1}$ is the Stokes automorphism,  with $\cS_\pm$ the lateral Borel resummation. This Stokes automorphism encodes the discontinuity across the Stokes line. In the Exact WKB picture, the turning points are associated with saddle points of an integral representation of the Airy function. These saddle points are branch points of a 3-sheeted Riemann surface: as one analytically continues $x$ around a turning point, the integration contour moves along this Riemann surface and its asymptotic expansion changes abruptly across Stokes lines corresponding to passing between sheets. The subleading terms added to the asymptotic expansion across Stokes lines provide an explanation for the imaginary parts in the non-perturbative corrections needed to cancel the  Borel ambiguities of the perturbative series, as encoded in the DDP formula. 

The Exact WKB quantization condition in Eq.~(\ref{eq:conclusion_quantization}) can be solved order by order in $\hbar$ and in $e^{-S_I/\hbar}$. At each order in the WKB expansion, new elliptic integrals are needed. Fortuitously, these integrals can all be expressed algebraically in terms of leading-order period integrals $S_P^0$ and $\partial_E S_P^0$ or $S_N^0$ and $\partial_E S_N^0$ with coefficients that are rational functions of $E$. This beautiful simplifying feature is a consequence of the Picard--Fuchs equation satisfied by the period integrals, which in turn is a consequence of the elliptic-curve structure of the classical energy shell: its first de~Rham cohomology is two-dimensional (there are two independent cycles around a torus). Once the period integrals are done, the perturbative energy spectrum follows. For the non-perturbative corrections there is a complication: as $E\to 0$ the turning points collide, and the WKB expansion breaks down and one must sum an infinite tower of terms to get $S_N$.  The leading tower resums into $\ln\Gamma(E/\hbar + \frac{1}{2})$, a result that can be derived independently from the Weber equation.  The subleading tower, whose coefficients have a Bernoulli-number structure matching the asymptotic expansion of the digamma function, can be resummed in closed form. Remarkably, the digamma cancels against the remainder, which comes from expanding $S_N^0$ with the two-loop perturbative energy correction. The full $\cO(\hbar^2)$ coefficient of $S_N$ is therefore a polynomial in $\kappa$.  A similar cancellation occurs at $\cO(\hbar^3)$ where the trigamma function arising from the subleading tower also cancels against a similar remainder term. 

Once the quantization condition is solved, one finds that odd instanton sectors ($E_1, E_3, \ldots$) produce the parity splitting while even sectors ($E_2, E_4, \ldots$) produce common energy shifts. We derived the energy corrections in closed form at leading $\hbar$ through $E_4$ for arbitrary level $N$, with the multi-instanton coefficients written as polynomials in $\Sigma = -\ln(8/\hbar) + \psi(\kappa+\tfrac12) - i\pi$ multiplied by $\hbar\fv^n$, with explicit $\psi'(\kappa+\tfrac12)$ and $\psi''(\kappa+\tfrac12)$ structure appearing at $E_3$ and $E_4$. Leveraging those results, we also showcased how the resurgent structure of the energy trans-series is effectively uncovered using alien calculus. The alien calculus reveals that the even sublattice ($E_P, E_2, E_4, \ldots$) and odd sublattice ($E_1, E_3, \ldots$) are completely disconnected. All Borel singularities of $E_P$ at $t = 2nS_I$ for $n \ge 1$ resum into a single logarithm:
\begin{equation*}
  \sum_{n=1}^\infty \lambda^{2n}\,\Delta_{2n}(E_P) \;=\; \frac{F}{2\pi i}\,\ln(1+\VN)\,,
\end{equation*}
where $F = \partial_\kappa E_P$. To make the alien calculus arguments less abstract, we explicitly verified the resurgent cancellation at the 2 and 4 instanton levels. At the 4-instanton level, the resurgent constraint is encoded in the alien calculus as
\begin{equation*}
  \frac{1}{2}\,\im[\Delta_4 E_P] + \frac{1}{2}\,\im[\Delta_2 E_2] + \im[E_4] = 0\,,
\end{equation*}
where we computed each term as an explicit function of $\hbar$ and verified that their sum vanishes.

\subsubsection*{Path integral summary}

On the path integral side, the starting point is the partition function $Z(T)$ and its parity-twisted counterpart $\widetilde{Z}(T)$ computed using the Euclidean Feynman path integral. The path integral is over all real paths, but this sum can be decomposed into a sum over Lefschetz thimbles, each associated to a complex saddle point of the action. 
These saddle points can be found exactly: they are doubly-periodic Weierstrass elliptic functions, classified by a pair of integers $(k,k')$ counting windings around the perturbative and non-perturbative cycles of an elliptic curve. Of these saddles, only the real ones, with $k'=0$, have nonzero intersection numbers in the Picard--Lefschetz decomposition, so only these are involved in the sum over saddles. The vanishing of $\eta_{k,k'}$ for $k'\ne0$ follows from a short lemma: $\im S$ is conserved along the gradient flow, so the unstable thimble of any complex saddle cannot intersect the real-path cycle $\GR$. This argument excludes even the self-conjugate $k'=k$ branch, where the saddle action happens to be real but the saddle itself is still complex-valued. Although the surviving saddles are real, the thimbles passing through them are complex and the path integral is only real once all the thimbles are summed over. 
The complex saddles, although not summed over, are important for the Stokes phenomenon. For example, the complex $(1,1)$ saddle sits at a Stokes point of the thimble associated with the real $(1,0)$ saddle corresponding to an instanton-anti-instanton pair.

For the path integral picture to reproduce the spectrum established by Exact WKB, it is critical to work at finite $T$. Taking $T\to\infty$, as in the dilute instanton gas, allows the ground state energy to be recovered, but cannot recover the excited state spectrum. The finite $T$ path integral is challenging, but not intractable. The 1-loop path integral around the elliptic saddles can be computed exactly in closed form because the fluctuation determinant is that of a Lam\'e operator with known spectrum. For the anti-periodic single-instanton saddle this gives a closed-form twisted partition function $\widetilde Z_1(T)$ valid at any finite $T$, from which the leading non-perturbative splittings $\Delta_N$ for all excited states fall out by Taylor expansion in $u = e^{-T}$ with no further resummation. Beyond 1-loop, systematic corrections around the perturbative saddles are calculable, but around the instanton saddles the calculations are very difficult. We therefore combine the finite-$T$ exact calculations around the perturbative saddles with leading-$T$ corrections around the instanton saddles. 

The geometric picture of resurgence on the path-integral side, summarized in our companion paper~\cite{DersySchwartz:2026}, is most concrete at $n=2$. In the $n=2$ sector at large $T$, the quasi-zero-mode space is one-dimensional, with a single complex variable $\alpha$ parametrizing the instanton--anti-instanton separation. The effective action has two saddles in the fundamental domain: the real $(1,0)$ saddle at $\alpha = T/2$ and the complex $(1,1)$ saddle at $\alpha = T/2 + i\pi$. The thimble through $(1,0)$ runs vertically toward $(1,1)$, bends at the Stokes point, and then runs horizontally; its imaginary part comes from the vertical segment between the two saddles. The lateral Borel resummation of the perturbative ($n=0$) series traces out the same vertical segment with opposite orientation, so the imaginary contributions cancel geometrically. For general $n$ the picture is the same with the QZM space promoted to $n-1$ dimensions: the thimble integral reduces exactly to a Meijer $G$-function whose imaginary part can be extracted at large $T$, and the leading asymptotic series of the $(n-2)$ sector is reproduced by a real-cycle integral over a codimension-one face of the $(n-1)$-simplex, whose lateral Borel resummation cancels the leading $T$ imaginary piece of the $n$-thimble.

At $n=4$ this geometric picture becomes richer. There are three quasi-zero modes corresponding to two pair-annihilation modes ($x_1, x_2$) and one breathing mode ($y$). The downward flow structure reveals two independent channels to $n=2$ and one channel to $n=0$, in precise correspondence with the $\Delta_2 Z_2$ and $\Delta_4 Z_0$ alien derivatives, mirroring the Exact WKB alien calculus. Despite using the large-$T$ approximation for these results, the path integral does not reduce to the dilute instanton gas: it gives a complex trans-series including $\ln \hbar$ factors in each instanton sector, a geometric picture of the resurgent structure, and a way to compute the leading asymptotic behavior in each instanton sector. Most importantly, in contrast to the dilute gas, the framework is systematically improvable. The higher-order calculations are daunting (e.g. Feynman diagrams on instanton backgrounds with the Lam\'e Green's function) but well-defined.

\subsubsection*{Complementarity}

Exact WKB and the path integral systematically compute different objects. Exact WKB constructs the wavefunction order by order in $\hbar$ and in $\lambda = e^{-S_I/\hbar}$, with the elliptic integrals demanded by the WKB expansion of the quantum momentum reduced via Picard--Fuchs to combinations of $S_P^0$ and $\partial_E S_P^0$ with rational coefficients in $E$. The path integral systematically computes the partition function $Z(T)$, with Euclidean time $T$ as a parameter from the start. Neither object is directly the spectrum: Exact WKB requires imposing the quantization condition $(1+\VP)^2 = -\VN$; the path integral requires extracting poles of the resolvent from $Z(T)$. An important difference is that the partition function $Z(T)$ already carries a trans-series in $\lambda$ before any spectral information is extracted, and the entire resurgent apparatus (BZJ cancellations, Stokes phenomena at complex saddles, the Lefschetz-thimble decomposition) lives in it directly. The wavefunction has no trans-series of its own; the resurgent constraints in Exact WKB (the DDP relation $\SA\VP = \VP(1+\VN)$, alien calculus) are formal relations on Voros symbols with no spectral content until quantization is imposed. In the path integral the trans-series is upstream of quantization; in Exact WKB it is downstream.
That difference makes the resurgent picture concrete on the path-integral side. The cancellations have a geometric realization on finite-dimensional quasi-zero-mode manifolds where they can be drawn: the shared codimension-one collapse face of the $(n-1)$-simplex connecting the $n$-instanton thimble to the $(n-2)$-instanton asymptotic series is the BZJ cancellation made manifest. Nothing in Exact WKB exposes this geometry; the same cancellations are present, but only after the full alien-calculus machinery has been set up.

The way perturbation theory works on the two sides is very different. Perturbation theory for the path integral is Feynman diagrams. On an instanton background these are punishing: the propagator is a Lam\'e Green's function, and two-vertex and higher integrals on the elliptic curve become formidable. The $n=0$ sector is special: there the propagator is on a harmonic oscillator background, and the finite-$T$ loop sum $\Delta_V(T,\hbar)$ at each order in $\hbar$ is fixed by spectral inversion of the Bender--Wu perturbative spectrum at every level $N$. The explicit loop calculation gives an independent derivation of an object already determined by the spectrum. Around instanton backgrounds, progress beyond 1-loop has only been possible so far at large $T$. Exact WKB has no Feynman diagrams at all. The analogous structure is the Picard--Fuchs reduction of the $P_n$, which is much faster and gives the perturbative and non-perturbative parts of $S_N$ to arbitrary order in $\hbar$ with bounded effort. The only catch is that a resummation of all orders in this expansion is needed to get an additional order in the $\hbar$ expansion of the energies.

The two methods are remarkably complementary: each step opaque in one is transparent in the other. The geometric picture of resurgence is invisible in Exact WKB and obvious in the path integral. The systematic $\hbar$ expansion within a sector is algebraic in Exact WKB (via Picard--Fuchs) but combinatorially explosive in the path integral. The leading non-perturbative splitting for excited states requires resumming an infinite WKB tower into $\Gamma(\kappa+\tfrac12)$ on the WKB side, but falls out of the first subleading $e^{-T}$ correction on the path-integral side. The leading-$\hbar$ coefficient of $E_n$ at every $n$ is an $(n+1)$-st-order Taylor expansion in WKB and a single derivative of a Meijer-$G$ function in the path integral.

Both formalisms compute on the same elliptic curve. The Voros period $S_N^0(E)$ and the path-integral saddle action $S_N^0(\varepsilon) - \varepsilon T$ are Legendre transforms of each other at the classical level, with $T = \omega_N(\varepsilon)$ and $\varepsilon = -E$. This is why the same special functions reappear on both sides: the $\Gamma(\kappa+\tfrac12)$ that requires an infinite WKB pole-tower resummation on the energy side, and the same $\Gamma(\kappa+\tfrac12)$ extracted from $[u^N] e^{8u/\hbar}$ on the partition function side, are both pulled from the same singular limit of the elliptic curve, where the perturbative cycle collapses and the instanton period diverges. 

One observation that we have not been able to derive from first principles: each order of the WKB pole-tower resummation in $S_N$ appears to correspond to one loop on the path-integral side, with the same numerical coefficients. The leading Weber resummation into $\hbar\ln\Gamma(\kappa+\tfrac12)$ matches the one-loop determinant; the digamma cancellation at $\cO(\hbar^2)$ matches the two-loop vacuum bubbles, with the $3N(N+1)/4$ structure extracted from the finite-$T$ expansion of $Z_P$; the trigamma cancellation at $\cO(\hbar^3)$ matches the three-loop bubbles. This dictionary holds at every order we have checked. Whether it has a clean derivation beyond order-by-order matching is an open question.

\subsubsection*{What is new}

The single largest contribution of this paper is the synthesis itself. The trans-series structure of the double-well spectrum has been understood, in pieces and in various languages, for more than four decades. Zinn-Justin and Jentschura (ZJJ)~\cite{ZinnJustinJentschura} packaged the multi-instanton expansion as the solution of a transcendental equation involving $\Gamma(1/2- B(E,g))$, derived from path-integral arguments combined with extensive numerical checks. Once the equation is in hand the multi-instanton coefficients fall out by Taylor expansion, but the equation itself reads as a prescription rather than a derivation: it is not clear from the path-integral argument why this particular combination of $\Gamma$ functions and exponentials should be exact. The rigorous justification on the WKB side came later through Delabaere--Dillinger--Pham~\cite{DelabaereDillingerPham1993,DelabaerePham1999}, and the path-integral counterpart was developed piecemeal over the following decades.
That the ZJJ results are equivalent to what Exact WKB and the Picard--Lefschetz path integral deliver is not in itself a new claim. DDP supplied the exact-WKB justification, \'Ecalle's alien calculus formalized the relations between sectors, Picard--Lefschetz organized the thimble decomposition of the path integral, and several authors have worked pieces of the path-integral side. What this paper adds is the execution at significant depth.

We push the exact-WKB calculation to four instanton orders at leading $\hbar$ and to three orders in $\hbar$ for the one-instanton ground-state splitting; write out the alien-calculus tower $\Delta_{2n} E_P$ term by term; verify the imaginary-part cancellations explicitly at $\cO(\lambda^2)$ and $\cO(\lambda^4)$. We decompose the partition function into thimble integrals through the $n=2, 3, 4$ saddles in closed Meijer-$G$ form, compute the exact finite-$T$ one-loop fluctuation determinant on every periodic saddle, and extract the excited-state splittings $\Delta_N$ for all~$N$ by Taylor expansion of a closed-form $\widetilde Z_1(T)$. At each step the $\Gamma$ functions, digamma derivatives, and polygamma functions that appear in ZJJ as poles of the prescribed quantization condition emerge here from a definite operation: a Weber resummation, a Picard--Fuchs reduction, or a finite-dimensional thimble integral. Carrying out the translation far enough on both sides makes the agreement between formalisms a nontrivial cross-check, beyond what the principle alone provides.

Several of those calculations yield results we believe to be new. On the WKB side: closed-form leading-$\hbar$ expressions for $E_3^\pm$ and $E_4$ at arbitrary level $N$, with explicit $\psi'(\kappa+\tfrac12)$ and $\psi''(\kappa+\tfrac12)$ structure (Eqs.~\eqref{eq:E3_summary}--\eqref{eq:E4_summary});
% \footnote{The cosine ground-state analog of $E_4$ is in Zinn-Justin's 1981 paper~\cite{ZinnJustin1981}; the double-well, arbitrary-$N$ versions extend the ZJJ framework.};
 the $\cO(\hbar^2)$ coefficient of $E_1^\pm$ at arbitrary $N$ (Eq.~\eqref{eq:E1_explicit_app}); the algebraic recursion in $R_N = \VN'/\VN$, $R_P = \VP''/\VP'$ that delivers $E_n$ at any order in $\lambda$; the on-shell alien resummation $\sum_n \lambda^{2n}\Delta_{2n}(E_P) = (F/2\pi i)\ln(1+\VN)$ (Eq.~\eqref{eq:alien_resum}); the explicit $\cO(\lambda^4)$ imaginary-part cancellation $\im[E_4] + \tfrac12\im[\Delta_2 E_2] + \tfrac12\im[\Delta_4 E_P] = 0$ (Eq.~\eqref{eq:ImSum_lambda4}); the digamma and trigamma cancellations in the $\hbar^2$ and $\hbar^3$ subleading pole towers of $S_N$, leaving polynomial-in-$\kappa$ coefficients; the next-to-leading large-order coefficients $C_1(\kappa), C_2(\kappa)$ of the perturbative series in closed polynomial form (Eqs.~\eqref{eq:C1kappa_app}--\eqref{eq:C2kappa_app}); and the polylogarithm form of the leading Borel transform of $E_P$ at $\kappa = 1/2$ and $\kappa = 3/2$.

On the path-integral side, the main new result is the Picard--Lefschetz origin of the Bogomolny--Zinn-Justin mechanism~\cite{DersySchwartz:2026}: at every $n$, the imaginary ambiguity of the lateral Borel resummation of the $(n-2)$-instanton sector cancels geometrically against an explicit thimble integral on the $n$-instanton quasi-zero-mode simplex, with the cancellation realized at $n=2$ as the original BZJ pair and extended to general $n$ via the codimension-one collapse face of the $(n-1)$-simplex. Additional new results include: the factorized form of the partition function on each thimble (Eq.~\eqref{eq:Z_factored}), decomposing $Z_{k,k'}$ into saddle action, transverse fluctuation determinant, perturbative and instanton-loop corrections $\Delta_V$ and $\Delta_L$, and a finite-dimensional quasi-zero-mode thimble integral $\cI_{k,k'}$; the distinction between the partition-function trans-series, which carries the full resurgent structure directly, and the energy trans-series, which inherits it downstream through the spectral inversion; the closed-form exact one-loop twisted partition function $\widetilde Z_1(T)$ valid at all $T$ (Eq.~\eqref{eq:Ztilde1_exact}), with the Legendre-transformed action $S_N^0(\varepsilon) - \varepsilon T$, and the excited-state splittings $\Delta_N$ for all $N$ extracted from it by Taylor expansion in $u = e^{-T}$ with no resummation needed; the three-way cancellation of $\ln u$ terms among the classical action, the one-loop determinant, and the two-loop perturbative vacuum bubbles in Section~\ref{sec:NP_excited_state}; the Meijer-$G$ representation of the multi-instanton thimble integral $\cI_n^\pm = \cG_n(\mu_n^n e^{\pm n\pi i})$ for general $n$ and the closed-form large-$T$ polynomial expansion (Eqs.~\eqref{eq:gen_smallz}, \eqref{eq:gen_imag_largeT}); the master formula for the leading-$\hbar$ coefficient of $E_n$ at every $n$, written as a single derivative of a generating function (Eq.~\eqref{eq:Enclosedform}); the intersection-number lemma showing $\eta_{k,k'}=0$ for all complex saddles, with the subtle exclusion of the self-conjugate $k'=k$ branch where $\im S$ vanishes but the saddle is still complex-valued; and the recovery of the perturbative excited-state spectrum from the finite-$T$ expansion of $Z_P$, with the explicit identification of the $e^{-mT}$ corrections that the dilute instanton gas misses.

\subsubsection*{Outlook}
There is much more to be done. For the double well, we have verified, at low orders, the order-by-order matching between WKB pole-tower resummation order and path-integral loop count, but this feature should be provable generally. Such a derivation would unify the two formalisms much more directly. Although the Exact-WKB program for the energy spectrum is essentially complete and algebraic, the finite-$T$ instanton-background Feynman diagrams on the path-integral side are not; the Lam\'e Green's function makes this calculation explicit in principle, and closing the finite-$T$ program would directly produce the excited-state higher-loop corrections that currently require WKB. The action--Borel correspondence introduced in Section~\ref{sec:zerodim} for the 0D model gives a direct reconstruction of the action from the Borel transform; whether an analog exists in QM, with the role of the action played by the WKB momentum or the saddle action, would be a cleaner formulation of resurgence than the one provided by alien calculus. Beyond the symmetric double well, but still within quantum mechanics, the same Picard--Lefschetz/exact-WKB synthesis should apply to the cosine, the asymmetric double well, the periodic Mathieu problem, and $H_2^+$, several of which have been studied piecewise in the literature but have not received the same complete treatment.

Looking beyond quantum mechanics, the next target is quantum field theory, where instanton calculations often rely on dilute-gas or valley approximations~\cite{tHooft1976,BalitskyYung} and where the relation between saddles, asymptotic perturbation theory, and ambiguity cancellation is much less explicit~\cite{tHooft:1977xjm,Beneke:1998ui,Bhattacharya:2024hhh}. One additional complication in quantum field theory is that there are sources of factorial growth in perturbation theory beyond instanton--anti-instanton pairs, such as renormalons~\cite{tHooft:1977xjm,Beneke:1998ui} which can be viewed as saddle points of an effective action~\cite{Bhattacharya:2024hhh}, and possibly other objects (see e.g.~\cite{clingerman2025asymptoticbehaviordiagramclasses}). In such a formulation, Picard--Lefschetz theory would not merely justify an analytic-continuation prescription after the fact; it would determine which saddle families contribute~\cite{Behtash:2018voa}, how their quasi-zero modes should be integrated, and how their Stokes jumps cancel perturbative ambiguities.  Compactified gauge theories, where semiclassical saddles and their constituents can be controlled more explicitly~\cite{Kraan:1998pm,Lee:1998bb,Argyres:2012ka}, provide a natural testing ground for this idea.

The double well has served physics for over a century, from the hydrogen molecular ion to instanton physics in gauge theories.  It remains, as Coleman appreciated, the simplest system in which the full architecture of non-perturbative quantum mechanics can be exhibited.  We hope that the synthesis presented here, unifying Exact WKB and Picard--Lefschetz thimble decomposition in a single self-contained treatment, will serve as both a reference and a starting point for the next generation of problems.

%% file: sections/appendix_perturbative.tex
% \!TEX root = ../DoubleDoubleMain.tex
\section{Perturbative calculations for the double well}
\label{appendix:benderWu}

In this appendix, we review the computation of the perturbative energy spectrum of the double-well potential using both Rayleigh-Schr\"odinger perturbation theory (RSPT) and the Bender-Wu recursion method. Both approaches yield identical results; the Bender-Wu method is a reformulation of RSPT into efficient recursion relations that enable computation
to very high orders.

\subsection{Rayleigh-Schr\"odinger perturbation theory}
\label{subsec:RSPT}

To expand the double-well potential $V(x) = \frac{1}{8}(x^2 - 1)^2$ around the minimum at $x = -1$ we write $x = -1 + y$. Then the Hamiltonian becomes
\begin{equation}
H = -\frac{\hbar^2}{2}\frac{d^2}{dy^2} + \frac{1}{2}y^2 - \frac{1}{2}y^3 + \frac{1}{8}y^4
\end{equation}
We write $H = H_0 + V_y$ where $H_0 = -\frac{\hbar^2}{2}\frac{d^2}{dy^2} + \frac{1}{2}y^2$ is the harmonic oscillator and $V_y = -\frac{1}{2}y^3 + \frac{1}{8}y^4$ is the perturbation. The unperturbed eigenstates are $|N\rangle$ with energies $E_N = \hbar(N +1/2)$. We organize the perturbative energy as a power series in $\hbar$,
\begin{equation}
E_P(N) = \sum_{k=1}^\infty e_k(N)\,\hbar^k\,.
\label{eq:EP_series}
\end{equation}
The leading term is simply $e_1(N) = N+1/2$, which is the harmonic oscillator result.

Up to second-order in RSPT the energy shift is:
\begin{equation}
E_P(N) =  E_N +  \langle N | V_y | N \rangle + \sum_{m \neq N} \frac{|\langle m | V_y | N \rangle|^2}{E_N - E_m} \,.
\label{RSPT1}
\end{equation}
In the first correction, $\langle N|y^3|N\rangle$ vanishes by parity.  Using $y = \sqrt{\frac{\hbar}{2}}(a + a^\dagger)$, the diagonal matrix element of $y^4$ gives
\begin{equation}
\langle N | y^4 | N \rangle = \frac{3\hbar^2}{4}(2N^2 + 2N + 1) = \frac{3\hbar^2}{4}\left(2\kappa^2 + \frac{1}{2}\right) \,.
\end{equation}
where $\kappa = N + \frac{1}{2}$. In the second-order correction, the term $-\frac{1}{2}y^3$ connects $|N\rangle$ to $|N\pm 1\rangle$ and $|N\pm 3\rangle$, while $\frac{1}{8}y^4$ connects to $|N\pm 2\rangle$, $|N\pm 4\rangle$ and $|N\rangle$. After summing all contributions and simplifying, one obtains
\begin{equation}
e_2(\kappa) = -\frac{3\kappa^2}{4} - \frac{1}{16} \,.
\end{equation}
Both corrections in Eq.~\eqref{RSPT1} land at $\hbar^2$ because $y \sim \sqrt{\hbar}$: the first-order term gives $\langle y^4/8\rangle \sim \hbar^2$, while the second-order term is dominated by the cubic, $|y^3/2|^2/E \sim \hbar^3/\hbar = \hbar^2$ (the quartic contribution to second order, $|y^4/8|^2/E \sim \hbar^3$, is subleading).

\subsection{The Bender-Wu recursion}
\label{subsec:BenderWu}

While the RSPT calculation becomes increasingly tedious at higher orders, Bender and Wu~\cite{BenderWu, BenderWu2} reformulated RSPT into efficient recursion relations. We present the method in sufficient detail to be reproducible. First, consider a general anharmonic oscillator with polynomial potential:
\begin{equation}
H = -\frac{\hbar^2}{2}\frac{d^2}{dy^2} + \frac{1}{2}y^2 + \sum_{k=3}^{K} g_k y^k\,.
\end{equation}
The key insight is to write an Ansatz for the wave function, expanding it in the basis of wavefunctions of the harmonic oscillator:
\begin{equation}
\psi(y) = e^{-y^2/(2\hbar)} \sum_{j=0}^\infty B_j(\hbar)\, H_j\left(\frac{y}{\sqrt{\hbar}}\right) \,,
\label{eq:BWansatz}
\end{equation}
where $H_j$ are Hermite polynomials satisfying $H_j'' - 2z H_j' + 2j H_j = 0$. Since odd-power perturbations ($y^3$, $y^5$, \ldots) contribute at half-integer powers of~$\hbar$ through $y^k = \hbar^{k/2}z^k$ with $z = y/\sqrt{\hbar}$, the coefficients $B_j$ must be expanded in powers of $\hbar^{1/2}$:
\begin{equation}
B_j(\hbar) = \sum_{r=0}^\infty B_j^{(r)} \hbar^{r/2} \,.
\end{equation}
The energy, by contrast, has only integer powers of $\hbar$ (the half-integer coefficients vanish by parity), so $E_P$ has the form of Eq.~\eqref{eq:EP_series}.
Substituting the ansatz~\eqref{eq:BWansatz} into the Schr\"odinger equation and using the identities
\begin{align}
z H_j(z) &= \tfrac{1}{2}H_{j+1}(z) + j H_{j-1}(z) \\
H_j'(z) &= 2j H_{j-1}(z)
\end{align}
with $z = y/\sqrt{\hbar}$, one projects onto $H_j(z)$ using the orthogonality of the Hermite polynomials.  Setting $e_1 = N + \frac{1}{2} \equiv \kappa$ (the harmonic oscillator value), the coefficient of $\hbar^{r/2}$ gives a recursion for the $B_j^{(r)}$. For the state with quantum number $N$, we impose $B_N^{(0)} = 1$ and $B_j^{(0)} = 0$ for $j \neq N$. At order $r \geq 1$ the recursion takes the form
\begin{equation}
(N - j) B_j^{(r)} = \sum_{k=3}^{K} g_k \sum_\ell [z^k]_{j,\ell}\, B_\ell^{(r-k+2)}
- \sum_{p=2}^{\lfloor r/2 \rfloor + 1} e_p\, B_j^{(r - 2p + 2)}
\label{eq:BWrecursion}
\end{equation}
where $[z^k]_{j,\ell}$ denotes the connection coefficient defined by $z^k H_\ell(z) = \sum_j [z^k]_{j,\ell}\,H_j(z)$, computable from the recurrence $z\,H_j = \tfrac{1}{2}H_{j+1} + j\,H_{j-1}$. Coefficients $B_\ell^{(s)}$ with $s < 0$ are set to zero. When $j = N$, the left-hand side vanishes and the equation becomes a solvability condition that determines the energy coefficients. Writing $r = 2R$ for the even steps (at which $e_{R+1}$ enters):
\begin{equation}
e_{R+1} = \sum_{k=3}^{K} g_k \sum_\ell [z^k]_{N,\ell}\, B_\ell^{(2R-k+2)}
- \sum_{p=2}^{R} e_p\, B_N^{(2R-2p+2)} \,.
\end{equation}
This determines $e_{R+1}$ in terms of previously computed quantities. The odd steps ($r$ odd) determine the half-integer $B_j$ coefficients generated by the odd-power perturbations; these do not contribute to the energy.

For the double-well potential expanded around $y = 0$, we have only  $g_3 = -\frac{1}{2}$ and $g_4 = \frac{1}{8}$. The recursion~\eqref{eq:BWrecursion} proceeds as follows:
 \begin{itemize}
     \item At order $r=1$ ($\hbar^{1/2}$): Only the cubic term contributes, generating off-diagonal $B_j^{(1)}$ for $j = N \pm 1, N \pm 3$ via $[z^3]_{j,N}$. The energy is unaffected (odd step).
     \item At order $r=2$ ($\hbar^1$): Both terms contribute. The solvability condition at $j = N$ gives $e_2 = g_3\sum_\ell [z^3]_{N,\ell}\,B_\ell^{(1)} + g_4\,[z^4]_{N,N}$, reproducing $e_2 = -\frac{3\kappa^2}{4} - \frac{1}{16}$.
     \item At order $r=4$ ($\hbar^2$): The solvability condition gives $e_3 = -\frac{17\kappa^3}{16} - \frac{19\kappa}{64}$.
\end{itemize}

The power of the Bender-Wu method is that it continues algorithmically to arbitrary order. The \texttt{BenderWu} package of Sulejmanpasic and \"Unsal~\cite{BenderWuPackage} provides a versatile Mathematica implementation for general polynomial potentials, enabling computation of $\sim 100$ orders in seconds and $\sim 250$ orders in a few hours. The computation is purely symbolic, yielding exact rational coefficients as polynomials in $\kappa$.

The first several coefficients are:
\begin{align}
e_1(\kappa) &= \kappa \\
e_2(\kappa) &= -\frac{3\kappa^2}{4} - \frac{1}{16} \\
e_3(\kappa) &= -\frac{17\kappa^3}{16} - \frac{19\kappa}{64} \\
e_4(\kappa) &= -\frac{375\kappa^4}{128} - \frac{459\kappa^2}{256} - \frac{131}{2048} \\
e_5(\kappa) &= -\frac{10689\kappa^5}{1024} - \frac{23405\kappa^3}{2048} - \frac{22709\kappa}{16384}
\end{align}
These coefficients grow factorially at large order:
\begin{equation}
e_k(\kappa) \sim -\frac{\Gamma(k + \kappa)}{\pi (2S_I)^{k+\kappa}} \left( c_0 + \frac{c_1}{k} + \cdots \right) \,,
\end{equation}
where $2S_I = \frac{4}{3}$ is twice the instanton action (the Borel singularity location). This factorial growth with non-alternating signs signals a non-Borel-summable asymptotic series, connecting to the non-perturbative physics discussed in the main text.

%% file: sections/appendix_Pn_decomposition.tex
% \!TEX root = ../DoubleDoubleMain.tex
\section{The \texorpdfstring{$P_n$}{Pn} decomposition: proof and algorithm}
\label{appendix:PicardFuchs}
\label{app:cohomological-reduction}
\label{app:polynomial-master}
In Exact WKB, at the lowest order we need to compute elliptic integrals over the classical momentum $P_0 = \sqrt{2V(x)-2E}$. Subsequently, we need to integrate over the momenta $P_{2m}$ which arise from the recursive solution to the Riccati equation. These $P_{2m}$ are rational functions of $P_0(x,E)$ and $x$, and directly integrating over them would require computing a new elliptic integral at each order. The key simplifying feature which allows these integrals to be done algebraically is that the $P_{2m}$ can be written as
\begin{equation}
P_{2m}(x,E)= a_{2m}(E)\, P_0(x,E) + b_{2m}(E)\, \partial_E P_0(x,E) + \partial_x f_{2m}(x,E)
\label{eq:Pn_decomp_app}
\end{equation}
where $a_{2m}$ and $b_{2m}$ are rational functions of $E$ alone and $f_{2m}(x,E)$ is single-valued. 
Integrating~\eqref{eq:Pn_decomp_app} over a closed cycle~$\gamma$ then gives
\begin{equation}
S_\gamma^{2m} = -\frac{1}{2}\oint_\gamma P_{2m}(x,E)\, dx
= \big[a_{2m}(E) + b_{2m}(E)\partial_E\big] S_\gamma^{0} 
\label{eq:Sn_decomp}
\end{equation}
with the \emph{same} coefficients $a_{2m}$, $b_{2m}$ for every cycle. Thus for each cycle, only the leading period and its energy derivative are needed to obtain the even WKB corrections to all orders.

This appendix explains why the decomposition exists, how to compute it, and works out the lowest-order decompositions. The general result is due to Ba\c{s}ar, Dunne, and \"Unsal~\cite{BasarDunneUnsal2017}; what follows is a self-contained account adapted to our conventions. An implementation that computes $a_{2m}$, $b_{2m}$, and $f_{2m}$ to arbitrary order is included in an ancillary file on arXiv.org.
\subsection{Existence of the \texorpdfstring{$P_n$}{Pn} decomposition}

To prove the existence of Eq.~\eqref{eq:Pn_decomp_app}
we rely on two different approaches: a topological one (the genus of the classical curve) and an analytic one (the Picard-Fuchs equation). Together they explain why exactly \emph{two} basis functions appear.

The classical energy shell for the double-well potential $V(x) = \frac{1}{8}(x^2-1)^2$ defines a curve $y^2 =Q(x,E)$ where 
\begin{equation}
Q(x,E) =P_0(x,E)^2= 2V(x) - 2E = \tfrac{1}{4}(x^2-1)^2 - 2E\,,
\end{equation}
Since $Q$ is a quartic polynomial in $x$, $y^2=Q$ is an elliptic curve of genus~$g = 1$. The WKB coefficients $P_n$ are rational functions of $x$ and $y$ on this curve, and we are interested in the meromorphic differentials $\omega_n = P_n\, dx$. Two such differentials are equivalent for period integrals if they differ by an exact form $dF$. The equivalence classes form a vector space whose dimension is $2g = 2$ with natural basis
\begin{equation}
\Omega_1 = y\, dx = P_0\, dx \quad \text{and} \qquad
\Omega_2 = -\frac{dx}{y} = \partial_E P_0\, dx ,
\end{equation}
where we used $\partial_E P_0 = \partial_E(2V - 2E)^{1/2} = -1/P_0$. Since the cohomology is two-dimensional, any meromorphic differential $\omega_n$ must be a linear combination of $\Omega_1$ and $\Omega_2$ modulo an exact form:
\begin{equation}
\omega_n = a_n(E)\,\Omega_1 + b_n(E)\,\Omega_2 + dF_n\,.
\end{equation}
Translating back from differentials to functions gives~\eqref{eq:Pn_decomp_app}. 

It is useful to separate the ingredients in this argument.  The genus-one statement is a statement about the topology of the classical curve: a torus has two independent one-cycles, so its period data are two-dimensional.  This is why two basis periods are enough.  The detailed form of the Riccati solutions is not what drives the cohomology reduction.  The Riccati recursion is used only to ensure that each WKB correction $P_n\,dx$ is a meromorphic differential, equivalently a rational function of $x$ and $y=P_0$ times $dx$.  Once that is true, the genus-one cohomology forces its periods to reduce to the two basis periods above, with coefficients $a_n(E)$ and $b_n(E)$ that are rational functions of the energy.

The genus argument guarantees two terms, but one can also arrive at this conclusion analytically. The WKB recursion generates $P_{2m}$ as a polynomial in $x$ divided by an odd power of~$P_0$. After integration over a cycle and using the identity
\begin{equation}
\frac{1}{P_0^{2k-1}} = -\frac{1}{(2k-3)!!}\, \partial_E^k P_0\,,
\label{eq:inverse_power_identity}
\end{equation}
(which follows by repeated differentiation of $P_0 = (2V-2E)^{1/2}$ with respect to~$E$), the cycle integral becomes a differential operator in $E$ acting on~$S_\gamma$:
\begin{equation}
S_\gamma^{(2m)} = \mathcal{D}_E^{(m)}\, S_\gamma(E)\,.
\label{eq:Sn_diffop}
\end{equation}
The operator $\mathcal{D}_E^{(m)}$ involves $\partial_E^k$ up to $k = 3m$. To reduce it to first order, we need the \textbf{Picard-Fuchs equation}. For the double well, both the leading perturbative action $S_P^0(E) = \pi i E\,{}_2F_1(\frac{1}{4},\frac{3}{4},2;\,8E)$ and the non-perturbative $S_N^0(E) = 32^{-1/2}\pi  (1-8E)\,{}_2F_1(\frac{1}{4},\frac{3}{4},2;\,1-8E)$  satisfy
\begin{equation}
\boxed{E(1-8E)\,\frac{d^2 S}{dE^2} = \frac{3}{2}\,S} \,.
\label{eq:PF_doublewell}
\end{equation}
The two actions are the two linearly independent solutions, corresponding to the two cycles of the torus. This equation expresses $\partial_E^2 S$ as a rational multiple of~$S$:
\begin{equation}
\partial_E^2 S = \frac{3}{2E(1-8E)}\, S\,.
\end{equation}
Differentiating with respect to $E$ gives $\partial_E^3 S$ in terms of $S$ and $\partial_E S$, and by induction every $\partial_E^k S$ reduces to the form $\alpha_k(E)\,S + \beta_k(E)\,\partial_E S$, where $\alpha_k$ and $\beta_k$ are rational with poles only at $E = 0$ and $E = 1/8$. Substituting into~\eqref{eq:Sn_diffop} reduces $\mathcal{D}_E^{(m)}S$ to the two-term form~\eqref{eq:Sn_decomp}.

The fact that the Picard-Fuchs equation is \emph{second}-order is special to the double well (and more generally to the ``Chebyshev'' class of potentials~\cite{BasarDunneUnsal2017}). For a generic genus-$g$ potential, one would need $2g$~basis functions instead of two.

\subsection{Algorithmic decomposition}
\label{subsec:algorithm}

The cohomological reduction can be carried out directly at the level of polynomials, without ever integrating over cycles or applying the Picard-Fuchs equation. The key observation is that the Riccati recursion produces each even correction $P_{2m}$ in the form
\begin{equation}
P_{2m} = \frac{N_{2m}(x,E)}{P_0^k}
\end{equation}
where $k=6m-1$  and $N_{2m}$ is a polynomial in $x$ with $E$-dependent coefficients. Making the ansatz $f_{2m} = R(x,E)/P_0^{k-2}$ for the total-derivative piece and substituting into~\eqref{eq:Pn_decomp_app} gives, after multiplying through by $P_0^k$, the \emph{polynomial master equation}:
\begin{equation}
\boxed{N_{2m}(x,E) = a_{2m}\,(2V - 2E)^{(k+1)/2} - b_{2m}\,(2V - 2E)^{(k-1)/2} + R'(2V - 2E) - (k-2)\,R\,V'}
\label{eq:polynomial-master}
\end{equation}
where we used $P_0^2 = 2V - 2E$, $\partial_E P_0 = -1/P_0$, and $P_0 P_0' = V'$.

This is an identity between polynomials in $x$. The left-hand side is known from the Riccati recursion. The right-hand side involves the unknowns $a_{2m}(E)$, $b_{2m}(E)$, and the coefficients of $R(x,E)$. Expanding both sides in powers of $x$ and equating coefficients gives a finite linear system that determines all unknowns. The cohomological existence argument guarantees a solution.

In summary, obtaining the full expansion of $S_P$ and $S_N$ only requires knowledge of the coefficients $a_{2m}, b_{2m}$ at each even order~$2m$. Those are reconstructed following the given algorithm:
\begin{enumerate}
\item Compute $P_{2m}$ from the Riccati recursion, keeping the result in the form $N_{2m}/P_0^k$. The numerator is a polynomial in $x$ and $E$.
\item Set up the polynomial $R(x,E)$ with undetermined coefficients. The degree of $R$ is fixed by matching the highest power of~$x$ on both sides of~\eqref{eq:polynomial-master}.
\item Expand the master equation~\eqref{eq:polynomial-master} in powers of $x$ and solve the resulting linear system for $a_{2m}$, $b_{2m}$, and the coefficients of~$R$.
\end{enumerate}
This is entirely algebraic, with no mention of elliptic integrals or special functions. The genus-one topology enters only through the guarantee that the linear system is solvable; the algorithm itself is easily implemented. The poles of $a_{2m}(E)$ and $b_{2m}(E)$ occur only at $E = 0$ and $E = 1/8$, the energies where turning points coalesce and the elliptic curve degenerates. Away from these values the reduction is regular.

For example, from the Riccati recursion with $V = \frac{1}{8}(x^2-1)^2$, we have $P_2 = N_2/P_0^5$ with
\begin{equation}
N_2(x,E) = \tfrac{1}{32}\bigl[-2x^6 + 3x^4 - 24Ex^2 + 8E - 1\bigr]\,.
\end{equation}
The master equation~\eqref{eq:polynomial-master} with $k = 5$ reads
\begin{equation}
N_2 = a_2\,(2V-2E)^3 - b_2\,(2V-2E)^2 + R'(2V-2E) - 3\,R\,V'\,,
\end{equation}
where degree counting gives $\deg(R) = 9$. Expanding in powers of $x$ and solving the linear system yields
\begin{equation}
a_2(E) = \frac{12E - 1}{16E(1-8E)}\,, \qquad b_2(E) = \frac{1}{8}\,.
\label{eq:a2b2}
\end{equation}
A convenient explicit primitive for the total-derivative remainder is
\begin{equation}
f_2(x,E) = \frac{1}{32}\!\left[
\frac{10}{3}\frac{x^3-x}{P_0^3}
+ \frac{t_1 x + t_3 x^3}{P_0}
+ u_1\, x\, P_0
\right],
\label{eq:f2-explicit}
\end{equation}
with $t_1 = \frac{16E-1}{3E(8E-1)}$, $t_3 = \frac{1-12E}{3E(8E-1)}$, and $u_1 = \frac{2(12E-1)}{3E(8E-1)}$, and one may verify~\eqref{eq:Pn_decomp_app} directly by differentiation.

The same algorithm extends to arbitrary order. An ancillary file on arXiv.org implements this algorithm and has been verified through~$P_{24}$. For example,
\begin{equation}
a_4 = \frac{11280E^3 - 5370E^2 + 663E - 28}{7680\,E^3(8E-1)^3}\,,\qquad
b_4 = -\frac{360E^2 - 115E + 7}{1920\,E^2(8E-1)^2}\,,
\label{eq:a4b4}
\end{equation}
with analogous (lengthier) expressions for $a_6,\ldots,a_{10}$ and $b_6,\ldots,b_{10}$.

%% file: sections/appendix_Weber_v2.tex
% \!TEX root = ../DoubleDoubleMain.tex
\section{Resummation of the nonperturbative action at fixed \texorpdfstring{$\kappa$}{kappa}}
\label{appendix:Weber}
The expansion of the action in Eq.~\eqref{eq:Sn_decomp} using the Picard--Fuchs decomposition:
\begin{equation}
S_\gamma^{2m} = \big[a_{2m}(E) + b_{2m}(E)\partial_E\big] S_\gamma^{0} 
\end{equation}
is computed as a series in $\hbar$ from the Riccati expansion at fixed~$E$. The coefficients $a_{2m}(E)$ and $b_{2m}(E)$ are rational functions of~$E$ and have poles at $E=0$, as can be seen in Eqs.~\eqref{eq:a2b2} and~\eqref{eq:a4b4}. The perturbative action $S_P^0(E)$ has a zero at $E=0$, but the non-perturbative action $S_N^0(E)$ does not. Since $E\sim \hbar$ the poles in the $a_{2m}$ and $b_{2m}$ make all the terms in the expansion the same order in $\hbar$ and the series needs to be resummed. The cause of the failure is that $E=0$ is at the bottom of the wells, so the four classical turning points collide there in two pairs: $(x_1,x_2)\to -1$ and $(x_3,x_4)\to +1$. The perturbative cycles around $(x_1,x_2)$ and $(x_3,x_4)$ contract at these collisions and have smooth small-$E$ limits. The non-perturbative cycle runs from $x_2$ to $x_3$ and remains finite at leading order, but its endpoints become double turning points; this endpoint degeneration produces the pole tower in the higher WKB corrections to $S_N$.

Substituting $E = E_P  = \hbar\kappa+\cdots$ and re-expanding in~$\hbar$ at fixed~$\kappa$ we can write
\begin{equation}
  S_N(E_P,\hbar) = S_I + \sum_{k=1}^\infty \hbar^k\, s_k(\kappa,\hbar)\,,
  \label{eq:SN_hk}
\end{equation}
where $S_I = S_N^{(0)}(0) = \frac{2}{3}$ is the instanton action.  Each coefficient $s_k$ receives two types of contribution:
\begin{equation}
  s_k(\kappa,\hbar) = \tilde s_k(\kappa) + r_k(\kappa,\hbar)\,.
  \label{eq:hk_split}
\end{equation}
The pole tower $\tilde s_k(\kappa)$ sums the $(k{-}2m)$-th Laurent coefficient of $S_N^{(2m)}(E)$ at $E=0$ over all WKB orders, producing a divergent series in inverse powers of~$\kappa$.  The remainder $r_k$ collects finitely many contributions from the regular parts of $S_N^{(2m)}$ at low orders and from the perturbative energy re-expansion $E = \hbar\kappa + \hbar^2 e_2 + \cdots$; these are straightforward to compute.  The coefficients $s_k$ are functions of~$\kappa$, possibly with explicit $\ln\hbar$ dependence before cancellations.  The nontrivial step at each order is resumming~$\tilde s_k$: in practice we compute enough tower coefficients from the Picard--Fuchs recursion for $a_{2m}(E)$ and $b_{2m}(E)$ to identify the pattern as a known asymptotic expansion of a special function.

\subsection{Leading order: the Weber equation and \texorpdfstring{$\ln\Gamma$}{logGamma}}
\label{app:leading_Weber}
The resummation of the leading pole tower for $S_N$ can be done analytically, using the uniform-WKB approach of Dunne and \"Unsal~\cite{DunneUnsal2014WKB}. For higher pole towers we resort to identifying the series pattern and summing it.  Uniform WKB is based on the observation that the singular behavior of $S_N$ is due to turning points colliding. When turning points are isolated, expanding near them leads to a linear potential and the Airy function. This is how Exact WKB was constructed and detailed in Section~\ref{sec:exactWKB}. When two turning points collide, the local potential is instead quadratic and the local equation is the Weber equation and the solutions are not Airy functions but parabolic cylinder functions.

Setting $y = x+1$ and expanding $V \approx \tfrac12 y^2$ around the well bottom, the leading-order Schr\"odinger equation reads
\begin{equation}
-\frac{\hbar^2}{2}\psi''(y) + \frac{1}{2} y^2\,\psi(y) = \hbar\kappa\,\psi(y)\,,\qquad E = \hbar\kappa\,.
\label{eq:Weber_app}
\end{equation}
This is the Weber equation; its solutions are parabolic cylinder functions $D_\nu(\zeta)$ where $\nu = \kappa - 1/2$ and $\zeta = \sqrt{2/\hbar}\,y$. The bound states are at $\nu=N$, i.e. $\kappa = N+1/2$, for $N = 0,1,2,\ldots$, recovering the harmonic spectrum.  At large $|\zeta|$, $D_\nu$ has the two-term asymptotic expansion~\cite{DLMF}
\begin{equation}
D_\nu(\zeta) \;\sim\; \zeta^\nu\,e^{-\zeta^2/4} \;+\; e^{\pm i\pi\nu}\,\frac{\sqrt{2\pi}}{\Gamma(-\nu)}\,\zeta^{-\nu-1}\,e^{+\zeta^2/4}\,,
\label{eq:Dnu_resurgent}
\end{equation}
where the relative weight of the second saddle is set by the Stokes coefficient $\sqrt{2\pi}/\Gamma(-\nu)$ --- an intrinsic property of $D_\nu$, universal to any harmonic turning-point collision, independent of the global form of~$V$.

In uniform WKB~\cite{DunneUnsal2014WKB}, the double-well wave function is written with a coordinate change matching Weber to the global potential: $\psi(y) = D_\nu(u(y)/\sqrt\hbar)/\sqrt{u'(y)}$ where $u(y) = u_0(y) + \hbar\,u_1(y) + \cdots$.  Substituting this ansatz into the Schr\"odinger equation produces an exact nonlinear ODE for $u(y)$, solved recursively in~$\hbar$ via a Riccati-type expansion analogous to the standard-WKB recursion for $P(x)$.  At leading order, $u_0\,u_0' = 2\sqrt{2V}$, equivalently $u_0^2(y) = 4\int_0^y \sqrt{2V}\,dy'$; at the barrier midpoint $y=1$ (i.e.\ $x=0$), $u_0^2(1)/2 = S_I = 2/3$ is the instanton action between the wells.  Imposing parity at the midpoint and using~\eqref{eq:Dnu_resurgent} to balance the two pieces of $D_\nu$ produces, after using the pole of $\Gamma(-\nu)$ at integer $\nu = N$, a level splitting
\begin{equation}
\delta E^{(N)}_\pm \;=\; \pm\,\hbar\,\frac{(C/\hbar)^\kappa}{\sqrt{2\pi}\,\Gamma(\kappa+\frac{1}{2})}\,e^{-S_I/\hbar}\,\bigl(1+\cO(\hbar)\bigr)\,,
\label{eq:delta_E_struct}
\end{equation}
where $C$ is a matching constant independent of $\hbar$.  The exponential $e^{-S_I/\hbar}$ comes from the midpoint value of $u_0$, while $\Gamma(\kappa+1/2)$ comes from the pole expansion of the Stokes coefficient in~\eqref{eq:Dnu_resurgent}.  Via the exact-WKB quantization $\VP = -1 \pm i\sqrt{\VN}$, this translates into the same structural form for $\sqrt{\VN}$.

The constant $C$ is not fixed by the local Weber equation alone; it is a global matching constant.  Rather than computing the full uniform-WKB matching here, we fix it from the global classical action: at leading WKB order $\sqrt{\VN}^{(0)} = e^{-S_N^{(0)}(E)/\hbar}$, with
\begin{equation}
S_N^{(0)}(E) = \tfrac{2}{3} + E\bigl[\ln(E/8) - 1\bigr] + \cO(E^2)
\label{eq:SN0_local}
\end{equation}
from the Picard--Fuchs expansion (Appendix~\ref{appendix:PicardFuchs}).  At $E = \hbar\kappa$,
\begin{equation}
e^{-S_N^{(0)}(\hbar\kappa)/\hbar} \;=\; e^{-S_I/\hbar}\,(8/\hbar)^\kappa\,\kappa^{-\kappa}\,e^\kappa\,\bigl(1+\cO(\hbar)\bigr)\,,
\label{eq:SN0_expanded}
\end{equation}
and Stirling's formula $\sqrt{2\pi}/\Gamma(\kappa+\tfrac12)\sim\kappa^{-\kappa}\,e^\kappa$ at large $\kappa$ identifies the matching constant uniquely as $C=8$, giving
\begin{equation}
\boxed{\;\sqrt{\VN} \;=\; e^{-S_I/\hbar}\,\frac{\sqrt{2\pi}}{\Gamma(\kappa+\tfrac12)}\,\Bigl(\frac{8}{\hbar}\Bigr)^{\!\kappa}\,\bigl(1+\cO(\hbar)\bigr)\,.\;}
\label{eq:VN_Weber}
\end{equation}
Each factor has a transparent origin: $e^{-S_I/\hbar}$ from $\exp(-u_0^2(1)/(2\hbar))$ at the midpoint; $\sqrt{2\pi}/\Gamma(\kappa+\tfrac12)$ from the Stokes coefficient in~\eqref{eq:Dnu_resurgent}, intrinsic to $D_\nu$; $(8/\hbar)^\kappa$ pinned by the global $S_N^{(0)}$.  The $(1+\cO(\hbar))$ remainder collects everything beyond the local Weber model: anharmonic corrections to $V$, the global matching, and the perturbative energy re-expansion $E = \hbar\kappa + \hbar^2 e_2 + \cdots$.  These generate $s_2, s_3, \ldots$.

The result~\eqref{eq:VN_Weber} can be checked order by order in $\hbar$ by computing the leading $1/E^{2m-1}$ pole of $S_N^{(2m)}$ directly from Picard--Fuchs: writing
\begin{equation}
S_N^{(2m)}(E) = \frac{\alpha_{2m}}{E^{2m-1}} + \text{(less singular)}\,,\qquad
\alpha_{2m} = \tfrac{2}{3}\,\mathrm{Res}_{E=0}\!\bigl[E^{2m-2}\,a_{2m}(E)\bigr]\,,
\end{equation}
the residues evaluate to $\alpha_2 = -1/24$, $\alpha_4 = 7/2880$, $\alpha_6 = -31/40320$, $\alpha_8 = 127/215040$, $\ldots$, matching the Bernoulli identity $\alpha_{2m} = B_{2m}(\tfrac12)/[2m(2m-1)]$ (we have checked to $m=12$, i.e. up to $a_{24}$ and $b_{24}$).  This is precisely the coefficient pattern in Stirling's asymptotic for $\ln\Gamma(\kappa+\tfrac12)$,
\begin{equation}
\ln\frac{\Gamma(\kappa+\tfrac12)}{\sqrt{2\pi}} - \kappa\ln\kappa + \kappa \;\sim\; \sum_{m\ge 1} \frac{B_{2m}(\tfrac12)}{2m(2m-1)}\,\kappa^{1-2m}\,,
\label{eq:Stirling_check}
\end{equation}
confirming the resummation of all leading poles into the closed-form $\Gamma$.

Finally, taking $S_N = -\hbar\ln\sqrt{\VN}$ in~\eqref{eq:VN_Weber}:
\begin{equation}
s_1(\kappa,\hbar) = \underbrace{\ln\frac{\Gamma(\kappa+\tfrac12)}{\sqrt{2\pi}}}_{\tilde s_1(\kappa)} \,\underbrace{-\,\kappa\ln(8/\hbar)}_{r_1(\kappa,\hbar)}\,.
\label{eq:SN_logGamma}
\end{equation}
$\tilde s_1$ is the resummed leading-pole tower (the $\Gamma$ with its Bernoulli tail), and $r_1=-\kappa\ln(8/\hbar)$ is the $\hbar^1$ piece of the regular Taylor of $S_N^{(0)}$ not already inside Stirling\footnote{Starting from Eq.~\eqref{eq:Stirling_check}, the $\kappa-\kappa \ln \kappa$ term from resumming the leading-pole tower mostly cancels against the $\kappa \ln(\kappa \hbar /8)-\kappa$ term from the expansion of $S_N^{(0)}$ (see Eq.~\eqref{SNexpanded}). The remainder is the $-\kappa \ln (8/\hbar)$ contribution.}. 

Two remarks are in order before moving on.  First, this derivation never used a period integral.  The instanton action $S_I = \int_0^1 \sqrt{2V}\,dy = 2/3$ is an elementary integral of a polynomial, and the matching factor $(8/\hbar)^\kappa$ was fixed from the local Taylor expansion of $S_N^{(0)}$ at $E=0$, which is itself a series solution of the Picard--Fuchs equation $E(1-8E)S''=(3/2)S$ near its regular singular point.  The same is true at subleading orders: uniform WKB extends to higher $\hbar$ by solving the global ODE for $u(y) = u_0 + \hbar u_1 + \hbar^2 u_2 + \cdots$ to higher order, and the next coefficients $s_2, s_3, \ldots$ are extracted from $u_n(1)$ at the barrier midpoint without any cycle integral~\cite{DunneUnsal2014WKB}.  In Sections~\ref{appendix:subleading} and~\ref{appendix:hbar3} below we instead use the algorithmically simpler Picard--Fuchs/Riccati route via the $a_{2m}(E), b_{2m}(E)$, but in principle the entire trans-series can be built from the local Weber data and the coordinate change $u(y)$. Second, the same $(8/\hbar)^\kappa/\Gamma(\kappa+\tfrac12)$ structure arises in the path-integral derivation of the splitting $\Delta_N$ (Section~\ref{sec:NP_excited_state}, Eq.~\eqref{eq:DeltaN_action}) with no resummation and no concern for the colliding turning points at all: it arises from the finite-$T$ classical action around the exact $n=1$ instanton saddle, before any $\hbar$ expansion. Thus there are three independent and complementary routes to the same result: the local Weber equation, the Picard--Fuchs expansion, and the path integral.

\subsection{Order \texorpdfstring{$\hbar^2$}{hbar2}: Bernoulli patterns and the digamma function}
\label{appendix:subleading}
Beyond leading order, we do not use the Weber equation.  Instead, we compute the tower coefficients $a_{2m}$ and $b_{2m}$ directly and identify the Bernoulli-number structure of the resulting sequence to determine the closed-form resummed expressions. The subleading pole tower
\begin{equation}
  \tilde s_2(\kappa) = \sum_{m=2}^\infty c_m\,\kappa^{2-2m}
\end{equation}
starts at $S_N^{(4)}$: $c_m$ is the coefficient of $E^{2-2m}$ in $S_N^{(2m)}(E)$, since $\hbar^{2m}E^{2-2m}=\hbar^2\kappa^{2-2m}$.  The finite and logarithmic terms in $S_N^{(2)}$ are included in the remainder $r_2$.  For the genuine pole coefficients the logarithms from $a_{2m}\,S_N^{(0)}$ cancel those from $b_{2m}\,\partial_E S_N^{(0)}$, so $\tilde s_2(\kappa)$ is a clean power series.  Computing from the Picard--Fuchs recursion gives
\begin{equation}
  c_m = \frac{3(2-4^{1-m})}{16m}\,B_{2m} - \frac{1}{32(m-1)}\,B_{2m-2}\!\left(\tfrac{1}{2}\right),
  \label{eq:cm_formula}
\end{equation}
verified through $m=12$ (i.e., $c_2$ through $c_{12}$, computed from $a_4,b_4,\ldots,a_{24},b_{24}$).  To identify the closed form, we use the identity $B_{2\ell}(\frac{1}{2}) = (2^{1-2\ell}-1)\,B_{2\ell}$ to express everything in ordinary Bernoulli numbers, and compare with the asymptotic expansion of the digamma function
\begin{equation}
  \psi\!\left(\kappa+\tfrac{1}{2}\right) - \ln\kappa \sim -\sum_{\ell=1}^\infty \frac{B_{2\ell}(\frac{1}{2})}{2\ell}\,\kappa^{-2\ell}\,.
  \label{eq:digamma_asy}
\end{equation}
Multiplying by $(12\kappa^2+1)/16 = \frac{3}{4}\kappa^2 + \frac{1}{16}$ and collecting coefficients, the $\kappa^2$ factor shifts the summation index to produce the $B_{2m}$ contribution in~\eqref{eq:cm_formula}, while the constant produces the $B_{2m-2}(1/2)$ contribution.  The combination reproduces~\eqref{eq:cm_formula} exactly, giving
\begin{equation}
  \tilde s_2(\kappa) = \frac{12\kappa^2+1}{16}\left[\psi\!\left(\kappa+\tfrac{1}{2}\right) - \ln\kappa\right] - \frac{1}{32}\,.
  \label{eq:h2_tower}
\end{equation}

The remainder $r_2$ comes from expanding $S_N^{(0)}(E(\hbar))$ with the energy correction $E = \hbar\kappa + \hbar^2(-\frac{3}{4}\kappa^2 - \frac{1}{16}) + \cdots$ and extracting the finite part of $\hbar^2\,S_N^{(2)}$. It reads
\begin{equation}
  r_2(\kappa) = \frac{17}{8}\kappa^2 + \frac{11}{48} + \frac{12\kappa^2+1}{16}\bigl(\ln\kappa - \psi(\kappa+\tfrac{1}{2})\bigr)\,.
  \label{eq:r2}
\end{equation}
Adding $\tilde s_2 + r_2$, the $\psi - \ln\kappa$ terms cancel algebraically, and the full coefficient $s_2$ in~\eqref{eq:SN_hk} is simply a polynomial:
\begin{equation}
  \boxed{
  s_2(\kappa) = \frac{17}{8} \kappa^2 + \frac{19}{96}
  }\,.
    \label{eq:h2_result}
\end{equation}

\subsection{Order \texorpdfstring{$\hbar^3$}{hbar3}: digamma and trigamma}
\label{appendix:hbar3}

At $\cO(\hbar^3)$ the pole tower $\tilde s_3(\kappa) = \sum_{m=2}^{\infty} d_m\,\kappa^{3-2m}$ is again log-free (verified through $m=12$), so it is a clean series in odd inverse powers of~$\kappa$. The identification now requires two building blocks: the digamma expansion~\eqref{eq:digamma_asy} (which produces even inverse powers, shifted to odd by an odd-power polynomial prefactor) and the trigamma function $\psi'(\kappa+\frac{1}{2})$, whose asymptotic expansion
\begin{equation}
  \psi'\!\left(\kappa+\tfrac{1}{2}\right) \sim \frac{1}{\kappa} + \sum_{\ell=1}^\infty B_{2\ell}\!\left(\tfrac{1}{2}\right)\,\kappa^{-(2\ell+1)}
  \label{eq:trigamma_asy}
\end{equation}
produces odd inverse powers.  Multiplying~\eqref{eq:digamma_asy} by a cubic $\frac{35}{16}\kappa^3 + \frac{25}{64}\kappa$ gives two Bernoulli contributions at each order, while multiplying~\eqref{eq:trigamma_asy} by the quartic $(12\kappa^2+1)^2/512$ gives three.  Five basis functions are needed and five suffice to determine all coefficients, including the polynomial part:
\begin{equation}
  \tilde s_3(\kappa) = \left(\frac{35}{16}\kappa^3 + \frac{25}{64}\kappa\right)\!\left[\psi\!\left(\kappa+\tfrac{1}{2}\right) - \ln\kappa\right]
  + \frac{(12\kappa^2+1)^2}{512}\,\psi'\!\left(\kappa+\tfrac{1}{2}\right)
  - \frac{9}{32}\kappa^3 - \frac{11}{96}\kappa\,.
  \label{eq:h3_closed}
\end{equation}
The polynomial $-\frac{9}{32}\kappa^3 - \frac{11}{96}\kappa$ is fixed by the requirement that $\tilde s_3(\kappa) \to 0$ as $\kappa\to\infty$ (matching the tower, whose terms are all negative powers of~$\kappa$): the $\psi'$ term would otherwise produce $\cO(\kappa^3)$ growth, and the next terms from both $\psi-\ln\kappa$ and $\psi'$ would produce $\cO(\kappa)$ growth.  The polynomial cancels both.  The full formula is verified by matching asymptotic expansions through $\kappa^{-21}$ (eleven coefficients, six over-determined beyond the five-parameter fit), and predicts the tower coefficients at all higher orders.

The regular terms needed at this order are the $E^3$ coefficient of $S_N^{(0)}$ and the $E^1$ coefficient of $S_N^{(2)}$:
\begin{equation}
  S_N^{(0)}(E) = \cdots + E^3\left(\frac{35}{16}\ln\frac{E}{8} + \frac{59}{8}\right) + \cdots,
  \qquad
  S_N^{(2)}(E) = \cdots + E\left(\frac{25}{64}\ln\frac{E}{8} + \frac{605}{384}\right) + \cdots.
  \label{eq:S0S2_h3_input}
\end{equation}
Setting $E=\hbar\kappa$ gives $(\frac{35}{16}\kappa^3+\frac{25}{64}\kappa)\ln\frac{\kappa\hbar}{8}+\frac{59}{8}\kappa^3+\frac{605}{384}\kappa$.  As at $\cO(\hbar^2)$, we must also include the energy correction from $E_P = \hbar\kappa + \hbar^2 e_2 + \hbar^3 e_3 + \cdots$, with $e_2=-(12\kappa^2+1)/16$ and $e_3=-(68\kappa^3+19\kappa)/64$.  At $\cO(\hbar^3)$ this contributes three additional pieces:
\begin{itemize}
\item $e_3\bigl[\psi(\kappa+\tfrac{1}{2})-\ln(8/\hbar)\bigr]$ from expanding $\hbar\ln\Gamma(E_P/\hbar+\frac{1}{2})$,
\item $\frac{e_2^2}{2}\,\psi'(\kappa+\tfrac{1}{2})$ from the second-order Taylor expansion of $\ln\Gamma$,
\item $e_2\,\frac{d s_2^{\rm raw}}{d\kappa}$ from the $\cO(\hbar)$ correction to $\partial_E S_N$.
\end{itemize}
Here $s_2^{\rm raw} = [(1+12\kappa^2)/16][\psi(\kappa+\tfrac{1}{2})+\ln(\hbar/8)] + \frac{17}{8}\kappa^2+\frac{19}{96}$ denotes the $\hbar^2$ coefficient before the energy re-expansion (i.e.\ at $E=\hbar\kappa$).  For the full remainder, including all energy-correction pieces, we find
\begin{equation}
  r_3(\kappa,\hbar) =
  \left(\frac{35}{16}\kappa^3 + \frac{25}{64}\kappa\right)\!
  \left[\ln\kappa - \psi\!\left(\kappa+\tfrac{1}{2}\right)\right]
  - \frac{(12\kappa^2+1)^2}{512}\,\psi'\!\left(\kappa+\tfrac{1}{2}\right)
  + \frac{67}{16}\kappa^3 + \frac{503}{384}\kappa\,.
  \label{eq:r3}
\end{equation}
Adding $\tilde s_3 + r_3$, \emph{all} special functions cancel: the digamma cancels between the tower's $(\psi-\ln\kappa)$ and the remainder's $(\ln\kappa-\psi)$, and the trigamma cancels between the tower and the energy correction.  The full coefficient $s_3$ in~\eqref{eq:SN_hk} is therefore a pure polynomial, just like $s_2$:
\begin{equation}
  \boxed{
  s_3(\kappa) = \frac{125}{32}\,\kappa^3 + \frac{153}{128}\,\kappa
  }\,.
  \label{eq:h3_full}
\end{equation}

%% file: sections/appendix_Lame_propagator.tex
% !TEX root = ../DoubleDoubleMain.tex

%%%%%%%%%%%%%%%%%%%%%%%%%%%%%%%%%%%%%%%%%%%%%%%%%%%%%%%%%%%%%%%%%%%%%%%%%%%%%%%
%%%%%.    APPENDIX: LAMÉ EQUATION SPECTRUM                              %%%%%%%
%%%%%%%%%%%%%%%%%%%%%%%%%%%%%%%%%%%%%%%%%%%%%%%%%%%%%%%%%%%%%%%%%%%%%%%%%%%%%%%

\section{Spectrum of the Lam\'e equation}
\label{appendix:Lame}

In Section~\ref{sec:morse} we identified the Lam\'e operator as characterizing fluctuations around the exact instanton saddles. In this appendix we review some aspects of its spectrum that are relevant to the body of the paper: the band-edge eigenfunctions and eigenvalues, the reduction to the P\"oschl--Teller spectrum at $T=\infty$, and the Floquet counting that fixes the Morse index of the multi-instanton saddles.

%%%%%%%%%%%%%%%%%%%%%%%%%%%%%%%%%%%%%%%%%%%%%%%%%%%%%%%%%%%%%
\subsection{The fluctuation operator as a Lam\'e equation}
\label{app:Lame_identification} 
%%%%%%%%%%%%%%%%%%%%%%%%%%%%%%%%%%%%%%%%%%%%%%%%%%%%%%%%%%%%%
For a single instanton, the exact saddle, with anti-periodic boundary conditions on a Euclidean circle of circumference $T$ is given in Eq.~\eqref{xCIfirst} as
\begin{equation}
x_\cI(t) = A_\sig\sn(u,\sig^2),\quad
u = \frac{t}{\sqrt{2(1+\sig^2)}}
  , \qquad
  A_\sig = \sqrt{\frac{2\sig^2}{1+\sig^2}}\,,
  \label{eq:H_exact_saddle}
\end{equation}
where $\sn(u,\sigma^2)$ is the Jacobi elliptic sine with modulus~$\sig$, and $K(\sig^2)=\int_0^{\pi/2}d\theta/\sqrt{1-\sig^2\sin^2\theta}$ is the complete elliptic integral of the first kind. Since $\sn(u+2K(\sigma^2),\sigma^2)=-\sn(u,\sigma^2)$ the quantization condition  is
\begin{equation}
  T = 2K(\sig^2)\,\sqrt{2(1+\sig^2)}\,.
  \label{eq:H_T_from_K}
\end{equation}
In the limit $\sig\to 1$, $K(\sig^2)\to\infty$, $T\to\infty$, and $\sn(u,1)=\tanh u$, recovering the dilute-gas $\tanh(t/2)$ instanton. The fluctuation operator around $x_\cI$ is $\opO_\cI = -\partial_t^2 + V''[x_\cI(t)]$ where
\begin{equation}
   V''(x_\cI(t))
  =  \frac{1}{2}(3x_\cI^2-1)
  =  \frac{1}{2}\!\left(\frac{6\sig^2}{1+\sig^2}\,\sn^2(u,\sig^2) - 1\right)\,.
  \label{eq:H_fluct_op}
\end{equation}
With the change of variable $u = t/\sqrt{2(1+\sig^2)}$, so that $\partial_t^2 = [2(1+\sig^2)]^{-1}\partial_u^2$, the eigenvalue equation $\opO_\cI\psi = \lambda\psi$ becomes
\begin{equation}
  -\frac{1}{2(1+\sig^2)}\,\psi''(u)
  + \frac{1}{2}\!\left(\frac{6\sig^2}{1+\sig^2}\,\sn^2(u,\sigma^2) - 1\right)\psi(u)
  = \lambda\,\psi(u)\,.
\end{equation}
Multiplying by $2(1+\sig^2)$ and rearranging:
\begin{equation}
  \boxed{\;
  -\psi''(u) + 6\sig^2\,\sn^2(u,\sig^2)\,\psi(u) = \mathcal{E}\,\psi(u)\,,
  \qquad
  \mathcal{E} = (1+\sig^2)(1+2\lambda)}\,,
  \label{eq:H_Lame}
\end{equation}
or equivalently,
\begin{equation}
  \lambda = \frac{\mathcal{E}-(1+\sig^2)}{2(1+\sig^2)}\,.
  \label{eq:H_lambda_from_E}
\end{equation}
Equation~\eqref{eq:H_Lame} is the \textbf{Lam\'e equation} with index $n_L=2$ (since $n_L(n_L+1)=6$) in Jacobi form.  The Lam\'e equation is one of the classical exactly solvable equations in mathematical physics, and its spectral theory is completely known~\cite{DLMF,WhittakerWatson}. It is sometimes written in the equivalent Weierstrass form $\opO_\cI = -\partial_t^2 + 6\,\tilde\wp(t) + \tfrac{1}{2}$, obtained via $x^2 = 4\tilde\wp+2/3$ with invariants $\tilde g_2 = \tfrac{1}{12}-2\varepsilon$, $\tilde g_3 = -\tfrac{1}{216}-\tfrac{\varepsilon}{3}$.

The same derivation applies to fluctuations about any real multi-instanton saddle $x_{k,0}$ with $k\ge 1$. Each such saddle traverses the same classical orbit as the single instanton $k$ times, so $x_{k,0}^2(t) = A_\sigma^2 \,\mathrm{sn}^2(u,\sigma^2)$ has exactly the same functional form, and the eigenvalue equation reduces to the same Lam\'e equation~\eqref{eq:H_Lame}. What changes is the quantization condition relating $\sigma$ to $T$: for $x_{k,0}$,
\begin{equation}
  T = 2k\,\omega_N(\varepsilon) = 4k\,K(\sigma^2)\sqrt{2(1+\sigma^2)}\,,
\end{equation}
in place of $T = 2K(\sigma^2)\sqrt{2(1+\sigma^2)}$ for the single instanton, so the saddle covers $2k$ Lam\'e fundamental cells. The corresponding boundary condition is periodic for $k\ge 1$ (rather than anti-periodic for the single instanton); how this selects which Lam\'e modes appear in the spectrum is handled by the Floquet counting of Section~\ref{app:Floquet_counting}.

%%%%%%%%%%%%%%%%%%%%%%%%%%%%%%%%%%%%%%%%%%%%%%%%%%%%%%%%%%%%%
\subsection{Band-edge spectrum}
\label{app:Lame_spectrum}
%%%%%%%%%%%%%%%%%%%%%%%%%%%%%%%%%%%%%%%%%%%%%%%%%%%%%%%%%%%%%

For the Lam\'e equation with index $n_L=2$, exactly $2n_L{+}1=5$ eigenfunctions are \emph{algebraic} (polynomial in the Jacobi functions) and periodic or anti-periodic with half-period $2K(\sigma^2)$. These five solutions are exact eigenfunctions of $\opO_{k,0}$ around every real multi-instanton saddle, since their Floquet multipliers $\rho=\pm 1$ satisfy the periodicity condition $\rho^{2k}=1$ for any $k$ (with $\sigma$ taking its $k$-dependent value). 
For $k=1$, they are the lowest five eigenvalues. They are \emph{band-edge} solutions that separate the continuous Bloch-Floquet spectrum into three allowed bands and two forbidden gaps. For $k\ge 2$, the condition $\rho^{2k}=1$ admits additional interior Floquet phases $\rho=e^{i\pi j/k}$ with $j=1,\ldots,k-1$, giving extra eigenvalues that are interleaved within each band and that are not algebraic in $\sigma$.

The five band-edge eigenfunctions and eigenvalues, verified by direct substitution into~\eqref{eq:H_Lame}, are
\begin{alignat}{3}
  &\text{(i)}  &&\quad \psi = \operatorname{cn}(u, \sigma^2)\,\operatorname{dn}(u, \sigma^2)\,,
    &&\qquad \mathcal{E}_1 = 1+\sig^2\,,
    \label{eq:H_cndn}\\
  &\text{(ii)}  &&\quad \psi = \sn(u, \sigma^2)\,\operatorname{dn}(u, \sigma^2)\,,
    &&\qquad \mathcal{E}_2 = 1+4\sig^2\,,
    \label{eq:H_sndn}\\
  &\text{(iii)} &&\quad \psi = \sn(u, \sigma^2)\,\operatorname{cn}(u, \sigma^2)\,,
    &&\qquad \mathcal{E}_3 = 4+\sig^2\,,
    \label{eq:H_sncn}\\
  &\text{(iv)} &&\quad \psi = 1 + B_-\,\sn^2(u,\sigma^2)\,,
    &&\qquad \mathcal{E}_- = 2(1+\sig^2)+2\sqrt{1-\sig^2+\sig^4}\,,
    \label{eq:H_poly_minus}\\
  &\text{(v)} &&\quad \psi = 1 + B_+\,\sn^2(u,\sigma^2)\,,
    &&\qquad \mathcal{E}_+ = 2(1+\sig^2)-2\sqrt{1-\sig^2+\sig^4}\,,
    \label{eq:H_poly_plus}
\end{alignat}
where $B_\pm = -(1+\sig^2)\pm\sqrt{1-\sig^2+\sig^4}$ are the roots of $B^2+2(1+\sig^2)B+3\sig^2=0$, obtained by matching coefficients of $1$, $\sn^2$, and $\sn^4$ in~\eqref{eq:H_Lame} for the polynomial ansatz $\psi=1+B\sn^2$. The eigenvalues of $\opO_\cI$ are related to the Lam\'e eigenvalues by~\eqref{eq:H_lambda_from_E}.

Under $u\to u+2K(\sigma^2)$, which corresponds to $t\to t+T$ in the original time variable, the Jacobi functions transform as $\sn\to -\sn$, $\operatorname{cn}\to -\operatorname{cn}$, $\operatorname{dn}\to\operatorname{dn}$. Thus the first two modes (i) and (ii) above are anti-periodic, while the remaining three modes (iii), (iv), and (v) are periodic. Single-instanton fluctuations satisfy anti-periodic boundary conditions in~$t$, so only the anti-periodic sector is relevant. The two anti-periodic band edges give:
\begin{itemize}
\item \emph{Zero mode}: The eigenfunction $\psi^{(i)} = \operatorname{cn}(u, \sigma^2)\,\operatorname{dn}(u, \sigma^2) \propto \dot{x}_\cI(t)$ is the translation mode, and has $\lambda_0=0$ as expected.
\item \emph{First excited state}: The second band edge solution is $\psi^{(ii)}(t) \propto \sn(u, \sigma^2)\,\operatorname{dn}(u, \sigma^2)$ with $\lambda_1 = 3\sig^2/[2(1+\sig^2)]$. This is the finite-$T$ generalization of the P\"oschl--Teller bound state at $\lambda=3/4$.
\end{itemize}
All higher anti-periodic eigenvalues lie in Band~III, above $\mathcal{E} = 2(1{+}\sig^2)+2\sqrt{1{-}\sig^2{+}\sig^4}$, and are positive. In particular, the spectrum of $\opO_\cI$ in the anti-periodic sector is non-negative, confirming that $x_\cI$ is a true saddle.

%%%%%%%%%%%%%%%%%%%%%%%%%%%%%%%%%%%%%%%%%%%%%%%%%%%%%%%%%%%%%
\subsection{Recovery of the P\"oschl--Teller spectrum at \texorpdfstring{$T=\infty$}{T=infinity}}
\label{app:PT_recovery}
%%%%%%%%%%%%%%%%%%%%%%%%%%%%%%%%%%%%%%%%%%%%%%%%%%%%%%%%%%%%%

As a check, we verify that the Lam\'e spectrum reduces to the $n_L=2$ P\"oschl--Teller (PT) spectrum in the $\sig\to 1$ ($T\to\infty$) limit.  Recall that this spectrum comprises one bound state, a zero mode, and a continuum.
As $\sig\to1$:
\begin{equation}
  \sn(u,\sig^2)\to\tanh u,\quad
  \operatorname{cn}(u,\sig^2)\to\operatorname{sech} u,\quad
  \operatorname{dn}(u,\sig^2)\to\operatorname{sech} u\,.
  \label{eq:H_k1_limits}
\end{equation}
The Lam\'e equation becomes $-\psi''+6\tanh^2\!u\;\psi = \mathcal{E}\psi$, i.e.\ $-\psi''+(6-6\operatorname{sech}^2u)\,\psi = \mathcal{E}\psi$, equivalent to the $n_L=2$ PT equation shifted by $6$, with spectral parameter $p^2 = \mathcal{E}-6$. Tracking the five modes of \eqref{eq:H_cndn}--\eqref{eq:H_poly_plus} in this limit: modes (ii) and (iii) both reduce to $\tanh u\,\operatorname{sech} u$ with $\mathcal{E}_{2,3}\to 5$ ($\lambda=3/4$), the PT bound state — gap~II closes as these two modes merge into a single eigenfunction. Mode (iv), $1+B_-\sn^2\to 1-3\tanh^2 u$, has $\mathcal{E}_-\to 6$, which is the onset of the continuum ($\lambda=1$, $p^2=0$). Modes (i) and (v) reduce to $\operatorname{sech}^2 u$ with $\mathcal{E}_1,\mathcal{E}_+\to 2$ and collapse onto the PT zero mode at $\lambda=0$, so gap~I closes as well. The resulting three-level spectrum ($\lambda=0$, $3/4$, continuum from $1$) is the standard PT result.

%%%%%%%%%%%%%%%%%%%%%%%%%%%%%%%%%%%%%%%%%%%%%%%%%%%%%%%%%%%%%
\subsection{Floquet counting and Morse index}
\label{app:Floquet_counting}
%%%%%%%%%%%%%%%%%%%%%%%%%%%%%%%%%%%%%%%%%%%%%%%%%%%%%%%%%%%%%

The number of negative eigenvalues of the Lam\'e operator (the Morse index of the saddle) for a multi-instanton saddle on a circle is determined by Floquet (Bloch) theory~\cite{MagnusWinkler}. The monodromy matrix $M(\lambda)$ propagates two independent solutions of the Lam\'e equation~\eqref{eq:H_Lame} across one fundamental cell of length $2K(\sigma^2)$:
\begin{equation}
  \begin{pmatrix} \psi(2K) \\ \psi'(2K) \end{pmatrix}
  = M(\lambda)
  \begin{pmatrix} \psi(0) \\ \psi'(0) \end{pmatrix}.
\end{equation}
Since $\det M(\lambda) = 1$,  the eigenvalues $\rho_{1,2}$ of $M$ (the Floquet multipliers) satisfy $\rho_1\rho_2 = 1$, so we write $\rho_{1,2} = e^{\pm i\theta}$. Bounded (Bloch) solutions exist when both lie on the unit circle, meaning $\theta \in \RR$. The Hill discriminant
\begin{equation}
  \Delta(\lambda) \equiv \operatorname{tr} M(\lambda) = 2\cos\theta
\end{equation}
ranges between $-2$ and $+2$ inside each allowed band, with $\Delta = +2$ at the periodic (P) band edge and $\Delta = -2$ at the anti-periodic (AP) edge. Within Band~I, $\Delta(\lambda)$ decreases monotonically from $+2$ at the bottom to $-2$ at the top.

For an $n=2k$ periodic saddle, the total period $T = 4kK\sqrt{2(1+\sigma^2)}$ covers $2k$ fundamental cells, so true periodicity $\psi(u+2k\cdot 2K)=\psi(u)$ requires $\rho^{2k}=1$:
\begin{equation}
  \rho = e^{i\pi j/k},\qquad j=0,1,\ldots,2k-1.
\end{equation}
The allowed eigenvalues are determined by
\begin{equation}
  \Delta(\lambda) = 2\cos\frac{\pi j}{k}\,.
  \label{eq:H_Bloch_condition}
\end{equation}
Within Band~I, the non-degenerate solutions correspond to $j=0$ ($\Delta=+2$, bottom of Band~I, most negative $\lambda$) and $j=k$ ($\Delta=-2$, top of Band~I, the zero mode at $\lambda=0$). The remaining values $(j, 2k-j)$ for $j=1,\ldots,k-1$ pair up to give $k-1$ doubly-degenerate eigenvalues with $\lambda<0$. The total count of negative eigenvalues is
\begin{equation}
  1 + 2(k-1) = 2k-1\,,
\end{equation}
giving the Morse index $\mu_{k,0}=2k-1=n-1$ quoted in the main text.

For the perturbative saddles $x_*(t)=\pm 1$ ($k=0$), the operator $\opO_0=-\partial_t^2+1$ has only positive eigenvalues, so $\mu=0$. For the single instanton with anti-periodic boundary conditions ($n=1$), only the AP band-edge $\Delta=-2$ in Band~I is allowed; this is the zero mode at $\lambda=0$, leaving no negative eigenvalues, so again $\mu=0$.

%% file: sections/appendix_loop_corrections.tex
% !TEX root = ../DoubleDoubleMain.tex

%%%%%%%%%%%%%%%%%%%%%%%%%%%%%%%%%%%%%%%%%%%%%%%%%%%%%%%%%%%%%%%%%%%%%%%%%%%%%%%
%%%%%.    APPENDIX: HIGHER-LOOP CORRECTIONS TO THE INSTANTON DENSITY    %%%%%%%
%%%%%%%%%%%%%%%%%%%%%%%%%%%%%%%%%%%%%%%%%%%%%%%%%%%%%%%%%%%%%%%%%%%%%%%%%%%%%%%

\section{Perturbative corrections around an instanton background}
\label{appendix:loop_corrections}

In this appendix we compute the loop correction factor $\Delta_L(\hbar)$ coming from connected vacuum bubbles in an instanton background, at leading order in the limit $T \to \infty$. We work mostly around the single-instanton background, and then explain how the same localized calculation repeats around each core in an $n$-instanton background, up to $\cO(e^{-T/n})$ overlap corrections. The Feynman rules involve the cubic and quartic action vertices, the collective-coordinate Jacobian source, and the instanton-background propagator $\prop_I$. We evaluate the diagrams at 2-loop order, confirming the results of W\"ohler--Shuryak~\cite{Wohler:1994pg}, and quote Escobar-Ruiz, Shuryak, and Turbiner~\cite{Escobar-Ruiz:2015nsa} for the 3-loop result.

%%%%%%%%%%%%%%%%%%%%%%%%%%%%%%%%%%%%%%%%%%%%%%%%%%%%%%%%%%%%
\subsection{Feynman rules}
\label{app:single_instanton_setup}
%%%%%%%%%%%%%%%%%%%%%%%%%%%%%%%%%%%%%%%%%%%%%%%%%%%%%%%%%%%%

At leading order in the $T\to\infty$ limit, the single instanton is the tanh kink $x_I(t) = \tanh(t/2)$. We decompose a general path as
\begin{equation}
  x(t) = x_I(t-t_0) + \xi(t-t_0)\,,
  \label{eq:path_decomp}
\end{equation}
with $t_0$ the collective coordinate and $\xi$ the transverse fluctuation. With $V(x) = \frac{1}{8}(x^2-1)^2$ and using $\delta S/\delta x|_{x_I} = 0$ to drop the linear term, the action expansion is
\begin{equation}
  S[x_I + \xi] =  S_I \;+\; \int_{-\infty}^\infty\!dt\,\bigg[\frac{1}{2}\xi\big(-\partial_t^2 + V''[x_I]\big)\xi \;+\; \frac{1}{2}x_I(t)\,\xi(t)^3 \;+\; \frac{1}{8}\xi(t)^4\bigg]\,,
  \label{eq:action_expansion}
\end{equation}
giving the kinetic operator $\opO_I = -\partial_t^2 + V''[x_I]$ which leads to the $\xi$ propagator and the cubic and quartic vertices.

To compute corrections at two loops and higher, we need to also include new vertices coming from the change to collective coordinates. The normalized zero mode is $\psi_0(t) = \dot x_I(t)/\sqrt{S_I}$ and $\xi$ must be constrained to be orthogonal to it.  We can impose this orthogonality by inserting into the path integral a Faddeev--Popov constraint\footnote{Since we are dealing with a single-instanton background, in which the path crosses the origin once, we do not need to further correct this identity with the collective coordinate fix factor (see the discussion below Eq.~\eqref{eq:Z_factored} and~\cite{CollectiveCoordinateFix}).}
\begin{equation}
  1 = \int_{-\infty}^{\infty} dt_0\,\delta \Big(F[x;t_0]\Big)\,\big|\partial_{t_0}F\big|\,,
  \label{eq:FP_identity}
\end{equation}
where
\begin{equation}
  F[x;t_0]  = \int_{-\infty}^\infty dt\,\psi_0(t-t_0)\xi(t-t_0) = \int_{-\infty}^\infty dt\,\psi_0(t-t_0)\,[x(t) - x_I(t-t_0)]\,.
\end{equation}
The $\delta$-function allows us to integrate over fluctuations in the $\psi_0$ direction trivially, leaving a $dt_0$ collective-coordinate integral. The Jacobian is
\begin{align}
  \partial_{t_0}F &= \int_{-\infty}^\infty dt\,\big\{\psi_0(t-t_0)\,\dot x_I(t-t_0) - \dot\psi_0(t-t_0) \xi(t-t_0) \big\}\\
&=\sqrt{S_I} -\frac{1}{\sqrt{S_I}} \!\int_{-\infty}^\infty dt\, \ddot x_I(t)\,\xi(t)\,,
\end{align}
where we integrated by parts in the last term. We can turn this second term into a set of interactions by writing
\begin{align}
  \partial_{t_0}F &=\sqrt{S_I} \exp \ln \bigg[1-\frac{1}{S_I}\int_{-\infty}^\infty dt\, \ddot x_I(t)\,\xi(t)\bigg]\\
  &=\sqrt{S_I} \exp \bigg[-\frac{1}{\hbar}\int_{-\infty}^\infty dt\,\frac{\hbar}{S_I}\,\ddot x_I(t)\,\xi(t) \;-\;\frac{1}{\hbar}\,\frac{\hbar}{2 S_I^2}\bigg(\int_{-\infty}^\infty dt\,\ddot x_I(t)\,\xi(t)\bigg)^{\!2} + \cO(\xi^3)\bigg]\,.
\end{align}
The leading constant $\sqrt{S_I}$ pairs with the Gaussian normalization to give the standard $\sqrt{S_I/(2\pi\hbar)}$ collective-coordinate Jacobian. The new interactions are suppressed by $\hbar$, so they only come in at two loops and higher. The first term appears as a local tadpole source in the Lagrangian: it emits a single $\xi$-line with profile $\ddot x_I(t)$, weighted by $\hbar/S_I$. The quadratic and higher pieces contribute only at three loops and beyond.

The $\xi$ propagator $\prop_I(t_1,t_2)$ is a Euclidean Green's function for the P\"oschl--Teller fluctuation operator $\opO_I = -\partial_t^2 + \frac{1}{2}(3\tanh^2(t/2) - 1)$ on the subspace orthogonal to the zero mode. Thus, it satisfies
\begin{equation}
  \opO_I\,\prop_I(t_1,t_2) \;=\; \delta(t_1-t_2) \;-\; \psi_0(t_1)\psi_0(t_2)\,,
  \label{eq:Pi_def}
\end{equation}
where $\psi_0(t) = \dot x_I(t)/\sqrt{S_I} = \sqrt{3/8}\,\operatorname{sech}^2(t/2)$, together with decay at infinity. The propagator was first written by Olejn\'ik~\cite{Olejnik:1989id} and corrected by W\"ohler--Shuryak~\cite{Wohler:1994pg}:
\begin{multline}
  \prop_I(t_1,t_2) \;=\; \prop_0(t_1,t_2)\!\left[2 - s_1 s_2 + \frac{1}{4}|s_1-s_2|(11 - 3 s_1 s_2)
  + (s_1-s_2)^2\right] \\
  + \frac{3}{16}(1-s_1^2)(1-s_2^2)\!\left[\ln\!\big(2\,\prop_0(t_1,t_2)\big) - \frac{11}{3}\right],
  \label{eq:Pi_inst_full}
\end{multline}
where $s_1 = \tanh(t_1/2)$, $s_2 = \tanh(t_2/2)$, and the SHO building block is
\begin{equation}
  \prop_0(t_1,t_2) \;=\; \frac{1}{2}\,e^{-|t_1-t_2|} \;=\; \frac{1}{2}\,\frac{1 - |s_1-s_2| - s_1 s_2}{1 + |s_1-s_2| - s_1 s_2}\,.
  \label{eq:Pi0_xy}
\end{equation}
One can check that $\prop_I$ satisfies Eq.~\eqref{eq:Pi_def} and vanishes at $\infty$, so it is the unique correct solution. At coincident points $(t_1=t_2)$,
\begin{equation}
  \prop_I(t,t) \;=\; \tfrac{1}{2} \;+\; \tfrac{1}{2}\operatorname{sech}^2(t/2) \;-\; \tfrac{11}{16}\operatorname{sech}^4(t/2)\,,
  \qquad
  \prop_0(t,t) \;=\; \tfrac{1}{2}\,.
  \label{eq:Pi_coincident}
\end{equation}
Note that $\prop_I(t,t) - \prop_0(t,t)$ is localized around the instanton core, decaying as $\operatorname{sech}^2(t/2)$ at large $|t|$.

%%%%%%%%%%%%%%%%%%%%%%%%%%%%%%%%%%%%%%%%%%%%%%%%%%%%%%%%%%%%
\subsection{Two-loop graphs}
\label{app:four_diagrams}
%%%%%%%%%%%%%%%%%%%%%%%%%%%%%%%%%%%%%%%%%%%%%%%%%%%%%%%%%%%%
Using the cubic, quartic, and tadpole vertices, there are four connected vacuum bubble topologies at two-loop order. To isolate the instanton-background piece that goes into $\DL$, we subtract the perturbative contribution from each diagram, leading to
\begin{align}
  G_4
  \;&=\; \vcenter{\hbox{\FeynmanQuarticVertexInst}} \;-\; \vcenter{\hbox{\FeynmanQuarticVertex}}
  \;=\; -\frac{3\hbar}{8}\!\int\!dt\,\Big[\prop_I(t,t)^2 - \prop_0(t,t)^2\Big]\,,
  \label{eq:a1_expr}\\[3pt]
  G_{3a}
  \;&=\; \vcenter{\hbox{\FeynmanCubicBubbleInst}} \;-\; \vcenter{\hbox{\FeynmanCubicBubble}}
  \;=\; \frac{3\hbar}{4}\!\int\!dt_1\,dt_2\,\Big[x_I(t_1)x_I(t_2)\,\prop_I(t_1,t_2)^3 - \prop_0(t_1,t_2)^3\Big]\,,
  \label{eq:b11_expr}\\[3pt]
  G_{3b}
  \;&=\; \vcenter{\hbox{\FeynmanCubicSunsetInst}} \;-\; \vcenter{\hbox{\FeynmanCubicSunset}}
  \;=\; \frac{9\hbar}{8}\!\int\!dt_1\,dt_2\,\Big[x_I(t_1)x_I(t_2)\,\prop_I(t_1,t_1)\prop_I(t_1,t_2)\prop_I(t_2,t_2) \nonumber\\
  &\hspace{8cm}- \prop_0(t_1,t_1)\prop_0(t_1,t_2)\prop_0(t_2,t_2)\Big]\,,
  \label{eq:b12_expr}\\[3pt]
  G_1
  \;&=\; \vcenter{\hbox{\FeynmanCubicJacobian}}
  \;=\; \frac{9\hbar}{4}\!\int\!dt_1\,dt_2\,x_I(t_1)\,\ddot x_I(t_2)\,\prop_I(t_1,t_1)\prop_I(t_1,t_2)\,.
  \label{eq:c1_expr}
\end{align}
Note that the $G_1$ tadpole has no subtraction since there is no tadpole around the perturbative saddles. The $G_1$ prefactor $9\hbar/4$ comes from $3\hbar/(2 S_I)$ with $S_I = 2/3$.

The integrals are most efficiently performed by the change of variables $(t_1,t_2) \to (s_1,s_2) = (\tanh(t_1/2),\tanh(t_2/2))$, which maps $\R^2$ onto $[-1,1]^2$ and turns $\prop_0$ and $\prop_I$ into rational and rational-plus-logarithmic functions. The Jacobian is $dt_1\,dt_2 = \frac{4\,ds_1\,ds_2}{(1-s_1^2)(1-s_2^2)}$. After this substitution, the perturbative subtractions cancel boundary divergences and each integrand becomes a manifestly integrable expression on $[-1,1]^2$. Because the result is finite, the simplest approach is to just compute the integrals numerically with high precision and then identify the resulting number as a rational. We do this for all four diagrams, confirming the results of W\"ohler--Shuryak~\cite{Wohler:1994pg}. 

\paragraph{Figure-8 graph $G_4$ in detail.}
Before turning to numerics, we first illustrate the approach by obtaining an analytical result. For the simplest of the four diagrams, $G_4$, the propagators are at coincident times so the integral is one-dimensional. Substituting $u = 1 - s_1^2 = \operatorname{sech}^2(t/2)$ in Eq.~\eqref{eq:Pi_coincident},
\begin{equation}
  \prop_I(t,t)^2 - \prop_0(t,t)^2 \;=\; \frac{u}{2} - \frac{7}{16}u^2 - \frac{11}{16}u^3 + \frac{121}{256}u^4\,.
  \label{eq:fig8_integrand}
\end{equation}
The remaining $t$-integral can be done directly:
\begin{equation}
  \int\!dt\,u^k \;=\; 2\!\int_{-1}^1\!(1-s_1^2)^{k-1}\,ds_1 \;=\; \frac{2\sqrt{\pi}\,  \Gamma(k)}{\Gamma(k+1/2)} \;=\; \frac{2\cdot 4^k\,(k-1)!\,k!}{(2k)!} \;=\; 4,\;\frac{8}{3},\;\frac{32}{15},\;\frac{64}{35}\quad(k=1,2,3,4).
\end{equation}
Plugging in,
\begin{equation}
  \int\!dt\,\big[\prop_I(t,t)^2 - \prop_0(t,t)^2\big]
  \;=\; 2 - \frac{7}{6} - \frac{22}{15} + \frac{121}{140} \;=\; \frac{97}{420}\,,
\end{equation}
and therefore
\begin{equation}
  G_4 \;=\; -\frac{3\hbar}{8}\cdot\frac{97}{420} \;=\; -\frac{97}{1120}\,\hbar\,,
  \label{eq:a1_value}
\end{equation}
is the figure-8 evaluated from scratch in our conventions.

\paragraph{The 2D integrals.}
For $G_{3a}, G_{3b}, G_1$ the integrands depend on two times. After the substitution $(s_1,s_2)$, the propagator $\prop_0$ has a kink along the diagonal $s_1=s_2$ (from the $|s_1-s_2|$ factor) but is smooth in each of the regions $s_1>s_2$ and $s_1<s_2$. By the $t_1\leftrightarrow t_2$ symmetry, $G_{3a}$ and $G_{3b}$ split into two equal contributions; $G_1$ is not symmetric ($t_1$ carries the cubic vertex, $t_2$ the Jacobian source) so we keep both regions explicitly. In each smooth subregion the integrand is a sum of rational functions and rational$\,\times \log$ pieces, which can be integrated term-by-term against the Jacobian $4\,ds_1\,ds_2/[(1-s_1^2)(1-s_2^2)]$.

We evaluated the integrals numerically with adaptive 2D quadrature, splitting the domain on the diagonal to remove the kink. The results, to 12 significant digits and shown alongside the rational form they identify, are
\begin{equation}
\begin{aligned}
  G_{3a} \;&=\; -0.063095238095\ldots\,\hbar \;=\; -\frac{53}{840}\,\hbar\,,\\
  G_{3b} \;&=\; -0.104464285714\ldots\,\hbar \;=\; -\frac{117}{1120}\,\hbar\,,\\
  G_1 \;&=\; -1.224999999\phantom{99}\ldots\,\hbar \;=\; -\frac{49}{40}\,\hbar\,.
\end{aligned}
  \label{eq:diagram_numerics}
\end{equation}
The same rationals appear diagram-by-diagram in W\"ohler--Shuryak~\cite{Wohler:1994pg} (in different conventions, so up to a factor of $2/3$), providing independent confirmation that our setup of vertices, propagator, and Wick combinatorics is consistent.

Summing the four diagrams gives the single-instanton 2-loop correction:
\begin{equation}
  \Delta_L(\hbar) \;=\; -\frac{97}{1120}\,\hbar \;-\; \frac{53}{840}\,\hbar \;-\; \frac{117}{1120}\,\hbar \;-\; \frac{49}{40}\,\hbar \;+\; \cO(\hbar^2)
  \;=\; -\frac{71}{48}\,\hbar + \cO(\hbar^2)\,.
  \label{eq:DL_numerical}
\end{equation} Two structural features of this result deserve mention. First, all four diagrams contribute negative numbers, so the 2-loop correction systematically reduces the magnitude of the leading-order single-instanton contribution. Second, the Jacobian tadpole $G_1$ supplies nearly $83\%$ of the total. This dominance persists at three loops, reflecting the fact that the collective-coordinate Jacobian source is a localized object integrated against unconstrained transverse fluctuations, with no canceling perturbative subtraction.

\paragraph{Three loops.}
The same procedure can be carried to higher loops. At three loops there are eighteen connected diagrams, including six new tadpole-source contributions analogous to $G_1$. Their evaluation involves three- and four-dimensional integrals, some of which produce irrational $\zeta$-function pieces in individual diagrams. Remarkably, these irrational contributions cancel in the sum~\cite{Escobar-Ruiz:2015nsa}, leaving the rational
\begin{equation}
  \Delta_L^{(3\text{-loop})}(\hbar) \;=\; \frac{\hbar^2}{S_I^2}\!\left(B_2 - \frac{1}{2}B_1^2\right)
  \;=\; -\frac{315}{128}\,\hbar^2\,,
  \label{eq:DL_3loop}
\end{equation}
with $B_1 = -71/72$ and $B_2 = -6299/10368$.

Because the perturbative subtractions localize every diagram to the instanton core, the integrals remain finite at any loop order. An alternative to the analytic evaluation used in~\cite{Wohler:1994pg,Escobar-Ruiz:2015nsa} is to compute each diagram numerically to high precision and identify its value via an integer-relation algorithm such as PSLQ, fitting against a basis $\{1,\zeta(2),\zeta(3),\ldots\}$. This bypasses the rapidly growing analytic work and would be the natural route past three loops. 

%%%%%%%%%%%%%%%%%%%%%%%%%%%%%%%%%%%%%%%%%%%%%%%%%%%%%%%%%%%%
\subsection{The factor of \texorpdfstring{$n$}{n} in the \texorpdfstring{$n$}{n}-instanton sector}
\label{app:DIG_factorization}
%%%%%%%%%%%%%%%%%%%%%%%%%%%%%%%%%%%%%%%%%%%%%%%%%%%%%%%%%%%%

At large $T$, the $n$-instanton propagator decomposes onto the SHO piece plus single-instanton localized corrections,
\begin{equation}
  \prop_{k,k'}(t_1,t_2) \;=\; \prop_0(t_1,t_2) \;+\; \sum_{a=1}^{n}\big[\prop_a(t_1,t_2) - \prop_0(t_1,t_2)\big] \;+\; \cO(e^{-T/n})\,,
\end{equation}
where each $\prop_a$ is the single-instanton propagator translated to its instanton center. Subtracting the perturbative bubbles in each diagram (as in Section~\ref{app:four_diagrams}) removes the $\prop_0$ piece, and the exponential localization of each $\prop_a - \prop_0$ to its core means every diagram in the subtracted sum splits into $n$ independent single-instanton contributions plus exponentially small overlap terms. The Jacobian-tadpole source on each of the $n$ instantons contributes the same way. Summing over which instanton each diagram lives on gives an overall factor of $n$, and exponentiating connected bubbles puts the $n$ into the exponent:
\begin{equation}
  e^{\sum (\text{connected bubbles})_{n\text{-instanton}}}
  \;=\; e^{n\,\Delta_L(\hbar)} \;+\; \cO(e^{-T/n})\,.
  \label{eq:DLn_DIG}
\end{equation}
The argument holds at every loop order: the same single-instanton-localized topology gets repeated $n$ times, with cross-core terms suppressed by $\cO(e^{-T/n})$. Combining the 2-loop result of Eq.~\eqref{eq:DL_numerical} with the 3-loop result of Eq.~\eqref{eq:DL_3loop} for any $n$-instanton background, we only need
\begin{equation}
  \boxed{\;\;
  \Delta_L(\hbar) \;=\; -\frac{71}{48}\,\hbar \;-\; \frac{315}{128}\,\hbar^2 \;+\; \cO(\hbar^3)\,.
  \;\;}
  \label{eq:nDL_numeric}
\end{equation}
For the untwisted trace only the even real sectors $n=2,4,6,\ldots$ contribute, while the same per-instanton factor also applies to the odd sectors appearing in the twisted trace. 

The $\cO(e^{-T/n})$ corrections from inter-instanton overlaps and from the exact-instanton (Lam\'e) propagator vs.\ its tanh limit are not computed here and remain an open problem. They break the simple proportionality $n\Delta_L$.

%% file: sections/appendix_transseries.tex
% \!TEX root = ../DoubleDoubleMain.tex
\section{Trans-series coefficients for the double well}
\label{appendix:terms}

This appendix collects, in one place, the explicit form of the trans-series for the energy eigenvalues of the double-well potential $V(x) = \tfrac{1}{8}(x^2-1)^2$ derived in the body of the paper. We give the closed-form expressions for the perturbative energy~$E_P$ and the first four non-perturbative sectors~$E_1^\pm, E_2, E_3^\pm, E_4$, the large-order growth of $E_P$ together with the corresponding Borel singularity structure, and the traditional re-expansion in $\hbar$ and $\ln(8/\hbar)$ for comparison with the older literature.

The exact energy is organised as a formal trans-series in the tunneling factor $\lambda = e^{-S_I/\hbar}$ with instanton action $S_I = 2/3$ (Eq.~\eqref{eq:transseries_summary}):
\begin{equation}
  E \;=\; E_P \;+\; \lambda\,E_1^\pm \;+\; \lambda^2\,E_2 \;+\; \lambda^3\,E_3^\pm \;+\; \lambda^4\,E_4 \;+\; \cdots
  \label{eq:transseries_app}
\end{equation}
For the $N$-th level we set $\kappa = N+\tfrac{1}{2}$. Each $E_n$ is itself a formal asymptotic series in~$\hbar$ whose coefficients depend on~$\kappa$. The $\pm$ on the odd-instanton sectors distinguishes even- and odd-parity states: odd sectors $E_1, E_3, \ldots$ are proportional to odd powers of $E_1^\pm$ and produce splittings, while even sectors $E_2, E_4, \ldots$ depend only on even powers of $E_1^\pm$ and produce common shifts of both parity states.

The closed-form expressions are built from two ingredients: the resummed instanton prefactor $\fv$, in Eq.~\eqref{eq:fv_summary}, and the Stokes parameter $\Sigma$ in Eq.~\eqref{eq:Sigmadef}:
\begin{align}
  \fv(E_P) &\;=\; \frac{\sqrt{\VN(E_P)}}{\lambda} \;=\; \frac{\sqrt{2\pi}}{\Gamma\!\left(\kappa+\tfrac{1}{2}\right)}\left(\frac{8}{\hbar}\right)^{\!\kappa}\bigl(1 + O(\hbar)\bigr) \,, \label{eq:fv_app}\\[4pt]
  \Sigma &\;=\; -\ln\!\frac{8}{\hbar} \;+\; \psi\!\left(\kappa+\tfrac{1}{2}\right) \;-\; i\pi \;\equiv\; \Sigma_R - i\pi \,,
  \label{eq:Sigma_app}
\end{align}
with $\psi=\Gamma'/\Gamma$ the digamma function. All imaginary parts in the trans-series are carried by the $-i\pi$ piece of~$\Sigma$ and conspire with the lateral Borel resummation ambiguity of $E_P$ to make the spectrum real.

For the 1-instanton sector $E_1^\pm$ we give the closed form explicitly through $O(\hbar^2)$; the ground-state and first-excited splittings are tabulated through the same order in Eqs.~\eqref{eq:splitE0inst}--\eqref{eq:splitE1inst}. For $E_2, E_3, E_4$ we have computed only the leading order in~$\hbar$ in closed form; the higher-$\hbar$ corrections in each non-perturbative sector are in principle accessible from the recursion in Eqs.~\eqref{eq:E2result}--\eqref{eq:E4result} together with $S_P, S_N$ to higher orders, but become rapidly unwieldy because of the $1/E^{2m-1}$ poles in the Picard--Fuchs coefficients (see the closing paragraph of Section~\ref{sec:pertexp}).

\subsection{Perturbative sector \texorpdfstring{$E_P$}{EP}}
\label{app:EPsector}

Solving $\VP = -1$ order by order in $\hbar$ (Section~\ref{sec:pertexp}), or equivalently the Bender--Wu recursion of Appendix~\ref{appendix:benderWu}, yields the perturbative energy as a formal power series $E_P = \sum_{j\ge 1} e_j(\kappa)\,\hbar^j$. Through five orders,
\begin{multline}
  E_P \;=\; \hbar\,\kappa \;-\; \hbar^2\,\frac{1 + 12\kappa^2}{16}
  \;-\; \hbar^3\,\frac{19\kappa + 68\kappa^3}{64}
  \;-\; \hbar^4\,\frac{131 + 3672\kappa^2 + 6000\kappa^4}{2048} \\
  \;-\; \hbar^5\!\left(\frac{22709\kappa}{16384} + \frac{23405\kappa^3}{2048} + \frac{10689\kappa^5}{1024}\right) \;+\; \cdots
  \label{eq:EP_app_series}
\end{multline}
The first four terms reproduce Eq.~\eqref{eq:EP_summary}; the $\hbar^5$ term is from the Bender--Wu output in Appendix~\ref{appendix:benderWu}, which extends straightforwardly to several hundred orders. Each $e_j(\kappa)$ is a polynomial in~$\kappa$ of degree $j$ containing only even (odd) powers of $\kappa$ when $j$ is even (odd), as expected from $\hbar \to -\hbar$, $\kappa \to -\kappa$ symmetry.

The series~\eqref{eq:EP_app_series} is asymptotic; the coefficients grow factorially at the rate set by the leading Borel singularity at $t = 2S_I = \tfrac{4}{3}$. The singularity at $t = S_I = \tfrac{2}{3}$ is absent (a consequence of the parity-doubled spectrum), so every sector of the trans-series grows at worst as $(3/4)^j j!$. Applying the standard dispersion relation to the closed form $\Delta_2 E_P = \hbar\,\fv^2(E_P)/[2 S_P'(E_P)]$ (Eq.~\eqref{eq:Delta2_EP}), expanded using $S_P$ in Eq.~\eqref{SPexpanded} and $\fv$ via $S_N$ in Eq.~\eqref{SNh2}, gives the large-order behavior to two sub-leading orders in $1/j$:
\begin{equation}
  e_j(\kappa) \;\sim\; C_0(\kappa)\frac{\Gamma(j + 2\kappa - 1)}{(2S_I)^{j + 2\kappa - 1}}
  \!\left[1 + \frac{C_1(\kappa)}{j} + \frac{C_2(\kappa)}{j^2} + \cdots\right] \,,
  \label{eq:ej_largeorder_app}
\end{equation}
with
\begin{align}
  C_0(\kappa) &\;=\; -\,\frac{8^{2\kappa}}{2\pi\,\Gamma(\kappa + \tfrac{1}{2})^2} \,,
  \label{eq:C0kappa_app}\\[3pt]
  C_1(\kappa) &\;=\; -\,\frac{204\kappa^2 + 72\kappa + 19}{36} \,,
  \label{eq:C1kappa_app}\\[3pt]
  C_2(\kappa) &\;=\; \frac{41616\kappa^4 + 22752\kappa^3 - 25944\kappa^2 - 15912\kappa - 3743}{2592} \,.
  \label{eq:C2kappa_app}
\end{align}
The leading prefactor is fixed by $\im E_2|_{\text{leading}} = 8^{2\kappa}\,\hbar^{1-2\kappa}/[2\Gamma(\kappa+1/2)^2]$ (Eq.~\eqref{eq:E2_leading_app}); the $1/j$ and $1/j^2$ corrections come from the $\cO(\hbar)$ and $\cO(\hbar^2)$ corrections to $\fv^2/S_P'$ at $E = E_P$. Explicitly, for the ground state
\begin{equation}
  e_j\!\left(\frac{1}{2}\right) \;\sim\; -\,\frac{4}{\pi}\,\frac{(j-1)!}{(2S_I)^{j}}\!\left[1 - \frac{53}{18\,j} - \frac{3185}{648\,j^{2}} + O(1/j^{3})\right] \,,
  \label{eq:ej_groundstate_app}
\end{equation}
and for the first excited state
\begin{equation}
  e_j\!\left(\frac{3}{2}\right) \;\sim\; -\,\frac{256}{\pi}\,\frac{(j+1)!}{(2S_I)^{j+2}}\!\left[1 - \frac{293}{18\,j} + \frac{50371}{648\,j^{2}} + O(1/j^{3})\right] \,.
  \label{eq:ej_firstexcited_app}
\end{equation}

The structure of the Borel singularity is most cleanly read off from the Borel transform itself. Using $B(t) = \sum_{j\ge 1} e_j(\kappa)\,t^j/j!$ and summing at leading large-order via the integral representation $\Gamma(j+2\kappa-1) = \int_0^\infty s^{j+2\kappa-2}e^{-s}ds$, gives the leading singular behavior
\begin{equation}
  B(t) \;=\; C_0(\kappa)\,\frac{\Gamma(2\kappa-1)}{(2S_I)^{2\kappa-1}}\!\left(1 - \frac{t}{2S_I}\right)^{\!1-2\kappa}\!\!\left[1 + \frac{C_1(\kappa)}{2\kappa-2}\!\left(1 - \frac{t}{2S_I}\right) + \frac{C_2(\kappa)}{(2\kappa-2)(2\kappa-3)}\!\left(1 - \frac{t}{2S_I}\right)^{\!2} + \cdots\right]\,.
  \label{eq:Borel_lead_app}
\end{equation}
The leading Borel singularity at $t = 2S_I$ is of order $2\kappa - 1$ and each subsequent $C_k$ is one power less singular. As $\kappa$ approaches physical half-integer values, all the singularities become poles or branch cuts of classical polylogarithms. Explicitly, for the ground state ($\kappa = 1/2$),
\begin{equation}
  B(t)\big|_{\kappa = 1/2} \;=\; \frac{4}{\pi}\,\ln\!\left(1 - \frac{t}{2S_I}\right) + \frac{106}{9\pi}\,\operatorname{Li}_2\!\left(\frac{t}{2S_I}\right) + \frac{3185}{162\pi}\,\operatorname{Li}_3\!\left(\frac{t}{2S_I}\right) + \cdots \,,
  \label{eq:Borel_groundstate_app}
\end{equation}
while for the first excited state ($\kappa = 3/2$),
\begin{equation}
  B(t)\big|_{\kappa = 3/2} \;=\; -\,\frac{144}{\pi}\!\left(1 - \frac{t}{2S_I}\right)^{\!-2} + \frac{2344}{\pi}\!\left(1 - \frac{t}{2S_I}\right)^{\!-1} + \frac{79646}{9\pi}\ln\!\left(1 - \frac{t}{2S_I}\right) + \cdots \,.
  \label{eq:Borel_firstexcited_app}
\end{equation}

The next Borel singularity at $t = 4S_I$ is suppressed by $(2S_I/4S_I)^{j} = (1/2)^{j}$ relative to the leading one and contributes a sub-asymptotic correction. It corresponds to growth in the perturbative series completely fixed by the alien calculus constraint in Eq.~\eqref{eq:Delta2n_EP}:
\begin{equation}
e_j^{(4S_I)}(\kappa) \;=\; +\,2\pi^2\,C_0(\kappa)^2\,
\frac{\Gamma(j+4\kappa-1)}{(4S_I)^{j+4\kappa-1}}\,
\bigl(1+O(j^{-1})\bigr)\,.
\end{equation}
This contribution would be invisible to any fit to the numerical coefficients, since exponential suppression $(1/2)^j$ outpaces any polynomial $1/j^k$ for large $j$.

\subsection{Non-perturbative sectors \texorpdfstring{$E_1^\pm,\, E_2,\, E_3^\pm,\, E_4$}{E1,E2,E3}}
\label{app:nonperturbative_sectors}

At $\cO(\lambda)$, the parity-split quantization condition $1+\VP = \pm i\sqrt{\VN}$ (Eqs.~\eqref{odddown}--\eqref{evendown}) expanded around $E = E_P$ gives the exact-in-$S_N$ form $E_1^\pm = \pm i\hbar\,e^{-(S_N - S_I)/\hbar}/(2S_P'(E_P))$ (Eq.~\eqref{E1formula}). Plugging in $S_P$ from Eq.~\eqref{SPexpanded} and $S_N$ from Eq.~\eqref{SNh2} and re-expanding in $\hbar$,
\begin{multline}
  E_1^\pm \;=\; \pm\,\frac{\hbar}{\sqrt{2\pi}\,\Gamma(\kappa+\tfrac{1}{2})}
  \!\left(\frac{8}{\hbar}\right)^{\!\kappa}
  \!\bigg[\,1 \;-\; \!\left(\frac{3\kappa}{2} + \frac{17\kappa^2}{8} + \frac{19}{96}\right)\!\hbar \\
  \;+\; \!\left(\frac{289\kappa^4}{128} - \frac{23\kappa^3}{32} - \frac{2125\kappa^2}{768} - \frac{115\kappa}{128} - \frac{5111}{18432}\right)\!\hbar^2 \;+\; O(\hbar^3)\,\bigg] \,.
  \label{eq:E1_explicit_app}
\end{multline}
This is the explicit form of Eq.~\eqref{eq:E1_summary}.

At $\cO(\lambda^3)$ the squared quantization condition $(1+\VP)^2 + \VN = 0$ gives (Eq.~\eqref{eq:E2result})
\begin{equation}
  E_2 \;=\; \tfrac{1}{2}\,(E_1^\pm)^2\!\left[\frac{\VN'}{\VN} - \frac{\VP''}{\VP'}\right]
  \;=\; -\,\frac{(E_1^\pm)^2}{2}\!\left[\frac{2(S_N' - S_P')}{\hbar} + \frac{S_P''}{S_P'}\right] \,.
  \label{eq:E2_master_app}
\end{equation}
Substituting $S_P, S_N$ to leading order yields Eq.~\eqref{eq:E2_summary},
\begin{equation}
  E_2 \;=\; -\,\frac{\hbar\,\fv^2}{4\pi^2}\,\Sigma\,(1 + O(\hbar))
  \;=\; \frac{\hbar\,\fv^2}{4\pi^2}\!\left[\ln\!\frac{8}{\hbar} - \psi\!\left(\kappa+\tfrac{1}{2}\right) + i\pi\right]\!(1 + O(\hbar)) \,,
  \label{eq:E2_leading_app}
\end{equation}
which displays both the logarithm in~$\hbar$ characteristic of two-instanton configurations and the imaginary part $\operatorname{Im}\,E_2 = +\hbar\,\fv^2/(4\pi)$ that cancels the lateral-Borel ambiguity in~$E_P$ (Section~\ref{sec:alien}). 

Continuing the Taylor expansion of the quantization condition to $\cO(\lambda^4)$ and defining $R_N \equiv \VN'/\VN, R_P \equiv \VP''/\VP'$ (Eq.~\eqref{eq:RNRP}), one finds (Eq.~\eqref{eq:E3result})
\begin{equation}
  E_3 \;=\; \frac{(E_1^\pm)^3}{24}\!\left[9R_N^2 - 18\,R_N R_P + 8R_P^2 + 6R_N' - 4R_P'\right] \,,
  \label{eq:E3_master_app}
\end{equation}
where primes denote $\partial_E$ and all quantities are evaluated at $E = E_P$. Inserting the leading-order forms of $S_P, S_N$ gives Eq.~\eqref{eq:E3_summary},
\begin{equation}
  E_3^\pm \;=\; \pm\,\frac{\hbar\,\fv^3}{48\pi^3}\!\left[9\Sigma^2 + \pi^2 - 3\,\psi'\!\left(\kappa+\tfrac{1}{2}\right)\right]\!(1+O(\hbar)) \,.
  \label{eq:E3_leading_app}
\end{equation}
Since $E_3^\pm$ depends on the sign of $E_1^\pm$, this sector contributes to the parity splitting.

At $\cO(\lambda^5)$ the quantization condition gives (Eq.~\eqref{eq:E4result})
\begin{align}
  E_4 &\;=\; \frac{(E_1^\pm)^4}{24}\Big[\,8R_N^3 - 24\,R_N^2 R_P + 22\,R_N R_P^2 - 6R_P^3 \notag\\
  &\hphantom{\;=\; \frac{(E_1^\pm)^4}{24}\Big[\,} + 12\,R_N R_N' - 8\,R_N R_P' - 12\,R_N' R_P + 7\,R_P R_P' + 2R_N'' - R_P''\,\Big] \,,
  \label{eq:E4_master_app}
\end{align}
and at leading order in $\hbar$ (Eq.~\eqref{eq:E4_summary})
\begin{equation}
  E_4 \;=\; -\,\frac{\hbar\,\fv^4}{96\pi^4}\!\left[16\Sigma^3 + 4\Sigma\pi^2 - 12\Sigma\,\psi'\!\left(\kappa+\tfrac{1}{2}\right) + \psi''\!\left(\kappa+\tfrac{1}{2}\right)\right]\!(1+O(\hbar)) \,.
  \label{eq:E4_leading_app}
\end{equation}
Since this depends only on $(E_1^\pm)^4$, the 4-instanton sector contributes a common shift of both parity states.

The same recursion extends to all orders in~$\lambda$ at leading $\hbar$ by elementary Taylor expansion of $(1+\VP)^2 + \VN = 0$. At each new order $\cO(\lambda^{n+1})$, the equation is linear in $E_n$ and is solved algebraically once $E_1, \ldots, E_{n-1}$ are known; each $E_n$ comes out as a polynomial of degree $n-1$ in $\Sigma$ multiplied by $\hbar\,\fv^n$, with coefficients that involve $\psi$, $\psi'$, $\psi''$, $\pi^2$, and Riemann zeta values evaluated at $\kappa + \tfrac{1}{2}$. In contrast, going to higher orders in $\hbar$ within a single sector requires summing the entire pole tower in the Picard--Fuchs coefficients (see Section~\ref{sec:pertexp}).

\subsection{Traditional trans-series expansion in \texorpdfstring{$\hbar$}{hbar} and \texorpdfstring{$\ln(8/\hbar)$}{log(8/hbar)}}
\label{app:traditional}

The expressions for the energies can be re-expanded in powers of $\hbar$, $e^{-S_I/\hbar}$ and $\ln(8/\hbar)$ to give the trans-series in its traditional form. Each instanton sector $E_n$ scales as $\hbar\,\fv^n$ and, for $n \ge 2$, contains powers of $\ln(8/\hbar)$ up to $[\ln(8/\hbar)]^{n-1}$ from expanding $\Sigma^{n-1}$, giving
\begin{multline}
  E \;=\; \sum_{j=1}^\infty e_j(\kappa)\,\hbar^j
  \;\pm\; \frac{e^{-S_I/\hbar}\,\hbar\,\fv}{2\pi}\!\sum_{j=0}^\infty e_j^{(1)}(\kappa)\,\hbar^j
  \;+\; \frac{e^{-2S_I/\hbar}\,\hbar\,\fv^2}{4\pi^2}\!\left[\sum_{j=0}^\infty e_j^{(2)}(\kappa)\,\hbar^j \;+\; \ln\!\frac{8}{\hbar}\sum_{j=0}^\infty c_{j,1}^{(2)}(\kappa)\,\hbar^j\right] \\
  \pm\; \frac{e^{-3S_I/\hbar}\,\hbar\,\fv^3}{8\pi^3}\!\left[\sum_{j=0}^\infty e_j^{(3)}(\kappa)\,\hbar^j + \ln\!\frac{8}{\hbar}\sum_{j=0}^\infty c_{j,1}^{(3)}(\kappa)\,\hbar^j + \!\left(\ln\!\frac{8}{\hbar}\right)^{\!2}\!\sum_{j=0}^\infty c_{j,2}^{(3)}(\kappa)\,\hbar^j\right] \\
  +\; \cdots \;+\;
  (\pm)^n\,\frac{e^{-nS_I/\hbar}\,\hbar\,\fv^n}{(2\pi)^n}\!\left[\sum_{j=0}^\infty e_j^{(n)}(\kappa)\,\hbar^j + \sum_{k=1}^{n-1}\!\left(\ln\!\frac{8}{\hbar}\right)^{\!k}\sum_{j=0}^\infty c_{j,k}^{(n)}(\kappa)\,\hbar^j\right] \;+\; \cdots
  \label{eq:transseries_traditional_app}
\end{multline}
where $(\pm)^n$ gives $+$ for even $n$ and $\pm$ for odd $n$; $e_j^{(n)}$ denotes the non-logarithmic tower in the $n$-instanton sector, and $c_{j,k}^{(n)}$ denotes the tower multiplying $\ln^k(8/\hbar)$. Writing $\sigma \equiv \psi(\kappa+1/2) - i\pi$, the leading coefficients in each sector follow from Eqs.~\eqref{eq:E1_explicit_app}, \eqref{eq:E2_leading_app}, \eqref{eq:E3_leading_app}, \eqref{eq:E4_leading_app}. At $n=1$:
\begin{equation}
  e_0^{(1)} = 1 \,, \qquad
  e_1^{(1)} = -\frac{19 + 144\kappa + 204\kappa^2}{96} \,,\quad
  e_2^{(1)} = \frac{289\kappa^4}{128} - \frac{23\kappa^3}{32} - \frac{2125\kappa^2}{768} - \frac{115\kappa}{128} - \frac{5111}{18432} \,.
  \label{eq:e21_app}
\end{equation}
At $n=2$:
\begin{equation}
e_0^{(2)} \;=\; -\sigma\,,\quad  c_{0,1}^{(2)} \;=\; 1\,.
  \label{eq:e0c0_2inst_app}
\end{equation}
At $n=3$, 
\begin{equation}
  e_0^{(3)} \;=\; \frac{1}{6}\!\left[9\sigma^2 + \pi^2 - 3\,\psi'\right]\,, \qquad
  c_{0,2}^{(3)} \;=\; \frac{3}{2} \,, \qquad
  c_{0,1}^{(3)} \;=\; -3\sigma \,;
  \label{eq:ecd_3inst_app}
\end{equation}
and at $n=4$,
\begin{equation}
\begin{aligned}
  e_0^{(4)} &= -\,\frac{1}{6}\!\left[16\sigma^3 + 4\pi^2\sigma - 12\,\psi'\sigma + \psi''\right] \\
  c_{0,3}^{(4)} &\;=\; \frac{8}{3} \,, \qquad
  c_{0,2}^{(4)} \;=\; -8\sigma \,, \qquad
  c_{0,1}^{(4)} \;=\; 8\sigma^2 + \frac{2\pi^2}{3} - 2\,\psi' \,.
\end{aligned}
\label{eq:ecde_4inst_app}
\end{equation}
These expressions can be compared to the trans-series of Zinn-Justin and Jentschura~\cite{ZinnJustinJentschura1, ZinnJustinJentschura2} via the change of variables
\begin{equation}
  \xi \;=\; \frac{e^{-S_I/\hbar}\,\fv}{2\pi} \;=\; \frac{1}{N!\,\sqrt{2\pi}}\!\left(\frac{8}{\hbar}\right)^{\!N+1/2}\!e^{-S_I/\hbar}\bigl(1+O(\hbar)\bigr) \,,
  \label{eq:xi_ZJ_app}
\end{equation}
in terms of which their expansion uses $\xi^n$ in place of $e^{-nS_I/\hbar}\fv^n/(2\pi)^n$ at the $n$-instanton order.